\documentclass[twoside,11pt]{Latex/Classes/PhDthesisPSnPDF}

\usepackage{amsfonts,amsmath,braket,soul,color,bbm,mathtools}
\usepackage{amssymb,dsfont,amsbsy,amsthm,tensor,graphicx,epstopdf}
\usepackage{caption,subcaption,multicol,textcomp,wrapfig,enumerate}

\newcommand{\tr}{\mathrm{tr}}

\newcommand{\be}{\begin{equation}}
\newcommand{\ee}{\end{equation}}
\newcommand{\beq}{\begin{eqnarray}}
\newcommand{\eeq}{\end{eqnarray}}
\newcommand{\pr}{^{\prime}}

\DeclareMathAlphabet\mathbfcal{OMS}{cmsy}{b}{n}

\let\oldsqrt\sqrt
\def\sqrt{\mathpalette\DHLhksqrt}
\def\DHLhksqrt#1#2{%
\setbox0=\hbox{$#1\oldsqrt{#2\,}$}\dimen0=\ht0
\advance\dimen0-0.2\ht0
\setbox2=\hbox{\vrule height\ht0 depth -\dimen0}%
{\box0\lower0.4pt\box2}}

\newcommand{\BbbR}{\mathbb{R}}
\newcommand{\BbbZ}{\mathbb{Z}}

\newcommand{\cavlength}{\delta}

\newtheorem{theorem}{Theorem}[section]
\newtheorem{defn}{Definition}[section]

\newtheorem{innercustomthm}{Postulate}
\newenvironment{post}[1]
  {\renewcommand\theinnercustomthm{#1}\innercustomthm}
  {\endinnercustomthm}

\ifpdf  
    \pdfcatalog { /PageMode (/UseOutlines)
                  /OpenAction (fitbh)  }
\fi

\title{Localised Systems in Relativistic Quantum Information}

  \author{\href{pmxal3@nottingham.ac.uk}{Antony Richard Lee}}
  \collegeordept{\href{http://www.nottingham.ac.uk/mathematics/index.aspx}{School of Mathematical Sciences}}
  \university{\href{http://www.nottingham.ac.uk/}{University Of Nottingham}}

  \crest{\includegraphics[width=4cm]{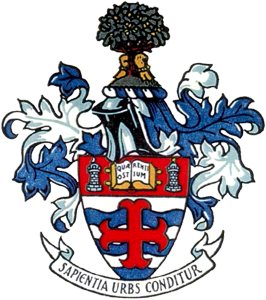}}

\degree{Doctor of Philosophy (PhD)\\ \vspace{5mm} {\normalfont\small Principal Assessor: Prof. John Barrett} \\{\normalfont\small External Assessor: Dr. Dan Browne}}
\degreedate{}

\setcitestyle{square}

\begin{document}

\renewcommand\baselinestretch{1.2}
\baselineskip=18pt plus1pt


\maketitle  


\frontmatter

\begin{acknowledgements}
\addcontentsline{toc}{chapter}{Acknowledgement}

When people say time flies when you are having fun they are not lying. However, they do not always emphasise how much faster time can appear to be moving when under the impending doom of finishing a large project, such as ones PhD thesis. That is not to say, though, the experience is not without its rewards. During the last four or so years, I have had the privilege to travel to places and speak to people quite out of the reach of normal day to day experiences. These two things alone are enough to make me humble and honoured to have been allowed to do the things I have done and to thank all those I have met along the way.

First has to be my principal supervisor Ivette Fuentes. Her friendship and enthusiasm has affected my work in an indescribable amount of ways, all of which have made me not only a better scientist but a better person. I will never quite understand why you took me under your wing in the way that you did but I am eternally grateful to you for doing so.

I would also like to thank my second supervisor Gerardo Adesso. It was magnificent learning the wonders of Gaussian state information theory from him and for all those fantastic games of squash we played over the past few years.

Next in line for my thanks are the many people I had the pleasure of collaborating with here in Nottingham (in alphabetical order): David Bruschi, Andrzej Dragan, Nicolai Friis, Jorma Louko and Carlos Sab\'{i}n. Working with you very much brought out the best out of me in every respect. I'd like to extend my thanks to the rest of the Quantum Information family in Nottingham (in no particular order this time and also missing some names): Jason Doukas, Mehdi Ahmadi, Anglea White, Valentina Baccetti, Luke Westwood, Giannis Kogia, Karishma Hathila, Lu\'{i}s Barbado, Davide Girolami, Sammy Ragy, Merlijn van Horssen, M\u{a}d\u{a}lin Gu\c{t}\u{a}, Ben Aaronson, Fernando Aguayo, C\u{a}t\u{a}lin C\u{a}tan\u{a}, Katarzyna Macieszczak and George Jaroszkiewicz. I'd like also to mention Prof. Viacheslav Belavkin who sadly passed away in the winter of 2012. Without him the entire Quantum Information group in Nottingham would probably not be still here today.

Further afield, I managed to to engage in many interesting, dynamic and fruitful discussions with a vast body of people such as Miguel Montero, Marco del Rey, Achim Kempf, Jonathan Oppenheim, Juan Le\'{o}n, Rob Mann, Marcus Huber, G\"{o}ran Johansson and many many more people who I cannot fit in here. Those who should be listed and are not will certainly know who they are and I apologise for missing you out.

A special thanks goes out to Eduardo Mart\'{i}n-Mart\'{i}nez whose persistent attempts to kill either myself or him on any continent we landed on are a constant source of entertainment. My appreciation to the ``maths-crew". Our little excursions and distractions helped me keep it together over the years.

During my time as a PhD student in Nottingham, I was blessed with the opportunity to organise a small conference aimed at postgraduate researchers titled ``Quantum Fields, Gravity and Information" (QFGI 2013 for short). Along side my co-organisers David Bruschi, Nicolai Friis and Sara Tavares, it was single handedly, at the time, the worst week of my life but has borne the most incredible results not only scientifically but on a personal level also. Thanks for all the hard work guys. One tip to those reading however, don't organise a conference when trying to write a thesis.

Finally I would like to emphasise my thanks and love to Sara. Although you have only been a constant in my life for a short time, believe me when I say your effects will stay with my for the rest of my life. For the times I felt like giving up, you were the only one who I needed to help me through. Thank you, thank you, thank you.\\

Ant

\end{acknowledgements}

\begin{abstracts}
\addcontentsline{toc}{chapter}{Abstract}

This thesis collects my own and collaborative work I have been involved with finding \emph{localised systems} in quantum field theory that are useful for quantum information. It draws from many well established physical theories such as quantum field theory in curved spacetimes, quantum optics and Gaussian state quantum information. The results are split between three chapters. 

For the first results, we set-up the basic framework for working with quantum fields confined to cavities. By considering the real Klein-Gordon field, we describe how to model the non-uniform motion of a rigid cavity through spacetime. We employ the use of \emph{Bogoliubov transformations} to describe the effects of changing acceleration. We investigate how entanglement can be generated within a single cavity and the protocol of quantum teleportation is affected by non-uniform motion.

The second set of results investigate how the Dirac field can be confined to a cavity for quantum information purposes. By again considering Bogoliubov transformations, we thoroughly investigate how the entanglement shared between two cavities is affected by non-uniform motion. In particular, we investigate the role of the Dirac fields charge in entanglement effects. We finally analyse a ``one-way-trip" of one of the entangled cavities. It is shown that different types of Dirac field states are more robust against motion than others.

The final results look at using our second notion of localisation, \emph{Unruh-DeWitt} detectors. We investigate how allowing for a ``non-point-like" spatial profile of the Unruh-DeWitt detector affects how it interacts with a quantum field around it. By engineering suitable detector-field interactions, we use techniques from symplectic geometry to compute the dynamics of a quantum state beyond commonly used perturbation theory. Further, the use of Unruh-DeWitt detectors in generating entanglement between two distinct cavities will be investigated.

\end{abstracts}


\setcounter{secnumdepth}{3} 
\setcounter{tocdepth}{3}    
\tableofcontents            

\mainmatter

\renewcommand{\chaptername}{} 

\chapter*{Publications}\label{chapter:publications}
\addcontentsline{toc}{chapter}{Publications}

This thesis is based on work presented in the following publications.

\begin{enumerate}[I]
	\item ``Kinematic entanglement degradation of fermionic cavity modes"~\cite{friis2012}
	\item ``Motion-generated quantum gates and entanglement resonance"~\cite{bruschi2012-2}
	\item ``Relativistic Quantum Teleportation with Superconducting Circuits"~\cite{friis2013-2}
	\item ``Spatially extended Unruh-DeWitt detectors for relativistic quantum information"~\cite{lee2012}
	\item ``Time evolution techniques for detectors in relativistic quantum information"~\cite{bruschi2013-2}
	\item ``Generating entanglement between two-dimensional boxes in uniform acceleration"~\cite{leeunpub}
\end{enumerate}

\listoffigures

\chapter*{Part I}
\addcontentsline{toc}{chapter}{Part I}

\chapter{Introduction}

The past decade has seen the emergence of a new, and rapidly growing, field of physics called \emph{relativistic quantum information}. This field looks at how concepts from \emph{quantum information} can be implemented and used within the framework of \emph{quantum field theory}. The principle motivation behind relativistic quantum information is: \emph{how does motion and gravitation affect entanglement?} A multifaceted question, it requires substantial theoretical considerations to establish well posed and specific problems. However, before doing so, it is useful to contextualise relativistic quantum information in terms of its two constituent parts, namely, quantum information and quantum field theory.

Quantum information asks questions about the storage, manipulation, processing and use of information in a quantum system~\cite{jaeger2007}. A fundamental question in quantum information is: \emph{can quantum systems be used to improve communications and computation?} The past twenty years have seen a huge body of work describing exactly how information stored in quantum systems can used to improve the classical description of information~\cite{jaeger2007,nielsen2010,diosi2011}. Remarkable advances worth mentioning are the protocol of quantum teleportation, quantum cryptography and quantum computation. Their mathematical framework have allowed people to invent exciting new concepts such as completely secure communication and even the so-called \emph{quantum computer}. The common theme between all these aspects of quantum information is the concept of \emph{quantum entanglement}. Entanglement is considered as one of the most fundamental properties of quantum physics. A consequence of the superposition principle (or equivalently the tensor product structure of a Hilbert space), it allows for quantum systems to contain correlations which are, in some sense, stronger than systems which just contain classical correlations. Entanglement, which was initially thought of a problematic aspect of quantum mechanics~\cite{einstein1935}, has now become the corner stone of quantum information. Through the examples given above, it has single-handedly allowed us to surpass classical information theory and has also made us question fundamental concepts of how information should be viewed. Thus, the investigation of entanglement and its thorough understanding is a fundamental, and largely unresolved, problem in modern theoretical physics.

The second theory underpinning relativistic quantum information is quantum field theory. Quantum field theory is the merger of principles from relativity theory and quantum mechanics. Unlike standard quantum mechanics, which describes systems with a fixed particle number, quantum field theory describes the interaction of systems where particle number can vary. In particular, it describes particles as excitations of more fundamental objects known as \emph{fields}. Typical examples of quantum fields are the Dirac field (which describes electrons) and the electromagnetic field (which describes photons). Quantum field theory revolutionised our understanding of how fundamental processes occurred and allowed us to describe the quantum theory of light, particle creation and predicted the (hopefully) recently discovered Higg's particle~\cite{higgsdiscovery}. Quantum field theory currently provides us with our best predictive theory for the interaction of fields in the presence of a gravitational force. It is therefore a natural framework for relativistic quantum information to work within.

As previously mentioned, relativistic quantum information's main aim is to answer questions about the overlap of relativity and the manipulation of information stored in quantum systems. More precisely, standard quantum information does not consider the effects of a system's motion through spacetime or the influence of changes in gravitation. In other words, spacetime is flat and relativistic considerations are negligible. So, if we take into account these more general scenarios, how does quantum information and its description change? This is precisely what relativistic quantum information is attempting to answer. Besides its obvious theoretical appeal, relativistic quantum information also has very real and concrete experimental motivation. Current technology is becoming increasingly accurate and is starting to step into the realm where relativistic effects are of great consequence. If we are to implement new quantum information ideas, such as communication over long distances and between the Earth and orbiting satellites, we need to have a fundamental grasp of how relativity affects entanglement~\cite{ursin2009,rideout2012}. 

The time, effort and creativity of people working in relativistic quantum information has culminated in two excellent review articles~\cite{peres2004,alsing2012}, which I advise any one interested in relativistic quantum information theory to read. 

To give further context to relativistic quantum information, we briefly review its early contributions. The first investigations into relativistic quantum information are attributed to Czachor~\cite{czachor1997}. Czachor showed that corrections to the violation of Bell's inequalities depended on the velocity between two massive particles. Following this, Peres~and~Terno~\cite{peres2002} emphasised the need for a possible reformulation of quantum information concepts in light of relativity theory. Further investigations went on to analyse definite momentum state entanglement as seen by different inertial observers~\cite{alsing2002}, how the reduced states of a bipartite systems transform under Lorentz transformations~\cite{GingrichAdami2002} and even considered interactions between spin-$1/2$ particles~\cite{PachosSolano2003}.

Inspired by the so-called \emph{Unruh effect}~\cite{unruh1976}, which predicts accelerated observers always measure a non-zero temperature around them, Alsing and Milburn~\cite{alsing2003,alsing2004} analysed teleportation for uniformly accelerated observers. The next natural step taken was then to study entanglement where the spacetime itself contains curvature, such as in the vicinity of a black hole. It was shown by Fuentes and collaborators that initial entanglement would be degraded in these settings and that entanglement is also observer dependent~\cite{funentes-schuller2005,martinmartinez2009}. Very recently, multi-particle entanglement has also been the subject of much investigation with results for momentum-spin entanglement between inertial observers~\cite{FriisBertlmannHuberHiesmayr2010,huber2011} and accelerated observers~\cite{friis2012-4}.

However, early results in relativistic quantum information relied on what is known as \emph{global mode entanglement}. This is entanglement that is shared between idealised plane wave wavefunctions which are spread out over \emph{all} spacetime. Essentially, they are states of particles which permeate the entire universe and are \emph{totally delocalised}. While having strong theoretical motivation, these delocalised states are difficult to measure physically. Physically well motivated systems, therefore, should be states that are localised to some finite region of spacetime. This would allow ``real" observers to store and manipulate quantum information in a more realistic way. To continue making progress, we must find \emph{localised systems in relativistic quantum information.}

In this thesis, we investigate two very promising candidates for localising systems in relativistic quantum information: set-ups involving \emph{cavities} and spatially confined quantum mechanical objects known as \emph{particle detectors}. The benefit of such considerations is clear. Cavities are commonly realised in quantum optics experiments and offer an ideal system for manipulation. Therefore, providing a flexible framework to make predictions with cavities will be of great use to both theorists and experimentalists. On the other hand, the benefit of using spatially confined objects (particle detectors) is that they can model atoms or other point-like systems. They, therefore, offer another system which can be manipulated in a local manner.

To be explicit, the two issues we will address in this thesis are,
\begin{enumerate}
	\item To construct quantum fields ``localised" to a finite region of spacetime by the use of cavities. Introduce a flexible framework in which to pose well motivated questions and investigate how quantum information is affected by relativistic considerations.
	
	\item To introduce a model of ``particle detector" which allows us to mathematically model interactions with a quantum field in a simple manner. We also want to introduce new tools to allow the investigation of non-perturbative quantum information within the framework of quantum field theory.
\end{enumerate}

This thesis is organised as follows: Part I introduces the mathematical tools needed to derive the results presented in this thesis. We start with Chapter~(\ref{chapter:qmqi}) and a basic introduction of quantum mechanics and quantum information such as its mathematical description and basic properties. We define entanglement and some useful methods of quantifying it. Finally we describe the paradigmatic protocol of \emph{quantum teleportation}. 

In Chapter~(\ref{chapter:qft}) we introduce \emph{canonical} quantum field theory. That is, quantum field theory where the notion of a particle can be well defined and entanglement can be thoroughly analysed. In particular, we will review the canonical quantisation of the Klein-Gordon field (spin-$0$) in Minkowski spacetime (which describes inertial observers) and what is known as \emph{Rindler} spacetime (which describes uniformly accelerated observers). We go on to relate the Minkowski spacetime treatment to the Rindler spacetime via what are known as \emph{Bogoliubov transformations}. Bogoliubov transformations are the standard way of relating different observers in quantum field theory and serve as the fundamental mathematical building blocks of Chapters~(\ref{chapter:bosons},\,\ref{chapter:moving-cavities}). Having discussed inertial and accelerated observers, we describe the Unruh effect and its implications for particle content between different observers. Continuing, we quantise and Dirac field (spin-$1/2$) in Minkowski and Rindler coordinates for our discussion of Fermionic entanglement in Chapter~(\ref{chapter:moving-cavities}). We end Chapter~(\ref{chapter:qft}) by introducing the Unruh-DeWitt particle detector model. The Unruh-DeWitt detector model is an operational way of defining what a particle is. In essence, it defines a particle as something that makes a detectors state change or, in other words, ``click". We describe the usual quantity of interest of an Unruh-DeWitt detector, the \emph{transition rate}. The transition rate of an Unruh-DeWitt detector essentially tells us about the probability of finding the detector in a given state and also how often it is ``clicking" per unit time.

Finally, Chapter~(\ref{chapter:cv}) introduces \emph{continuous variable} quantum mechanics and details how the special class of \emph{Gaussian} quantum states can be elegantly represented in the language of phase space and symplectic geometry. Gaussian states are useful for our purposes as they allow us to link quantum field theory and quantum information in a very elegant way. They also have a very broad set of possible experimental implementations which could be useful for future verifications of theoretical work. We define how to compute our measures of entanglement for Gaussian states and explain how the field of linear quantum optics and Gaussian quantum information can be related to quantum field theory.

Part II presents new results in relativistic quantum information. In Chapter~(\ref{chapter:bosons}), we describe mathematically cavities for Klein-Gordon fields. We demonstrate how the cavity's modes become entangled when moved through spacetime. This motion implements an \emph{entangling gate} and we discuss its implications for quantum computing. For the Klein-Gordon field, we also investigate the paradigmatic quantum information protocol of quantum teleportation. We show that motion through spacetime degrades the entanglement resource for the teleporation protocol. Further, we identify how to correct for the degradation by performing local operations and fine tuning the trajectory of the cavity. Finally, we introduce an experimental set-up to test our results using cutting-edge circuit quantum electrodynamics technology.

In Chapter~(\ref{chapter:moving-cavities}), we give a pedagogical presentation of the Dirac field contained within a cavity. Unlike the Klein-Gordon field, the Dirac field requires more complicated boundary conditions which, consequently, require extra effort to implement properly. Dirac fields also allow us to consider different classes of entangled states which are fundamentally different from Klein-Gordon field entangled states. Starting with two cavities which initially share an entangled state, we thoroughly investigate how motion affects the entanglement as one of the cavities move through spacetime. We find that the charge of the Dirac particles directly contributes to the observed degradation affects and that certain states are more robust against acceleration than others. We also look at a ``one-way trip" trajectory where one of the entangled systems departs to a, possibly distant, region of space.

Chapter~(\ref{chapter:fat-detectors}) explores how our second localised system, the Unruh-DeWitt detector, can be used for quantum information. We aim at developing new detector models which are more realistic and simpler to treat mathematically so that they can be used in relativistic quantum information processing. It will be shown that more physically realistic models of particle detectors in quantum field theory, which account for spatial size, modify the standard Unruh effect. In particular, the state of the detector can differ from a canonical thermal state in a significant way. Therefore, in principle, the temperature seen by an observer will not be directly proportional to their acceleration. We investigate specially engineered detector-field interactions that allow us to take advantage of well known tools from symplectic geometry and Gaussian state quantum information to analyse the evolution of quantum states. We show how to derive equations that determine the evolution of a state non-perturbatively. As an example, we analyse the state of a stationary detector coupled to a quantum field. Finally, we combine both cavities and particle detectors into a scenario where entanglement can be generated between two spatially separated systems. We consider a scenario where two cavities, one inertial and one accelerated, are initially not entangled. By passing an Unruh-DeWitt detector through each cavity, we show that the acceleration of one of the cavities degrades the entanglement generated between them. However, we find that only for extremely high accelerations is the generated entanglement degraded by a significant amount. This robustness could be used as a base for experimental verifications of predictions in relativistic quantum information.

Part III serves as an area for conclusions, final remarks and appendices. We summarise the results of the thesis and point out their interesting consequences and relevant physical interpretation. Continuing, a short discussion of ongoing work and very near future projects will be given. In particular, we look at how the results of the thesis can be extended in a natural way. Finally, a very speculative view of possible future directions for relativistic quantum information is given.

For reference and readability, a list of notational conventions has been made at the end of the thesis~(\ref{chapter:notation}).

\chapter{Quantum Information}\label{chapter:qmqi}

Quantum Information is the study of how information can be stored, manipulated and processed in quantum systems. A central question in quantum information is: \emph{how does quantum information differ from classical information?} Common tasks in quantum information involve communication~\cite{nielsen2010}, computer algorithms~\cite{mosca2008} and cryptography~\cite{gisin2002}. A common link that underpins many of these tasks is a quantum property known as \emph{entanglement}. Entanglement is viewed as the main resource for quantum information and is considered one of the basic aspects of quantum theory~\cite{schrodinger1935}. Two prominent examples of how entanglement provided improvements over classical information theory are Shor's so-called \emph{quantum factoring algorithm}~\cite{shor1994} and the paradigmatic protocol of quantum teleportation~\cite{bennett1993}. Shor's algorithm is a method of factoring an integer $N$ into its prime factors. Using entanglement, it offers a significant reduction in the time needed to compute the factorisation when compared to the best known classical algorithms. The teleportation protocol, which will be reviewed in more detail in Section~(\ref{sec:teleportation-qmqi}), is a method of exploiting entanglement to send quantum information efficiently. Therefore, understanding how to quantify, manipulate and use the entanglement contained within a system is a central question in quantum information. Given these few motivating examples, entanglement has been the subject of a vast body of work and has become a well-founded mathematical discipline in its own right~\cite{bruss2001,plenio2007}.

After its initial burst of interest, people turned to experimentally verifying the theoretical predictions of entanglement. This culminated in the physical realisation of \emph{Shor's quantum factoring algorithm}~\cite{vandersypen2001} and the quantum teleportation protocol~\cite{boschi1998}. Currently, experimental investigations of entanglement are being pushed to their very limit in terms of what is physically possible. This is exemplified by the group of Zeilinger who exchanged quantum information, in the form of photons, over a distance of 143~km~\cite{ma2012} and the group of Rempe, in the form of two-level atoms, over a distance of 21~m~\cite{nolleke2013}. Given the great success of such experiments, people have explored the possibility of using entanglement for communication via satellites orbiting the Earth and over large distances~\cite{villarosi2008,rideout2012}. In these new, more extreme, environments, discrepancies between theoretical descriptions and physical observation can become problematic. To guarantee the effective implementation of quantum communication over long distances, a thorough understanding of the effects on entanglement due to the motion of the satellites and the gravitational field of the Earth is vital. Relativistic quantum information is, therefore, perfectly suited to guide us through the problem of studying entanglement for communication through spacetime. 

This chapter is structured as follows: we briefly review the basic mathematical concepts needed to introduce quantum mechanics and quantum information. We define the postulates of quantum mechanics for what are known as \emph{pure} and \emph{mixed} states. Then, the definition of entanglement and a discussion of how to quantify it for different types of states is given. Finally, we illustrate the use of entanglement through the paradigmatic protocol of quantum teleportation.

For those who would like to delve deeper into quantum mechanics and quantum information, an excellent introductory text is Di\'{o}si~\cite{diosi2011} and for a more substantial, but extremely pedagogical, treatment is Nielsen and Chuang~\cite{nielsen2010}.

\section{Pure State Quantum Mechanics}

To begin, we define the basic mathematical structure that quantum mechanics is based on:
\begin{defn}
A Hilbert space $H$ is a normed, complex inner product space which is complete with respect to the inner product $\langle \psi|\phi \rangle\in\mathbb{C}$ for $\ket{\psi},\ket{\phi}\in H$.
\end{defn}
Hilbert spaces are the mathematical foundations of quantum mechanics and so we define our first postulate~\cite{nielsen2010}:
\begin{post}{1}\label{post1}
A physical pure quantum state is represented by a vector (also known as a ray) $\ket{\psi}$ in a Hilbert space $H$.
\end{post}
The state vectors $\ket{\psi}$ are known as \emph{pure states}. These vectors are normalised to unity i.e. $\langle\psi|\psi\rangle=+1$. The simplest example of a pure quantum state is known as a \emph{qubit}. A qubit is a two-dimensional quantum object which lives in the Hilbert space $\mathbb{C}^{2}$. By two-dimensional we mean that its state can be written as a linear superposition of two orthogonal quantum states. We can, of course, define any basis we want to represent the orthogonal quantum states. A common notation to represent the two states of the qubit is $\ket{0}$ to denote the ``ground state" and $\ket{1}$ to denote the ``excited state". Note that other possible nomenclature for the two states of a qubit are ``up/down" or ``on/off", among many others. Defining $\ket{0}$,$\ket{1}$ to be an \emph{orthonormal} basis, we can write a general pure qubit state as
\begin{eqnarray}
\ket{\psi}=a\ket{0}+b\ket{1},
\end{eqnarray}
where $a,b\in\mathbb{C}$ and $|a|^{2}+|b|^{2}=+1$. We will return to qubits when we speak about entanglement in Section~(\ref{sec:entanglement}).

When dealing with quantum systems, a suitable theory of describing measurements is essential. Quantum measurements are described by so-called \emph{measurement operators}. These are operators which act on the Hilbert space $H$ of a state $\ket{\psi}\in H$. Measurement operators form the central part of our second postulate of quantum mechanics~\cite{nielsen2010}:
\begin{post}{2}\label{post2}
Quantum measurements are described by a set of measurement operators $\hat{M}_{n}$. Each measurement operator has with it an associated measurement outcome $m_{n}$, where the probability of obtaining the measurement outcome is given by
\begin{eqnarray}
p_{n}=\bra{\psi}\hat{M}_{n}^{\dag}\hat{M}_{n}\ket{\psi},
\end{eqnarray}
with the state immediately afterwards reducing to
\begin{equation}
\ket{\psi'}=\frac{1}{\sqrt{p_{n}}}\hat{M}_{n}\ket{\psi},
\end{equation}
and the measurement operators must satisfy the completion equation
\begin{equation}
\sum_{n}\hat{M}_{n}^{\dag}\hat{M}_{n}=\hat{I}.
\end{equation}
\end{post}
Note in the above we have used $\hat{I}$ to denote the identity operator of a Hilbert space $H$, i.e. the operator that acts trivially on a quantum state. The statement of the completion equation comes from the fact that probabilities of measurements should sum to one. In the special case where the measurement observables are \emph{Hermitian}, i.e. $\hat{M}_{n}^{\dag}=\hat{M}_{n}$, and satisfy $\hat{M}_{m}\hat{M}_{n}=\delta_{mn}\hat{M}_{n}$, the measurement observables are said to be \emph{projective measurements}. Given a general observable, represented by a Hermitian operator $\hat{A}$, we can use projective measurements to decompose it as
\begin{equation}
\label{eqn:eigendecomp}
\hat{A}=\sum_{n}a_{n}\hat{P}_{n},
\end{equation}
where we have denoted a projective measurement as $\hat{P}_{n}$. This is known as the spectral decomposition of a Hermitian operator and the eigenvalues $a_{n}$ are called the spectrum of $\hat{A}$ and represent the possible measurement outcomes of the observer. Hermitian operators have the special property that their spectrum contains only real entries i.e. $a_{n}\in\mathbb{R}$. Projective measurements also allow us to write the \emph{expectation value} of an observable $\hat{A}$ in a particularly simple away:
\begin{defn}
The expectation value of an observable $\hat{A}$ for a given state $\ket{\psi}$ is defined as
\begin{equation}
\begin{aligned}
\langle\hat{A}\rangle&=\bra{\psi}\hat{A}\ket{\psi},\\
&=\sum_{n}a_{n}p_{n},
\end{aligned}
\end{equation}
where $a_{n}$ are the eigenvalues of $\hat{A}$ and have associated with them the probabilities\, $p_{n}=\bra{\psi}\hat{P}_{n}\ket{\psi}$.
\end{defn}
A fundamental result of measuring observables in quantum mechanics is the \emph{Heisenberg uncertainty principle}. This says that given multiple copies of a state $\ket{\psi}$, the standard deviation of two observables $\hat{A}$ and $\hat{B}$ when measured has to satisfy~\cite{heisenberg1927}
\begin{eqnarray}
\label{eqn:heisenberg-uncert}
\mathrm{Var}(\hat{A})\mathrm{Var}(\hat{B})\ge\frac{1}{4}\left|\langle [\hat{A},\hat{B}] \rangle\right|^{2}.
\end{eqnarray}
where $\mathrm{Var}(\hat{\mathcal{O}}):=\langle\hat{\mathcal{O}}^{2}\rangle-\langle\hat{\mathcal{O}}\rangle^{2}$ is the variance of an operator $\hat{\mathcal{O}}$.

Now that we have defined quantum states and measurements, it would be useful to know how a quantum system evolves in time, i.e. what are its \emph{dynamics?} This is done via \emph{Schr\"{o}dinger's} equation which tells us how a given Hermitian operator $\hat{H}$ evolves a quantum system~\cite{nielsen2010}:
\begin{post}{3}\label{post3}
For an isolated system, the dynamics of a state are governed by the Schr\"{o}dinger equation
\begin{eqnarray}
\label{eqn:schrodinger}
i\partial_{t}\ket{\psi(t)}=\hat{H}\ket{\psi(t)},
\end{eqnarray}
where $\hat{H}$ is a Hermitian operator and corresponds to the total energy of the system and $t$ is the time coordinate of the system. 
\end{post}
Dynamics written as in Eq.~(\ref{eqn:schrodinger}) are referred to as the Schr\"{o}dinger picture, in which the states evolve in time and the operators do not. As it represents the total energy of the system, the operator $\hat{H}$ is defined as the Hamiltonian of the quantum system in full analogy with classical dynamics. Equivalently, we can formulate the dynamics in what is known as the \emph{Heisenberg picture}. This is given by the Heisenberg equation which, for a given observable $\hat{A}$, reads
\begin{eqnarray}
\frac{d}{dt}\hat{A}(t)=i\left[\hat{H}(t),\hat{A}(t)\right]+\partial_{t}\hat{A}(t).
\end{eqnarray}
In this picture, the operators of observables evolve in time, not the state. We shall, however, work in what is known as the \emph{interaction picture}~\cite{greiner2007}. Consider a Hamiltonian $\hat{H}$ which can be split into a time independent and a time dependent term i.e.
\begin{eqnarray}
\hat{H}(t)=\hat{H}_{0}+\hat{H}_{1}(t).
\end{eqnarray}
We can define a new state as
\begin{eqnarray}
\ket{\psi_{I}(t)}:=e^{i\hat{H}_{0}t}\ket{\psi(t)},
\end{eqnarray}
where $\ket{\psi(t)}$ is the solution to Eq.~(\ref{eqn:schrodinger}). It can be shown the state $\ket{\psi_{I}(t)}$ obeys
\begin{eqnarray}
\label{eqn:schwinger}
i\partial_{t}\ket{\psi_{I}(t)}=\hat{H}_{I}(t)\ket{\psi_{I}(t)},
\end{eqnarray}
where we have defined the new operator $\hat{H}_{I}(t)$ as
\begin{equation}
\hat{H}_{I}(t):=e^{i\hat{H}_{0}t}\hat{H}_{1}(t)e^{-i\hat{H}_{0}t}.
\end{equation}
Eq.~(\ref{eqn:schwinger}) is nothing more than the Schr\"{o}dinger equation transformed to the interaction picture and is referred to as the Schwinger-Tomonaga equation~\cite{greiner2007}. The interaction picture is useful as it associates all trivial dynamics due to the free Hamiltonian to the states. This allows us to consider only the interaction Hamiltonian of our system when computing the dynamics of a state. From now on, we shall assume that we are always in the interaction picture and therefore drop any \emph{I} subscripts. The general solution to Eq.~(\ref{eqn:schwinger}) is
\begin{eqnarray}
\label{eqn:schrodingersolution}
\ket{\psi(t)}=\hat{U}\ket{\psi(0)},
\end{eqnarray}
where $\ket{\psi(0)}$ is the initial state of the system and the operator $\hat{U}$ is known as the \emph{evolution operator} for a given Hamiltonian and is defined as
\begin{eqnarray}
\label{eqn:evo-op-def}
\hat{U}=\overleftarrow{T}e^{-i\int d\tau \hat{H}(\tau)}.
\end{eqnarray}
Here, $\overleftarrow{T}$ denotes the \emph{time ordering operator}~\cite{greiner2007}. The reason for introducing the time ordering operator is that, in general, the Hamiltonian that governs the dynamics of a system does not commute with itself at different times. We therefore have to take this non-commutativity into account when solving quantum dynamics. We shall compute explicit examples of Eq.~(\ref{eqn:evo-op-def}) in Section~(\ref{sec:time-ordering}).

To end the pure quantum state section we discuss the notion of \emph{composite systems}. So far, we have only seen the discussion of a single Hilbert space denoted by $H$. However, the quantum mechanics of a single system, like a qubit, is quite trivial. We want to describe \emph{physical} situations which occur in nature such as the collision of two particles or the interaction of two clouds of gas. Thus, we need a concept of multiple systems. This can be easily accomplished by extending our definition of a Hilbert space to include multiple spaces.
\begin{defn}
For a set of $N$ quantum subsystems, each described by a Hilbert space $H_{j}$, the Hilbert space for the whole quantum system is defined as
\begin{eqnarray}
H=\bigotimes_{j=1}^{N}H_{j},
\end{eqnarray}
where $\otimes$ is the tensor product of the individual Hilbert spaces.
\end{defn}
Tensor products are a way of combining two vector spaces such that the resulting space is also a vector space. Linear operators and inner products of a subspace $H_{j}$ are mapped to linear operators and inner products on the larger space $H$.  This construction is important so that the postulates of quantum mechanics can be applied in a natural way to composite systems. To illustrate the tensor product, we show how it is used to combine states from two individual Hilbert spaces:
\begin{defn}
Given two independent quantum states $\ket{\psi_{A}}\in H_{A}$ and $\ket{\psi_{B}}\in H_{B}$, we define a combined state $\ket{\psi_{AB}}$ which lives in $H_{AB}=H_{A}\otimes H_{B}$ via the tensor product as
\begin{eqnarray}
\ket{\psi_{AB}}=\ket{\psi_{A}}\otimes\ket{\psi_{B}},
\end{eqnarray}
where $\otimes$ is the tensor product.
\end{defn}
Further, we can use the tensor product to write the most general pure state of the Hilbert space $H_{AB}=H_{A}\otimes H_{B}$:
\begin{defn}
For a set of possible (orthonormal) quantum states $\lbrace\ket{\psi_{A}^{j}}\rbrace$ and $\lbrace\ket{\psi_{B}^{k}}\rbrace$ which belong to the Hilbert spaces $H_{A}$ and $H_{B}$ respectively, the whole state of the Hilbert space $H_{AB}=H_{A}\otimes H_{B}$ is defined as
\begin{equation}
\label{eqn:tensor-state}
\ket{\psi_{AB}}=\sum_{j,k}c_{jk}\ket{\psi_{A}^{j}}\otimes\ket{\psi_{B}^{k}}.
\end{equation}
where $c_{ij}\in\mathbb{C}$ are complex amplitudes which satisfy $\sum_{j,k}|c_{jk}|^{2}=+1$.
\end{defn}
Note that the tensor product of a set of pure states is again pure. However, composite systems allow us to explore the concept of states on just a small subspace of the full space $H$. If the state of the whole system is pure, does it necessarily imply the state of a subsystem is also pure? In other words, does the expression~(\ref{eqn:tensor-state}) describe the most general state? The short answer is, of course, no. We can generalise pure states to what are known as \emph{mixed states}. We shall review them next.

\section{Mixed State Quantum Mechanics}

We have just seen the postulates of quantum mechanics for pure states. There is, however, a more convenient description of quantum states which generalises the notion of pure states. This description uses a tool known as the \emph{density operator}. The density operator (or sometimes density matrix) is a linear operator $\hat{\rho}$ on a Hilbert space which describes the general state of a system. In the following, we will denote a general composite system pure state as $\ket{\psi_{i}}$, i.e. the subscript $i$ does \emph{not} represent a subsystem in general.

Consider a quantum system which can be in a number of possible quantum states $\ket{\psi_{i}}$. This can arise in a very natural manner from the uncertainty of knowing whether a quantum system is in on state or another. This is in direct analogy with classical statistical mechanics where the state of a system is describe as a \emph{statistical ensemble} of accessible configurations. By combining the pure states with themselves as an \emph{outer product}, we can express the state of the system as
\begin{eqnarray}
\hat{\rho}=\sum_{i}\omega_{i}\ket{\psi_{i}}\bra{\psi_{i}},
\end{eqnarray}
where $\omega_{i}$ are statistical weights which give the probability of the system being in the state $\ket{\psi_{i}}$. The normalisation of a mixed state is easily expressed as
\begin{eqnarray}
\mathrm{tr}(\hat{\rho})=+1 \Rightarrow \sum_{i}\omega_{i}=+1,
\end{eqnarray}
where we have denoted the \emph{trace} of a linear operator as $\mathrm{tr}(\cdot)$. We can also write a given pure state $\ket{\psi}$ in its density operator form as
\begin{eqnarray}
\hat{\rho}=\ket{\psi}\bra{\psi}.
\end{eqnarray}
We can easily distinguish between pure and mixed states by the property of \emph{idempotence}. We therefore write
\begin{subequations}
\begin{align}
\hat{\rho}^{2}&=\hat{\rho}\Rightarrow \mathrm{pure~state},\\
\hat{\rho}^{2}&\not=\hat{\rho}\Rightarrow \mathrm{mixed~state}.
\end{align}
\end{subequations}
To be more precise, we can define the \emph{purity} of a state as
\begin{eqnarray}
\mu(\hat{\rho})=\mathrm{tr}\hat{\rho}^{2},
\end{eqnarray}
which obtains its maximum value of $+1$ when a state is pure i.e. $\mu(\ket{\phi}\bra{\phi})=+1$. We can now reformulate our postulates of quantum mechanics in the language of density matrices:
\begin{post}{1}
A physical quantum system is represented by a positive semi-define operator of trace one on a Hilbert space represented as
\begin{eqnarray}
\hat{\rho}=\sum_{i}\omega_{i}\ket{\psi_{i}}\bra{\psi_{i}}.
\end{eqnarray}
\end{post}
\begin{post}{2}
Quantum measurements are described by a set of measurement operators $M_{n}$. Each measurement operator has with it an associated measurement outcome $m_{n}$, where the probability of obtaining the measurement outcome is given by
\begin{eqnarray}
p_{n}=\mathrm{tr}\left(\hat{M}_{n}\hat{\rho}\hat{M}_{n}^{\dag}\right),
\end{eqnarray}
with the state immediately afterwards reducing to
\begin{eqnarray}
\hat{\rho}'=\frac{\hat{M}_{n}\hat{\rho}{M}_{n}^{\dag}}{\mathrm{tr}\left(\hat{M}_{n}\hat{\rho}\hat{M}_{n}^{\dag}\right)},
\end{eqnarray}
and the measurement operators must satisfy the completion equation
\begin{equation}
\sum_{n}\hat{M}_{n}^{\dag}\hat{M}_{n}=\hat{I}.
\end{equation}
\end{post}
\begin{post}{3}
The dynamics of an initial quantum state $\hat{\rho}(0)$ are governed by the Schr\"{o}dinger equation~(\ref{eqn:schrodinger}) and induce the \emph{unitary evolution}
\begin{eqnarray}
\hat{\rho}(t)=\hat{U}(t)\hat{\rho}(0)\hat{U}^{\dag}(t),
\end{eqnarray}
where $\hat{U}$ is evolution operator defined via Eq.~(\ref{eqn:evo-op-def}).
\end{post}
Although we have mathematically defined mixed states, we would like to understand physically how they arise. We can do this by considering the concept of a \emph{partial trace}. We first explain the partial trace for a composite system of two subsystems. The definitions extend in a natural way to composite systems of many subsystems.

Consider the two Hilbert spaces $H_{A}$ and $H_{B}$. The full Hilbert space of the composite system is $H_{AB}=H_{A}\otimes H_{B}$. We can write a general state of $H_{AB}$ as
\begin{eqnarray}
\hat{\rho}_{AB}=\sum_{k}\omega_{k}\ket{\psi_{AB}^{k}}\bra{\psi_{AB}^{k}},
\end{eqnarray}
where $\ket{\psi_{AB}^{k}}\bra{\psi_{AB}^{k}}$ are possible states of the composite system. The partial trace maps a density matrix $\hat{\rho}_{AB}\in H_{AB}$ to a density matrix acting on one of the subsystems $H_{A}$ or $H_{B}$. In other words, it is a way of obtaining the state of a single subsystem, removing any information about unwanted subsystems.
\begin{defn}
The partial trace of the state $\hat{\rho}_{AB}\in H_{AB}=H_{A}\otimes H_{B}$ over the subsystem $H_{B}$ is defined as the map
\begin{eqnarray}
\mathrm{tr}_{B}:\mathrm{trace}(H_{AB})\rightarrow\mathrm{trace}(H_{A}),
\end{eqnarray}
where $\mathrm{trace}(H)$ denotes the space of all trace class operators that live in the Hilbert space $H$ i.e. those which have a finite trace.
\end{defn}
In terms of density matrices the partial trace is written as
\begin{eqnarray}
\mathrm{tr}_{B}(\hat{\rho}_{AB})=\hat{\rho}_{A}.
\end{eqnarray}
Concretely, given two bases $\lbrace\ket{\psi_{A}^{j}}\rbrace\in H_{A}$ and $\lbrace\ket{\psi_{B}^{m}}\rbrace\in H_{B}$, we can define a general linear operator $\hat{\mathcal{O}}_{AB}$ on $H_{AB}=H_{A}\otimes H_{B}$ as
\begin{eqnarray}
\hat{\mathcal{O}}_{AB}=\sum_{jk,mn}\mathcal{O}_{jkmn}\ket{\psi_{A}^{j}}\ket{\psi_{B}^{m}}\bra{\psi_{A}^{k}}\bra{\psi_{B}^{n}}.
\end{eqnarray}
The partial trace of an operator, with respect to the subsystem $H_{B}$, is defined as
\begin{equation}
\begin{aligned}
\mathrm{tr}_{B}\hat{\mathcal{O}}_{AB}&=\sum_{l}\bra{\psi_{B}^{l}}\hat{\mathcal{O}}_{AB}\ket{\psi_{B}^{l}},\\
&=\sum_{jk,mn}\mathcal{O}_{jkmn}\ket{\psi_{A}^{j}}\bra{\psi_{A}^{k}}\cdot\langle \psi_{B}^{l}|\psi_{B}^{m}\rangle\langle \psi_{B}^{n}|\psi_{B}^{l} \rangle ,\\
&=\sum_{jk,m}\mathcal{O}_{jkmm}\ket{\psi_{A}^{j}}\bra{\psi_{A}^{k}},
\end{aligned}
\end{equation}
where we have used in the last line $\langle \psi_{B}^{l}|\psi_{B}^{m}\rangle=\delta_{lm}$ and performed the relevant summations. Note that we are left with an operator which acts purely on $H_{A}$. The operator obtained after partial tracing is also known as the \emph{reduced operator}. The partial trace, as previously mentioned, can be used with a state $\hat{\rho}_{AB}$ to obtain a \emph{reduced state} $\hat{\rho}_{A}$. Moreover, the partial trace preserves the positive semi-definite and unit trace properties so that the reduced state is still a physical quantum state.  Therefore, it is known as a \emph{trace preserving, completely positive map}~\cite{diosi2011,nielsen2010}. The partial trace map naturally extends to composite systems of any number of subsystems and can be applied to any subsystem. 

To get a feeling of how the partial trace works, we shall define a family of pure states which live in a composite space of two systems. States which are defined in terms of two subsystems only are known as \emph{bipartite}. Consider the following states which live in the composite space $H_{AB}=H_{A}\otimes H_{B}$:
\begin{subequations}
\label{eqn:bellstates}
\begin{align}
\ket{\phi^{\pm}}_{AB}=\frac{\ket{0}_{A}\ket{0}_{B}\pm\ket{1}_{A}\ket{1}_{B}}{\sqrt{2}},\\
\ket{\psi^{\pm}}_{AB}=\frac{\ket{0}_{A}\ket{1}_{B}\pm\ket{1}_{A}\ket{0}_{B}}{\sqrt{2}}.
\end{align}
\end{subequations}
These are pure, bipartite states and are known as the \emph{Bell states}. They posses the property that when tracing over one of the subsystems, the resulting reduced state is proportional to the identity operator. Expressing the Bell states in terms of their density operator, e.g. $\ket{\phi^{\pm}}\bra{\phi^{\pm}}$, and performing the partial trace with respect to the Hilbert space $H_{B}$ we obtain, for all four Bell states,
\begin{eqnarray}
\hat{\rho}_{A}=\frac{1}{2}\hat{I}_{A},
\end{eqnarray}
where $\hat{I}_{A}$ is the identity operator on the Hilbert space $H_{A}$. Notice that the reduced state of the Bell states are mixed, i.e. $\hat{\rho}_{A}^{2}\ne\hat{\rho}_{A}$. Thus, we can obtain mixed states from pure states. Therefore we can physically interpret mixed states as states where information has been \emph{lost} in some way. Finally, we mention that the Bell states posses another property that will be the main theme of the work presented here. They are \emph{entangled} between the systems $H_{A}$~and~$H_{B}$.

\section{Entanglement}\label{sec:entanglement}

Entanglement theory is concerned with quantifying entanglement. As previously mentioned, entanglement has many applications and so we briefly review how to define and quantify it. The central question in quantum entanglement theory is: \emph{given an arbitrary state $\hat{\rho}$, how can we determine if it is entangled or not?} Given that this question encompasses a deep and large breadth of work, we shall limit ourselves to only the most essential concepts needed for the work presented here. Our starting point will be to define what it means to \emph{not} be entangled, i.e. what is known as separability.
\begin{defn}
Separable state: the state of a composite system is said to be separable if, and only if, it can be written as a tensor product of individual subspace states~\cite{peres1996}.
\end{defn}
More precisely, let $H=\bigotimes_{j=1}^{N}H_{j}$ be the total Hilbert space of $N$-subsystems. An arbitrary mixed state is separable if it takes the form
\begin{eqnarray}
\hat{\rho}=\sum_{k}\omega_{k}\bigotimes_{j}\hat{\rho}_{j}^{k},
\end{eqnarray}
where $\hat{\rho}_{j}^{k}$ are the reduced states of the subsystems of unit trace and the $\omega_{k}$ are statistical weights which sum to unity i.e. $\sum_{k}\omega_{k}=+1$. Note that this definition also includes pure separable states. We can now define what it means to be entangled:
\begin{defn}
Entangled state: A state is entangled if, and only, if it is not separable. 
\end{defn}
Entanglement has a defining property of that it cannot be created via what are known as \emph{local operations and classical commutations} (LOCC). This means that by performing operations (which can be, for example, unitary transformations, completely positive maps or measurements) on a single subsystem of a state and communicating any information via a classical method, such as sending a laser signal, one cannot increase the entanglement contained within the system. These operations can, however, decrease the entanglement in the state.

\section{Entanglement Measures}

Here we review the entanglement measures that will be used in this thesis. For some excellent review articles on entanglement theory, the reader is advised to see~\cite{bruss2001,vedral2002,plenio2007,horodecki2009} and references therein. In essence, entanglement theory attempts to identify and quantify the amount of entanglement inherent in physical systems. One of the main reasons for investigating quantum correlations is that they can be used to implement protocols that would be otherwise impossible using classical systems. Of particular interest, which we shall discuss in later sections, is the protocol of quantum teleporation.

\subsection{Von Neumann Entropy}

In classical information theory, Shannon~\cite{shannon1948} defined a measure of the uncertainty associated with a random variable. Following this idea, Von Neumann~\cite{neumann1932} defined the quantum analogy as a measure of how \emph{mixed} a state is. It is fundamental in studying general quantum (and classical) correlations and forms the basis of many entanglement measures.
\begin{defn}
The Von Neumann entropy of a quantum state $\hat{\rho}$ is defined as
\begin{eqnarray}
\mathrm{S}(\hat{\rho}):=-\mathrm{tr}\left(\hat{\rho}\log_{2}\hat{\rho}\right).
\end{eqnarray}
\end{defn}
Using the eigendecomposition, see Eq.~(\ref{eqn:eigendecomp}), of a positive semi-definite, Hermitian operator we can write the Von Neumann entropy as a function of the eigenvalues of $\hat{\rho}$ as
\begin{eqnarray}
\label{enq:vne}
\mathrm{S}(\hat{\rho}):=-\sum_{j}\lambda_{j}\log_{2}\lambda_{j},
\end{eqnarray}
where the $\lambda_{j}$ are eigenvalues of $\hat{\rho}$. Note that the Von Neumann entropy is zero for pure states, i.e. $\text{S}(\ket{\psi}\bra{\psi})=0$. This can be seen by performing a power expansion of the Von Neumann entropy in terms of $\hat{\rho}$ and using the defining pure state property $\hat{\rho}^{2}=\hat{\rho}$.

The Von Neumann entropy can be used to determine the quantum correlations of pure bipartite states, i.e. states which are pure and only contain two subsystems. This can be seen by writing a pure bipartite state in what is known as its \emph{Schmidt decomposition}~\cite{nielsen2010}. Consider the bipartite Hilbert space $H_{AB}=H_{A}\otimes H_{B}$ where $\text{dim }H_{A}=n$,~$\text{dim }H_{B}=m$ and $n\le m$. We can always construct orthonormal bases for these Hilbert spaces $\lbrace\ket{\psi^{A}_{j}}\rbrace_{j=1}^{n}$ and $\lbrace\ket{\psi^{B}_{j}}\rbrace_{j=1}^{m}$ respectively such that for any $\ket{\psi}\in H_{AB}$ we have
\begin{eqnarray}
\label{eqn:schmidt}
\ket{\psi}=\sum_{j=1}^{n}\alpha_{j}\ket{\psi^{A}_{j}}\ket{\psi^{B}_{j}},
\end{eqnarray}
where $\alpha_{j}\ge 0$ are known as the Schmidt coefficients and $\sum_{j}\alpha_{j}^{2}=+1$. Using the Schmidt decomposition, we can write the reduced density matrices for the subsystems as
\begin{subequations}
\begin{align}
\hat{\rho}_{A}&=\sum_{j=1}^{n}\alpha_{j}^{2}\ket{\psi^{A}_{j}}\bra{\psi^{A}_{j}},\\
\hat{\rho}_{B}&=\sum_{j=1}^{n}\alpha_{j}^{2}\ket{\psi^{B}_{j}}\bra{\psi^{B}_{j}}.
\end{align}
\end{subequations}
The reduced states are diagonal and, moreover, have the same spectrum. It is clear to see that if the original Schmidt decomposition had only one non-zero $\alpha_{j}$, then it would have been a separable state. In other words for Schmidt coefficients where $d=1$ a pure bipartite state is not entangled. Thus for $d>1$ we have a non-separable state and know that the state is entangled. We can quantify the degree of entanglement in the state by computing the mixedness of the reduced states. As each reduced state is diagonal with the same spectrum, we can quantify the mixedness using just one of the reduced states. We therefore introduce our first entanglement measure, the \emph{entropy of entanglement}.
\begin{defn}
The entropy of entanglement for a state $\hat{\rho}_{AB}$ is defined as
\begin{eqnarray}
\label{eqn:entropy-of-entanglement}
\mathrm{E}(\hat{\rho}_{AB}):=\text{S}(\hat{\rho}_{A})=\text{S}(\hat{\rho}_{B}),
\end{eqnarray}
where $\mathrm{tr}_{B}(\hat{\rho}_{AB})=\hat{\rho}_{A}$ and $\mathrm{tr}_{A}(\hat{\rho}_{AB})=\hat{\rho}_{B}$ are the reduced states of the system $H_{A}$ and $H_{B}$ respectively and $S(\cdot)$ is the Von Neumann entropy.
\end{defn}
In particular, we can use the Schmidt decomposition to write
\begin{equation}
\label{eqn:entropy-of-entanglement-schmidt}
\mathrm{E}(\hat{\rho})=-\sum_{j}\alpha_{j}^{2}\log_{2}\alpha_{j}^{2}.
\end{equation}
The entropy of entanglement is \emph{the} measure for pure bipartite quantum correlations~\cite{adesso2006}. To illustrate the entropy of entanglement, we can use it to quantify the entanglement of the Bell states~(\ref{eqn:bellstates}). As they are pure and bipartite, we can use the entropy of entanglement to fully quantify the quantum correlations present. We have seen previously the reduced state of all four Bell states is $1/2\,\hat{I}$. Thus using the definition of the entropy of entanglement in terms of an operator's Schmidt coefficients~(\ref{eqn:entropy-of-entanglement-schmidt}), we find
\begin{eqnarray}
\mathrm{E}(\ket{\psi^{\pm}})=\mathrm{E}(\ket{\phi^{\pm}})=+1.
\end{eqnarray}
We notice that a pure, bipartite state whose reduced density matrices are $1/2\,\hat{I}$ maximises the entropy of entanglement. Such states are therefore known as \emph{maximally entangled}. However, in more complicated scenarios, such as composite system mixed states and states involving more than two subsystems, a general measure of entanglement is unknown~\cite{plenio2007}. One reason for this is that there is no analogue of the Schmidt decomposition for mixed states. It is therefore reasonable to look for alternative quantifiers. In what follows, we will introduce what are known as quantum \emph{negativity measures}. These are useful as they can be used to quantify entanglement in mixed states.

\subsection{Negativity Measures}\label{sec:negs}

As previously stated, a general ordering of entanglement measures is unknown. Given this, a vast body of work has been developed to find measures that are not only computable but are also known to bound any entanglement contained within a system. Here we introduce the \emph{negativity} and \emph{logarithmic negativity} for this purpose. They rely on a criterion for state separability known as the \emph{positive partial transpose} (PPT) criterion. To define this criterion, we need to define the \emph{partial transposition map}.

The partial transposition map is most easily shown with an example. Given a bipartite Hilbert space $H_{AB}=H_{A}\otimes H_{B}$, we can define an orthonormal basis $\lbrace\ket{\psi^{A}_{i}}\ket{\psi^{B}_{j}}\rbrace$ (see~(\ref{eqn:schmidt})). Thus we can write a general mixed state on $H_{AB}$ as
\begin{eqnarray}
\hat{\rho}=\sum_{ij,kl}\rho_{ij,kl}\ket{\psi^{A}_{i}}\ket{\psi^{B}_{j}}\bra{\psi^{A}_{k}}\bra{\psi^{B}_{l}}.
\end{eqnarray}
We define the partial transpose of the operator $\hat{\rho}$ on the Hilbert space $H_{AB}=H_{A}\otimes H_{B}$ with respect to, say, the subspace $H_{B}$ by interchanging the indices of the $B$ subspace:
\begin{eqnarray}
\hat{\rho}^{\text{tp}_{B}}:=\sum_{ij,kl}\rho_{ij,kl}\ket{\psi^{A}_{i}}\ket{\psi^{B}_{l}}\bra{\psi^{A}_{k}}\bra{\psi^{B}_{j}}.
\end{eqnarray}
We can of course partially transpose with respect to whatever subspace we choose and extend the definition to incorporate as many partitions as are necessary. We can now define the PPT criterion~\cite{peres1996}.
\begin{defn}
The PPT criterion for a bipartite state: For a given mixed state $\hat{\rho}_{AB}$, if its partial transposition has negative eigenvalues, then it is necessarily entangled, i.e.
\begin{eqnarray}
\hat{\rho}_{AB}^{\mathrm{tp}_{B}}\not\ge 0\Rightarrow \hat{\rho}_{AB}\,\mathrm{\,\,is\,\,entangled}.
\end{eqnarray}
\end{defn}
This definition is \emph{not} sufficient in general~\cite{vidal2002}. Only for states which live in the spaces $\mathbb{C}^{2}\otimes\mathbb{C}^{2}$ and $\mathbb{C}^{2}\otimes\mathbb{C}^{3}$ is the PPT criterion \emph{sufficient}~\cite{plenio2007}. This means that an entangled state might have a positive partial transpose. In other words
\begin{eqnarray}
\hat{\rho}_{AB}^{\mathrm{tp}_{B}}\ge 0 \not\Rightarrow \hat{\rho}_{AB}\,\mathrm{\,\,is\,\,separable}.
\end{eqnarray}
States which are entangled and have a positive partial transpose are known as \emph{bound entangled states}~\cite{horodecki1998}. In other words, they have zero distillable entanglement. The final definition we need before the negativities can be defined is the \emph{trace norm} of an operator. The trace norm of an operator $\hat{A}$ is defined as~\cite{conway2000}
\begin{eqnarray}
\label{eqn:tracenorm1}
\|\hat{A}\|_{1}:=\text{tr} \sqrt{\hat{A}^{\dag}\hat{A}}.
\end{eqnarray}
In the case where $\hat{A}$ is a normal operator, i.e. $[\hat{A},\hat{A}^{\dag}]=0$ , the trace norm is equal to the sum of the magnitudes of its eigenvalues, i.e.
\begin{equation}
\label{eqn:tracenorm2}
\|\hat{A}\|_{1}=\sum_{j}|\lambda_{j}|,
\end{equation}
where we have denoted $\lambda_{j}$ as the eigenvalues of $\hat{A}$. Grouping the positive and negative eigenvalues together, we can cast Eq.~(\ref{eqn:tracenorm1}) into the form
\begin{eqnarray}
\label{eqn:tracenorm3}
\|\hat{A}\|_{1}=\left|\sum_{k}\lambda_{k}^{+}\right|-\left|\sum_{l}\lambda_{l}^{-}\right|,
\end{eqnarray}
where $\lambda_{k}^{+}$ and $\lambda_{l}^{-}$ denote the positive and negative eigenvalues respectively of $\hat{A}$. We can now define the \emph{quantum negativity}~\cite{vidal2002}.
\begin{defn}
The quantum negativity of a state $\hat{\rho}$ is defined as
\begin{eqnarray}
\mathcal{N}(\hat{\rho}):=\frac{\|\hat{\rho}^{\text{tp}_{B}}\|_{1}-1}{2}
\end{eqnarray}
where $\|\cdot\|_{1}$ is the trace norm of a Hilbert space operator. 
\end{defn}
As $\hat{\rho}^{\text{tp}_{B}}$ is a Hermitian operator, and therefore a normal operator, the trace norm can be replaced by the sum of the magnitudes of its eigenvalues. We also know that $\text{tr}\hat{\rho}^{\text{tp}_{B}}=+1$ and hence using Eq.~(\ref{eqn:tracenorm3}) we can write the negativity as
\begin{eqnarray}
\mathcal{N}(\hat{\rho})=\left|\sum_{l}\lambda^{-}_{l}\right|.
\end{eqnarray}
In other words, the negativity is the absolute value of the sum of all negative eigenvalues of $\hat{\rho}^{\text{tp}_{B}}$. It is also said to quantify how much $\hat{\rho}^{\text{tp}_{B}}$ ``fails to be positive"~\cite{vidal2002}. The reason we use the partial transposition of a state in the definition of entanglement is due to the fact if a state was separable then transposing one if its subsystems would not affect its positivity~\cite{peres1996}.

Finally we define another well-known and well-studied negativity based measure: the \emph{logarithmic negativity}~\cite{vidal2002}.
\begin{defn}
We define the logarithmic negativity as
\begin{eqnarray}
\mathcal{E}_{\mathcal{N}}(\hat{\rho})=\log_{2}\|\hat{\rho}^{\text{tp}_{A}}\|_{1}
\end{eqnarray}
where $\|\cdot\|_{1}$ denotes the trace norm of an operator.
\end{defn}
Having defined entanglement and how to quantify it, we now illustrate its usefulness via a quantum information protocol: quantum teleportation.

\section{Quantum Teleportation}\label{sec:teleportation-qmqi}

Quantum teleportation is a protocol that uses quantum entanglement to transfer an unknown state from one party to another. It relies on the fact that entanglement cannot be created by LOCC. The seminal paper of Bennett et. al.~\cite{bennett1993} was the first to show how to exploit the entanglement shared between systems held by two parties. The usual names of the two participants in quantum information are Alice (A) and Bob (B). However during the course of writing this thesis, we discussed with Bob and he decided not to participate in any teleportation procedures. This is due to the extreme nature of incorporating relativity and quantum mechanics. Luckily, Rob, Bob's relativistic cousin, was happy to step in and help. Therefore, we will refer to Rob as Alice's partner for quantum information tasks in relativistic quantum information. We will continue to refer to Rob with the subscript (B) to avoid possible notation clashes later. The first quantum teleportation scheme showed how to send an unknown pure state from Alice to Rob. The original teleportation protocol involved qubits. Many generalisations and adaptations were then created that involved mixed states and systems that were not qubits, such as continuous variable Gaussian states~\cite{fiurasek2002}. The generalisation to continuous variables will be the subject of Chapter~(\ref{chapter:bosons}), Section~(\ref{sec:teleportation-results}).

Consider the maximally entangled qubit state shared between Alice and Rob
\begin{eqnarray}
\ket{\phi^{+}}_{AB}=\frac{\ket{0,0}_{AB}+\ket{1,1}_{AB}}{\sqrt{2}}.
\end{eqnarray}
Note we have defined $\ket{0,0}_{AB}\equiv\ket{0}_{A}\ket{0}_{B}$ etc. We wish to transfer an unknown pure qubit state, carried by Alice, to Rob. We can use the entanglement held by Alice and Rob to perform this transfer. It is this scenario that is known as quantum teleportation. The full state of the system can be written as
\begin{eqnarray}
\label{eqn:teleportationstate}
\ket{\Psi}_{ABC}=\ket{\phi^{+}}_{AB}\ket{\alpha}_{C},
\end{eqnarray}
where $\ket{\alpha}_{C}=a\ket{0}_{C}+b\ket{1}_{C}$ is the unknown qubit state and $|a|^{2}+|b|^{2}=+1$. Using the definitions of the Bell states~(\ref{eqn:bellstates}), we can rewrite~(\ref{eqn:teleportationstate}) as
\begin{equation}
\begin{aligned}
\ket{\Psi}_{ABC}&=\hspace{7mm}\ket{\phi^{+}}_{AC}\Big(a\ket{0}_{B}+b\ket{1}_{B}\Big)+\ket{\phi^{-}}_{AC}\Big(a\ket{0}_{B}-b\ket{1}_{B}\Big)\\
&\hspace{7mm}+\ket{\psi^{+}}_{AC}\Big(a\ket{1}_{B}+b\ket{0}_{B}\Big)+\ket{\psi^{-}}_{AC}\Big(a\ket{1}_{B}-b\ket{0}_{B}\Big)
\end{aligned}.
\end{equation}
Next, Alice performs a local measurement on the two qubits in her possession. Once this measurement is performed, Alice will possess two \emph{classical} bits of information and know what state Rob's qubit will have collapsed to. For example if Alice measured $\ket{\psi^{-}}$, she would know that Rob's qubit is $a\ket{1}_{B}-b\ket{0}_{B}$. Thus to obtain the original, still unknown, qubit she communicates to Rob classically that he needs to perform the unitary rotation $\boldsymbol{U}=-i\boldsymbol{\sigma}_{2}$, where $\boldsymbol{\sigma}_{2}$ is the second Pauli matrix. Rob's qubit will now be in the unknown state $a\ket{0}_{B}+b\ket{1}_{B}$. Hence, for every measurement Alice might wish to perform, she can classically communicate the result and an associated unitary operation such that Rob's qubit can be transformed into the unknown state $\ket{\alpha}$. Interestingly, we have transferred the state $\ket{\alpha}$ from Alice to Rob without \emph{any} loss of information. In this case we say the \emph{fidelity} of the teleportation was one. In any teleporation protocol, the figure or merit is usually classified by how close the input state and the teleported output state are i.e. the fidelity of the two states. We can quantify this  fidelity by computing the overlap of the input state and the output state. For an arbitrary mixed input state $\hat{\rho}_{1}$ and output state $\hat{\rho}_{2}$ we define the fidelity as 
\begin{equation}
\mathcal{F}(\hat{\rho}_{1},\hat{\rho}_{2})=\mathrm{tr}\left[\sqrt{\sqrt{\hat{\rho}_{2}}\hat{\rho}_{1}\sqrt{\hat{\rho}_{2}}}\right].
\end{equation}
This simple example serves as a demonstration of how teleportation works and the usefulness of entangled systems. As we have discussed the basics of quantum theory and quantum information, it will be useful to review another physical theory which will play an important role in the following work: quantum field theory.
\chapter{Quantum Field Theory}\label{chapter:qft}

After the revolutionary success of quantum mechanics in its early days, it became interesting to understand how, and if, quantum mechanics is compatible with relativity theory. Early attempts to unify these two theories considered finding a relativistic version of Schr\"{o}dinger's equation. One of the first attempts to replace Schr\"{o}dinger's equation was put forward by Klein~\cite{klein1926} and Gordon~\cite{gordon1926} in the late $1920$'s. This resulted in the so-called Klein-Gordon equation and describes spinless particles. This attempted to generalise Schr\"{o}dinger's wave equation to be compatible with concepts from relativity such as invariance under changes from one inertial observer to another. However, as we will see, it had severe drawbacks. The next great step was that of Dirac~\cite{dirac1928} in $1928$ with the equation he introduced to describe an electron, i.e. spin-$1/2$ particles. Both equations succeeded in bringing together principles of relativity and quantum theory but were plagued by problems of predicting solutions with \emph{negative energy}. These negative energy solutions, it would turn out, pointed towards a more profound understanding of nature. Negative energy solutions were disturbing as the Klein-Gordon and Dirac fields could not be interpreted as a single particle with physical energy levels, as was the case for the Schr\"{o}dinger equation. Given this problem, it led people to ask: \emph{how can the Klein-Gordon and Dirac equations be interpreted?} It was finally realised that we could use these problematic solutions not as the negative energy eigenfunctions of a particle but as \emph{positive} energy eigenfunctions of an \emph{anti-particle}. Thus, the notion of single particle mechanics could not be maintained and lead to the interpretation that the physical objects described by these relativistic quantum wave equations were not single particles, but were in fact \emph{fields}. This new description of quantum systems also allowed for a previously unaccounted phenomenon, \emph{particle creation}. In other words, the solutions to the Klein-Gordon and Dirac equation were not energy eigenfunctions of a single particle, but were actually describing excitations of a field which can be interpreted as particles~\cite{srednicki2007}. Therefore, not only did the new theory bring relativity and quantum mechanics together, but also allowed for the creation and annihilation of particles. Thus, a need to move away from the concept of single particles and towards considering true field equations was motivated. These early works, which combined relativity and quantum mechanics through field equations, gave birth to \emph{quantum field theory}. This was all, however, in \emph{flat} spacetime.

Given the great success of quantum field theory in flat spacetime, people naturally asked how to extend the theory to more generic \emph{curved} spacetimes. In particular, a very natural question arose: \emph{is the notion of a particle well defined in general spacetimes?} A na\"{i}ve, and understandable, first answer could be: \emph{why not?} However, this seemingly innocent question turned out to be a stumbling block for quantum field theory in curved spacetimes. It also inspired people to ask how different observers, who travel on different trajectories, would perceive a field around them. Thus, there was a need to find a consistent method of defining particles, for \emph{different} observers. The ideal situation would be to find a universal definition of what a particle is and hope that it does not change as an observer started to travel through spacetime. However, in a generic spacetime, where conserved quantities cannot be assumed to exist in general, there is no consistent way of defining a notion of particle along every point of a generic observers trajectory~\cite{alsing2012}. This is problematic as these observers would not be able to meaningfully describe the field around them. In light of this problem, what approach can be taken? We need situations where we can \emph{consistently} define a notion of particle. Moreover, we need to be able to do this such that the definition does not change in \emph{time}. This definition would help in our concept of particles. Such a method is based upon what are known as \emph{Killing vector fields}. Killing vector fields are a way of identifying the symmetries of a spacetime and can be used to define a consistent notion of particle. As we will see, allowing an observer to meaningfully describe the particles of a field in a consistent way will help us address the question of how two \emph{different} observers describe particle content.

Once the notion of Killing vector fields was applied to quantum field theory in curved spacetimes, people could address the question of ``observer-dependent" particle content. The standard method of relating quantum states as described by different observers as they travel along different Killing vector fields is through maps known as \emph{Bogoliubov transformations}. We will discuss Bogoliubov transformations in more detail in Section~(\ref{sec:bogo-qft}) but essentially they are a way to relate one set of solutions of a field equation to another. Using Bogoliubov transformations, a landmark contribution to the question of a spacetime's particle content was given by Hawking~\cite{hawking1975}. Starting with a scenario where the state of a field in the distant past, as described by some inertial observer, was in the vacuum, he showed that a collapsing black hole would emit radiation as perceived by an observer in the distant future. Thus, the two different observers describe a different particle content for the field. This observation then lead Unruh to consider the particle content difference between an inertial observer in flat spacetime and a uniformly accelerated one. From this, Unruh predicted the so-called \emph{Unruh effect}~\cite{unruh1976}. In short, the Unruh effect tells us that a uniformly accelerated observer will perceive a uniform ``sea" of particles. This turns out to be exactly the same as a stationary observer at rest in a thermal bath of particles. As will be shown in Chapters~(\ref{chapter:bosons})~and~(\ref{chapter:moving-cavities}), the setting of Unruh, describing inertial and accelerated observers, is particularly useful for modelling the motion of an observer through spacetime. We can approximate an arbitrary observer's trajectory by combining periods of inertial motion with periods of accelerated motion. Therefore studying the quantisation of a field in terms of both inertial and accelerated observers is essential.

This chapter is organised as follows: we introduce the basic objects of differential geometry, and in particular Killing vector fields, which are essential for our studies of inertial and accelerated observers in quantum field theory. We then quantise the Klein-Gordon field in coordinates suitable for inertial~(\ref{sec:kg-mink-quant}) and accelerated~(\ref{app:rindler-coords}) observers. We go on to define the Bogoliubov transformation~(\ref{sec:bogo-qft}) between two sets of solutions to a quantum field, in particular the Klein-Gordon field, and note its relation to the particle content between different observers. Continuing, we give a short review of the quantisation of the Dirac field~(\ref{sec:dirac-field}) for inertial and accelerated observers. Finally, we discuss what is known as the \emph{Unruh-DeWitt detector model}~(\ref{sec:qft-unruh-detector}). The Unruh-DeWitt model gives us a flexible tool to investigate how particles of a field will be viewed by observers on different trajectories through spacetime.

\section{Geometry}

In this section, we assume that the reader has an intuitive picture of manifolds, geometry and notions common to relativity. An excellent introductory text for differential geometry and general relativity is Carrol~\cite{carroll2004}.

Differential geometry underpins the entire of Einstein's theory of general relativity~\cite{carroll2004}. The starting point consists of modelling physical spacetime as a \emph{smooth manifold} equipped with patches of \emph{local coordinates}. Essentially, manifolds are objects that, when viewed in a sufficiently small region, resemble flat space. The local coordinates are the coordinate systems used by an observer who is measuring space and time in this sufficiently small region.

Thus, considering a smooth manifold with a patch of local coordinates $x^{\mu}=(t,\boldsymbol{x})$, where $t$ is the time coordinate and $\boldsymbol{x}$ are the spatial coordinates, we can define the line element 
\begin{eqnarray}
\label{eqn:metric}
ds^{2}=g_{\mu\nu}dx^{\mu}dx^{\nu},
\end{eqnarray}
where $g^{\mu\nu}$ are the components of the \emph{metric tensor} and the indices run over $\mu,\nu=0,1,2,3$. Here, we have used the Einstein summation convention $\sum_{j}x_{j}x^{j}\rightarrow x_{j}x^{j}$. In other words, repeated indices are implicitly summed over. The line element of a coordinate patch is the infinitesimal distance between two different points. It can therefore be used to define a notion of distance. The metric tensor $g$ is used not only to define notions of length on a manifold but can also be used to define angles between vectors.

For the following discussion in this thesis, we will restrict ourselves to the case where the metric can be decomposed into the form
\begin{eqnarray}
\label{eqn:our-metric}
ds^{2}=-N(t,\boldsymbol{x})^{2}dt^{2}+G_{ab}(t,\boldsymbol{x})dx^{a}dx^{b},
\end{eqnarray}
where the indices $a,b$ run over $1,2,3$ and the signature of the metric is $(-,+,+,+)$. In general, the functions $N(t,\boldsymbol{x})$ and $G_{ab}(t,\boldsymbol{x})$ depend on all spacetime coordinates. As mentioned earlier, a fundamentally important concept in the canonical quantisation of fields is that of \emph{Killing vector fields}. A vector field $K=K^{\nu}\partial_{\nu}$ on a manifold is called Killing if it satisfies the equation~\cite{carroll2004}
\begin{eqnarray}
\nabla_{\mu}K_{\nu}+\nabla_{\nu}K_{\mu}=0,
\end{eqnarray}
where $\nabla_{\nu}$ is the covariant derivative in the manifold. Killing vector fields allow us to identify trajectories for observers which can define quantities which do not change in time~\cite{carroll2004}. In quantum field theory, quantities that are preserved in time are usually associated with inner products for field modes. We identify time conserved quantities via what are known as \emph{timelike} Killing vector fields.
\begin{defn}
A vector field $K$ is said to be timelike with respect to a metric tensor $g$ if $g(K,K)<0$.
\end{defn}
As mentioned before, the Klein-Gordon and Dirac fields admit what are known as positive and negative energy solutions with respect to their inner products. This is problematic as the inner product for the vector space we construct from these solutions will not be positive definite and therefore will \emph{not} be a Hilbert space~\cite{wald1994}. We therefore need a systematic way of splitting these two sets of solutions up into their corresponding positive and negative parts. This is achieved by the following definition:
\begin{defn}\label{def:particles}
Let $K$ be a timelike Killing vector field and $\lbrace\phi_{\boldsymbol{k}}(t,\boldsymbol{x})\rbrace$ be a set of solutions to a field equation. We define \emph{positive} and \emph{negative} energy solutions respectively as solutions to the eigenfunction equation
\begin{subequations}
\label{eqn:positive-negative-def}
\begin{align}
-iK\phi_{\boldsymbol{k}}(t,\boldsymbol{x})&=+\omega_{\boldsymbol{k}}\phi_{k}(t,\boldsymbol{x})\Rightarrow \mathrm{positive},\\
-iK\phi_{\boldsymbol{k}}(t,\boldsymbol{x})&=-\omega_{\boldsymbol{k}}\phi_{k}(t,\boldsymbol{x})\Rightarrow \mathrm{negative},
\end{align}
\end{subequations}
where $\omega_{\boldsymbol{k}}>0$ denotes the eigenvalue of the solution $\phi_{\boldsymbol{k}}(t,\boldsymbol{x})$.
\end{defn}
From this we define \emph{particles} to be associated with the modes with $+\omega_{k}$ and \emph{anti-particles} with modes with $-\omega_{k}$. The definition~(\ref{def:particles}) has two physical motivations. The first is that Killing vector fields generate trajectories for observers through spacetime who will be able to define conserved quantities. The second is that for a given observer following a timelike Killing vector field, any Lorentz transformation will leave the timelike property of the Killing vector field unchanged. Thus, the observer will have a constant notion of particle as they travel through spacetime. We can perform a quantisation procedure on the individual solution spaces and thus obtain a well defined Hilbert space~\cite{wald1994}.

As mentioned in the introductory section, we will be analysing the Klein-Gordon and Dirac field as described by two different observers. The first will be an inertial observer who describes the spacetime via \emph{Minkowski coordinates}. The second observer will be a uniformly, \emph{linearly} accelerated observer. That is, an observer whose proper acceleration (the acceleration measured it their rest frame) is constant in time. Uniformly accelerated observers are most conveniently described by the so-called \emph{Rindler} coordinates. First, however, we introduce the Klein-Gordon field and quantise it for Minkowski coordinates to get a feel for how canonical quantisation of a field works.

\section{Klein-Gordon Field}

The Klein-Gordon field is arguably the simplest relativistically invariant field equation we can write as it does not contain spinorial or vectorial components. It corresponds to spin-$0$ particles and can be either real or complex valued. Physically, excitations of the Klein-Gordon field can be realised as composite particles such as the Pions or the Kaons.

\subsection{Field Equation Definition}
We define the Klein-Gordon equation as~\cite{birrell1982}
\begin{eqnarray}
\label{eqn:covariant-kg-eqn}
\frac{1}{\sqrt{-g}}\partial_{\mu}\left(g^{\mu\nu}\sqrt{-g}\partial_{\nu}\right)\phi -m^{2}\phi =0,
\end{eqnarray}
where $\phi$ is the Klein-Gordon field, $m$ denotes the mass of the field and $g=\mathrm{det}(g^{\mu\nu})$. Given two solutions, $\phi_{\boldsymbol{k}},\phi_{\boldsymbol{k}'}$, of the Klein-Gordon equation, we can define a \emph{covariant} mode product on the space of Klein-Gordon solutions as~\cite{crispino2008}
\begin{eqnarray}
\label{eqn:kg-inner-product}
\left(\phi_{\boldsymbol{k}},\phi_{\boldsymbol{k}'}\right)\equiv -i\int_{\Sigma}d\boldsymbol{x}\sqrt{+G}N^{-1}\left(\bar{\phi}_{\boldsymbol{k}}\partial_{t}\phi_{\boldsymbol{k}'}-\phi_{\boldsymbol{k}'}\partial_{t}\bar{\phi}_{\boldsymbol{k}}\right),
\end{eqnarray}
where $\Sigma$ is a hypersurface of constant time i.e. $t=\mathrm{const}$, $G=\mathrm{det}(G_{ab})$ and the functions $G_{ab}$, $N$ are as defined in~(\ref{eqn:our-metric}). The mode product~(\ref{eqn:kg-inner-product}) can be conveniently rewritten in terms of the fields \emph{conjugate momentum} $\Pi(t,\boldsymbol{x})$. Defining~\cite{crispino2008}
\begin{eqnarray}
\label{eqn:kg-conjugate-momentum}
\Pi_{\boldsymbol{k}}(t,\boldsymbol{x})=\sqrt{+G}N^{-1}\partial_{t}\phi_{\boldsymbol{k}}(t,\boldsymbol{x}),
\end{eqnarray}
we can write the inner product~(\ref{eqn:kg-inner-product}) as
\begin{eqnarray}
\left(\phi_{\boldsymbol{k}},\phi_{\boldsymbol{k}'}\right)\equiv -i\int_{\Sigma}d\boldsymbol{x} \left(\bar{\phi}_{\boldsymbol{k}}\Pi_{\boldsymbol{k}'}-\bar{\Pi}_{\boldsymbol{k}}\phi_{\boldsymbol{k}'}\right).
\end{eqnarray}
The identification of a field's conjugate momentum will prove useful when performing canonical quantisation.

\subsection{Minkowski Quantisation}\label{sec:kg-mink-quant}

In standard flat $(3+1)$-dimensional Minkwoski spacetime, we can define the coordinates $(t,\boldsymbol{x})$ where the components of the metric tensor are $g^{\mu\nu}=\mathrm{diag}(-1,1,1,1)$ and so the line element~(\ref{eqn:metric}) takes the form
\begin{eqnarray}
ds^{2}=-dt^{2}+dx_{a}dx^{a}.
\end{eqnarray}
where $a$ runs over $1,2,3$. The resulting expression for the Klein-Gordon equation~(\ref{eqn:covariant-kg-eqn}) is
\begin{eqnarray}
-\partial_{t}\partial_{t}\phi(t,\boldsymbol{x}) +\partial^{a}\partial_{a}\phi(t,\boldsymbol{x})-m^{2}\phi(t,\boldsymbol{x})=0.
\end{eqnarray}
The standard basis of plane wave solutions is defined as
\begin{eqnarray}
\label{eqn:kg-plane-waves}
\phi_{\boldsymbol{k}}^{M}(t,\boldsymbol{x})=N_{\boldsymbol{k}}e^{-i\omega_{\boldsymbol{k}} t+\boldsymbol{k}\cdot\boldsymbol{x}},
\end{eqnarray}
where $N_{\boldsymbol{k}}$ is a normalisation constant. Note that for the modes~(\ref{eqn:kg-plane-waves}) to satisfy the Klein-Gordon equation the parameter $\omega_{\boldsymbol{k}}$ needs to satisfy the dispersion relation
\begin{eqnarray}
\omega_{\boldsymbol{k}}^{2}=\boldsymbol{k}\cdot\boldsymbol{k}+m^{2},
\end{eqnarray}
which is the well known mass-energy relation from special relativity and so we can identify the parameter $\omega_{\boldsymbol{k}}$ with the energy of a particle with momentum $\boldsymbol{k}$ and mass $m$. Thus, assuming the field is real, i.e. $\bar{\phi}(t,\boldsymbol{x})=\phi(t,\boldsymbol{x})$, we can expand the full solution to the Klein-Gordon equation as
\begin{eqnarray}
\label{eq:minkowski-mode-field-expansion}
\phi(t,\boldsymbol{x})=\int d\boldsymbol{k}\left(a_{\boldsymbol{k}}\phi_{\boldsymbol{k}}^{M}(t,\boldsymbol{x})+\mathrm{c.c.}\right),
\end{eqnarray}
where the $a_{\boldsymbol{k}}$ are arbitrary complex functions. Note that the integration measure in Eq.~\eqref{eq:minkowski-mode-field-expansion} is \emph{not} Lorentz invariant. Other authors choose to have an integration measure which is manifestly Lorentz invariant as to make the Lorentz invariance of the field explicit. For details see, for example,~\cite{srednicki2007}. So far this is a purely classical field. From Eq.~(\ref{eqn:kg-conjugate-momentum}) the conjugate momentum of the field is~\cite{birrell1982,srednicki2007}
\begin{eqnarray}
\Pi(t,\boldsymbol{x})=\partial_{t}\phi(t,\boldsymbol{x}).
\end{eqnarray}
In standard Minkowski coordinates the Klein-Gordon inner product~(\ref{eqn:kg-inner-product}) becomes
\begin{eqnarray}
\label{eqn:kg-minkowski-inner-product}
\left(\phi_{\boldsymbol{k}}^{M},\phi_{\boldsymbol{k}'}^{M}\right)= -i\int_{\Sigma}d\boldsymbol{x}\left(\bar{\phi}_{\boldsymbol{k}}^{M}\partial_{t}\phi_{\boldsymbol{k}'}^{M}-\phi_{\boldsymbol{k}'}^{M}\partial_{t}\bar{\phi}_{\boldsymbol{k}}^{M}\right),
\end{eqnarray}
where we can choose $\Sigma$ to be the hypersurface $t=0$. Here we choose $N_{\boldsymbol{k}}=1/\sqrt{2\omega_{\boldsymbol{k}} (2\pi)^{3}}$ such that the inner products of the modes are normalised as
\begin{subequations}
\begin{align}
\label{eqn:kg-mode-normalisations1}\left(\phi_{\boldsymbol{k}}^{M}(x),\phi^{M}_{\boldsymbol{k}^{\prime}}(x)\right)&=\delta(\boldsymbol{k}-\boldsymbol{k}^{\prime}),\\
\label{eqn:kg-mode-normalisations2}\left(\bar{\phi}^{M}_{\boldsymbol{k}}(x),\bar{\phi}^{M}_{\boldsymbol{k}^{\prime}}(x)\right)&=-\delta(\boldsymbol{k}-\boldsymbol{k}^{\prime}),\\
\label{eqn:kg-mode-normalisations3}\left(\phi^{M}_{\boldsymbol{k}}(x),\bar{\phi}^{M}_{\boldsymbol{k}^{\prime}}(x)\right)&=0.
\end{align}
\end{subequations}
Equation~(\ref{eqn:kg-mode-normalisations2}) is problematic as it says the norm of the complex conjugates of the modes, i.e. $\bar{\phi}^{M}_{\boldsymbol{k}}(x)$, are negative. Thus the inner product is not positive definite for all solutions of the Klein-Gordon equation. We can, however, separate these problematic modes via the timelike Killing vector $K=\partial_{t}$. We notice that the modes and their complex conjugates satisfy Eq.~(\ref{eqn:positive-negative-def}):
\begin{subequations}
\begin{align}
i\partial_{t}\phi^{M}_{\boldsymbol{k}}(t,\boldsymbol{x})&=+\omega_{\boldsymbol{k}}\phi_{\boldsymbol{k}}^{M}(t,\boldsymbol{x}),\\
i\partial_{t}\bar{\phi}^{M}_{\boldsymbol{k}}(t,\boldsymbol{x})&=-\omega_{\boldsymbol{k}}\bar{\phi}^{M}_{\boldsymbol{k}}(t,\boldsymbol{x}).
\end{align}
\end{subequations}
Therefore we can naturally split the solution space of the Klein-Gordon field into its positive and negative parts. This allows quantise the field and construct a proper Hilbert space for the Klein-Gordon field.

\emph{Canonical quantisation} is achieved by introducing the equal time \emph{canonical commutation relations} (CCR's)~\cite{birrell1982,srednicki2007}
\begin{eqnarray}
\left[\hat{\phi}(t,\boldsymbol{x}),\hat{\Pi}(t,\boldsymbol{x}^{\prime})\right]&=&\delta(\boldsymbol{x}-\boldsymbol{x}^{\prime}),
\end{eqnarray}
with all other equal time commutation relations vanishing. Thus, the field solution $\phi(t,\boldsymbol{x})$ is promoted to the operator $\hat{\phi}(t,\boldsymbol{x})$ and has the consequence of promoting the arbitrary functions $\lbrace a_{\boldsymbol{k}},a_{\boldsymbol{k}}^{\dag}\rbrace$ to operators satisfying
\begin{eqnarray}
\left[\hat{a}_{\boldsymbol{k}},\hat{a}_{\boldsymbol{k}^{\prime}}^{\dag}\right]&=&\delta(\boldsymbol{k}-\boldsymbol{k}^{\prime}),
\end{eqnarray}
where, again, all other commutators vanish. These are, of course, the same commutation relations as the standard quantum harmonic oscillator. Thus, we can interpret the field as a collection of \emph{decoupled} harmonic oscillators. This implies that the Klein-Gordon field is \emph{Bosonic} in nature. 

We call the $\phi^{M}_{\boldsymbol{k}}$ modes of \emph{positive energy} with respect to the timelike Killing vector $K=\partial_{t}$ and the complex conjugates $\bar{\phi}^{M}_{\boldsymbol{k}}$ modes of \emph{negative energy} with respect to $K=\partial_{t}$. We therefore identify positive energy modes with \emph{particles} and negative energy modes with \emph{anti-particles}. Note that excitations of the field, whether it be associated with a particle or anti-particle, \emph{always}, physically, have positive energy. This is a simple example of how timelike Killing vector fields are useful in quantum field theory on curved spacetimes.

As we have quantised the Klein-Gordon field, we expect to have an associated vector space in which to define our states. We do this by defining the so-called \emph{vacuum state} of a quantum field which, for the Minkowski Klein-Gordon field, is defined as
\begin{eqnarray}
\hat{a}_{\boldsymbol{k}}\ket{0}_{M}=0\,\forall\,\boldsymbol{k}.
\end{eqnarray}
The vacuum state is interpreted as being the state with no particles, i.e. it is the ground state of the field. It is normalised as $\langle 0|0 \rangle=+1$ and spans the zero particle Hilbert space $\mathbb{C}$. The set of all mode operators $\lbrace\hat{a}_{\boldsymbol{k}},\hat{a}_{\boldsymbol{k}}^{\dag}\rbrace$ allow us to define a general \emph{single} particle state as
\begin{eqnarray}
\ket{\psi}_{1}=\int_{\boldsymbol{k}}d\boldsymbol{k}\alpha_{\boldsymbol{k}}\hat{a}_{\boldsymbol{k}}^{\dag}\ket{0},
\end{eqnarray}
where $\int_{k}d\boldsymbol{k}|\alpha_{\boldsymbol{k}}|^{2}=+1$. We denote the Hilbert space of all possible single particle states as $H$ and refer to it as the \emph{single particle sector}. A similar construction can be done for the space of all \emph{two} particle states and we denote this as $\mathrm{sym}(H^{\otimes 2})$, where the notation $\mathrm{sym}(\cdot)$ denotes symmetrisation of the tensor product over Bosonic states. Combining all possible particle sectors, we can construct the full Fock space for a Bosonic system~\cite{fock1932}:
\begin{eqnarray}
\label{eqn:fock-space}
\mathbb{F}=\mathbb{C} \oplus H \oplus \mathrm{sym}(H^{\otimes 2}) \oplus \mathrm{sym}(H^{\otimes 3}) \oplus \ldots
\end{eqnarray}
For each mode there is a natural basis known as the Fock basis defined as
\begin{eqnarray}
\ket{n_{\boldsymbol{k}}}=\frac{1}{\sqrt{n!}}\hat{a}_{\boldsymbol{k}}^{\dag n}\ket{0},
\end{eqnarray}
where $n_{\boldsymbol{k}}$ denotes the number of particles in mode $\boldsymbol{k}$ and which, when combined with all other mode bases, spans the entire Fock space.

To summarise, we quantised the Klein-Gordon field in standard Minkowski coordinates making use of the timelike Killing vector $K=\partial_{t}$ which corresponded to inertial observers. This allowed us to separate the space of solutions into positive and negative parts which allowed for a proper quantisation procedure. As alluded to earlier, the accelerated observers also play a role in field quantisation. We will see that accelerated observers admit their own timelike Killing vector fields and quantisation procedure. However we first discuss how to define accelerated observers and motivate why they are useful.

\subsection{Rindler Quantisation}\label{app:rindler-coords}

As previously mentioned, Unruh investigated how uniformly, linearly accelerated observers would perceive a quantum field around them. Uniformly accelerated observers offer a useful way to analyse non-trivial effects without the need for going to complicated curved spacetime scenarios. These special types of observers are best described using what are known as \emph{Rindler coordinates}. The Rindler coordinates were first introduced by Rindler~\cite{rindler1966} to study uniformly accelerated observers in special relativity. Uniformly accelerated observers have many interesting consequences, the most important of which for our purposes is the inequivalent particle content described by an inertial observer and an accelerated one. Our primary motivation for using Rindler coordinates is that they allow us to describe uniformly accelerated motion in spacetime. In Chapter~(\ref{chapter:moving-cavities}), we combine trajectories of inertial and accelerated motion to model the non-uniform trajectory of a \emph{cavity}, a term to be defined later, through spacetime.

\begin{figure}[t]
\centering
\includegraphics[width=0.5\linewidth]{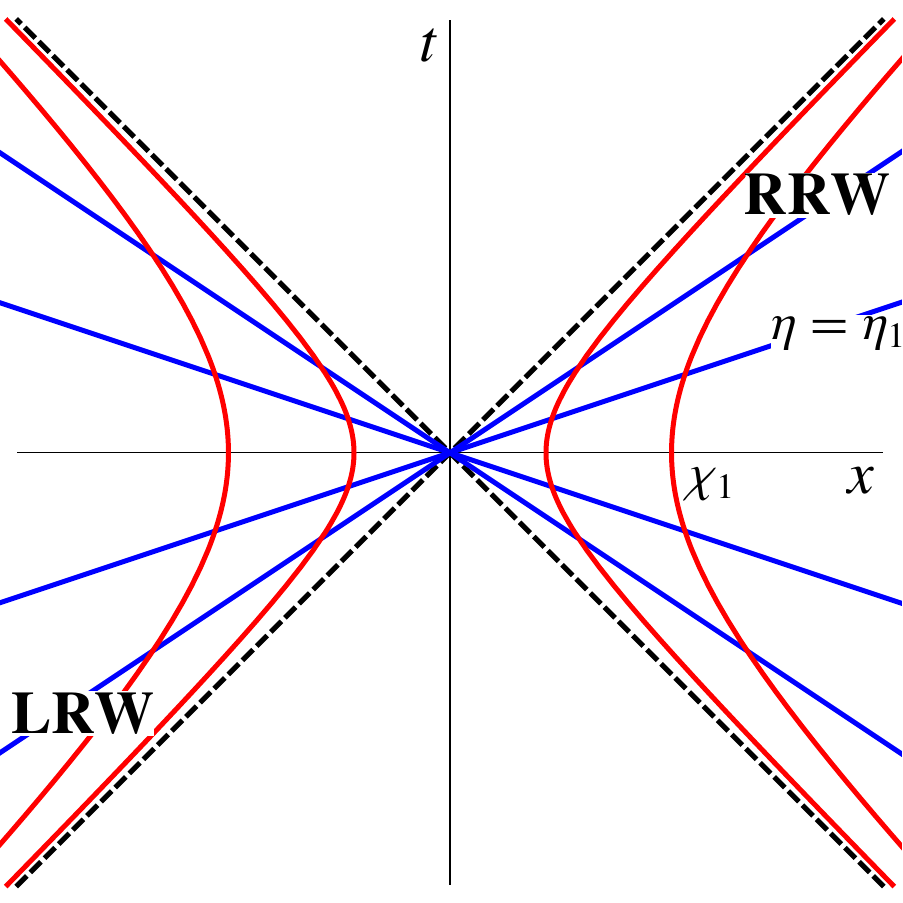}
\caption[Schematic diagram of Rindler trajectories]{Schematic diagram of Rindler trajectories. The solid red lines are the trajectories of stationary Rindler observers i.e. $\chi=\mathrm{const}$. They have a constant proper acceleration as observed by an inertial observer. The solid blue lines are hypersurfaces of constant Rindler coordinate time $\eta$ and only intersect a given accelerated trajectory once. The dashed black lines are the Rindler horizons.}
\label{fig:rindlercoords}
\end{figure}

The standard Rindler coordinates, which we denote as $(\eta,\boldsymbol{\chi})$, are defined via~\cite{takagi1986}
\begin{subequations}
\label{eqn:rindler-coord-def}
\begin{align}
t&=\chi\sinh(\eta),\\
x&=\chi\cosh(\eta),
\end{align}
\end{subequations}
where $(t,x)$ are the usual $(1+1)$-dimensional Minkowski spacetime coordinates and all other coordinates are unchanged. The ``spatial" coordinate $\chi$ is defined to be strictly positive ($\chi>0$) while the ``temporal" coordinate $\eta$ can take any real value ($\eta\in\mathbb{R}$). The Rindler coordinates $(\eta,\chi)$ only cover a part of Minkowski spacetime. This region is bounded by the asymptotes $x>|t|$, see Fig.~(\ref{fig:rindlercoords}), and is referred to as the \emph{right Rindler wedge}. We can define a second set of coordinates which covers the so-called \emph{left Rindler wedge} as
\begin{subequations}
\label{eqn:rindler-coord-def2}
\begin{align}
t&=\chi\sinh(\eta),\\
x&=-\chi\cosh(\eta).
\end{align}
\end{subequations}
The two coordinate patches~(\ref{eqn:rindler-coord-def})~and~(\ref{eqn:rindler-coord-def2}) only cover part of Minkowski spacetime, see Fig.~(\ref{fig:rindlercoords}). However, as we will see later, they are enough for our purposes of quantising the Klein-Gordon and Dirac field. To get a feel for the Rindler coordinates properties, we can write the usual $(1+1)$ flat spacetime metric as
\begin{eqnarray}
ds^{2}=-dt^{2}+dx^{2}=-\chi^{2}d\eta^{2}+d\chi^{2},
\end{eqnarray}
This line element is valid in both right and left Rindler wedges. An observer who is described as being at a fixed point in the Rindler coordinates, i.e. $\chi=\mathrm{const}$, travels through Minkowski spacetime with a \emph{constant} proper acceleration. To see this, consider a particle moving along the trajectory $\chi=1/\beta$ for some fixed $\beta>0$, as viewed in the right Rindler wedge. For this trajectory, the line element in Rindler coordinates becomes
\begin{eqnarray}
\label{eqn:rindlerpropertime}
ds^{2}=-\beta^{-2}d\eta^{2}.
\end{eqnarray}
The proper time, $\tau$, for a particle on any worldine is defined by the relation $ds^{2}=-d\tau^{2}$. From~(\ref{eqn:rindlerpropertime}) we find the differential equation
\begin{eqnarray}
d\tau^{2}=\beta^{-2}d\eta^{2}\Rightarrow \eta(\tau)=\beta\tau,
\end{eqnarray}
where we have assumed $\eta(0)=0$. Thus, as parametrised by its proper time, the particles worldline can be written as
\begin{eqnarray}
\label{eqn:rindler-parametrisation}
\eta(\tau)=\beta\tau,\,\,\chi(\tau)=1/\beta.
\end{eqnarray}
We can also parametrise the original Minkowski coordinates via the particles proper time as
\begin{equation}
t(\tau)=1/\beta\cosh(\beta\tau),\,\,x(\tau)=1/\beta\sinh(\beta\tau).
\end{equation}
Combining these two expressions we can find how an inertial observer would describe the trajectory of an accelerated observer. The result is
\begin{eqnarray}
\label{eqn:rindler-traj-inertial}
x(t)=\sqrt{\beta^{-2}-t^{2}}.
\end{eqnarray}
Thus, an inertial observer sees an accelerated observer on a \emph{hyperbolic} trajectory. As can be seen in Fig.~(\ref{fig:rindlercoords}), the Rindler trajectories are asymptotically bound by the lines $x=|t|$. These asymptotic bounds are known as the \emph{Rindler horizons}. They form a natural barrier for the Rindler observers such that an observer travelling in the right Rindler wedge is \emph{causally disconnected} from an observer travelling in the left wedge. This means a particle emitted by an observer in one wedge will never reach the observer in the other wedge. As a consequence, a Rindler observer in the right wedge \emph{cannot} transfer information to a Rindler observer in the left wedge. We can also relate the Minkowski coordinates to the Rindler coordinates for constant time slices $\eta=\eta_{1}$. This gives
\begin{eqnarray}
\label{eqn:rindler-time-inertial}
x(t)=\frac{t}{\tanh(\eta_{1})}.
\end{eqnarray}
From this equation, we see that different constant Rindler time slices foliate the spacetime in a unique way. That is to say, no two time slices cross at any point (other than the spacetime origin), see Fig.~(\ref{fig:rindlercoords}). The components of the particles acceleration vector can be written as
\begin{subequations}
\label{eqn:accn-components}
\begin{align}
\alpha^{0}=\frac{d^{2}}{d\tau^{2}}t(\tau)&=\beta\cosh(\beta\tau),\\
\alpha^{1}=\frac{d^{2}}{d\tau^{2}}x(\tau)&=\beta\sinh(\beta\tau).
\end{align}
\end{subequations}
The proper acceleration of a particle is defined as $|\alpha|=\sqrt{-g_{\mu\nu}\alpha^{\mu}\alpha^{\nu}}$. Using the acceleration components~(\ref{eqn:accn-components}) we find that the proper acceleration of a stationary Rindler observer is
\begin{eqnarray}
|\alpha |=\beta>0 .
\end{eqnarray}
Therefore, as expected, the proper acceleration for a Rindler observer is constant and equal to $\beta$. Finally, we identify the timelike Killing vector field which is associated with the Rindler observers. The required Killing vector field is
\begin{eqnarray}
K=\partial_{\eta}=x\partial_{t}+t\partial_{x}.
\end{eqnarray}
Thus, we expect the solutions to the Klein-Gordon field to be classified according to $K=\partial_{\eta}$. In the following, we show how to quantise the field in the right Rindler wedge only, as quantisation in the left wedge follows the exact same methodology. 

In Rindler coordinates, the Klein-Gordon equation~(\ref{eqn:covariant-kg-eqn}) takes the form
\begin{eqnarray}
\label{eqn:kg-equation-rindler}
-\partial_{\eta\eta}\phi +\left(\chi\partial_{\chi}\chi\partial_{\chi}+\chi^{2}\partial^{a}\partial_{a}\right)\phi -\chi^{2}m^{2}\phi =0,
\end{eqnarray}
for $a=2,3$ and the inner product~(\ref{eqn:kg-inner-product}) reads
\begin{eqnarray}
\label{eqn:kg-rindler-inner-product}
\left(\phi_{\boldsymbol{k}}^{R},\phi_{\boldsymbol{k}'}^{R}\right)_{R}&=& -i\int_{\Sigma} d\boldsymbol{\chi}\frac{\bar{\phi}_{\boldsymbol{k}}^{R}\partial_{\eta}\phi_{\boldsymbol{k}'}^{R}-\phi_{\boldsymbol{k}'}^{R}\partial_{\eta}\bar{\phi}_{\boldsymbol{k}}^{R}}{\chi},
\end{eqnarray}
and we choose $\Sigma$ to be the $\eta=0$ hypersurface. Note we have also used, in slight conflict with the Minkowski modes, the notation $\boldsymbol{k}=(\Omega,\boldsymbol{k}_{\perp})$ where $\Omega$ is a strictly positive parameter ($\Omega>0$) identified with a Rindler particle's energy and $\boldsymbol{k}_{\perp}\in\mathbb{R}^{2}$ denotes the momentum of a particle in all other spatial dimensions orthogonal to $\chi$. Following the Minkowski coordinate procedure, we define the solution to the Rindler coordinate Klein-Gordon equation in the right Rindler wedge to be 
\begin{eqnarray}
\label{eqn:kg-bessel-waves}
\phi(\eta,\boldsymbol{\chi})=\int d\boldsymbol{k}\left(A_{\boldsymbol{k}}\phi_{\boldsymbol{k}}^{R}(\eta,\boldsymbol{\chi})+\mathrm{c.c.}\right),
\end{eqnarray}
where
\begin{eqnarray}
\phi_{\boldsymbol{k}}^{R}(\eta,\boldsymbol{\chi}):=N_{\Omega}K_{i\Omega}\left(m\chi\right)e^{-i\Omega\eta +i\boldsymbol{k}_{\perp}\cdot\boldsymbol{y}_{\perp}},
\end{eqnarray}
with $N_{\Omega}=\sqrt{\frac{\sinh(\pi\Omega)}{4\pi^{2}}}$ and $K_{i\Omega}\left(m\chi\right)$ are modified Bessel functions of the second kind~\cite{NIST:DLMF}. Note that here the integration measure in~\eqref{eqn:kg-bessel-waves} is defined as $d\boldsymbol{k}:=d\Omega d\boldsymbol{k}_{\perp}$ and integration of $\Omega$ is in the interval $[0,\infty[$. As expected, we can separate the Rindler modes naturally via the timelike Killing vector field $K=\partial_{\eta}$. We find that the Rindler modes satisfy the eigenvalue equation
\begin{subequations}
\begin{align}
i\partial_{\eta}\phi_{\boldsymbol{k}}^{R}(\eta,\boldsymbol{\chi})&=+\Omega\phi_{\boldsymbol{k}}^{R}(\eta,\boldsymbol{\chi}),\\
i\partial_{\eta}\bar{\phi}_{\boldsymbol{k}}^{R}(\eta,\boldsymbol{\chi})&=-\Omega\bar{\phi}_{\boldsymbol{k}}^{R}(\eta,\boldsymbol{\chi}).
\end{align}
\end{subequations}
Thus, we again identify the modes $\phi_{\boldsymbol{k}}^{R}(\eta,\boldsymbol{\chi})$ with particles and the $\bar{\phi}_{\boldsymbol{k}}^{R}(\eta,\boldsymbol{\chi})$ with anti-particles. As in the Minkowski coordinate case, we impose equal time canonical commutation relations on the field and its conjugate momentum to arrive at the relations
\begin{eqnarray}
\left[\hat{A}_{\boldsymbol{k}},\hat{A}_{\boldsymbol{k}'}^{\dag}\right]=\delta(\Omega-\Omega')\delta(\boldsymbol{k}_{\perp}-\boldsymbol{k'}_{\perp}).
\end{eqnarray}
We can define the right Rindler wedge vacuum state as (where the subscript $r$ indicates the \emph{right} Rindler region only)
\begin{eqnarray}
\hat{A}_{\boldsymbol{k}}\ket{0}_{r}=0\,\,\forall\,\,\boldsymbol{k},
\end{eqnarray}
and from here we can define a Fock basis in which to describe our Rindler Fock space. As pointed out earlier, the Rindler coordinates only cover a portion of Minkowski spacetime. However, we can extend the coordinate definition to include the left Rindler wedge, see Fig.~(\ref{fig:rindlercoords}). Even though this still does not fully cover all of Minkowski spacetime, it can be shown that quantising the field in the left and right wedges and analytically extending the modes to the rest of Minkowski spacetime is sufficient to represent the Klein-Gordon field on the \emph{whole} of flat spacetime~\cite{unruh1976,takagi1986,crispino2008}. Therefore, the whole Klein-Gordon field quantised via the Rindler coordinates is
\begin{eqnarray}
\hat{\phi}(\eta,\boldsymbol{\chi})=\sum_{\epsilon=r,l}\int d\boldsymbol{k} \hat{A}_{\boldsymbol{k},\epsilon}\phi_{\boldsymbol{k},\epsilon}^{R}(\eta,\boldsymbol{\chi})+\mathrm{h.c.}
\end{eqnarray}
where we have denoted the \emph{right} Rindler modes and operators with $\epsilon=r$ and \emph{left} Rindler modes with $\epsilon=l$. Finally, we note that the \emph{full} Rindler Klein-Gordon vacuum state is defined as
\begin{eqnarray}
\hat{A}_{\boldsymbol{k},\epsilon}\ket{0}_{r}\otimes\ket{0}_{l}=\hat{A}_{\boldsymbol{k},\epsilon}\ket{0}_{R}=0\,\forall\,\epsilon,\,\boldsymbol{k},
\end{eqnarray}
where we have defined $\ket{0}_{r}\otimes\ket{0}_{l}=\ket{0}_{R}$. We have seen two examples of observers who travel along different timelike Killing vector fields, i.e. inertial observers who follow trajectories generated by $\partial_{t}$ and accelerated observers who follow trajectories generated by $\partial_{\eta}$. As mentioned before, we intend to use these two types of observer to model the motion of quantum states through spacetime. A natural question to ask is how two sets of field solutions are related. Expressing field modes and operators in terms of another set of solutions allows us to transform a state of one observer to another. These transformations are known as \emph{Bogoliubov transformations}.

\section{Bogoliubov Transformations}\label{sec:bogo-qft}

Quite simply, a Bogoliubov transformation is a change of basis from one set of mode solutions to another that preserves the commutation (or anti-commutation) relations of the field operators. In other words, it is a unitary transformation from one field representation to another. Here we illustrate the process of computing Bogoliubov transformations for a real Klein-Gordon field. For simplicity and convenience when presenting our results later, we shall work with mode solutions which have a discrete spectrum. To obtain the continuous case, one needs to replace the summations with integrals and the appropriate measure.

We have seen two different, but equivalent, quantisations of the real Klein-Gordon field~\cite{unruh1976,crispino2008}. One was done in the usual Minkowski coordinates and the other in the Rindler coordinates.  For simplicity, we shall work in the $(1+1)$-dimensional case. However, all concepts can be extended to higher dimensional spacetimes naturally.

Abstractly, we can write the two field expansions as
\begin{subequations}
\label{eqn:kg-bogo}
\begin{align}
\hat{\phi}(t,x)&=\sum_{k}\left[\hat{a}_{k}\phi_{k}^{M}(t,x)+\hat{a}_{k}^{\dag}\bar{\phi}_{k}^{M}(t,x)\right],\\
\hat{\phi}(\eta,\chi)&=\sum_{k,\epsilon}\left[\hat{A}_{k,\epsilon}\phi_{k,\epsilon}^{R}(\eta,\chi)+\hat{A}_{k,\epsilon}^{\dag}\bar{\phi}_{k,\epsilon}^{R}(\eta,\chi)\right],
\end{align}
\end{subequations}
where the notation is as in the previous sections. Following~\cite{birrell1982,crispino2008,takagi1986}, we can use the completeness of both sets to write
\begin{subequations}
\label{eqn:kg-bogo-ansatz-1}
\begin{align}
\phi_{k',\epsilon}^{R}(t,x)&=\sum_{k}\left[\alpha_{k'k}^{\epsilon}\phi_{k}^{M}(t,x)+\beta_{k'k}^{\epsilon}\bar{\phi}_{k}^{M}(t,x)\right],\\
\bar{\phi}_{k',\epsilon}^{R}(t,x)&=\sum_{k}\left[\bar{\alpha}_{k'k}^{\epsilon}\bar{\phi}_{k}^{M}(t,x)+\bar{\beta}_{k'k}^{\epsilon}\phi_{k}^{M}(t,x)\right],
\end{align}
\end{subequations}
where the modes $\phi_{k',\epsilon}^{R}$ have been written in terms of the standard Minkowski modes and coordinates. The coefficients $\alpha_{k'k}^{\epsilon}$ and $\beta_{k'k}^{\epsilon}$ are known as \emph{Bogoliubov coefficients}. Using the orthonormality of the $\phi_{k}^{M}(t,x)$ modes~(\ref{eqn:kg-mode-normalisations1}), we can take Klein-Gordon field inner products of~(\ref{eqn:kg-bogo-ansatz-1}) to obtain~\cite{birrell1982}
\begin{subequations}
\label{eqn:kg-bogo-def}
\begin{align}
\alpha_{k'k}^{\epsilon}&:=\left(\phi_{k',\epsilon}^{R}(t,x),\phi_{k}^{M}(t,x)\right)\left|_{t=0}\right. ,\\
\beta_{k'k}^{\epsilon}&:=-\left(\phi_{k',\epsilon}^{R}(t,x),\bar{\phi}_{k}^{M}(t,x)\right)\left|_{t=0}\right. ,
\end{align}
\end{subequations}
where, as the notation suggests, the inner product, in Minkowski coordinates, is taken over the $t=0$ hypersurface for convenience. From now on, we drop the $\epsilon$ index as all considerations for the left and right Rindler wedges are the same.

It is worth mentioning the overall structure and representation of the Bogoliubov transformations. From~(\ref{eqn:kg-bogo-ansatz-1}) we can collect the modes $\phi_{k'}^{R},\bar{\phi}_{k'}^{R}$ and $\phi_{k'}^{M},\bar{\phi}_{k'}^{M}$ into column vectors and represent the coefficients~(\ref{eqn:kg-bogo-def}) as matrices such that we can write~(\ref{eqn:kg-bogo}) as
\begin{eqnarray}
\label{eqn:kg-bogo-ansatz-matrix}
\left(\begin{array}{cc}\boldsymbol{\phi}^{R}\\ \bar{\boldsymbol{\phi}}^{R}\end{array}\right)=\left(\begin{array}{cc}\boldsymbol{\alpha} & \boldsymbol{\beta}\\ \bar{\boldsymbol{\beta}} & \bar{\boldsymbol{\alpha}}\end{array}\right)\left(\begin{array}{cc}\boldsymbol{\phi}^{M}\\ \bar{\boldsymbol{\phi}}^{M}\end{array}\right).
\end{eqnarray}
Further, the corresponding transformation for the mode operators goes as
\begin{subequations}
\label{eqn:kg-bogo-ops}
\begin{align}
\label{eqn:kg-bogo-annih}\hat{A}_{k'}&=\sum_{k}\left[\bar{\alpha}_{k'k} \hat{a}_{k}-\bar{\beta}_{k'k} \hat{a}_{k}^{\dag}\right],\\
\hat{A}^{\dag}_{k'}&=\sum_{k}\left[\alpha_{k'k} \hat{a}_{k}^{\dag}-\beta_{k'k} \hat{a}_{k}\right],
\end{align}
\end{subequations}
which has the matrix form
\begin{eqnarray}
\label{eqn:kg-bogo-ops-matrix}
\left(\begin{array}{cc}\hat{\boldsymbol{A}}\\ \hat{\bar{\boldsymbol{A}}}\end{array}\right)=\left(\begin{array}{cc}\bar{\boldsymbol{\alpha}} & -\bar{\boldsymbol{\beta}}\\ -\boldsymbol{\beta} & \boldsymbol{\alpha}\end{array}\right)\left(\begin{array}{cc}\hat{\boldsymbol{a}}\\ \hat{\bar{\boldsymbol{a}}}\end{array}\right).
\end{eqnarray}
As the Bogoliubov transformation implements a unitary transformation of the field operators it must preserve the commutation relations of the field. In the case of the Klein-Gordon field this amounts to the two \emph{Bogoliubov identities}
\begin{subequations}
\label{eqn:kg-bogo-identities}
\begin{align}
\sum_{n}\left[\alpha_{np}\bar{\alpha}_{nq}-\beta_{np}\bar{\beta}_{nq}\right]&=\delta_{pq},\\
\sum_{n}\left[\alpha_{np}\beta_{nq}-\beta_{np}\alpha_{nq}\right]&=0.
\end{align}
\end{subequations}
These identities can be written in matrix form as
\begin{subequations}
\begin{align}
\boldsymbol{\alpha}\boldsymbol{\alpha}^{\dag}-\boldsymbol{\beta}\boldsymbol{\beta}^{\dag}&=\boldsymbol{I},\\
\boldsymbol{\alpha}\boldsymbol{\beta}^{\text{tp}}&=\left(\boldsymbol{\alpha}\boldsymbol{\beta}^{\text{tp}}\right)^{\text{tp}}.
\end{align}
\end{subequations}
However, it should be noted that the analysis is not restricted to real Klein-Gordon fields. We can perform the same type of expansion using Dirac fields for, say, Minkowski and Rindler modes and compute the Bogoliubov coefficients. However, we postpone theses computations until Chapter~(\ref{chapter:moving-cavities}) where we analyse Fermionic cavity entanglement in depth.

It is interesting to note that in the special case $\beta_{kk'}=0$, the transformation of the annihilation operator in Eq.~(\ref{eqn:kg-bogo-annih}) only contains other annihilation operators. That is to say, if the Bogoliubov transformation takes annihilation operators to annihilation operators (and of course creation operators to creation operators) then they define the same vacuum. This can be expressed as
\begin{eqnarray}
\hat{a}_{k}\ket{0}=\hat{A}_{k}\ket{0}=0.
\end{eqnarray}
Thus the particle content of the vacuum is unchanged. However, in the more general case $\beta_{kk'}\not= 0$ the two vacua do not coincide and the two observers describe a \emph{different} particle content. This is exactly the case for inertial and accelerated observers and is the foundation of the Unruh effect~\cite{unruh1976}.

\subsection{Unruh Effect}

We now give a brief discussion of the original Unruh effect~\cite{unruh1976}. For a very thorough and pedagogical derivation, the reader is encouraged to read the seminal paper of Takagi~[\cite{takagi1986}, Section (2.8)]. In short, the Unruh effect says: an accelerated observer views the Minkowski vacuum as a thermal state with temperature proportional to their acceleration.

We restrict ourselves to the $(1+1)$ Klein-Gordon field. Our starting point for the Unruh effect is the operator transformations~(\ref{eqn:kg-bogo-ops}) which we state again for clarity
\begin{eqnarray*}
\hat{A}_{k',\epsilon}&=&\sum_{k}\left[\bar{\alpha}_{k'k}^{\epsilon} \hat{a}_{k}-\bar{\beta}_{k'k}^{\epsilon} \hat{a}_{k}^{\dag}\right],\\
\hat{A}^{\dag}_{k',\epsilon}&=&\sum_{k}\left[\alpha_{k'k}^{\epsilon} \hat{a}_{k}^{\dag}-\beta_{k'k}^{\epsilon} \hat{a}_{k}\right].
\end{eqnarray*}
A reasonable question to ask is how are the vacua of the two modes related? It turns out that, because of the linearity of the Bogoliubov transformation, the preservation of the commutation relations between Minkowski and Rindler mode operators and the normalisation of the Minkowski vacuum, we can write the result~\cite{takagi1986}
\begin{eqnarray}
\ket{0}_{M}=\bigotimes_{m}N_{m}e^{r_{m}\hat{A}^{\dag}_{m,r}\hat{A}^{\dag}_{m,l}}\ket{0}_{R},
\end{eqnarray}
where $\ket{0}_{M}$ is the Minkowski vacuum state, $\ket{0}_{R}=\ket{0}_{r}\otimes\ket{0}_{l}$ is the Rindler vacuum state, $N_{m}=\sqrt{1-r_{m}^{2}}$, $r_{m}=e^{-\pi\Omega_{m}}$, $\Omega_{m}$ is a dimensionless Rindler frequency, $\hat{A}^{\dag}_{m,r}/\hat{A}^{\dag}_{m,l}$ are the Rindler creation operators for right/left Rindler wedges and $m$ labels the mode of the field. This complicated expression can be more simply understood when we restrict our view to a single Rindler mode~$m$. Replacing the exponential map with its power series definition we obtain
\begin{eqnarray}
\label{eqn:unruh-state}
N_{m}e^{r_{m}\hat{A}^{\dag}_{m,r}\hat{A}^{\dag}_{m,l}}\ket{0}_{R}=\sqrt{1-e^{-2\pi\Omega_{m}}}\sum_{n=0}^{\infty}e^{-n\pi\Omega_{m}}\ket{n_{m},n_{m}}_{R},
\end{eqnarray}
where $\ket{n_{m},n_{m}}_{R}:=(1/n!)(\hat{A}^{\dag}_{m,r})^{n}(\hat{A}^{\dag}_{m,l})^{n}\ket{0}_{R}$. The result is that the Minkowski vacuum state, when expressed in terms of the Rindler modes, is composed of a superposition of right and left Rindler particles. By inspection, we can see that the state~(\ref{eqn:unruh-state}) cannot be factored into a product between a state in the right Rindler region and a state in the left Rindler region. It is, therefore, an entangled state between the right and left Rindler modes. However, an observer in the right Rindler wedge cannot access the modes in the left Rindler wedge, due to the Rindler horizons. Therefore, the observer in the right Rindler wedge needs to trace over the left Rindler modes leaving the reduced state
\begin{eqnarray}
\label{eqn:thermal-state}
\hat{\rho}_{m,r}=(1-e^{-2\pi\Omega_{m}})\sum_{n}e^{-2n\pi\Omega_{m}}\ket{n_{m}}_{rr\!\!}\bra{n_{m}}.
\end{eqnarray}
where $\ket{n_{m}}_{rr\!\!}\bra{n_{m}}$ are the Fock states in the right Rindler wedge only. The state~(\ref{eqn:thermal-state}) has the form of a canonical thermal state but there is no direct link to a temperature. As a heuristic step, we consider the case where the dimensionless Rindler coordinate time is parametrised via the proper time of an observer with proper acceleration $\beta$. From Eq.~(\ref{eqn:rindler-parametrisation}), we can write the Rindler coordinate time as
\begin{eqnarray}
\eta(\tau)=\beta\tau ,
\end{eqnarray}
where $\tau$ is the proper time of the accelerate observer. When inserted into the phase of a Rindler mode, we find
\begin{eqnarray}
\eta\Omega_{m} &=& \beta\tau\Omega_{m} .
\end{eqnarray}
From here, we can identify the ``physical" energy of a Rindler particle as
\begin{eqnarray}
E_{m}=\beta\Omega_{m} .
\end{eqnarray}
Next, we can relate the dimensionless frequency $\Omega_{m}$ to the temperature as measured by an observer travelling along the timelike Killing vector $K=\partial_{\eta}$. The relation between the Killing vector and the temperature at a point is given by the Ehrenfest-Tolman relation~\cite{rovelli2010},
\begin{eqnarray}
\label{eqn:Ehrenfest-Tolman}
T\sqrt{-g_{\mu\nu}K^{\mu}K^{\nu}}=\mathrm{const},
\end{eqnarray}
where $T$ is the temperature at a spacetime point and $K^{\mu}$ are the components of the timelike Killing vector an observer is flowing along. For an observer \emph{fixed} at the Rindler position $\chi=1/\beta$, the Ehrenfest-Tolman relation~(\ref{eqn:Ehrenfest-Tolman}) reduces to
\begin{eqnarray}
\frac{T}{\beta}=\mathrm{const.}
\end{eqnarray}
Thus, choosing the constant of proportionality to be $\mathrm{const}=1/2\pi$, we can identify Unruh temperature as
\begin{equation}
\label{eqn:unruh-temp}
T=\beta/2\pi ,
\end{equation}
which finally gives us the relation
\begin{equation}
2\pi\Omega_{m}=E_{m}/T.
\end{equation}
Remarkably, this gives the state~(\ref{eqn:thermal-state}) the exact form of a canonical thermal state~\cite{unruh1976,takagi1986}. Thus, the Minkowski vacuum is an entangled state which, when restricted to the right Rindler wedge, is a thermal state whose temperature is given by~(\ref{eqn:unruh-temp}). This is the Unruh effect. This result of the left and right Rindler wedges sharing entanglement is one of the original inspirations for the field of relativistic quantum information. However, this result was based entirely on the well defined concept of particles for Minkowski and Rindler observers. As we have mentioned before, such a concept is not guaranteed to exist in a general spacetime. This is because timelike Killing vector fields do not exist a generic spacetime. This leads one to ask if there is another way of talking about the particle content of a field as described by different observers. 

An operational definition of what a particle is was first put forward by Unruh~\cite{unruh1976}. This operational definition used what is now known as a ``particle detector". This was meant in the sense that a particle detector is a quantum mechanical object which responds to the presence of excitations of a field. This model came to be known as the Unruh-DeWitt detector. Unruh-DeWitt detectors will be the foundations of the results in Chapter~(\ref{chapter:fat-detectors}) and will be introduced in Section~(\ref{sec:qft-unruh-detector}). However, we first finish our introduction of the quantum fields that will be used in Chapter~(\ref{chapter:moving-cavities}). We therefore review the Dirac field.

\section{Dirac Field}\label{sec:dirac-field}

The Dirac equation, as mentioned earlier, describes a field of spin-$1/2$ particles in a relativistic manner. It merges special relativity with quantum mechanics for Fermions, such as electrons and neutrinos, and correctly accounts for their spin. As we will be analysing entanglement between Fermionic modes of a cavity in Chapter~(\ref{chapter:moving-cavities}), we shall briefly review the theory for free fields.

\subsection{Field Equation Definition}

In the following analysis, we restrict ourselves to the $(1+1)$-dimensional case. The Dirac equation in its covariant form reads~\cite{crispino2008}
\begin{eqnarray}
\label{eqn:diracequation}
-i\boldsymbol{\gamma}^{\mu}_{R}\nabla_{\mu}\psi +m\psi=0,
\end{eqnarray}
where $\boldsymbol{\gamma}_{R}^{\mu}=e_{\alpha}^{\mu}\boldsymbol{\gamma}^{\alpha}$ are the curved spacetime Dirac matrices, $\nabla_{\mu}=\partial_{\mu}+\Gamma_{\mu}$ is the covariant derivative, $\Gamma_{\mu}=1/8 [\boldsymbol{\gamma}^{\alpha},\boldsymbol{\gamma}^{\alpha}]e_{\alpha}^{\lambda}\nabla_{\mu}e_{\beta\lambda}$ are the spin-connections and $e_{\alpha}^{\lambda}$ are frame fields of the spacetime at hand. The matrices $\boldsymbol{\gamma}^{\mu}$ are $4\times 4$ and satisfy
\begin{eqnarray}
\label{eqn:cliffordalgebra}
\lbrace\boldsymbol{\gamma}^{\mu},\boldsymbol{\gamma}^{\nu}\rbrace =2\eta^{\mu\nu}\boldsymbol{I},
\end{eqnarray}
where $\lbrace\cdot,\cdot\rbrace$ is the anti-commutator of two matrices, $\eta^{\mu\nu}$ is the Minkowski metric tensor and $m$ denotes the bare bass of the field. The solutions to Eq.~(\ref{eqn:diracequation}) are naturally four component vectors, known as \emph{spinors}. The conserved inner product, given two solutions to the Dirac equation $\psi_{k}$, $\psi_{k'}$, in its covariant form reads~\cite{boulanger2006,crispino2008}
\begin{eqnarray}
\label{eqn:dirac-inner-product}
\left(\psi_{k},\psi_{k'}\right):=
\int_{\Sigma}d\Sigma_{\mu}\psi_{k}^{\dag}\,\boldsymbol{\gamma}^{0}_{R}\boldsymbol{\gamma}^{\mu}_{R}\,\psi_{k'}
\end{eqnarray}
where $\Sigma$ is a constant time hypersurface and $d\Sigma_{\mu}$ is an appropriate spacetime measure we are integrating with. In much the same way as the Klein-Gordon field, we canonically quantise the Dirac field in both Minkowski and Rindler coordinates.

\subsection{Minkowski Quantisation}

It should first be noted that the choice of the matrices $\boldsymbol{\gamma}^{\mu}$ is not unique, i.e. we can pick any set of matrices which satisfy the algebra relations~(\ref{eqn:cliffordalgebra}). Thus we choose whatever representation is useful for our purposes. We define the $\boldsymbol{\gamma}^{\mu}$ to be
\begin{eqnarray}
\boldsymbol{\gamma}^{0}&=&
\begin{pmatrix}
  \boldsymbol{I} & \boldsymbol{0}  \\
  \boldsymbol{0} & -\boldsymbol{I}
 \end{pmatrix},\\
\boldsymbol{\gamma}^{k}&=&
\begin{pmatrix}
  \boldsymbol{0} & \boldsymbol{\sigma}_{k}  \\
  -\boldsymbol{\sigma}_{k} & \boldsymbol{0}
 \end{pmatrix},
\end{eqnarray}
where $\boldsymbol{\sigma}_{k}$ are the Pauli matrices. The solutions to the Dirac equation are
\begin{eqnarray}
\label{eqn:dirac-mink-basis}
\psi^{\pm}_{k,\sigma}(t,z)=\frac{1}{\sqrt{4\pi\omega}}e^{\mp i\omega t+ikz}\left[\begin{array}{c}
\pm\sqrt{\omega\pm m}\Lambda_{\sigma} \\ \frac{\sigma k}{\sqrt{\omega\pm m}}\Lambda_{\sigma}
\end{array}\right],
\end{eqnarray}
where we have defined $\Lambda_{+}=(1,0)$ and $\Lambda_{-}=(0,1)$ and $\sigma=+,-$ denotes the \emph{spin} of a particle. Notice that, as was expected, the Minkowski basis~(\ref{eqn:dirac-mink-basis}) satisfies
\begin{eqnarray}
\mp i\partial_{t}\psi^{\pm}_{k,\sigma}=\pm\omega\psi^{\pm}_{k,\sigma}.
\end{eqnarray}
This again allows us to split up our field equation solutions into their positive and negative frequency parts allowing for quantisation. The normalisation of the modes has been computed via the Minkowski inner product~(\ref{eqn:dirac-inner-product})
\begin{eqnarray}
\left(\psi_{\omega,\sigma},\psi_{\omega',\sigma'}\right)=\int_{\mathbb{R}}dz\,\psi_{\omega,\sigma}^{\dag}\boldsymbol{\gamma}^{0}\boldsymbol{\gamma}^{3}\psi_{\omega',\sigma'},
\end{eqnarray}
on the $t=0$ hypersurface. Further, $\omega>0$ is the Minkowski Dirac particle energy which satisfies the mass-energy relation
\begin{equation}
\omega^{2}=k^{2}+m^{2}.
\end{equation}
We expand the full solution to the Dirac field as~\cite{crispino2008}
\begin{eqnarray}
\psi(t,z)=\sum_{\sigma=\pm}\int dk\left(b_{k,\sigma}\psi_{k,\sigma}^{+}(t,z)+\bar{d}_{k,\sigma}\psi_{-k,-\sigma}^{-}(t,z)\right),
\end{eqnarray}
where $b_{k,\sigma}$ and $d_{k,\sigma}$ are arbitrary complex functions. Quantisation is achieved by imposing equal time \emph{anti-commutation} relations (CAR's) on the field and its Hermitian conjugate. The CAR's are imposed component by component on the field such that
\begin{eqnarray}
\label{eqn:minkowski-car}
\left\lbrace \hat{\psi}_{\alpha}(t,z),\hat{\psi}_{\beta}^{\dag}(t,z')\right\rbrace=\delta_{\alpha\beta}\delta(z-z'),
\end{eqnarray}
with all other anti-commutators vanishing. Note that in the above equation, $\psi_{\alpha}$ denotes the $\alpha$ component of the \emph{full} Dirac field and not a mode solution. This quantisation then implies the anti-commutation relations for the mode operators
\begin{subequations}
\begin{align}
\left\lbrace \hat{b}_{k,\sigma},\hat{b}_{k',\sigma'}^{\dag}\right\rbrace &= \delta(k-k')\delta_{\sigma\sigma'},\\
\left\lbrace \hat{d}_{k,\sigma},\hat{d}_{k',\sigma'}^{\dag}\right\rbrace &= \delta(k-k')\delta_{\sigma\sigma'},
\end{align}
\end{subequations}
with all other anti-commutators vanishing. After quantisation, we associate the modes $\psi_{k,\sigma}^{+}$ with \emph{particles} and $\psi_{k,\sigma}^{-}$ with \emph{anti-particles}. The Minkowski vacuum state of the Dirac field is defined by
\begin{eqnarray}
\hat{b}_{k,\sigma}\ket{0}_{M}=\hat{d}_{k,\sigma}\ket{0}_{M}=0\,\,\forall\,\, k,\sigma,
\end{eqnarray}
with Fock basis elements are
\begin{subequations}
\label{eqn:dirac-fock-basis}
\begin{align}
\ket{1_{k,\sigma}}^{+}&:=\hat{b}_{k,\sigma}^{\dag}\ket{0}_{M},\\
\ket{1_{k,\sigma}}^{-}&:=\hat{d}_{k,\sigma}^{\dag}\ket{0}_{M},
\end{align}
\end{subequations}
where the $\pm$ superscript denotes particle/ anti-particle states. This basis defines another Fock space but due to the anti-commutation relations no more than one particle, of momentum $k$ with spin $\sigma$, can occupy the state at any one time. This Dirac field Fock space is called the \emph{anti-symmetrised} Fock space. Next is the standard procedure for quantising the Dirac field in Rindler coordinates.

\subsection{Rindler Quantisation}

To quantise the free Dirac field we follow~\cite{crispino2008}. In Rindler coordinates $(\eta,\chi)$, the massive Dirac equation reads
\begin{eqnarray}
\label{eqn:rindler-dirac-equation}
i\partial_{\eta}\psi_{\Omega,\sigma}=\left(\boldsymbol{\gamma}^{0}m\chi-i\boldsymbol{\alpha}_{3}/2-i\boldsymbol{\alpha}_{3}\chi\partial_{\chi}\right)\psi_{\Omega,\sigma},
\end{eqnarray}
where $\boldsymbol{\alpha}_{j}\equiv\boldsymbol{\gamma}^{0}\boldsymbol{\gamma}^{j}$, $\Omega>0$ is a particle's energy and $\sigma$ denotes spin. Note that the parameter $\Omega$ is a \emph{dimensionless} energy. This is due to the definition of the Rindler coordinates~(\ref{eqn:rindler-coord-def}). To relate it to the \emph{physical} energy of a particle, one must take into account the proper acceleration of a Rindler observer. We will see this when we review the Unruh effect in section~(\ref{sec:qft-unruh-detector}). Continuing, in our particular representation of the Dirac matrices, the normalised mode solutions read~\cite{crispino2008}
\begin{eqnarray}
\label{eqn:dirac-free-modes}
\psi_{\Omega,\sigma}=N_{\Omega}e^{-i\Omega\eta}\left(
\mathrm{K}_{i\Omega +\sigma/2}(m\chi)\left[\begin{array}{c}\Lambda_{\sigma} \\ -\sigma\Lambda_{\sigma}\end{array}\right]
+i\mathrm{K}_{i\Omega -\sigma/2}(m\chi)\left[\begin{array}{c}\Lambda_{\sigma} \\ \sigma\Lambda_{\sigma}\end{array}\right]
\right),
\end{eqnarray}
where $N_{\Omega}=\sqrt{\frac{m\cosh(\pi\Omega)}{2\pi^{2}}}$ and the $\Lambda_{\pm}$ are as in the previous section. Again, the Rindler modes satisfy the equation
\begin{subequations}
\begin{align}
-i\partial_{\eta}\psi_{\pm\Omega,\sigma}&=\pm\Omega\psi_{\pm\Omega,\sigma}.
\end{align}
\end{subequations}
In the right Rindler wedge the inner product reads
\begin{eqnarray}
\left(\psi_{\Omega,\sigma},\psi_{\Omega',\sigma'}\right)=\int_{\mathbb{R}^{+}}d\chi\,\psi_{\Omega,\sigma}^{\dag}\boldsymbol{\gamma}^{0}\boldsymbol{\gamma}^{3}\psi_{\Omega',\sigma'},
\end{eqnarray}
where integration is over $\chi>0$ due to the definition of the right Rindler wedge coordinates. We express the full solution to the Dirac equation in terms of the modes~(\ref{eqn:dirac-free-modes}) and impose anti-commutation relations. As in the Minkowski case, this is imposed component wise for the Dirac field as
\begin{eqnarray}
\label{eqn:rindler-car}
\left\lbrace \psi_{\alpha}(\eta,\chi),\psi_{\beta}^{\dag}(\eta,\chi')\right\rbrace=\delta_{\alpha\beta}\delta(\chi-\chi').
\end{eqnarray}
This is equivalent to imposing the following CAR's for the mode operators
\begin{subequations}
\begin{align}
\left\lbrace \hat{b}_{\Omega,\sigma},\hat{b}_{\Omega',\sigma'}^{\dag} \right\rbrace &= \delta(\Omega-\Omega')\delta_{\sigma\sigma'},\\
\left\lbrace \hat{d}_{\Omega,\sigma},\hat{d}_{\Omega',\sigma'}^{\dag} \right\rbrace &= \delta(\Omega-\Omega')\delta_{\sigma\sigma'},
\end{align}
\end{subequations}
where $\lbrace \hat{b}_{\Omega,\sigma},\hat{b}_{\Omega,\sigma}^{\dag}\rbrace$ are the annihilation/creation operators for particle excitations and $\lbrace \hat{d}_{\Omega,\sigma},\hat{d}_{\Omega,\sigma}^{\dag}\rbrace$ are the annihilation/creation operators for anti-particle excitations. The vacuum state associated with the Rindler Dirac field is defined as
\begin{eqnarray}
\hat{b}_{\Omega,\sigma}\ket{0}_{R}=\hat{d}_{\Omega,\sigma}\ket{0}_{R}=0\,\,\forall\,\,\Omega,\,\sigma,
\end{eqnarray}
and a Fock basis is constructed as in the previous subsections, see~(\ref{eqn:dirac-fock-basis}). Finally, we expand the field in the right Rindler wedge as
\begin{eqnarray}
\hat{\psi}(\eta ,\chi)=\sum_{\sigma=\pm}\int_{0}^{\infty}d\Omega \left(\hat{b}_{\Omega,\sigma}\psi_{\Omega,\sigma}(\eta ,\chi)+\hat{d}_{\Omega,\sigma}^{\dag}\psi_{-\Omega,-\sigma}(\eta ,\chi)\right)
\end{eqnarray}
A similar procedure can be done to quantise the Dirac field in the left Rindler wedge. We can therefore expand the whole flat spacetime Dirac field in terms of the left and right Rindler modes~\cite{unruh1976,crispino2008}. This concludes our discussion of how canonical field quantisation is implemented. As we have seen, the notion of particle, and a constant definition of such a concept, relied heavily on the ability to identify timelike Killing vector fields. We now move to an operational method of investigating an observer's perception of the field around them: the Unruh-DeWitt particle detector model.

\section{Unruh-DeWitt Detector}\label{sec:qft-unruh-detector}

\subsection{Unruh-DeWitt detector: Definition}

We end the quantum field theory chapter by introducing a model used extensively in curved spacetimes: the Unruh-DeWitt detector model. It corresponds to a point like object whose internal degrees of freedom interact with a quantum field. Examples of physical realisations are electric dipoles interacting with an electromagnetic field~\cite{haffner2008,miller2005,guzman2006}, flux qubits in circuit QED~\cite{peropadre2010,gambetta2011} and trapped ions in optical lattices~\cite{thompson1990}. They provide an operational definition for particles in scenarios where a clear definition of particle does not exist, such as spacetimes without global timelike Killing vectors. They are therefore useful for analysing how different types of observer motion can affect the observation of quantum states. We follow \cite{unruh1976,DeWitt1979,schlicht2004} by defining the simplest case of Unruh-DeWitt detector, a monopole operator interacting with a real Klein-Gordon field.

We work in the interaction picture, see Eq.~(\ref{eqn:schwinger}). The Unruh-DeWitt detector model is defined via the action
\begin{equation}
\label{eqn:takagi-action}
S=\int dV\mathcal{M}(t,\boldsymbol{x})\phi(t,\boldsymbol{x}),
\end{equation}
where $dV$ is the spactime manifold volume element written in terms of the \emph{general} coordinates $(t,\boldsymbol{x})$. Here, $\mathcal{M}(t,\boldsymbol{x})$ is a two-level quantum system that describes the Unruh-DeWitt detector and $\phi(t,\boldsymbol{x})$ is the real Klein-Gordon field. We introduce a set of coordinates which are the natural coordinates used by an observer comoving with the detector, denoted by $(\tau,\boldsymbol{\zeta})$. Such coordinates are adequately described by the so-called \emph{Fermi-Walker coordinates}~\cite{misner1973}. Essentially, Fermi-Walker coordinates describe the spatial dimensions (i.e. those which are orthogonal to the temporal dimension) at each instant along a pointlike trajectory. The definitions for the Fermi-Walker coordinates can be found in ``Gravitation"~\cite{misner1973} however a very nice discussion can be found in~\cite{schlicht2004}. Consequently, we write the Unruh-DeWitt detector action~(\ref{eqn:takagi-action}) as
\begin{eqnarray}
S=\int d\tau d\boldsymbol{\zeta}\sqrt{-g(x[\tau,\boldsymbol{\zeta}])}\mathcal{M}(x[\tau,\boldsymbol{\zeta}])\phi(x[\tau,\boldsymbol{\zeta}]),
\end{eqnarray}
where we have parametrised the coordinates $(t,\boldsymbol{x})$ via $(\tau,\boldsymbol{\zeta})$. In the scenarios we will consider, we
assume the detector can be written in the form
\begin{eqnarray}
\mathcal{M}(x[\tau,\boldsymbol{\zeta}])=f(\tau,\boldsymbol{\zeta})M(\tau),
\end{eqnarray}
where $M(\tau)$ describes the internal quantum degrees of freedom of the detector via
\begin{eqnarray}
\label{eqn:internal-degrees-of-freedom}
M(\tau):=\sigma_{-}e^{-i\tau\Delta}+\sigma_{+}e^{+i\tau\Delta},
\end{eqnarray}
and $f(\tau,\boldsymbol{\xi})$ is known as the detector's \emph{spatial profile}. Here $\sigma_{+}$ and $\sigma_{-}$ denote the raising and lowering operators of the detector's internal energy state and $\Delta$ denotes the internal energy gap between the detector's two states. The Hamiltonian for the system is given by
\begin{eqnarray}
\label{eqn:takagi-lagrangian}
H:=\int d\boldsymbol{\zeta}\sqrt{-g(x[\tau,\boldsymbol{\zeta}])}f(\tau,\boldsymbol{\zeta})M(\tau)\cdot\phi(x[\tau,\boldsymbol{\zeta}]).
\end{eqnarray}
Noting the detectors' internal degrees of freedom~(\ref{eqn:internal-degrees-of-freedom}) are spatially independent, we can write the Hamiltonian as
\begin{eqnarray}
H=M(\tau)\cdot\tilde{\phi}(\tau),
\end{eqnarray}
where we have now defined the \emph{smeared field operator}
\begin{eqnarray}
\label{eqn:smeared-field-operator}
\tilde{\phi}(\tau):=\int d\boldsymbol{\zeta}\sqrt{-g(x[\tau,\boldsymbol{\zeta}])}f(\tau,\boldsymbol{\zeta})\phi(x[\tau,\boldsymbol{\zeta}]).
\end{eqnarray}
Physically, the smeared field operator corresponds to the \emph{effective} field a detector couples to. In particular, the detector will not couple to all modes in a uniform manner but will interact with each mode with a different strength. This object will be the main focus of our investigations in Chapter~(\ref{chapter:fat-detectors}), Sections~(\ref{sec:free-detector})~and~(\ref{sec:time-ordering}). Having identified the smeared field operator for Unruh-DeWitt detectors, we shall define the usual quantity associated with a detector moving through spacetime: its \emph{transition rate}.

\subsection{Unruh-DeWitt detector: Transition Rate}

The standard quantity to analyse when using the Unruh-DeWitt detector is its \emph{transition rate}. This is due, as we will see, to its explicit, non-trivial, dependence on the state of the field and the trajectory of the detector. Following~\cite{schlicht2004}, we define it as the probability per unit time the detector will undergo a transition from one state to another when probing the \emph{Minkowski} vacuum. To find the exact form of the transition rate, we need to solve the full dynamics of the field-detector interaction governed by the Hamiltonian~(\ref{eqn:takagi-lagrangian}). This is usually a formidable task, thus we turn to other methods of calculating quantum field theory quantities. Most works use perturbation theory and calculate the first order contribution. Thus, working to first order in perturbation theory, we define the transition rate of the detector probing the Minkowski vacuum as~\cite{schlicht2004}
\begin{eqnarray}
\label{eqn:transition-rate}
\dot{F}_{\tau}(\Delta):=2\int ds\, \mathrm{Re}\left[e^{-i\tau\Delta}W(\tau,\tau-s)\right],
\end{eqnarray}
where we have defined the Wightman function
\begin{eqnarray}
W(\tau,\tau'):=\bra{0}\tilde{\phi}(\tau)\tilde{\phi}(\tau')\ket{0},
\end{eqnarray}
with $\ket{0}$ being the Klein-Gordon Minkowski vacuum and $\tilde{\phi}(\tau)$ the smeared field operator~(\ref{eqn:smeared-field-operator}). For a derivation of this expression see~\cite{schlicht2004}. From the transition rate~(\ref{eqn:transition-rate}), we can compute how an Unruh-DeWitt detector perceives the Minkowski vacuum while undergoing different forms of motion. For example, a point-like detector, whose spatial profile is $f(\tau,\boldsymbol{\xi})=\delta(\boldsymbol{\xi}-\boldsymbol{x})$ for some fixed space point $\boldsymbol{x}$, which is on a stationary inertial trajectory while probing the $(3+1)$-dimensional \emph{massless} Klein-Gordon vacuum has the transition rate~\cite{schlicht2004}
\begin{eqnarray}
\dot{F}_{\tau}(\Delta)=-\frac{1}{2\pi}\Delta\Theta(-\Delta),
\end{eqnarray}
where $\Theta(\Delta)$ is the usual Heaviside theta function. This result is interpreted as the detector remaining unexcited when initially in its ground state ($\Delta>0$) and undergoing spontaneous emission when in its excited state ($\Delta<0$). This can be expected as a stationary detector should not be reacting in a significant way when perturbing the vacuum state around it. A scenario of more importance, however, occurs when the detector undergoes uniform acceleration. This type of motion, as we have seen previously, is described by a stationary Rindler trajectory and the resulting transition rate is given by~\cite{schlicht2004}
\begin{eqnarray}
\label{eqn:accelerated-response}
\dot{F}_{\tau}(\Delta)=\frac{1}{2\pi}\frac{\Delta}{\beta}\frac{1}{e^{2\pi\Delta/\beta}-1}.
\end{eqnarray}
Interestingly, the final multiplicative factor in~(\ref{eqn:accelerated-response}) is a \emph{Bose-Einstein distribution} when we make the identification with a temperature $T=\beta/2\pi$ with $\beta$ playing the role of the proper acceleration of the detector. The significance of this result is that it shows, to first order in perturbation theory, that a uniformly accelerated detector's internal degrees of freedom will respond as if immersed in a thermal bath. This can be related to the Unruh effect which predicts that the Minkowski vacuum state as perceived by an observer travelling with proper acceleration $\beta$ will observe a thermal quantum state of temperature $T\propto\beta$.

The Unruh-DeWitt detector has been an indispensable tool in relativistic quantum information for extracting entanglement from the vacuum state by swapping it from field to detectors~\cite{reznik2005} and also as a way to store information in a quantum field for later use~\cite{olson2011}. In Chapter~(\ref{chapter:fat-detectors}) we propose novel techniques to solve the evolution for these detectors beyond perturbation theory. We do this by considering detectors with Harmonic oscillator degrees of freedom. Harmonic oscillators are an example of systems with \emph{continuous variable} degrees of freedom. Continuous variable systems have been extensively studied within quantum information~\cite{braunstein2005}. In particular, the field of \emph{Gaussian state} quantum information has been at the centre of a large body of work~\cite{laurat2005,weedbrook2012}. Continuous variables systems include \emph{Bosonic} fields, such as the Klein-Gordon field. This connection will prove to be useful when analysing entanglement in quantum field theory. Thus, we review the basics of continuous variable Gaussian quantum information next.

\chapter{Continuous Variables}\label{chapter:cv} 

A system of great interest in quantum field theory is the electromagnetic field~\cite{cerf2007,puri2001,scully1999,walls2008}. Once quantised, the electromagnetic field can be thought of as a collection of decoupled \emph{quantum harmonic oscillators}, much in the same way as the Klein-Gordon field in the previous chapter. These harmonic oscillators correspond to the so-called \emph{modes} of the electromagnetic field. We refer to a field mode as a solution to a field equation governing a system. Quantum optics studies the interaction of the electromagnetic field with other systems such as atoms or other fields. A system of particular interest is that of \emph{optical cavities}. This area of research has seen many successes in probing the fundamental laws of quantum mechanics and has recently received the Nobel prize awarded to Haroche and Wineland for ``ground-breaking experimental methods that enable measuring and manipulation of individual quantum systems"~\cite{nobelprize2012}. Fields that are contained in cavities can be used to study strong quantum mechanical effects between different modes of the field~\cite{JohanssonJohanssonWilsonNori2010,WilsonDynCasNature2012}. To explain such experimental set-ups the theory of continuous variables was developed. 

Continuous variable systems involve physical observables which have a continuous spectrum. Simple examples of these are the position and momentum operators of standard quantum mechanics. Another well known system where continuous variable operators arise are Bosonic fields. Thus, our starting point will be the definition of Bosonic operators and the Hilbert space they live in.

We call a set of operators $\lbrace\hat{x}_{k},\hat{p}_{k}\rbrace$, where in general $k\in\mathbb{Z}$, \emph{Bosonic} if they satisfy the canonical commutation relations
\begin{eqnarray}
\left[\hat{x}_{k},\hat{p}_{k^{\prime}}\right]=i\,\delta_{kk^{\prime}}.
\end{eqnarray}
Note for a continuous parameter $k\in\mathbb{R}$ the Kronecker delta above would be replaced with a Dirac delta distribution $\delta(k-k')$. The operators are of course nothing more than the usual quantum harmonic oscillators of standard quantum mechanics. They can be related to the field mode operators, like those of the Klein-Gordon field, via
\begin{equation}
\label{eqn:quadrature-operators-def}
\begin{aligned}
\hat{x}_{k}&:=\frac{1}{\sqrt{2}}\left(\hat{a}_{k}+\hat{a}_{k}^{\dag}\right),\\
\hat{p}_{k}&:=\frac{-i}{\sqrt{2}}\left(\hat{a}_{k}-\hat{a}_{k}^{\dag}\right).
\end{aligned}
\end{equation}
These transformations imply the fundamental commutation relations $[\hat{a}_{k},\hat{a}_{k'}^{\dag}]=\delta_{kk'}$. In the previous chapter, we saw how the Klein-Gordon field could be written as a collection of decoupled oscillators. Therefore, the continuous variables formalism is naturally applicable to analyse this system. 

Note that the annihilation operators $\hat{a}_{k}^{\dag}$ define a vacuum state
\begin{eqnarray}
\hat{a}_{k}^{\dag}\ket{0}=0\,\forall\,k,
\end{eqnarray}
from which we can construct the usual Bosonic Fock space basis of Section~(\ref{sec:kg-mink-quant}). From this basis we can construct any state which lives in a Fock space. A state of interest is the so called \emph{coherent state}. A single mode coherent state describes, ideally, a laser beam of frequency $k$~\cite{glauber1963}. We can construct a coherent state using the Weyl \emph{displacement operator}
\begin{eqnarray}
\label{eqn:weyloperator}
\hat{D}_{k}(\alpha_{k})=e^{-\bar{\alpha}_{k}\hat{a}_{k}+\alpha_{k}\hat{a}_{k}^{\dag}},
\end{eqnarray}
where $\alpha_{k}\in\mathbb{C}$ is an arbitrary complex parameter. We define a general coherent state as
\begin{eqnarray}
\label{eqn:coherent-state}
\ket{\alpha_{k}}\equiv\hat{D}_{k}(\alpha_{k})\ket{0}.
\end{eqnarray}
Note that a trivial consequence of Eq.~(\ref{eqn:coherent-state}) is that for $\alpha_{k}=0$ a coherent state coincides with the vacuum state. Useful properties of the coherent states are that they are the eigenstates of the annihilation operators i.e.
\begin{eqnarray}
\label{eqn:coherent-property}
\hat{a}_{k}\ket{\alpha_{k}}=\alpha_{k}\ket{\alpha_{k}},
\end{eqnarray}
and minimise the Heisenberg uncertainty relation~(\ref{eqn:heisenberg-uncert})
\begin{eqnarray}
\mathrm{Var}(\hat{x}_{k})\mathrm{Var}(\hat{p}_{k})=\frac{1}{4},
\end{eqnarray}
where $\mathrm{Var}(\hat{\mathcal{O}}):=\langle\hat{\mathcal{O}}^{2}\rangle-\langle\hat{\mathcal{O}}\rangle^{2}$ is the variance of an observable for a given state. Coherent states are the simplest of what are known as \emph{Gaussian states}. This special class of continuous variable states are central in the forthcoming chapters and so a brief overview of their description and quantification will be illuminating.

\section{Gaussian States}\label{sec:gaussian-states}

In this section we define a subset of continuous variable states known as \emph{Gaussian states} \cite{simon1994,giedke2003,braunstein2005,weedbrook2012}. They are of great physical relevance since they are easy to access and manipulate experimentally. Moreover, they admit an elegant mathematical description which allows one to compute exact expressions for the quantification of entanglement, which are otherwise unattainable.

In the continuous variables formalism, we collect the set of $2N$ Bosonic operators into the vector
\begin{eqnarray}
\label{eqn:gerardo-basis}
\hat{\boldsymbol{X}}=\left( \hat{x}_{1},\hat{p}_{1},\ldots ,\hat{x}_{N},\hat{p}_{N}\right),
\end{eqnarray}
where the \emph{quadrature} operators $\hat{x}_{m}$, $\hat{p}_{m}$ are defined in~(\ref{eqn:quadrature-operators-def}). In this notation, the commutation relations of the quadrature operators $\hat{X}_{j}$ can be neatly written as
\begin{eqnarray}
\left[\hat{X}_{m},\hat{X}_{n}\right]=i\Omega_{mn},
\end{eqnarray}
where the matrix $\boldsymbol{\Omega}$ is a \emph{symplectic form} associated with the mode operators and reads~\cite{birrell1982}
\begin{eqnarray}
\label{eqn:real-symplectic-form}
\boldsymbol{\Omega}=\bigoplus\limits_{k=1}^{N}\boldsymbol{\omega}=\left( 
\begin{array}{ccc}
\boldsymbol{\omega} & & \\
 & \ddots & \\
 & & \boldsymbol{\omega}
\end{array}
\right),
\hspace{5mm}
\boldsymbol{\omega}:=\left(\begin{array}{cc}
0 & 1\\
-1 & 0
\end{array}\right).
\end{eqnarray}
Next, we define the characteristic function of a quantum state $\hat{\rho}$ as
\begin{eqnarray}
\label{eqn:characteristic-function}
\chi(\boldsymbol{\xi}):=\mathrm{tr}\left[\hat{\rho} e^{i\hat{\boldsymbol{X}}^{\mathrm{tp}}\boldsymbol{\Omega}\boldsymbol{\xi}}\right],
\end{eqnarray}
where $\boldsymbol{\xi}\in\mathbb{R}^{2N}$. Through a Fourier transformation, we can relate a state's characteristic function to its \emph{Wigner function}~\cite{braunstein2005,weedbrook2012}
\begin{eqnarray}
W(\boldsymbol{X})&=&\frac{1}{(2\pi)^{2N}}\int_{\mathbb{R}^{2N}}d^{2N}\boldsymbol{\xi}e^{-i\boldsymbol{X}^{\mathrm{tp}}\boldsymbol{\Omega}\boldsymbol{\xi}}\chi(\boldsymbol{\xi}),
\end{eqnarray}
where $\boldsymbol{X}\in\mathbb{R}^{2N}$ represents the classical values associated with $\hat{\boldsymbol{X}}$. A state's Wigner function represents it uniquely (hence so does it's characteristic function)~\cite{wigner1932}. A Gaussian state is defined as a quantum state which is fully characterised by its \emph{first} and \emph{second} moments only. The first moments $\boldsymbol{d}$ of a state are defined as
\begin{eqnarray}
d_{j}:=\left\langle\hat{X}_{j}\right\rangle,
\end{eqnarray}
and the second moments $\boldsymbol{\Gamma}$ (also known as a \emph{covariance matrix}) are
\begin{eqnarray}
\label{eqn:cov-mat-def}
\Gamma_{ij}=\left\langle\hat{X}_{i}\hat{X}_{j}+\hat{X}_{j}\hat{X}_{i}\right\rangle -2\left\langle\hat{X}_{i}\right\rangle\left\langle\hat{X}_{j}\right\rangle.
\end{eqnarray}
We note that the covariance matrix $\boldsymbol{\Gamma}$ is a symmetric positive-definite matrix. Therefore for Gaussian states, the characteristic and Wigner functions reduce to the form
\begin{subequations}
\begin{align}
\chi(\boldsymbol{\xi})&=e^{-\frac{1}{4}\boldsymbol{\xi}^{\mathrm{tp}}\boldsymbol{\Omega}\boldsymbol{\Gamma}\boldsymbol{\Omega}^{\mathrm{tp}}\boldsymbol{\xi}-i\left(\boldsymbol{\Omega}\boldsymbol{d}\right)^{\mathrm{tp}}\boldsymbol{\xi}},\\
W(\boldsymbol{X})&=\frac{1}{\pi^{N}}\frac{1}{\sqrt{\mathrm{det}(\boldsymbol{\Gamma})}}e^{-(\boldsymbol{X}-\boldsymbol{d})^{\mathrm{tp}}\boldsymbol{\Gamma}^{-1}(\boldsymbol{X}-\boldsymbol{d})}.
\end{align}
\end{subequations}
Note that our above definitions differ from many other good references for Gaussian state quantum information in the sense that strictly our covariance matrix ``twice" the one defined by other authors. We have also chosen the natural unit convention $\hbar\rightarrow+1$ which again other authors choose differently. It is therefore prudent to be aware of which conventions a particular article has chosen.

Continuing, as an example, we compute the first moments and covariance matrix of a coherent state $\ket{\alpha_{k}}$. Using the definitions of the quadrature operators~(\ref{eqn:quadrature-operators-def}) and the property~(\ref{eqn:coherent-property}), we find
\begin{eqnarray}
\label{eqn:cov-mat-coherent}
\boldsymbol{d}=\sqrt{2}\left(\begin{array}{c}\mathrm{Re}(\alpha_{k}) \\ \mathrm{Im}(\alpha_{k})\end{array}\right),\,\,\boldsymbol{\Gamma}=\boldsymbol{I}.
\end{eqnarray}
Remarkably, the covariance matrix for a coherent state is identical for all coherent parameters $\alpha_{k}$ and is just the identity matrix.

The characteristic function~(\ref{eqn:characteristic-function}) represents the state of a physical system. The density matrix represents a physical state if, and only if, it is positive definite. For Gaussian states this means
\begin{eqnarray}
\mathbf{\boldsymbol{\Gamma}}+i\mathbf{\Omega}\ge 0.
\end{eqnarray}
Gaussian states can, of course, be pure or mixed. We can easily define pure and mixed Gaussian states by
\begin{eqnarray}
\label{eqn:gaussian-purity}
\mathrm{det}(\boldsymbol{\Gamma})=\left\{ 
  \begin{array}{l l}
    +1 \Rightarrow \mathrm{pure}\\
    < 1 \Rightarrow \mathrm{mixed}
  \end{array} \right. .
\end{eqnarray}
Now that we have defined the basic quantities that represent a Gaussian state, one may ask: \emph{how are unitary transformations represented?} Unitary transformations on a Hilbert space correspond to real \emph{symplectic transformations} on the first and second moments as
\begin{equation}
\label{eqn:symp-transformation}
\hat{U}^{\dag}\hat{\rho}\hat{U}\rightarrow\left\{ 
  \begin{array}{l l}
    \boldsymbol{d}'=\boldsymbol{S}\boldsymbol{d} \\
    \boldsymbol{\Gamma}'=\boldsymbol{S}\boldsymbol{\Gamma}\boldsymbol{S}^{\mathrm{tp}}
  \end{array} \right. ,
\end{equation}
where $\boldsymbol{S}$ is a \emph{symplectic matrix} which corresponds to the action of $\hat{U}$ on the state $\hat{\rho}$. This simple transformation rule only holds, however, for unitary transformations whose exponents are, at most, quadratic in the mode operators $\lbrace \hat{a}_{k},\hat{a}_{k}^{\dag}\rbrace$. Such unitary transformations preserve the Gaussian nature of the states. Gaussian state quantum information is built around these symplectic transformations. It will therefore be useful to introduce some of their properties.

\section{Symplectic Geometry}

Symplectic geometry has its foundations firmly rooted in classical mechanics. However, as stated earlier, it also has a place in quantum theory with profound and deep consequences. For an excellent introductory text of classical symplectic geometry see Berndt~\cite{berndt2001} and for its application to quantum mechanics see Gosson~\cite{gosson2006}. For a particularly useful summary of symplectic geometry and its use in continuous variables quantum systems see Arvind et. al.~\cite{arvind1995}.

\subsection{Definitions}

The group of real symplectic matrices is defined by
\begin{eqnarray}
\label{eqn:symplectic-def-1}
\boldsymbol{S}\boldsymbol{\Omega}\boldsymbol{S}^{\mathrm{tp}}=\boldsymbol{\Omega},
\end{eqnarray}
where $\boldsymbol{\Omega}$ is the symplectic form defined via Eq.~(\ref{eqn:real-symplectic-form}). We denote this group by $\mathrm{Sp}(2N,\mathbb{R})$ and so define
\begin{eqnarray}
\mathrm{Sp}(2N,\mathbb{R})=\left\lbrace\boldsymbol{S}| \boldsymbol{S}\boldsymbol{\Omega}\boldsymbol{S}^{\mathrm{tp}}=\boldsymbol{\Omega}\right\rbrace .
\end{eqnarray}
Note that symplectic matrices are always square ($2N\times 2N$), invertible matrices with determinant $\mathrm{det}(\boldsymbol{S})=+1$. Given the arrangement of operators in the basis of~(\ref{eqn:gerardo-basis}), we decompose the symplectic matrix into the block form
\begin{eqnarray}
\boldsymbol{S}=\left(
\begin{array}{cccc}
\boldsymbol{s}_{11} & \boldsymbol{s}_{12} & \cdots & \boldsymbol{s}_{1N} \\
\boldsymbol{s}_{21} & \boldsymbol{s}_{22} &  &  \\
\vdots &  & \ddots &  \\
\boldsymbol{s}_{N1} &  &  & \boldsymbol{s}_{NN}
\end{array}
\right),
\end{eqnarray}
where the $2\times 2$ sub-block $\boldsymbol{s}_{mn}$ represents the transformation between the modes $m$ and $n$. A very special property is that any symmetric positive-definite matrix can be diagonalised by a symplectic matrix. This is the \emph{Williamson normal form} theorem~\cite{williamson1937}.

\subsection{Williamson Normal Form}

Williamson showed that any symmetric positive-definite matrix can be put into a diagonal form via a symplectic transformation. Its main use is in finding the \emph{symplectic spectrum} of an arbitrary state characterised by a covariance matrix $\boldsymbol{\Gamma}$. This statement is formalised in the following theorem~\cite{williamson1937}:
\begin{theorem}
Let $\boldsymbol{\Gamma}$ be a $2N\times 2N$ positive-definite matrix. Then there exists a unique $\boldsymbol{S}\in\mathrm{Sp}(2N,\mathbb{R})$ that diagonalises $\boldsymbol{\Gamma}$ such that\\
\begin{eqnarray*}
\boldsymbol{\boldsymbol{\Gamma}}=\boldsymbol{S}^{\mathrm{tp}}\bigoplus\limits_{k=1}^{N}
\left(
\begin{array}{cc}\nu_{k} & 0 \\ 0 & \nu_{k} 
\end{array}
\right)\boldsymbol{S}
\end{eqnarray*}
\end{theorem}
A proof of this theorem can be found in the excellent text~\cite{gosson2006}. We can collect the $N$ eigenvalues $\nu_{k}$ into $\boldsymbol{\nu}=\mathrm{diag}(\nu_{1},\ldots,\nu_{N})$ (either a diagonal matrix or vector). $\boldsymbol{\nu}$ is known as the \emph{symplectic spectrum} of $\boldsymbol{\Gamma}$. As a consequence of this result, we can characterise the purity~(\ref{eqn:gaussian-purity}) of a state by rewriting its determinant as
\begin{eqnarray}
\mathrm{det}(\boldsymbol{\Gamma})=\prod_{k}\nu_{k}^{2}.
\end{eqnarray}
The Williamson normal form is relevant for our purposes since any physical Gaussian state, represented by a covariance matrix $\boldsymbol{\Gamma}$, is a positive-definite matrix . In particular we can use it to find the symplectic spectrum of the covariance matrix $\boldsymbol{\Gamma}$. In practice, however, it is usually much more convenient to obtain the spectrum $\boldsymbol{\nu}$ from the relation~\cite{gosson2006}
\begin{equation}
\label{eqn:symplectic-spectrum}
\boldsymbol{\nu}=\mathrm{Eig}_{+}\left(i\boldsymbol{\Omega}\boldsymbol{\Gamma}\right),
\end{equation}
where $\mathrm{Eig}_{+}\left(\boldsymbol{A}\right)$ denotes the diagonal matrix of \emph{positive} eigenvalues of the matrix $\boldsymbol{A}$. Knowing the symplectic spectrum of a given covariance matrix is very powerful. As we shall see, it will allow us to cast our entanglement measures into functions of $\boldsymbol{\nu}$.
A natural next step would be to link, in a concrete way, the relationship between a unitary operator and its symplectic representation. However, to do so we benefit from reviewing the different bases we can choose to write a symplectic matrix in.

\subsection{Representations of $\mathrm{Sp}(2N,\mathbb{R})$}\label{sec:symplectic-representations}

From the previous section, a second representation of $\mathrm{Sp}(2N,\mathbb{R})$ from~(\ref{eqn:symplectic-def-1}) is given by the transformation
\begin{eqnarray}
\left(\begin{array}{c}
\hat{x}_{1}\\
\vdots \\
\hat{x}_{N}\\
\hat{p}_{1}\\
\vdots\\
\hat{p}_{N}
\end{array}\right)\equiv
\boldsymbol{L}
\left(\begin{array}{c}
\hat{x}_{1}\\
\hat{p}_{1}\\
\vdots\\
\vdots\\
\hat{x}_{N}\\
\hat{p}_{N}
\end{array}\right),
\end{eqnarray}
where $\boldsymbol{L}$ is a basis changing matrix. This is, of course, nothing more than a rearrangement of the original basis~(\ref{eqn:gerardo-basis}). We will refer to the vector
\begin{eqnarray}
\label{eqn:quadrature-basis}
\hat{\boldsymbol{Y}}:=\left(\hat{x}_{1},\ldots,\hat{x}_{N},\hat{p}_{1},\ldots,\hat{p}_{N}\right),
\end{eqnarray}
as the \emph{quadrature basis} vector.

The benefit of this basis is that the symplectic form and any symplectic matrix now takes a block form (with an abuse of notation calling the symplectic form and symplectic matrices with respect to this new basis again $\boldsymbol{\Omega}$ and $\boldsymbol{S}$)
\begin{eqnarray}
\boldsymbol{\Omega}=\left(
\begin{array}{cc}
\boldsymbol{0} & \boldsymbol{I} \\
-\boldsymbol{I} & \boldsymbol{0}
\end{array}
\right),
\hspace{5mm}
\boldsymbol{S}=\left(
\begin{array}{cc}
\boldsymbol{A} & \boldsymbol{B} \\
\boldsymbol{C} & \boldsymbol{D}
\end{array}
\right).
\end{eqnarray}
Note the above matrices are now $N\times N$ dimensional. We can use the defining symplectic relation~(\ref{eqn:symplectic-def-1}) to find the following conditions for $\boldsymbol{A},\boldsymbol{B},\boldsymbol{C}$ and $\boldsymbol{D}$:
\begin{subequations}
\begin{align}
\left(\boldsymbol{A}\boldsymbol{B}^{\mathrm{tp}}\right)^{\mathrm{tp}}&=\boldsymbol{A}\boldsymbol{B}^{\mathrm{tp}},\\
\left(\boldsymbol{C}\boldsymbol{D}^{\mathrm{tp}}\right)^{\mathrm{tp}}&=\boldsymbol{C}\boldsymbol{D}^{\mathrm{tp}},\\
\boldsymbol{A}\boldsymbol{D}^{\mathrm{tp}}&-\boldsymbol{B}\boldsymbol{C}^{\mathrm{tp}}=\boldsymbol{I}.
\end{align}
\end{subequations}
The corresponding expression for the Williamson normal form is then
\begin{eqnarray*}
\boldsymbol{\boldsymbol{\Gamma}}=\boldsymbol{S}^{\mathrm{tp}}\left(\begin{array}{cc}\boldsymbol{\nu} & \boldsymbol{0} \\ \boldsymbol{0} & \boldsymbol{\nu} \end{array}\right)\boldsymbol{S},
\end{eqnarray*}
where $\boldsymbol{\nu}$ is the previously defined symplectic spectrum associated with the covariance matrix $\boldsymbol{\Gamma}$. Using this representation, we can naturally transform to a new basis which is known as the \emph{complex form} of $\mathrm{Sp}(2N,\mathbb{R})$~\cite{arvind1995}. Note that this is not a ``complexification" of the group, it is simply a change of basis which is very convenient. It is essentially a transformation from the quadrature operators $\hat{x}_{j},\hat{p}_{j}$ to the mode operators $\hat{a}_{j},\hat{a}_{j}^{\dag}$ given by
\begin{eqnarray}
\left(\begin{array}{c}
\hat{a}_{1}\\
\vdots \\
\hat{a}_{N}\\
\hat{a}_{1}^{\dag}\\
\vdots\\
\hat{a}_{N}^{\dag}
\end{array}\right)\equiv
\boldsymbol{L}_{(c)}
\left(\begin{array}{c}
\hat{x}_{1}\\
\vdots \\
\hat{x}_{N}\\
\hat{p}_{1}\\
\vdots\\
\hat{p}_{N}
\end{array}\right),
\end{eqnarray}
where the basis changing matrix elegantly reads
\begin{eqnarray}
\boldsymbol{L}_{(c)}:=\frac{1}{\sqrt{2}}
\left(\begin{array}{cc}
\boldsymbol{I} & i\boldsymbol{I} \\
\boldsymbol{I} & -i\boldsymbol{I}
\end{array}\right).
\end{eqnarray}
For later convenience we denote the vector of mode operators as
\begin{eqnarray}
\label{eqn:mode-op-basis}
\hat{\boldsymbol{\xi}}:=\left(\hat{a}_{1},\ldots, \hat{a}_{N},\hat{a}^{\dag}_{1},\ldots, \hat{a}^{\dag}_{N}. \right)
\end{eqnarray}
In this representation, we can find the complex form of any matrix written in the quadrature basis~(\ref{eqn:quadrature-basis}) via
\begin{eqnarray}
\label{eqn:complex-form-embedding}
\boldsymbol{S}\rightarrow \boldsymbol{S}_{(c)}=\boldsymbol{L}_{(c)}\boldsymbol{S}\boldsymbol{L}_{(c)}^{\dag}.
\end{eqnarray}
Using this rule, we find that the complex form of the symplectic matrices are particularly aesthetically pleasing~\cite{arvind1995}:
\begin{eqnarray}
\label{eqn:complex-form-defs}
\boldsymbol{\Omega}_{(c)}=-i\boldsymbol{K},
\hspace{5mm}
\boldsymbol{K}:=\left(
\begin{array}{cc}
\boldsymbol{I} & \boldsymbol{0} \\
\boldsymbol{0} & -\boldsymbol{I}
\end{array}
\right),
\hspace{5mm}
\boldsymbol{S}_{(c)}=\left(
\begin{array}{cc}
\boldsymbol{\alpha} & \boldsymbol{\beta} \\
\bar{\boldsymbol{\beta}} & \bar{\boldsymbol{\alpha}}
\end{array}
\right).
\end{eqnarray}
In addition, the defining symplectic relation~(\ref{eqn:symplectic-def-1}) is replaced by
\begin{eqnarray}
\label{eqn:symplectic-relation-complex}
\boldsymbol{S}_{(c)}\boldsymbol{K}\boldsymbol{S}_{(c)}^{\dag}=\boldsymbol{K},
\end{eqnarray}
where we notice that the transposition operation has been promoted to a Hermitian conjugation due to the embedding~(\ref{eqn:complex-form-embedding}). Using~(\ref{eqn:symplectic-relation-complex}), we find that the conditions for $\boldsymbol{S}_{(c)}$ to be symplectic result in the expressions
\begin{subequations}
\begin{align}
\boldsymbol{\alpha}\boldsymbol{\alpha}^{\dag}-\boldsymbol{\beta}\boldsymbol{\beta}^{\dag}&=\boldsymbol{I},\\
\boldsymbol{\alpha}\boldsymbol{\beta}^{\text{tp}}&=\left(\boldsymbol{\alpha}\boldsymbol{\beta}^{\text{tp}}\right)^{\text{tp}}.
\end{align}
\end{subequations}
Remarkably, these are the same relations we found for the Klein-Gordon field Bogoliubov transformation in Section~(\ref{sec:bogo-qft}). This identification of Bogoliubov transformations and symplectic matrices allows us to use the full power of modern quantum optics and quantum information in our studies of quantum information in quantum field theory.

Finally, the Williamson normal form for the complex form of $\mathrm{Sp}(2N,\mathbb{R})$ reads
\begin{eqnarray*}
\boldsymbol{\boldsymbol{\Gamma}}_{(c)}=\boldsymbol{S}^{\dag}_{(c)}\left(\begin{array}{cc}\boldsymbol{\nu} & \boldsymbol{0} \\ \boldsymbol{0} & \boldsymbol{\nu} \end{array}\right)\boldsymbol{S}_{(c)},
\end{eqnarray*}
where $\boldsymbol{\nu}$ remains unchanged from the previous definitions and $\boldsymbol{\boldsymbol{\Gamma}}_{(c)}$ is the complex form of the covariance matrix $\boldsymbol{\boldsymbol{\Gamma}}$ defined via
\begin{eqnarray}
(\boldsymbol{\boldsymbol{\Gamma}}_{(c)})_{mn}=\langle\hat{\xi}_{m}\hat{\xi}_{n}^{\dag}+\hat{\xi}_{m}^{\dag}\hat{\xi}_{n}\rangle-2\langle\hat{\xi}_{n} \rangle\langle\hat{\xi}_{m}^{\dag}\rangle.
\end{eqnarray}
It should be noted that in the complex form, the symplectic spectrum~(\ref{eqn:symplectic-spectrum}) of a covariance matrix can be computed using
\begin{eqnarray}
\label{eqn:symplectic-spectrum-complex}
\boldsymbol{\nu}=\mathrm{Eig}_{+}(\boldsymbol{K}\boldsymbol{\Gamma}_{(c)}).
\end{eqnarray}
Of course, the complex form covariance matrix $\boldsymbol{\boldsymbol{\Gamma}}_{(c)}$ can be obtained from $\boldsymbol{\boldsymbol{\Gamma}}$ by the transformation rule~(\ref{eqn:complex-form-embedding}) which results in the block form
\begin{eqnarray}
\boldsymbol{\boldsymbol{\Gamma}}_{(c)}=\left(\begin{array}{cc}
\mathbfcal{V} & \mathbfcal{U} \\
\bar{\mathbfcal{U}} & \bar{\mathbfcal{V}}
\end{array} \right),
\end{eqnarray}
with the conditions $\mathbfcal{V}^{\dag}=\mathbfcal{V}$ and $\mathbfcal{U}^{\mathrm{tp}}=\mathbfcal{U}$. This implies that $\boldsymbol{\boldsymbol{\Gamma}}_{(c)}^{\dag}=\boldsymbol{\boldsymbol{\Gamma}}_{(c)}$.

In Chapter~(\ref{chapter:fat-detectors}), Sections~(\ref{sec:boson-resonance},\,\ref{sec:teleportation-results}), we will use the above notions of symplectic geometry to compute exactly the evolution of a quantum state. As examples, these techniques will be applicable to to solve systems which involve cavities moving through spacetime and multiple Unruh-DeWitt detectors coupled to a quantum field. To explain these techniques, we must first comment on the Lie algebra structure of the complex form of $\mathrm{Sp}(2N,\mathbb{R})$. The Lie algebra of the symplectic group will help us derive equations which govern the time evolution of a quantum state.

\section{Lie Algebra of $\mathrm{Sp}(2N,\mathbb{R})$}

An excellent text on Lie groups and algebras is Hall~\cite{hall2004}.

To begin, we define a set of Hermitian, $2N\times 2N$, linearly independent basis matrices $\boldsymbol{G}_{j}$, such that
\begin{eqnarray}
\mathfrak{sp}(2N,\mathbb{R})=\left\lbrace \boldsymbol{K}\boldsymbol{G}_{j}| \boldsymbol{G}_{j}^{\dag}=\boldsymbol{G}_{j}\right\rbrace .
\end{eqnarray}
$\mathfrak{sp}(2N,\mathbb{R})$ is known as the \emph{Lie algebra} of $\mathrm{Sp}(2N,\mathbb{R})$. We can link a Lie algebra with its group via the exponential map~\cite{hall2004}. The real symplectic group is \emph{connected} (though not simply connected) and is \emph{non-compact}. Being non-compact implies that not every symplectic matrix can be written as the exponential of a single matrix. However, as we will see, every element of the symplectic group \emph{can} be written as a product of exponentials. 

The matrices $\boldsymbol{K}\boldsymbol{G}_{j}$ are the infinitesimal generators of $\mathrm{Sp}(2N,\mathbb{R})$ and is a finite, closed algebra of dimension $N(2N+1)$. To ensure the correct properties of the symplectic group, the matrices $\boldsymbol{G}$ are necessarily of the form
\begin{eqnarray}
\boldsymbol{G}=\left(\begin{array}{cc}
\mathbfcal{X} & \mathbfcal{Y} \\
\bar{\mathbfcal{Y}} & \bar{\mathbfcal{X}}
\end{array} \right),
\end{eqnarray}
with the conditions $\mathbfcal{X}^{\dag}=\mathbfcal{X}$ and $\mathbfcal{Y}^{\mathrm{tp}}=\mathbfcal{Y}$. Note that the matrices $\boldsymbol{G}_{j}$ are not the most arbitrary type of Hermitian matrix (which for $2N\times 2N$ matrices has dimension $4N^{2}$) and have dimension $\mathrm{dim}(\mathbfcal{X})+\mathrm{dim}(\mathbfcal{Y})=N^{2}+2N+N(N-1)=N(2N+1)$ as stated earlier. A useful consequence of the algebra being closed is that we can decompose any symplectic matrix in the product decomposition
\begin{eqnarray}
\boldsymbol{S}_{(c)}=\prod_{j}e^{-ig_{j}\boldsymbol{K}\boldsymbol{G}_{j}},
\end{eqnarray}
where $g_{j}\in\mathbb{R}$ and the product runs over the $N(2N+1)$ independent symplectic generators. This decomposition will be useful to solve the time evolution of moving detectors in section~(\ref{sec:time-ordering}). There is, of course, an equivalent structure for each representation of $\mathrm{Sp}(2N,\mathbb{R})$, however, they are irrelevant for our purposes. The complex form of $\mathrm{Sp}(2N,\mathbb{R})$ allows us to consider the role of the mode operators $\lbrace a_{k},a_{k}^{\dag} \rbrace$ in Gaussian state quantum information. To gain a better grasp of this relationship, we shall briefly review some useful concepts from linear quantum optics.

\section{Linear Optics}

Now that we have defined Gaussian states, and their transformations through symplectic matrices, we want to relate unitary operators on a Hilbert space to their symplectic matrix counterpart. The answer to this can be found by considering concepts from \emph{linear optics}. The field of continuous variables is built on the power of techniques from linear optics. The backbone of these techniques are unitary transformations whose exponent is quadratic in field operators. Starting with the vector of mode operators~(\ref{eqn:mode-op-basis}), we can compactly write the mode operator commutation relations as
\begin{eqnarray}
\left[\hat{\xi}_{m},\hat{\xi}_{n}^{\dag}\right]=K_{mn},
\end{eqnarray}
where $K_{mn}$ are the components of $\boldsymbol{K}$ defined in~(\ref{eqn:complex-form-defs}). As stated in Section~(\ref{sec:gaussian-states}), symplectic transformations represent unitary operators that are \emph{quadratic} in mode operators. Any quadratic combination of field operators can be written as
\begin{eqnarray}
\label{eqn:quadraticoperator}
\hat{H}=\hat{\boldsymbol{\xi}}^{\dag}\cdot\mathbf{H}\cdot\hat{\boldsymbol{\xi}},
\end{eqnarray}
where
\begin{eqnarray}
\label{eqn:h-symplectic-def}
\mathbf{H}=\left(\begin{array}{cc} \mathbf{A} & \mathbf{B} \\ \bar{\mathbf{B}} & \bar{\mathbf{A}} \end{array}\right),
\end{eqnarray}
and $\mathbf{A}^{\dag}=\mathbf{A}$ and $\mathbf{B}^{\text{tp}}=\mathbf{B}$. $\mathbf{H}$ is a characterised by $N(2N+1)$ real parameters. We can now define the group of Gaussian persevering linear operators as
\begin{eqnarray}
\mathcal{U}=\left\lbrace\hat{U}|\hat{U}\hat{U}^{\dag}=\hat{U}^{\dag}\hat{U}=\hat{I}\right\rbrace .
\end{eqnarray}
From the structure of the $\hat{U}$ operators, it follows that $\mathcal{U}$ is the group of quadratic unitary operators i.e. their exponents are at most quadratic in mode operators. The Lie algebra structure of the linear operator group can be found in the standard way by differentiating the defining the unitary condition with respect to some group parameter $t$. Therefore the Lie algebra associated with the group of quadratic unitary operators is
\begin{eqnarray}
\mathfrak{u}=\left\lbrace \hat{H}|\hat{H}^{\dag}-\hat{H}=0 \right\rbrace .
\end{eqnarray}
So we find, as was expected, that $\hat{H}$ has to be a Hermitian operator, which will later be useful as it will be identified with the Hamiltonian of the system.

Transformations of this form take an arbitrary linear combination of field operators to another arbitrary linear combination of field operators. Mathematically we have, for complex coefficients $\alpha_{jk},\beta_{jk}$,
\begin{equation}
\begin{aligned}
\hat{U}^{\dag}\hat{a}_{k}\hat{U}&=&\sum_{j}\alpha_{jk}\hat{a}_{j}+\sum_{j}\beta_{jk}\hat{a}_{j}^{\dag},\\
\hat{U}^{\dag}\hat{a}_{k}^{\dag}\hat{U}&=&\sum_{j}\overline{\alpha}_{jk}\hat{a}_{j}^{\dag}+\sum_{j}\overline{\beta}_{jk}\hat{a}_{j},
\end{aligned}
\end{equation}
which can be written compactly as
\begin{eqnarray}
\hat{U}^{\dag}\left(\begin{array}{c}\hat{\boldsymbol{a}}\\ \hat{\boldsymbol{a}}^{\dag}\end{array}\right)\hat{U}=\left(\begin{array}{cc} \boldsymbol{\alpha} & \boldsymbol{\beta} \\ \bar{\boldsymbol{\beta}} & \bar{\boldsymbol{\alpha}} \end{array}\right) \left(\begin{array}{c} \hat{\boldsymbol{a}}\\ \hat{\boldsymbol{a}}^{\dag}\end{array}\right).
\end{eqnarray}
As these linear transformations must preserve the commutation relations, due to $\hat{U}$ begin unitary, we find the conditions
\begin{subequations}
\label{eqn:symplectic-def-bogo}
\begin{align}
\boldsymbol{\alpha}\boldsymbol{\alpha}^{\dag}-\boldsymbol{\beta}\boldsymbol{\beta}^{\dag}&=\boldsymbol{I},\\
\boldsymbol{\alpha}\boldsymbol{\beta}^{\text{tp}}&=\left(\boldsymbol{\alpha}\boldsymbol{\beta}^{\text{tp}}\right)^{\text{tp}}.
\end{align}
\end{subequations}
Remarkably, we find the conditions on $\boldsymbol{\alpha}$ and $\boldsymbol{\beta}$ are nothing more than the defining relations for the complex form of $\mathrm{Sp}(2N,\mathbb{R})$. Thus we can write (dropping the complex form subscript)
\begin{eqnarray}
\mathbf{S}=\left(\begin{array}{cc} \boldsymbol{\alpha} & \boldsymbol{\beta} \\ \bar{\boldsymbol{\beta}} & \bar{\boldsymbol{\alpha}} \end{array}\right).
\end{eqnarray}
The correspondence between unitary transformations (or linear transformations) and symplectic relations, i.e.
\begin{eqnarray}
\hat{U}^{\dag}\hat{\boldsymbol{\xi}}\hat{U}=\mathbf{S}\cdot\hat{\boldsymbol{\xi}},
\end{eqnarray}
allows us to use the power of symplectic geometry to calculate linear transformations of our systems. We would now like to construct a direct relation between the unitary operator $\hat{U}$ and its symplectic representation. Generally this will be a formidable task but in certain situations a simple relation between a Gaussian preserving unitary operator and its symplectic counterpart is achievable. Given a unitary operator which can be written as a single exponential (e.g. the time ordered integral can be performed trivially), the corresponding symplectic matrix can be written as single exponential also. Details of this can be found in appendix~(\ref{app:symplectic-matrix-derivation}). The result is~\cite{luis1995}
\begin{eqnarray}
\label{eqn:sym-from-h}
\boldsymbol{S}=e^{-i\boldsymbol{K}\boldsymbol{H}},
\end{eqnarray}
where $\boldsymbol{K}$ is the commutator component matrix~(\ref{eqn:complex-form-defs}) and $\boldsymbol{H}$ is defined via~(\ref{eqn:h-symplectic-def}). This correspondence between quadratic Hamiltonians and symplectic matrices is best understood through examples. Two prominent cases of unitary transformations in quantum optics are the \emph{beam splitter} and \emph{two-mode squeezing} operations.

\subsection{Beam Splitter}

Beam splitters are devices used in quantum optics to split light into two separate beams. Their most common use is in interferometers. Mathematically, we can describe a beam splitter as a unitary transformation between two modes of a quantum field. To this effect, we define the beam splitter Hamiltonian as
\begin{eqnarray}
\hat{H}=2ir\left(\hat{a}^{\dag}\hat{b}-\hat{a}\hat{b}^{\dag}\right).
\end{eqnarray}
Here, $\hat{a},\hat{a}^{\dag},\hat{b},\hat{b}^{\dag}$ denotes the operators of two different modes and $r$ is a real parameter. The matrix representation of this Hamiltonian is
\begin{eqnarray}
\boldsymbol{H}=ir\left(
\begin{array}{cc}
i\boldsymbol{\sigma}_{2} & \boldsymbol{0} \\
\boldsymbol{0} & -i\boldsymbol{\sigma}_{2}
\end{array} \right).
\end{eqnarray}
From here we can use the expression~(\ref{eqn:sym-from-h}) to find
\begin{equation}
\label{eqn:beam-splitter-matrix}
\boldsymbol{S}=\left(
\begin{array}{cc}
\boldsymbol{R}& \boldsymbol{0} \\
\boldsymbol{0} & \boldsymbol{R}
\end{array} \right),
\end{equation}
where the rotation matrix $\boldsymbol{R}$ is defined as
\begin{eqnarray}
\boldsymbol{R}=\left(\begin{array}{cc}
\cos(r) & \sin(r) \\
-\sin(r) & \cos(r)
\end{array}\right).
\end{eqnarray} 
The symplectic matrix~(\ref{eqn:beam-splitter-matrix}) is known as the beam splitter operation. Beam splitters, along with local phase rotations, constitute what are known as \emph{passive} transformations in quantum optics~\cite{arvind1995}. We note that, in the complex form of $\mathrm{Sp}(2N,\mathbb{R})$, the most general passive transformation can be written as
\begin{equation}
\boldsymbol{S}=\left(
\begin{array}{cc}
\boldsymbol{U}& \boldsymbol{0} \\
\boldsymbol{0} & \bar{\boldsymbol{U}}
\end{array} \right),
\end{equation}
where $\boldsymbol{U}\boldsymbol{U}^{\dag}=\boldsymbol{I}$ is a unitary matrix i.e. $\boldsymbol{U}\in\mathrm{U}(N)$ (not $\mathrm{SU}(N)$). Next we define the two-mode squeezing operation.

\subsection{Two-Mode Squeezing}

We define the two-mode squeezing Hamiltonian as
\begin{equation}
\label{eqn:parametric-down-conversion}
\hat{H}=2ir\left(\hat{a}^{\dag}\hat{b}^{\dag}-\hat{a}\hat{b}\right),
\end{equation}
where $\hat{a},\hat{a}^{\dag},\hat{b},\hat{b}^{\dag}$ denotes the operators of two different modes and $r$ is a real parameter. Hamiltonians of this form are used to describe particle creation in quantum optics through the process of \emph{parametric down conversion}~\cite{burnham1970}. To progress we would like to relate the Hamiltonian~(\ref{eqn:parametric-down-conversion}) to its symplectic counterpart. The matrix representation of $\hat{H}$ is given by
\begin{eqnarray}
\label{eqn:two-mode-squeezed-matrix}
\boldsymbol{H}=ir\left(
\begin{array}{cc}
\boldsymbol{0} & \boldsymbol{\sigma}_{1} \\
-\boldsymbol{\sigma}_{1} & \boldsymbol{0}
\end{array} \right),
\end{eqnarray}
where $\boldsymbol{\sigma}_{1}$ is the first Pauli matrix. Using~(\ref{eqn:two-mode-squeezed-matrix}) and the definition of a symplectic matrix~(\ref{eqn:sym-from-h}) we find
\begin{eqnarray}
\boldsymbol{S}=\left(
\begin{array}{cc}
\cosh(r)\boldsymbol{I} & \sinh(r)\boldsymbol{\sigma}_{1} \\
\sinh(r)\boldsymbol{\sigma}_{1} & \cosh(r)\boldsymbol{I}
\end{array} \right).
\end{eqnarray}
This symplectic matrix is known as the \emph{two-mode squeezing operator}. The two-mode squeezing operator is an example of a time independent operator. In general, Hamiltonians can be time-dependent. In this case, the unitary operator associated with a time dependent Hamiltonian relies on computing the Hamiltonian and its commutation relations at different times, see Eq.~(\ref{eqn:evo-op-def}). This is generally a formidable task. However, using the results presented in this section, we can explicitly construct the time dependent Hamiltonians unitary operator. This will be the foundations of the results presented in Section~(\ref{sec:time-ordering}).

The two-mode squeezing operator is an entanglement generating operation. This can be seen from calculating how the two-mode squeezing operator acts on a coherent state of two modes. Using the Weyl displacement operator~(\ref{eqn:weyloperator}), the two mode coherent state can be written as $\ket{\alpha_{A},\alpha_{B}}\equiv\hat{D}(\alpha_{A})\hat{D}(\alpha_{B})\ket{0}$ and is, by definition, separable. From the definition of a coherent state~(\ref{eqn:coherent-state}) we find that it is represented by the ($4\times 4$) covariance matrix
\begin{eqnarray}
\boldsymbol{\Gamma}=\boldsymbol{I}.
\end{eqnarray}
We can therefore use the transformation rule~(\ref{eqn:symp-transformation}) to find how the two-mode squeezing operator transforms a coherent state. Denoting the transformed state as $\boldsymbol{\Gamma}(r)=\boldsymbol{S}\boldsymbol{S}^{\dag}$ we find
\begin{eqnarray}
\boldsymbol{\Gamma}(r)&=&\left(
\begin{array}{cc}
\label{eqn:two-mode-squeezed-state}
\cosh(2r)\boldsymbol{I} & \sinh(2r)\boldsymbol{\sigma}_{1} \\
\sinh(2r)\boldsymbol{\sigma}_{1} & \cosh(2r)\boldsymbol{I}
\end{array} \right).
\end{eqnarray}
This state is called a \emph{two-mode squeezed state}. Two-mode squeezed states are pure bipartite states and so their entanglement can be fully described via the entropy of entanglement. However, we first need to define how our entanglement measures can be computed in the Gaussian state formalism.

\section{Measures of Entanglement}

In this section, we shall review the measures of entanglement that are important in continuous variable systems. The power of continuous variable and covariance matrices is made explicit by the simple form of many measures of entanglement. 

First, however, we will define the partial tracing for Gaussian states. As in standard quantum mechanics, this is most easily shown through an example. In the real form basis~(\ref{eqn:gerardo-basis}), a three mode state can be written as:
\begin{eqnarray}
\boldsymbol{d}=\begin{pmatrix}
\boldsymbol{d}_{A} \\ \boldsymbol{d}_{B} \\ \boldsymbol{d}_{C}
\end{pmatrix},\,\,\boldsymbol{\Gamma}_{ABC}=\begin{pmatrix}
\boldsymbol{\Gamma}_{A} & \boldsymbol{\Gamma}_{AB} & \boldsymbol{\Gamma}_{AC} \\
\boldsymbol{\Gamma}_{AB}^{\mathrm{tp}} & \boldsymbol{\Gamma}_{B} & \boldsymbol{\Gamma}_{BC} \\
\boldsymbol{\Gamma}_{AC}^{\mathrm{tp}} & \boldsymbol{\Gamma}_{BC}^{\mathrm{tp}} & \boldsymbol{\Gamma}_{C}
\end{pmatrix},
\end{eqnarray}
where $\boldsymbol{d}_{j}$ are the first moments of system-$j$, $\boldsymbol{\Gamma}_{j}$ is the variance matrix for system-$j$ and $\boldsymbol{\Gamma}_{jj'}$ are the covariance matrices between the systems-$j,j'$. In the above example, the reduced state between the $A,B$ subsystems is obtained by omitting all information about the $C$ subsystem. In other words, denoting the partial trace over the $C$ subsystem as $\tr_{C}$, we have
\begin{eqnarray}
\label{eqn:gaussian-partial-trace}
\tr_{C}(\boldsymbol{d})=\begin{pmatrix}
\boldsymbol{d}_{A} \\ \boldsymbol{d}_{B}
\end{pmatrix},\,\,\tr_{C}(\boldsymbol{\Gamma}_{ABC})=\begin{pmatrix}
\boldsymbol{\Gamma}_{A} & \boldsymbol{\Gamma}_{AB} \\
\boldsymbol{\Gamma}_{AB}^{\mathrm{tp}} & \boldsymbol{\Gamma}_{B}
\end{pmatrix}.
\end{eqnarray}
The Gaussian partial trace can naturally be extended to more modes and can be performed as many times as needed. It is, of course, a completely positive, trace preserving map like its Hilbert space counterpart and also preserves the Gaussian nature of the states. We can now define our entanglement measures.

\subsection{Von Neumann Entropy}

As in standard quantum information, the Von Neumann entropy can be used to quantify the entanglement between pure, bipartite Gaussian states~(\ref{enq:vne}). In the continuous variable formalism, the Von Neumann entropy can be written terms of the symplectic spectrum of a given covariance matrix $\boldsymbol{\Gamma}$ as
\begin{defn}
Let $\boldsymbol{\boldsymbol{\Gamma}}$ be the covariance matrix that represents a Gaussian state. The Von Neumann entropy of $\boldsymbol{\boldsymbol{\Gamma}}$ is defined as~\cite{holevo1999,serafini2004}
\begin{eqnarray}
\label{eqn:vne-Gaussian}
\mathrm{S}(\boldsymbol{\Gamma})&=&\sum_{k=1}^{N}f(\nu_{k}),
\end{eqnarray}
where $\nu_{k}$ are the symplectic eigenvalues of $\boldsymbol{\Gamma}$ and 
\begin{eqnarray}
f(x)\equiv \frac{x+1}{2}\log\left( \frac{x+1}{2} \right)-\frac{x-1}{2}\log\left( \frac{x-1}{2} \right).
\end{eqnarray}
\end{defn}
It should be noted the definition of the entropy of entanglement for a pure bipartite state remains the same as in~(\ref{eqn:entropy-of-entanglement}). All we need do is replace the method of computing the Von Neumann entropy for a Gaussian state with the above definition.

\subsection{Negativity Measures}\label{subsec:gasussian-negs}

The simplest measures that can be computed for mixed Gaussian sates are the negativity-type measures. They rely on the positive partial transpose (PPT) criterion~\cite{peres1996}. We therefore need to represent the partial transposition operation in terms of a covariance matrix. As we will be concerned only with bipartite systems, we restrict our definitions accordingly.
\begin{defn}
For a given bipartite Gaussian state, represented by a covariance matrix $\boldsymbol{\Gamma}$, we define its partial transposition with respect to one of the two subsystems as
\begin{eqnarray}
\tilde{\boldsymbol{\Gamma}}=\boldsymbol{P}\boldsymbol{\Gamma}\boldsymbol{P}
\end{eqnarray}
where we have defined the symplectic partial transposition matrix as, with respect to the complex form basis~(\ref{eqn:mode-op-basis}),
\begin{eqnarray}
\label{eqn:gaussian-pt}
\boldsymbol{P}=\left(\begin{array}{cccc}
1 & 0 & 0 & 0 \\
0 & 0 & 0 & 1 \\
0 & 0 & 1 & 0 \\
0 & 1 & 0 & 0 
\end{array}\right).
\end{eqnarray}
\end{defn}
It should be pointed our that the partial transposition matrix has a different form in each representation of $\mathrm{Sp}(2N,\mathbb{R})$. This has no physical consequence but for completeness we write explicitly their form in the $\hat{\boldsymbol{X}}$ basis~(\ref{eqn:gerardo-basis}) and $\hat{\boldsymbol{Y}}$ basis~(\ref{eqn:quadrature-basis}) as $\boldsymbol{P}_{\boldsymbol{X}}$ and  $\boldsymbol{P}_{\boldsymbol{Y}}$ respectively,
\begin{eqnarray}
\boldsymbol{P}_{\boldsymbol{X}}=\boldsymbol{I}\oplus\boldsymbol{\sigma}_{3},\,\,\boldsymbol{P}_{\boldsymbol{Y}}=\boldsymbol{I}\oplus\boldsymbol{\sigma}_{1}.
\end{eqnarray}
The transformation~(\ref{eqn:gaussian-pt}), with out loss of generality, partially transposes the Gaussian state with respect to the \emph{second} mode. We now have everything we need to quantify bipartite Gaussian state entanglement in terms of either the negativity or the logarithmic negativity~\cite{serafini2004}.
\begin{defn}
Let $\boldsymbol{\boldsymbol{\Gamma}}$ be the covariance matrix representing a bipartite Gaussian state. The negativity of  $\boldsymbol{\boldsymbol{\Gamma}}$ is given by
\begin{eqnarray}
\label{eqn:neg-gaussian}
\mathcal{N}(\boldsymbol{\boldsymbol{\Gamma}})=\frac{1}{2}\mathrm{max}\left[\frac{1-\tilde{\nu}}{\tilde{\nu}},0\right],
\end{eqnarray}
where $\tilde{\nu}$ is the smallest symplectic eigenvalue of the partial transposition of $\boldsymbol{\boldsymbol{\Gamma}}$.
\end{defn}
\begin{defn}
Let $\boldsymbol{\boldsymbol{\Gamma}}$ be the covariance matrix representing a bipartite Gaussian state. The logarithmic negativity of  $\boldsymbol{\boldsymbol{\Gamma}}$ is given by
\begin{eqnarray}
\label{eqn:log-neg-gaussian}
\mathcal{E}_{\mathcal{N}}(\boldsymbol{\Gamma})&=&\mathrm{max}\left[-\log(\tilde{\nu}),0\right],
\end{eqnarray}
where $\tilde{\nu}$ is the smallest symplectic eigenvalue of the partial transposition of $\boldsymbol{\boldsymbol{\Gamma}}$.
\end{defn}
\begin{figure}[t!]
\centering
\includegraphics[width=0.7\linewidth]{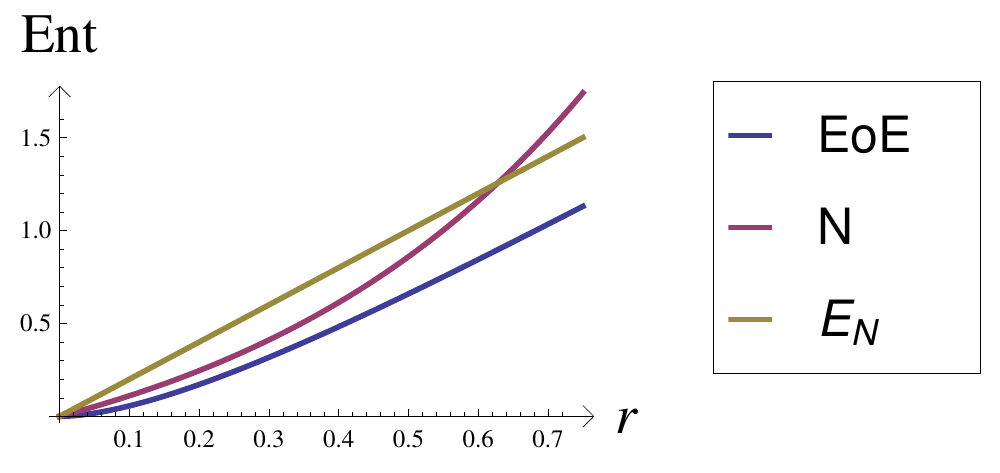}
\caption[Entanglement of Two-Mode Squeezed State]{Plot of different entanglement measures for the two-mode squeezed state as a function of the squeezing parameter $r$. The blue line is the entropy of entanglement, the purple line is the negativity and the mustard line is the logarithmic negativity.}
\label{fig:2ms-plots}
\end{figure}
These negativities will be used to quantify entanglement between localised systems in Sections~(\ref{sec:resonance})~and~(\ref{sec:teleportation-results}). As an example of how our entanglement measures work for Gaussian states, we compute the entropy of entanglement, negativity and logarithmic negativity for the two-mode squeezed state~(\ref{eqn:two-mode-squeezed-state}). To compute the entropy of entanglement we need to find the reduced state of either the first or second mode of the two-mode squeezed state. Using Eq.~(\ref{eqn:gaussian-partial-trace}), we can easily find the reduced state of a Gaussian state. In the case of a two-mode squeezed state the reduced state of the B (or A) system is
\begin{eqnarray}
\boldsymbol{\Gamma}_{B}(r)=\cosh(2r)\boldsymbol{I}.
\end{eqnarray}
The symplectic spectrum of $\boldsymbol{\Gamma}_{B}(r)$, computed via~(\ref{eqn:symplectic-spectrum-complex}), consists of the single eigenvalue~$\nu=\cosh(2r)$. Thus, using the definition of the Von Neumann entropy~(\ref{eqn:vne-Gaussian}), we find the entropy of entanglement for a two-mode squeezed state is
\begin{eqnarray}
\label{eqn:2mss-entanglement}
\mathrm{E}(\boldsymbol{\Gamma}(r))=\cosh^{2}(r)\log\cosh^{2}(r)-\sinh^{2}(r)\log\sinh^{2}(r).
\end{eqnarray}
For the negativity and logarithm negativity, we need to compute the smallest symplectic eigenvalue of the partially transposed state $\tilde{\boldsymbol{\Gamma}}(r)$~(\ref{eqn:two-mode-squeezed-state}). Using the partial transposition definition~(\ref{eqn:gaussian-pt}) and the symplectic spectrum rule~(\ref{eqn:symplectic-spectrum-complex}), we find the symplectic spectrum of the partially transposed two-mode squeezed state~(\ref{eqn:two-mode-squeezed-state}) is
\begin{eqnarray}
\boldsymbol{\nu}=\lbrace e^{+2r},e^{-2r} \rbrace .
\end{eqnarray}
Identifying the smallest partially transposed symplectic eigenvalue as $\tilde{\nu}=e^{-2r}$, the negativity~(\ref{eqn:neg-gaussian}) and logarithmic negativity~(\ref{eqn:log-neg-gaussian}) give respectively
\begin{subequations}
\begin{align}
\mathcal{N}&=\frac{e^{+2r}-1}{2},\\
\mathcal{E}_{\mathcal{N}}&=+2r.
\end{align}
\end{subequations}
We have plotted the three entanglement measures against each other in Fig.~(\ref{fig:2ms-plots}).

Interestingly, the \emph{squeezing parameter} $r$, which can take any real value, allows the entanglement of a two-mode squeezed state to be infinite. This can be traced back to the infinite dimensional nature of continuous variable systems. Two-mode squeezed states play a central role in Gaussian state quantum information and are used in continuous variable teleportation which discuss in what follows

\section{Gaussian State Teleportation}\label{sec:cv-teleportation}

Here we outline the Gaussian state teleportation protocol. Quantum teleportation forms a cornerstone of quantum information for qubits and therefore understanding how it passes over to Gaussian states is also of great importance. Gaussian states are particularly useful for our purposes as they have great experimental accessibility and are immediately applicable to the Klein-Gordon (or electromagnetic) field. This makes Gaussian state quantum information a natural choice to use in relativistic quantum information. Moreover, in Chapter~(\ref{chapter:bosons}), Section~(\ref{sec:teleportation-results}), we will consider relativistic effects on moving cavities. The original continuous variable teleportation protocol was devised by Braunstein \emph{et.al.}~\cite{braunstein1998}. We follow~\cite{adesso2005,mari2008} for its application to Gaussian states.

For our discussion, it is more convenient to work in the $\hat{\boldsymbol{X}}$ basis~(\ref{eqn:gerardo-basis}). The teleportation procedure for Gaussian states is as follows. Alice and Bob meet and prepare an entangled state, represented by the covariance matrix 
\begin{eqnarray}
\boldsymbol{\Gamma}=\begin{pmatrix}
\boldsymbol{A} & \boldsymbol{C} \\
\boldsymbol{C}^{\mathrm{tp}} & \boldsymbol{B}
\end{pmatrix},
\end{eqnarray}
between a mode controlled by Alice and a mode controlled by Rob. Rob then leaves taking his part of the entangled state and undergoes some form of motion. During this time, Alice obtains (by some means) an \emph{unknown} single mode coherent state. She wishes to teleport the unknown state to Rob using the entanglement shared between them. First, she performs a \emph{beam splitting} operation on her mode and the unknown state in her possession. Subsequently, Alice performs what is known as a \emph{homodyne} measurement of her mode and the input state to obtain classical bits, in analogy with the qubit case, see Section~(\ref{sec:teleportation-qmqi}).\footnote{Homodyne detection of a single mode of light is a well-used experimental method. It uses a strong coherent state and beam splitter to measure different quadratures of the mode under inspection. For details see~\cite{leonhardt1997}} She finally communicates the measurement results via classical communication to Rob who then performs a final displacement operation on his mode to obtain the unknown input coherent state. 

As stated in Section~(\ref{sec:teleportation-qmqi}), the usual figure of merit for teleportation is the fidelity between the input state and the teleported output state. In our scheme of Gaussian teleportation, the initial state is an unknown coherent state, which is pure. Thus, the computation of the fidelity is between a pure input state and, possibly, a mixed output state. In terms of the original shared entanglement resource state $\boldsymbol{\Gamma}$, the fidelity in this case reads~\cite{mari2008}
\begin{eqnarray}
\mathcal{F}=\dfrac{2}{\sqrt{4+2\tr(\boldsymbol{N})+ \det(\boldsymbol{N})}},
\end{eqnarray}
where the matrix $\boldsymbol{N}$ is defined as
\begin{eqnarray}
\boldsymbol{N}=\boldsymbol{\sigma}_{3}\boldsymbol{A}\boldsymbol{\sigma}_{3}+\boldsymbol{\sigma}_{3}\boldsymbol{C}+\boldsymbol{C}^{\mathrm{tp}}\boldsymbol{\sigma}_{3}+\boldsymbol{B},
\end{eqnarray}
and $\boldsymbol{\sigma}_{3}$ is the third Pauli matrix. This expression will form the basis of Section~(\ref{sec:teleportation-results}) where we investigate how the motion of Bob affects the fidelity of Gaussian teleportation.

This concludes our mathematical preliminaries chapters. We can now present recent results regarding cavities moving through spacetime and how Unruh-DeWitt detectors can be used for quantum information purposes.

\chapter*{Part II}
\addcontentsline{toc}{chapter}{Part II}

\chapter{Moving Cavities for Relativistic Quantum Information: Bosons}\label{chapter:bosons}



\section{Introduction}\label{sec:boson-resonance}

In this chapter, we study the effects of gravity and motion on entanglement and how this affects quantum teleportation. Moreover, we show how to use relativistic motion to implement a quantum gate. In order to study the effects of motion in quantum information it is necessary to use quantum field theory. This will allow us to properly incorporate both quantum and relativistic effects. Our starting point is to find localised systems in quantum field theory to store information. The need to use localised systems comes from a very physical motivation. When using quantum systems, it is critical that one can prepare states and perform accurate measurements on them. This only makes sense when the system one is dealing with is confined to a finite region of spacetime. A quantum optical device called a \emph{cavity} is widely used to store quantum information in a localised manner. As such, we will consider cavities as our localised quantum systems which contain quantum fields. Cavities are best described as perfectly reflecting mirrors which trap quantum fields. Not only do they describe the localisation of a quantum system to a finite region of spacetime, they are also accessible to experiments. This fact lead to the Nobel prize for physics in 2012 being awarded to the pioneering efforts of Haroche (shared with Wineland)~\cite{nobelprize2012}. There are plans to use satellites for quantum communication and cryptography purposes along with testing the laws of nature~\cite{ursin2009,rideout2012}. As satellites are under the influence of both gravity and motion in orbit, a proper account of entanglement at these scales must be made.

Our main aim is to understand how does relativity affect quantum information and find ways to use relativistic effects to improve quantum information tasks. We will show that is its possible to generate observable amounts of entanglement via motion and that motion has observable effects on quantum information protocols. These two questions will be addressed in this chapter.

This chapter is organised into two sections. The first describes the basic set-up for a Klein-Gordon field confined to a cavity. We use this set-up to analyse the entanglement generated between modes of the cavity due to its motion through spacetime. We find that under certain conditions, the entanglement generated can be maximised via resonances produced by periodic motion. As such, the motion of the cavity implements a \emph{quantum entanglement gate}. Quantum gates form an important part of quantum computing~\cite{nielsen2010}. They are used to implement quantum computing circuits which can be used to run quantum algorithms. Therefore, cavities could open the door to implementing quantum gates via their motion through spacetime or, colloquially, ``gates by moving" (coined by Ivette Fuentes).

The second section studies the effect of motion on continuous variable teleportation. Quantum teleportation is a method of transferring quantum states and so has great significance in communications. If quantum communication is to be achieved between satellites and over large distances then the effects of motion and gravity must be accounted for. We analyse the fidelity of continuous variable quantum teleportation in the situation where one of the teleportation parties is moving through spacetime.

\section{Entanglement Resonances and Gates}

\subsection{Introduction}

Understanding how motion and gravity affect entanglement is a key feature in the implementation of new quantum information technologies, including quantum cryptography and teleportation, in space-based scenarios that are currently under investigation~\cite{villarosi2008}. Quantifying entanglement in situations where motion or gravitation have a significant role can also provide guidance for theories about the microscopic structure of spacetime, via the Hawking-Unruh effect and its connections to thermodynamics and statistical mechanics~\cite{hawking1975,unruh1976}. Recent work in relativistic quantum information has shown that non-uniform motion of a cavity creates entanglement between the cavity field modes~\cite{friis2012-2,friis2012-3}. In this section we show that by repeating any trajectory segment periodically it is possible to generate a resonance creating higher amounts of entanglement for small cavities and large accelerations or large cavities and small accelerations. Non-uniform motion induces transformations on the states which are equivalent to two-mode squeezing operations upto local unitary transformations. These transformations play the role of two-mode quantum gates in continuous variables systems~\cite{weedbrook2012}.

Finding suitable ways to store and process information in a quantum and relativistic setting is a main goal in the field of relativistic quantum information. Moving cavities are good candidates to store information~\cite{alphacentauri2012,friis2012,downes2011} since confined fields can be realised experimentally and observers can directly access their states by means of local operations. When a cavity is accelerated for a finite time the cavity modes are affected by the motion. A mismatch between the vacua at different times gives rise to the creation of particles which populate the modes~\cite{birrell1982}. The initial and final modes are related through Bogoliubov transformations which mix all frequency modes~\cite{alphacentauri2012,friis2012-2} generating entanglement~\cite{friis2012-2,friis2012-3}. Quantum correlations are created between all modes; however, higher degrees of entanglement are produced between oddly separated modes. As a consequence, maximally entangled states of modes confined in two different cavities degraded when the cavities undergo non-uniform motion~\cite{alphacentauri2012}.

We mention that two-mode squeezed states can be produced in a single cavity by the periodic repetition of any trajectory segment. Entanglement resonances occur when the frequency associated to the segment travel time is equal to the sum of two oddly separated cavity mode frequencies. We show analytically that for \textit{any\/} couple of oddly separated modes it is possible to find a segment travel time where the entanglement between such modes increases linearly with the number of repetitions. These resonances appear independently of the details of the trajectory though the amount of entanglement generated does depend on the trajectory itself.

As a concrete example we present a travel scenario which allows for simple analytical expressions. The travel segment which we repeat starts with a period of acceleration followed by inertial coasting. The cavity then accelerates in the same or opposite direction for a second period of time and finally undergoes a second period of inertial coasting. We find the conditions for the resonances to occur and investigate how the magnitude and the direction of the accelerations affect the final entanglement generated. In the special case where there is no inertial coasting and the accelerations alternate in direction, our sample trajectory reduces to the oscillatory motion that is often considered in the dynamical Casimir effect literature~\cite{dodonov1990}.

\subsection{Field Quantisation}\label{sec:resonance}

We consider a real scalar field $\phi$ of mass $m$ contained within a cavity in $(1+1)$-dimensional spacetime. The massless field can be treated as a special case of our study and the effect of additional transverse dimensions can be included as a positive contribution to $m$.  The cavity follows a worldtube which is composed of periods of inertial and uniformly accelerated motion.  We will start by describing the field within the cavity during such periods as seen by a co-moving observer. 

We wish to construct standing wave solutions of the field contained within a stationary cavity. The cavity walls are placed at $x=a$ and $x=b$ where $0< a < b$, so that $\delta=b-a$ is the length of the cavity. We impose Dirichlet boundary conditions on the field mode by requiring the field to vanish at the cavity walls. Using the free field planes wave basis~(\ref{eqn:kg-plane-waves}), a standing wave solution can be found by superimposing a left travelling mode with a right travelling mode as
\begin{eqnarray}
\phi_{n}(t,x)=A\phi_{k}(t,x)+B\phi_{-k}(t,x),
\end{eqnarray}
where we have added the subscript $n$ in anticipation of the momentum being discretised. Our boundary conditions are
\begin{eqnarray}
\phi_{n}(t,a)=\phi_{n}(t,b)=0,
\end{eqnarray}
and thus we find the positive frequency standing wave mode functions with respect to the time Killing vector field $\partial_t$ are:
\begin{subequations}
\begin{align}
\label{MinkowskiSolutions}
\phi^{M}_n (t,x) &= \frac{1}{\sqrt{\omega_{n}L}}\sin\left[\frac{n\pi}{L}(x-a)\right]e^{-i\omega_{n}t} \ ,
\\ \omega_{n} &= \sqrt{{(n\pi/L)}^2+m^2} \ ,
\end{align}
\end{subequations}
where $n\in\mathbb{N}$. Note that we have normalised the solutions with respect to the conserved inner product~(\ref{eqn:kg-minkowski-inner-product}).
 
Our investigation requires us to analyse how non-uniform motion through spacetime affects quantum entanglement. We do this by approximating non-uniform motion by periods of inertial and accelerated motion. As we will see in the next subsection~(\ref{sec:bogo-boson-graft}), this is done by relating Minkowski modes to Rindler modes via Bogoliubov transformations. By successively transforming from reference frame to another, we can build general trajectories which model the motion of a cavity through spacetime.

We employ Rindler coordinates $(\eta,\chi)$ to describe the field during periods of uniformly accelerated motion. The cavity walls are placed at $\chi=a$ and $\chi=b$ and the proper time and acceleration at the centre of the cavity are given by $\tau$ and $2/(b+a)$, respectively. We write the positive energy modes with respect to timelike Killing vector $\partial_{\eta}$ for the massive Klein-Gordon field~(\ref{eqn:kg-equation-rindler}) as
\begin{eqnarray}
\label{eq:general bessel solutions}
\phi_{\Omega}(\eta,\chi)=\left(A_{\Omega}\text{I}_{i\Omega}(m\chi)+B_{\Omega}\text{I}_{-i\Omega,}(m\chi)\right)e^{-i\Omega\eta}
\end{eqnarray}
where $\text{I}_{\pm i\Omega}(m\chi)$ are the modified Bessel functions of the first kind. Imposing the Dirichlet boundary conditions
\begin{eqnarray}
\phi_{\Omega}(\eta,a)=\phi_{\Omega}(\eta,b)=0,
\end{eqnarray}
we obtain the positive frequency Rindler standing wave modes solutions with respect to $\partial_{\eta}$
\begin{equation}
\label{RindlerSolutions2}
\phi^{R}_{n} (\eta,\chi) =  N_{k} \bigl[I_{i\Omega_{n}}(m\chi)I_{-i\Omega_{n}}(m x_{A}) - I_{-i\Omega_{n}}(m\chi)I_{i\Omega_{n}}(mx_{A})\bigr]e^{-i\Omega_{n}\eta}\ ,
\end{equation}
where $N_{n}$ are normalisation constants found by using the Rindler Klein-Gordon inner product~(\ref{eqn:kg-rindler-inner-product}). In the massless case the mode functions and the frequencies reduce to simple expressions (see~\cite{alphacentauri2012}). Note the frequencies $\Omega_{n}$ have acquired a subscript. This is due to the discretisation of the spectrum imposed by the cavity boundary conditions. For the massive scalar field, closed expressions do not exist for $\Omega_{n}$ or $N_{n}$ in contrast to the massless case. In the following, we will use perturbation approximations to the frequencies $\Omega_{n}$ and normalisation constants $N_{n}$ in terms of small perturbation parameter.

\subsection{Grafting Trajectory Segments}\label{sec:bogo-boson-graft}

\begin{figure}[t]
\begin{center}
\includegraphics[width=0.5\textwidth]{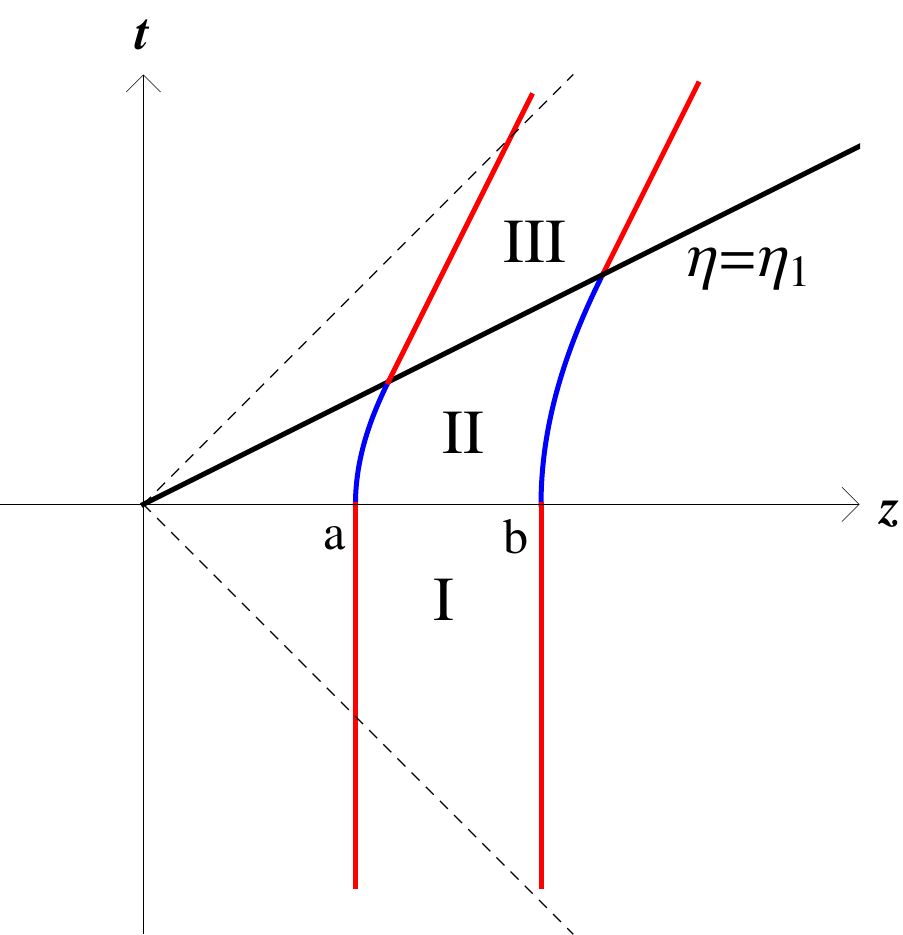}
\caption[Cavity Basic Building Block]{Space-time diagram of cavity motion is shown.
Rob's cavity is at rest initially (Region~$\mathrm{I}$),
then undergoes a period of uniform acceleration from $t=0$ to $\eta=\eta_{1}$
(Region $\mathrm{I\!I}$) and is thereafter
again inertial (Region $\mathrm{I\!I\!I}$).}
\label{fig:bogobogoboxes-boson}
\end{center}	
\end{figure}

Our procedure for modelling the motion of a cavity through spacetime involves three distinct regions, denoted as (I), (II) and (III). We call the first region, denoted by (I), the ``in" region i.e. the region where the cavity is initially at rest before it begins to move. The region denoted by (II) is where the cavity is undergoing uniform acceleration. After some finite period of time, the cavity stops accelerating and becomes inertial again in region (III), which we call the ``out" region. The Bogoliubov transformation which relates the ``in" region modes with the ``out" region modes allows us to compute the state of the cavity after it has undergone the non-uniform motion. For a schematic diagram of the travel scenario, see Fig.~(\ref{fig:bogobogoboxes-boson}).

Following the canonical quantisation procedure for fields, we write the quantised field operators in each region as
\begin{subequations}
\label{eq:regionsI-IIquantization-bosons}
\begin{align}
\mathrm{I}:\ \ \
\hat{\phi} &=	\sum\limits_{n=1}^{\infty} \bigl(\phi^{M}_{n} \hat{a}_{n} + \text{h.c.} \bigr),
\label{eq:region I quantization2}\\[2mm]
\mathrm{I\! I}:\ \ \
\hat{\phi} &=	\sum\limits_{n=1}^{\infty} \bigl(\phi^{R}_{n} \hat{A}_{n} + \text{h.c.} \bigr),
\label{eq:region II quantization2}\\[2mm]
\mathrm{I\! I\! I}:\ \ \
\hat{\phi} &=	\sum\limits_{n=1}^{\infty} \bigl(\tilde{\phi}^{M}_{n} \tilde{a}_{n} + \text{h.c.} \bigr),
\label{eq:region III quantization2}
\end{align}
\end{subequations}
with commutation relations
\begin{subequations}
\begin{align}
\mathrm{I}:&\ \ \
\left[\hat{a}_{n},\hat{a}_{m}^{\dag} \right] =	\delta_{nm}, \\
\mathrm{I\! I}:&\ \ \
\left[\hat{A}_{n},\hat{A}_{m}^{\dag} \right] =	\delta_{nm}, \\
\mathrm{I\! I\! I}:&\ \ \
\left[\tilde{a}_{n},\tilde{a}_{m}^{\dag} \right] =	\delta_{nm},
\end{align}
\end{subequations}
where in region (III) the modes have been adapted to new inertial coordinates, $(\tilde{t},\tilde{z})$, which are described naturally by an observer comoving with the cavity. 

We will work in the covariant matrix formalism which is applicable to systems consisting of discrete Bosonic modes as long as the analysis is restricted to Gaussian states, see Chapter~(\ref{chapter:cv}).

Changes from inertial to accelerated motion and vice versa are implemented by the action of Bogoliubov transformations.  Consider that at $t=0$ a cavity initially at rest begins to accelerate. From Section~(\ref{sec:bogo-qft}), we can compute the Bogoliubov transformation between the Rindler and Minkowski Klein-Gordon modes using the Klein-Gordon inner product evaluated at $t=0$~\cite{birrell1982,fabbri2005,crispino2008}. We can identify a small expansion parameter $h=2\delta/(b+a)$ to perform a perturbative analysis of the Bogoliubov transformations (i.e. $h\ll 1$). The Bessel functions in Eq.~(\ref{RindlerSolutions2}) are difficult to manipulate analytically. However, the coefficients can be computed using uniform asymptotic expansions of the Bessel functions~\cite{NIST:DLMF,dunster1990}. The results are
\begin{eqnarray}
\label{eqn:fundamental-bogos-matrix}
\boldsymbol{\alpha} &=& \boldsymbol{I} + \boldsymbol{\alpha}^{(1)}h + O(h^2),\\
\boldsymbol{\beta} &=& \boldsymbol{\beta}^{(1)}h + O(h^2),
\end{eqnarray}
where
\begin{subequations}
\label{eqn:fundamental-bogos}
\begin{align}
\alpha_{nn}^{(1)}&=0,\\
\alpha_{nm}^{(1)}&=\frac{\left[-1+(-1)^{m+n}\right]}{\pi^{4}(m^{2}-n^{2})^{3}}mn\left[\frac{(M^{2}+\pi^{2}n^{2})^{1/4}\left[\pi^{2}(n^{2}+3m^{2})+4M^{2}\right]}{(M^{2}+\pi^{2}m^{2})^{1/2}}\right.\\
&\hspace{5mm}+\left.\frac{(M^{2}+\pi^{2}m^{2})^{1/4}\left[\pi^{2}(m^{2}+3n^{2})+4M^{2}\right]}{(M^{2}+\pi^{2}n^{2})^{1/2}}\right],\notag\\
\beta_{nn}^{(1)}&=0,\\
\beta_{nm}^{(1)}&=\frac{\left[-1+(-1)^{m+n}\right]}{\pi^{4}(m^{2}-n^{2})^{3}}mn\left[\frac{(M^{2}+\pi^{2}n^{2})^{1/4}\left[\pi^{2}(n^{2}+3m^{2})+4M^{2}\right]}{(M^{2}+\pi^{2}m^{2})^{1/2}}\right.\\
&\hspace{5mm}-\left.\frac{(M^{2}+\pi^{2}m^{2})^{1/4}\left[\pi^{2}(m^{2}+3n^{2})+4M^{2}\right]}{(M^{2}+\pi^{2}n^{2})^{1/2}}\right]\notag,
\end{align}
\end{subequations}
Note that here $n,m\in\mathbb{Z}^{+}$ and $M:=\mu\delta/c^{2}$ is the ``dimensionless" mass of the field and $\mu$ is the bare mass of the scalar field. At this point a comment on the physical interpretation and relative size of $h$ is suitable. The dimensionless parameter $h$ is the product of the length of the cavity and the acceleration at the \emph{centre} of the cavity. Reinstating the appropriate factors of $c$ (the speed of light in a vacuum) gives us
\begin{equation}
h=\frac{2\delta}{c^{2}(b+a)}.
\end{equation}
We can immediately see that the factor of $c^{2}$ in the denominator implies that only for extremely large cavities or accelerations will $h>1$. As an example, for a cavity approximately $1$m in length, the allowed acceleration at the centre of the cavity for which $h<0.01$ is of the order $10^{14}\mathrm{ms}^{-2}$. Thus, the \emph{proper acceleration} at the centre our cavities can be extremely large.

Continuing, as stated in Section~(\ref{sec:bogo-qft}), we can collect the Bogoliubov coefficients into matrices which act on the cavity modes and operators. Denoting the matrix which transforms the modes as $\boldsymbol{V}$ and the matrix which transforms the mode operators as $\boldsymbol{Q}$, we can write the Bogoliubov transformations in the complex form of the symplectic group (see~(\ref{sec:symplectic-representations})), dropping the $(c)$ subscript, as
\begin{eqnarray}
\label{eqn:bogo-matrices-bosons}
\boldsymbol{V}=
\left(
\begin{array}{cc}
\boldsymbol{\alpha} & \boldsymbol{\beta} \\
\bar{\boldsymbol{\beta}} & \bar{\boldsymbol{\alpha}}
\end{array}
\right),\hspace{2mm}
\boldsymbol{Q}=
\left(
\begin{array}{cc}
\bar{\boldsymbol{\alpha}} & -\bar{\boldsymbol{\beta}} \\
-\boldsymbol{\beta} & \boldsymbol{\alpha}
\end{array}
\right).
\end{eqnarray}
Both $\boldsymbol{V}$ and $\boldsymbol{Q}$ are necessarily sympletic with associated symplectic form
\begin{eqnarray}
\boldsymbol{\Omega}=-i\boldsymbol{K},
\end{eqnarray}
where $\boldsymbol{K}$ is defined in Eq.~(\ref{eqn:complex-form-defs}). For periods of either inertial or uniform acceleration, the Hamiltonian that governs the dynamics of the cavity does not involve interactions between the cavity modes. Therefore, the Hamiltonian is nothing more than the free Hamiltonian of the field. This free Hamiltonian evolution induces phase rotations in each mode, which does not change the particle content or entanglement between modes. These phase rotations are represented via the symplectic matrix
\begin{eqnarray}
\boldsymbol{U}=\bar{\boldsymbol{G}}\oplus\boldsymbol{G},
\end{eqnarray}
where $\boldsymbol{G}=\mathrm{diag}\left(e^{i\theta_{1}},e^{i\theta_{2}},\ldots \right)$. The angles are given by $\theta_{n}=\omega_{n}t$ during coasting periods and  $\theta_{n}=\Omega_{n}\eta$ during acceleration. Once parametrised by the proper time at the centre of the cavity, the accelerated phase rotation angles read
\begin{eqnarray}
\Omega_{n}\eta = \omega_{n}\tau +O(h^{2}),
\end{eqnarray}
thus the phase rotations for both free inertial and accelerated motion take the same form to first order in $h$.

We can compose the Bogoliubov matrix $\boldsymbol{Q}$ with free evolution for propertime $\tau_{j}$ to create \emph{basic building blocks} of motion as
\begin{eqnarray}
\label{eqn:symplectic-building-block}
\boldsymbol{S}_{j}:=\boldsymbol{Q}^{-1}(h_{j})\boldsymbol{U}(\tau_{j})\boldsymbol{Q}(h_{j}).
\end{eqnarray}
The matrix $\boldsymbol{S}_{j}$ tells us how mode operators in region (I) are related to mode operators in region (III) after undergoing acceleration for some propertime $\tau_{j}$. The inverse of the matrix $\boldsymbol{Q}$ represents the transformation from accelerated to inertial i.e. the \emph{inverse transformation} of inertial to accelerated motion. Note that in the case $h=0$, $\boldsymbol{S}_{j}=\boldsymbol{U}(\tau_{j})$ representing a period of inertial motion and in the case $\tau_{j}=0$ the transformation is trivial i.e. there was no motion. Using~(\ref{eqn:symplectic-building-block}), a general travel \emph{segment} can be constructed as
\begin{eqnarray}
\label{eq:general-segment}
\boldsymbol{S}(\lbrace h_{j};\tau_{j}\rbrace):=\prod_{j}\boldsymbol{S}_{j}.
\end{eqnarray}
This matrix represents any number of basic building blocks and can be composed in an arbitrary way to approximate a general trajectory through spacetime. An interesting, and useful, consequence of this product decomposition is that the zero order contribution to~(\ref{eq:general-segment}) takes the form
\begin{eqnarray}
\boldsymbol{S}^{(0)}=\bar{\boldsymbol{G}}^{(0)}(T)\oplus\boldsymbol{G}^{(0)}(T),
\end{eqnarray}
where $T:=\sum_{j}\tau_{j}$ is the \emph{total} proper time of the travel segment. The propertime can be both the propertime while inertial (where it coincides with Minkwoski coordinate time $t$) or while uniformly accelerating.

While working within a single cavity, we can take advantage of the fact that the reduced state between two modes $k,k'$ only depends on the Bogoliubov transformation between those two modes. This is only true, however, while working to first order in $h$. We consider the state between two modes only for simplicity and illustration. One could choose to investigate multi-partite correlations and is the subject of~\cite{friis2012-4}. The derivation of a pure reduced state to first order in $h$ can be found in appendix~(\ref{app:Derivation of two-mode transformation}). After a single travel scenario we can write the reduced state between the modes $k,k'$ as
\begin{eqnarray}
\label{eq:one-travel-scenario}
\boldsymbol{\Gamma}_{kk'}=\boldsymbol{s}_{kk'}\boldsymbol{\Gamma}_{0}\boldsymbol{s}_{kk'}^{\dag},
\end{eqnarray}
where $\boldsymbol{\Gamma}_{0}$ is the initial state between the two modes $k,k'$ and Bogoliubov matrix takes the form
\begin{eqnarray}
\label{nico}
\boldsymbol{s}_{kk'}&=&
\left(
\begin{array}{cc}
\bar{\boldsymbol{A}} & -\bar{\boldsymbol{B}} \\
-\boldsymbol{B} & \boldsymbol{A}
\end{array} 
\right), 
\end{eqnarray}
with
\begin{eqnarray}
\boldsymbol{A}=
\left(
\begin{array}{cc}
A_{kk} & A_{kk'} \\
A_{k'k} & A_{k'k'}
\end{array} 
\right)\!\!, 
\boldsymbol{B}=
\left(
\begin{array}{cc}
B_{kk} & B_{kk'} \\
B_{k'k} & B_{k'k'}
\end{array} 
\right).
\end{eqnarray}
Here $\boldsymbol{A}$ and $\boldsymbol{B}$ represent the \emph{general} Bogoliubov transformations between the initial ``in" modes and the final ``out" modes after a composition of basic building block trajectories. Writing out the defining symplectic group relation to first order in $h$ gives us the following perturbative relations
\begin{subequations}
\begin{align}
\boldsymbol{s}^{(0)}_{kk'}\boldsymbol{\Omega}\boldsymbol{s}^{(0)\dag}_{kk'}&=\boldsymbol{\Omega},\\
\boldsymbol{s}^{(0)}_{kk'}\boldsymbol{\Omega}\boldsymbol{s}^{(1)\dag}_{kk'}+\boldsymbol{s}^{(1)}_{kk'}\boldsymbol{\Omega}\boldsymbol{s}^{(0)\dag}_{kk'}&=\boldsymbol{0},
\end{align}
\end{subequations}
which, given the structure of the symplectic matrix $\boldsymbol{S}$ , allows us to identify
\begin{subequations}
\label{eqn:order-h-relations}
\begin{align}
\bar{\boldsymbol{G}}^{(0)}\boldsymbol{A}^{(1)\mathrm{tp}}+\bar{\boldsymbol{A}}^{(1)}\boldsymbol{G}^{(0)}&=\boldsymbol{0},\\
\bar{\boldsymbol{G}}^{(0)}\boldsymbol{B}^{\dag (1)}&=\bar{\boldsymbol{B}}^{(1)}\bar{\boldsymbol{G}}^{(0)}.
\end{align}
\end{subequations}
We are interested in constructing trajectory segments which will be repeated to generate resonances. A segment contains any number of different basic building blocks and could be for example, a return voyage to Alpha-Centauri \cite{alphacentauri2012}.  We consider the cavity to be initially in the vacuum state, which is represented by the identity matrix $\boldsymbol{\Gamma}_{0}=\boldsymbol{I}$, see Eq.~(\ref{eqn:cov-mat-coherent}). 

Working to order $O(h)$, we find the reduced state~(\ref{eq:one-travel-scenario}) is
\begin{eqnarray}
\label{eqn:perturbed-state}
\boldsymbol{\Gamma}_{kk'}&=&
\boldsymbol{I}-2\left(
\begin{array}{cc}
\boldsymbol{0} & \bar{\boldsymbol{B}}^{(1)}\bar{\boldsymbol{G}}^{(0)} \\
\boldsymbol{B}^{(1)}\boldsymbol{G}^{(0)} & \boldsymbol{0}
\end{array}
\right)h,
\end{eqnarray}
where we have used Eq.~(\ref{eqn:order-h-relations}) to simplify our expressions. It can be easily seen that the reduced state is pure to $O(h^{2})$~\cite{friis2012-3} i.e. $\text{det}(\boldsymbol{\Gamma}_{kk'})=1+O(h^{2})$. As the state is pure and bipartite, it can be shown that it is equivalent to a two-mode squeezed state, upto local symplectic transformations on the modes $k$ and $k'$~\cite{laurat2005}. This implies, remarkably, that our transformation matrix~(\ref{nico}) is equivalent to a two-mode squeezing operator and is an entangling gate operation. Therefore, we have found a physical implementation of a quantum gate by ``shaking" a cavity through spacetime. Knowing this, we can now quantify the entanglement generated by the gate~(\ref{nico}).

The entanglement of a bipartite system in a pure state is quantified by the entropy of entanglement; however, this measure is not suitable in our perturbative regime because the first order contribution in the expansion of the entropy is of the form $\mathrm{S}\sim h\ln h$ which cannot appear as a term in a power series expansion. Fortunately, we can turn to the entanglement negativities for the quantification of the entanglement in our system. From Section~(\ref{subsec:gasussian-negs}), to quantify the negativity of a two-mode Gaussian state we need the smallest symplectic eigenvalue of the partially transposed state $\tilde{\boldsymbol{\Gamma}}_{kk'}$. As we are working with the complex form of the symplectic group, we can compute the partial transposition, with respect to the second mode, of $\boldsymbol{\Gamma}_{kk'}$ via
\begin{eqnarray}
\tilde{\boldsymbol{\Gamma}}_{kk'}=\boldsymbol{P}\boldsymbol{\Gamma}_{kk'}\boldsymbol{P},
\end{eqnarray}
where
\begin{eqnarray}
\boldsymbol{P}=\left(\begin{array}{cccc}
1 & 0 & 0 & 0 \\
0 & 0 & 0 & 1 \\
0 & 0 & 1 & 0 \\
0 & 1 & 0 & 0 
\end{array}\right).
\end{eqnarray}
The symplectic spectrum can be computed using~(\ref{eqn:symplectic-spectrum-complex}) and from it we can identify the smallest symplectic eigenvalue $\tilde{\nu}_{-}$.

As our unperturbed state is the identity matrix, the symplectic eigenvalues of the partial transpose are degenerate and read $\boldsymbol{\nu}=\lbrace 1,  1 \rbrace$. Thus, we need to find the corrections to the symplectic degenerate eigenvalue $\tilde{\nu}=+1$. Using standard degenerate perturbation theory~\cite{bransden2000}, and making explicit use of $B_{kk}^{(1)}=0$ and the perturbative identity
\begin{eqnarray}
B^{(1)}_{k'k}=+G_{k}^{(0)}\bar{G}_{k'}^{(0)}B^{(1)}_{kk'},
\end{eqnarray}
we find the symplectic eigenvalue $\tilde{\nu}$ is perturbed to
\begin{eqnarray}
\label{eqn:symplectic-eigenvalues}
\tilde{\nu}_{\pm}=1\pm 2|B^{(1)}_{kk'}|h+O(h^{2}).
\end{eqnarray}
Interestingly, to $O(h)$ the negativity and logarithmic negativity coincide exactly and so, using $\tilde{\nu}_{-}$ from Eqn.~(\ref{eqn:symplectic-eigenvalues}), we can write, after one travel scenario,
\begin{eqnarray}
\mathcal{N}(\boldsymbol{\Gamma}_{kk'})=|B^{(1)}_{kk'}|h+O(h^{2}).
\end{eqnarray}
This quantifies the entanglement in our system for our general travel scenarios. However, we are interested in generating the most entanglement possible for a given situation. We notice that when the commutator of the symplectic transformation vanishes, i.e.
\begin{eqnarray}
\label{eqn:bogo-resonance-condition}
\left[\boldsymbol{s}_{kk'},\boldsymbol{s}^{\dag}_{kk'}\right]=\boldsymbol{0},
\end{eqnarray}
to first order in $h$ the state after $N$ repetitions of the symplectic transformation~(\ref{eq:general-segment}) can be written as
\begin{eqnarray}
\boldsymbol{\Gamma}_{kk'}^{N}&=&\boldsymbol{I}+N\boldsymbol{\Gamma}_{kk'}^{(1)}h+O(h^{2}),
\end{eqnarray}
where $\boldsymbol{\Gamma}_{kk'}^{(1)}$ is the first order matrix of the state~(\ref{eqn:perturbed-state}). The partially transposed state is therefore given by
\begin{eqnarray}
\tilde{\boldsymbol{\Gamma}}_{kk'}^{N}&=&\boldsymbol{I}+N\tilde{\boldsymbol{\Gamma}}_{kk'}^{(1)}h+O(h^{2}).
\end{eqnarray}
Thus, the the only change in the computation of the smallest symplectic eigenvalue for the $N$ repetition comes as a multiplicative factor of $N$ in front of the first order correction to the state. Thus, we obtain
\begin{equation}
\label{eqn:resonance-symplectic-correction}
\tilde{\nu}_{N}^{(1)}:=N\tilde{\nu}_{-}^{(1)}.
\end{equation}
We call the vanishing of the commutator $\left[\boldsymbol{s}_{kk'},\boldsymbol{s}^{\dag}_{kk'}\right]=\boldsymbol{0}$ a \emph{resonance condition}. 

Using~(\ref{eqn:resonance-symplectic-correction}), it is straightforward to see, when on resonance,
\begin{equation}
\mathcal{N}(\boldsymbol{\Gamma}_{kk'}^{N})=N|B_{kk'}^{(1)}|h+O(h^{2}).
\end{equation}
In other words, the entanglement generated between two modes $k,k'$ grows linearly with the number of repetitions $N$ when the travel times of the scenario are picked such that we have a resonance. To stay within our pertubative regime, the condition $N|B_{kk'}^{(1)}|h\ll 1$ must be satisfied. It would be useful to know when the resonant situations occur. We proceed with a computation of the resonance commutator~(\ref{eqn:bogo-resonance-condition}). Using explicitly $B^{(1)}_{kk}=0$ and the perturbative Bogoliubov relations, we can write
\begin{eqnarray}
\left[\boldsymbol{s}_{kk'},\boldsymbol{s}^{\dag}_{kk'}\right]=
2\left(
\begin{array}{cc}
\boldsymbol{0} & \bar{\boldsymbol{C}} \\
\boldsymbol{C} & \boldsymbol{0}
\end{array} 
\right)h,\hspace{4mm}\boldsymbol{C}=\left(G_{k'}^{(0)}-\bar{G}_{k}^{(0)}\right)B_{kk'}^{(1)}.
\end{eqnarray}
Hence, we observe a resonance when
\begin{eqnarray}
\left(G_{k'}^{(0)}-\bar{G}_{k}^{(0)}\right)B_{kk'}^{(1)}=0.
\end{eqnarray}
Therefore, resonances occur when $B_{kk'}^{(1)}\neq0$ and the total proper time $T$ takes discrete values
\begin{equation}
T_{n} = \frac{2n\pi}{\omega_k + \omega_{k'}}\label{resonant:total:times}.
\end{equation}
We emphasise that the value of $T_n$ does not depend on the details of the travel scenario; however, the total amount of entanglement generated $\mathrm{N}(\boldsymbol{\Gamma}_{kk'}^{N})=N|B_{kk'}^{(1)}|h$ does depend on the specifics of the trajectory through the Bogoliubov coefficient.

\subsection{Cavity Resonances}

We now specialise to our sample travel scenario which corresponds to a cavity which is initially inertial, travels with proper acceleration $h$ for proper time $\tau_{1}$, coasts for proper time $\tau_{2}$, travels with proper acceleration $\lambda h$ for proper time $\tau_{1}$ and finally coasts for proper time $\tau_{2}$. Here $\lambda$ is a real parameter that indicates the magnitude of acceleration for the second period and the direction of acceleration as
\begin{subequations}
\begin{align}
\lambda &>0 \Rightarrow \text{same direction}\\
\lambda &<0 \Rightarrow \text{opposite direction}
\end{align}
\end{subequations} 
The evolution matrix can then be written in terms of the basic building blocks as
\begin{eqnarray}
\boldsymbol{S}=\boldsymbol{S}(0,\tau_{2})\boldsymbol{S}(\lambda h,\tau_{1})\boldsymbol{S}(0,\tau_{2})\boldsymbol{S}(h,\tau_{1}).
\end{eqnarray}
Note that the transformations with $h=0$ correspond to inertial motion.

Using this decomposition, and the relation $\boldsymbol{S}^{-1}\equiv\boldsymbol{K}\boldsymbol{S}^{\dag}\boldsymbol{K}^{\dag}$ to find the inverse of a symplectic matrix~\cite{arvind1995}, we find to first order 
\begin{equation}
\label{entanglement}
|B_{kk'}^{(1)}|=\beta_{kk'}^{(1)}|1-G_{k}^{(0)}(\tau_{1})G_{k'}^{(0)}(\tau_{1})||1+\lambda G_{k}^{(0)}(\tau_{1})G_{k'}^{(0)}(\tau_{2})G_{k}^{(0)}(\tau_{2})G_{k'}^{(0)}(\tau_{1})|.
\end{equation}
Note that $\beta_{kk'}^{(1)}$ are the first order off-diagonal elements of the fundamental Bogoliubov transformation~(\ref{eqn:fundamental-bogos}).

Substituting Eq. (\ref{resonant:total:times}) in Eq. (\ref{entanglement}) we find that the logarithmic negativity at resonant times is,
\begin{equation}
\label{final:general:casimir:beta}
\mathcal{N}(\boldsymbol{\Gamma}_{kk'}^{N})=N\beta_{kk'}|1-(-1)^{n}e^{-i(\omega_k+\omega_k^{\prime})\tau_{2}}||1+ (-1)^{n}\lambda|h.
\end{equation}
Note that the negativity vanishes when $n$ is even and the time of coasting is $\tau_{2}=2\pi m/(\omega_k+\omega_k^{\prime})$ and when $n$ is odd and $\tau_{2}=(2m+1)\pi/(\omega_k+\omega_k^{\prime})$ with $m\in\mathbb{N}$.  

In the case the accelerations have the same magnitude ($|\lambda |=+1$) maximal amounts of entanglement are generated when $n$ is even and  $\tau_{2}=\pi (2m+1)/(\omega_k+\omega_k^{\prime})$ when the accelerations have the same direction ($\lambda>0$) and when $n$ is odd and  $\tau_{2}=2\pi m/(\omega_k+\omega_k^{\prime})$ when the accelerations alternate direction  ($\lambda<0$). 

\begin{figure}[!ht]
\centering
\includegraphics[width=0.7\columnwidth]{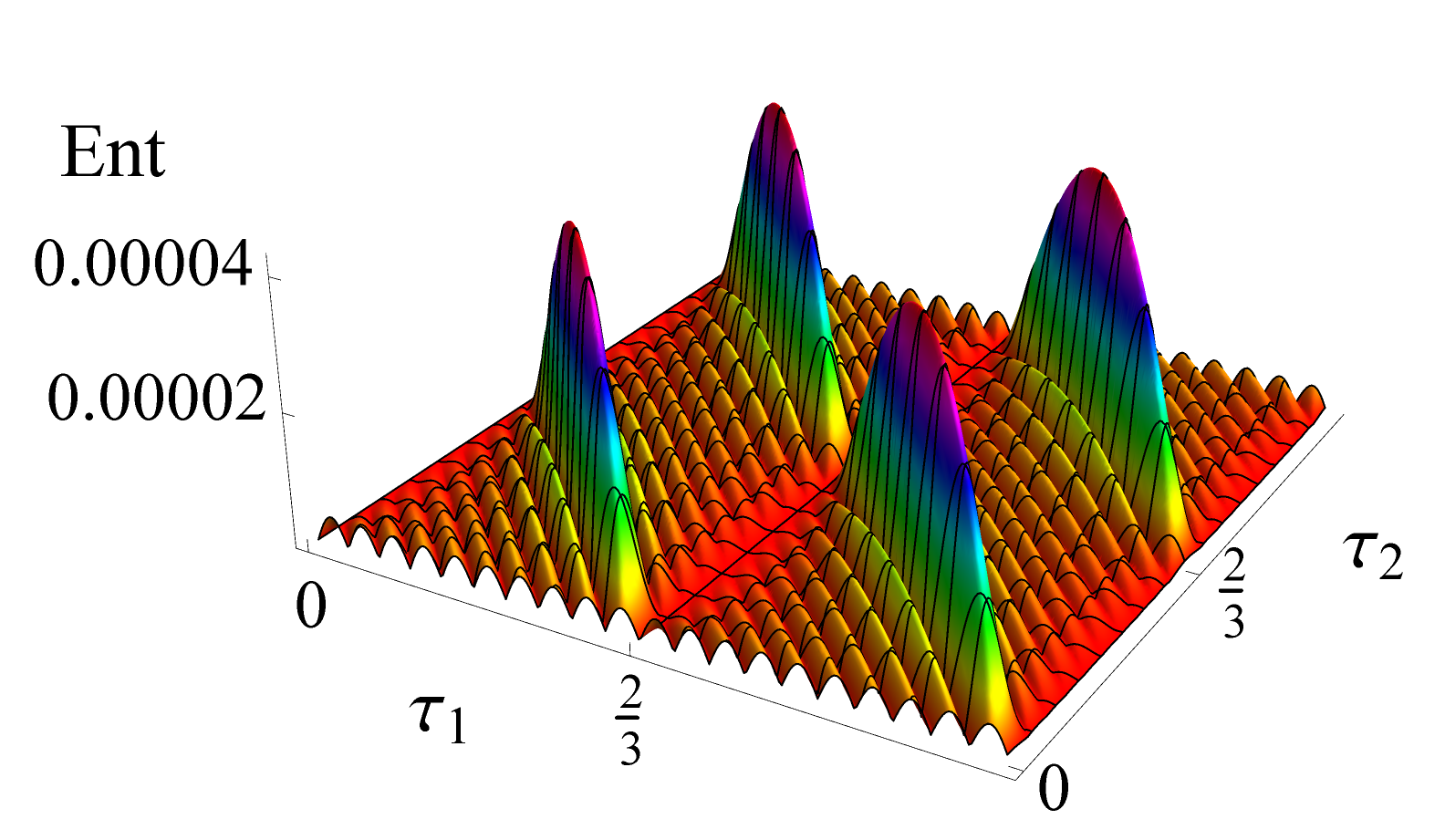}
\caption[Cavity Resonance]{\label{dupa} The correction to the symplectic eigenvalue $\tilde{\nu}^{(1)}_{N}$  after $N=5$ segment repetitions as a function of the proper time of acceleration~$\tau_{1}$ and the time of coasting $\tau_{2}$. We considered a cavity length of $L=1$, massless modes $k=1$ and $k'=2$ and accelerations $\sim 10^{-4}$.} 
\end{figure}

This behavior is shown in Fig.~(\ref{dupa}) where we plot the symplectic eigenvalue  $\tilde{\nu}^{(1)}_{N}$ after $N=5$ segment repetitions as a function of the proper time of acceleration~$\tau_{1}$ and the time of coasting $\tau_{2}$. We considered a cavity length of $L=1$, massless modes $k=1$ and $k'=2$ and accelerations $\sim 10^{-4}$. 
Interestingly, the special case of alternating acceleration directions ($\lambda=-1$) and no coasting ($\tau_{2}=0$) corresponds to the standard dynamical Casimir setting where the cavity oscillates periodically as a whole. A resonant enhancement of particle creation occurs in the dynamical Casimir effect \cite{dodonov1990} which was recently demonstrated in the laboratory in a superconducting circuit consisting of a coplanar transmission line with a tunable electrical length which produces an effective moving boundary \cite{WilsonDynCasNature2012}.

\subsection{Discussion}

We have shown that non-uniform motion can generate two-mode squeezing gates which produce observable amounts of entanglement.
Finding ways to create significant amounts of entanglement in relativistic settings is of great interest since entanglement is necessary for quantum communications and information processing  \cite{schutzhold2008}. Recent studies in relativistic quantum information show that small amounts of mode entanglement are created when a cavity undergoes non-uniform motion. \cite{alphacentauri2012,friis2012-3}. We show that particle creation and bipartite mode entanglement can be linearly enhanced by repeating any travel segment periodically. Via the equivalence principle, our results suggest that fluctuations of a gravitational field can produce entanglement. For example, consider a small cavity containing a Bosonic field in its vacuum state free falling in the presence of a gravitational field \cite{kothawala2013}. Entanglement between the modes is generated by suddenly holding the cavity at a fixed position against the action of the gravitational field. If the cavity's position changes periodically or the gravitational field fluctuates, the entanglement can be enhanced. However, the gates produced here are linear unitary transformations. To implement full quantum computing one would need to find ways of producing non-linear transformations which cannot be represented by symplectic matrices. Finally, we mention that there are currently very few implementations of quantum gates in relativistic quantum information. We therefore hope these results can pioneer quantum communication and even possible quantum computing using moving cavities.

\section{Teleportation between moving cavities}\label{sec:teleportation-results}

\subsection{Introduction}

\emph{How are quantum information tasks affected by relativistic motion?} This seemingly simple question has been intriguing researchers for more than a decade~\cite{peres2004} and it has been the cornerstone of an entire new field of physics, \emph{relativistic quantum information}, see Ref.~\cite{alsing2012} for a recent review. Besides its theoretical interest, the topic is increasingly gaining practical relevance as quantum information experiments are reaching relativistic regimes~\cite{rideout2012}. However, a satisfactory empirically testable framework to address this question has been missing.

The first attempts to answer this open question considered highly idealised situations where observers with constant, eternal accelerations analysed the entanglement between global modes of a quantum field \cite{alsing2003,funentes-schuller2005,bruschi2010}. Typically, these studies consider only the mathematical intricacies of the problem and contain little reference to realistic physical set-ups. However, recent theoretical work~\cite{downes2011}, including the previous section, has analysed relativistic entanglement in paradigmatic quantum optical scenarios such as cavity QED \cite{nobelprize2012}. In the relativistic case cavities move with accelerations that can arbitrarily vary in time.  As shown in the previous section, relativistic motion can also be used to implement quantum gates. 

On the other hand, the \emph{dynamical Casimir effect} has recently been demonstrated in a real experiment using superconducting circuits where the relativistic motion of an effective boundary condition was successfully achieved~\cite{WilsonDynCasNature2012}.

In this section we show that non-uniform accelerated motion has  effects on a paradigmatic quantum information protocol,
\emph{quantum teleportation}, in a framework at which the theoretical predictions can be tested in Earth-based experiments. Such experiments will be capable  of informing future space-based experiments. We focus on the effects of non-uniform motion on the fidelity of the standard protocol for quantum teleportation with continuous variables, see Fig.~\ref{fig:teleportation in motion}. We employ the powerful tools of quantum optics for Gaussian states. In this setting we can take advantage
of very recent experimental developments in circuit quantum electrodynamics~\cite{YouNori2011}.

Our main results are the following. We observe two distinguishable degradation effects on the fidelity of teleportation when we consider one of the parties undergoing non-uniform motion. The first is due to the time evolution of the field. The fidelity loss due to this effect can be corrected by applying local operations which depend on the proper time. The second degradation effect is solely due to acceleration and it can only be avoided by conveniently choosing the duration of accelerated motion.

\begin{figure}[t!]
\centering
\includegraphics[width=0.5\columnwidth]{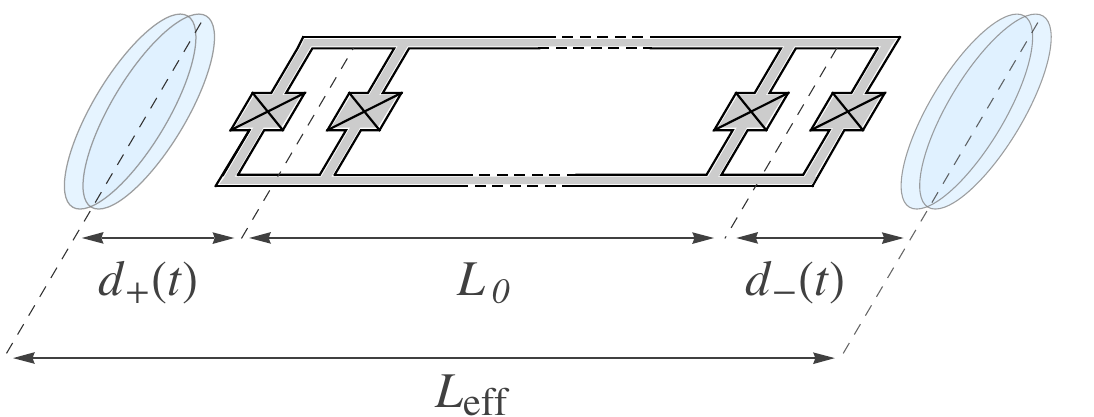}
\caption[SQUID Configuration]{\textbf{Sketch of the experimental setup:} A coplanar waveguide is interrupted by two \emph{superconducting quantum interference devices}
(SQUIDs) placed at a fixed distance $L_{0}$, creating a cavity of effective length $L_{\mathrm{eff}}=L_{0}+d_{+}(t)+d_{-}(t)$. The
time dependence of the distances $d_{\pm}(t)$ between the SQUIDs and the effective boundaries are controlled with external drive fields applied to
the superconducting circuits to simulate a cavity of constant length with respect to its rest frame.}
\label{fig:microwave cavity setup}
\end{figure}

Finally, we introduce our experimental setup, see Fig.~\ref{fig:microwave cavity setup}, to test our results using cutting-edge circuit QED technology. We generalise the dynamical Casimir effect idea of producing a single oscillating boundary condition and now describe the case of a rigid cavity moving with constant acceleration during a finite time interval.

\subsection{Physical Set-up}

We now introduce our model starting with the standard teleportation protocol in continuous variables systems which assumes Alice and Rob to be at rest at all times~\cite{mari2008}. The novelty of our approach will be to consider that Rob undergoes non-uniform relativistic motion before concluding the protocol. As we will be considering the teleportation scheme between Alice and Rob using a single mode of Alice and a single mode of Rob, it will be more convenient to work with the real representation of $\mathrm{Sp}(2N,\mathbb{R})$ introduced in Section~(\ref{sec:symplectic-representations}). We will therefore be working in the basis $\hat{\boldsymbol{X}}=(\hat{x}_{k},\hat{p}_{k},\hat{x}_{k'},\hat{p}_{k'})$ where $k$ denotes the mode controlled by Alice and $k'$ denotes the mode controlled by Rob.

Initially Alice and Rob are at rest and share a two-mode squeezed state~(\ref{eqn:two-mode-squeezed-state}) of a $(1+1)$-dimensional Klein-Gordon field contained within a cavity with squeezing parameter $r>0$. In the case for two modes, the quantum correlations of this state are characterized by its covariance matrix which, from Eq.~(\ref{eqn:cov-mat-def}), can be written as
\begin{eqnarray}
\label{eqn:initial-state-teleportation}
\boldsymbol{\Gamma}=
\begin{pmatrix} 
\boldsymbol{A} &  \boldsymbol{C} \\ 
\boldsymbol{C}^{\text{tp}} &  \boldsymbol{B} 
\end{pmatrix}
\end{eqnarray}
with the further requirements that $\boldsymbol{A}^{\mathrm{tp}}=\boldsymbol{A}$ and $\boldsymbol{B}^{\mathrm{tp}}=\boldsymbol{B}$. Alice wants to use the entanglement of this state as a resource to teleport an additional unknown coherent state to Rob. 

From Section~(\ref{sec:cv-teleportation}), given the covariance matrix $\boldsymbol{\Gamma}$, the fidelity of this protocol is~\cite{mari2008}
\begin{equation}
\label{eq:fidelity}
\mathcal{F} = \dfrac{2}{\sqrt{4+2\tr(\boldsymbol{N})+ \det(\boldsymbol{N})}}\,,
\end{equation}
where $\boldsymbol{N}=\boldsymbol{\sigma}_{3}\boldsymbol{A}\boldsymbol{\sigma}_{3}+\boldsymbol{\sigma}_{3}\boldsymbol{C}+\boldsymbol{C}^{\mathrm{tp}}\boldsymbol{\sigma}_{3}+\boldsymbol{B}$ and $\boldsymbol{\sigma}_{3}$ is the third Pauli matrix. In addition, Alice and Rob can use local operations and classical communication (LOCC) to improve the fidelity of the protocol without increasing the amount of shared entanglement. In particular, optimising over all \emph{local} Gaussian operations the upper bound to the optimal fidelity of teleportation can be expressed as~\cite{adesso2005,mari2008}
\begin{equation}
\label{eq:optimalfidelity}
\mathcal{F}_{\mathrm{opt}} \leq \dfrac{1}{1+\tilde{\nu}_{-}}\,,
\end{equation}
where $\tilde{\nu}_{-}$ is the smallest (positive) symplectic eigenvalue of the partial transpose of $\boldsymbol{\Gamma}$. The upper bound is achieved precisely if Alice and Rob share a two mode squeezed state~(\ref{eqn:two-mode-squeezed-state}), for which $\tilde{\nu}_{-}=\exp(-2r)$, $r>0$~\cite{adesso2005,mari2008}. Then $\mathcal{F}=\mathcal{F}_{\mathrm{opt}}=1/\bigl(1+\exp(-2r)\bigr)$. As we have not yet taken into account the observers' motion, it comes as no surprise that the optimal fidelity of teleportation only depends on the squeezing. In this case the initial entanglement is~(\ref{eqn:2mss-entanglement})
\begin{eqnarray}
\mathrm{E}(\boldsymbol{\Gamma})=\cosh^{2}(r)\log\cosh^{2}(r)-\sinh^{2}(r)\log\sinh^{2}(r).
\end{eqnarray}
In the limit $r\to 0$ the entanglement between Alice and Rob vanishes. It is interesting to see however that for $r\to\infty$ the entanglement shared between Alice and Rob is unbounded. This is a consequence of the infinite degrees of freedom of continuous variable states.

\begin{figure}[t!]
\centering
\includegraphics[width=0.5\columnwidth]{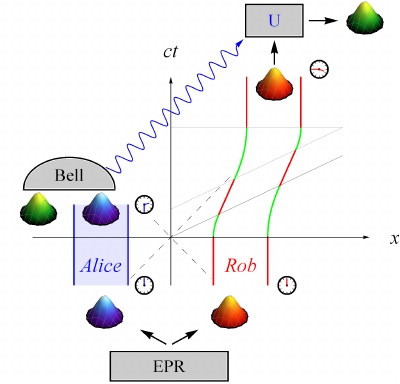}
\caption[Teleportation Set-up]{Alice and Rob initially share a two-mode squeezed cavity state that is produced by the EPR source. Consecutively, Rob's cavity undergoes non-uniform motion consisting of segments of constant acceleration (green hyperbolae) and inertial coasting (red, parallel lines). Alice, who remains inertial, sends the outcome of the Bell measurement on the input state and her mode of the entangled
state to Rob via a classical channel (blue, wavy arrow). Rob can then retrieve the teleported state by performing the appropriate unitary $U$. In addition to the standard protocol both Alice and Rob measure their respective proper times and perform local rotations to compensate for the phases accumulatied during the motion.}
\label{fig:teleportation in motion}
\end{figure}

Now let us consider the scenario that is sketched in Fig.~(\ref{fig:teleportation in motion}).
After the preparation of the initial state Rob's cavity undergoes non-uniform motion, consisting of periods of constant acceleration and inertial motion.  
As in the previous section, we model the evolution of the systems state by its symplectic evolution
\begin{eqnarray}
\tilde{\boldsymbol{\Gamma}}=\boldsymbol{S}\boldsymbol{\Gamma}\boldsymbol{S}^{\mathrm{tp}}.
\end{eqnarray}
The reduced covariance matrix $\tilde{\boldsymbol{\Gamma}}_{kk\pr}$ for two modes $k$ and $k\pr$ can be obtained from $\tilde{\boldsymbol{\Gamma}}$ via partial tracing.
If the motion is inertial for the time $t$ then $\boldsymbol{S}$ is simply composed of local rotations with angles $\omega_{k}t$ and $\omega_{k\pr}t$, where $\omega_{k}$ and $\omega_{k\pr}$ are the angular frequencies of the modes $k$ and $k\pr$ respectively. We let Rob's cavity accelerate for a proper time $\tau$ which is measured at the center of the rigid cavity. Then $\boldsymbol{S}$ is given in terms of the Bogoliubov coefficients $\alpha_{mn}$ and $\beta_{mn}$ that relate the mode functions of the inertial and accelerated cavity~\cite{friis2012-3}. The coefficients $\alpha_{mn}$ account for mode mixing, while $\beta_{mn}$ accounts for particle pair production.

Notice that, for cavity sizes $\sim 1$m, this approach can accommodate extremely large accelerations, as we will see in detail below. Moreover, there are no restrictions on the duration, covered distance or the achieved velocity of the motion. 

\subsection{Teleportation Fidelity}

Now we will focus on the effect of the motion on the fidelity of the teleportation protocol described above. Although the formalism allows us to consider arbitrary trajectories, we consider here the simplest case: Rob's motion is inertial apart from one finite interval of constant acceleration.
\begin{figure}[b!]
\hspace*{-0.4cm}
\includegraphics[width=0.5\linewidth]{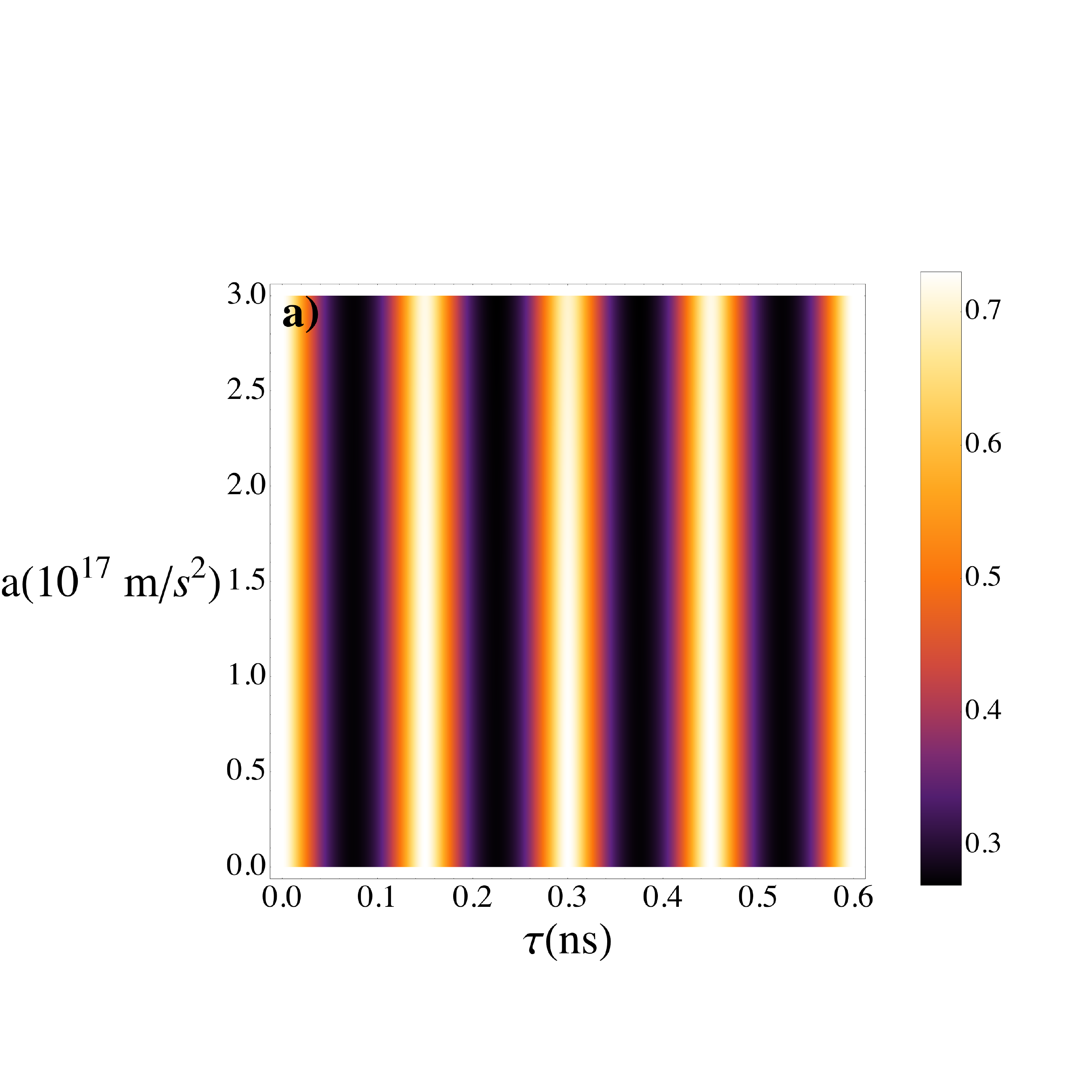}
\includegraphics[width=0.5\linewidth]{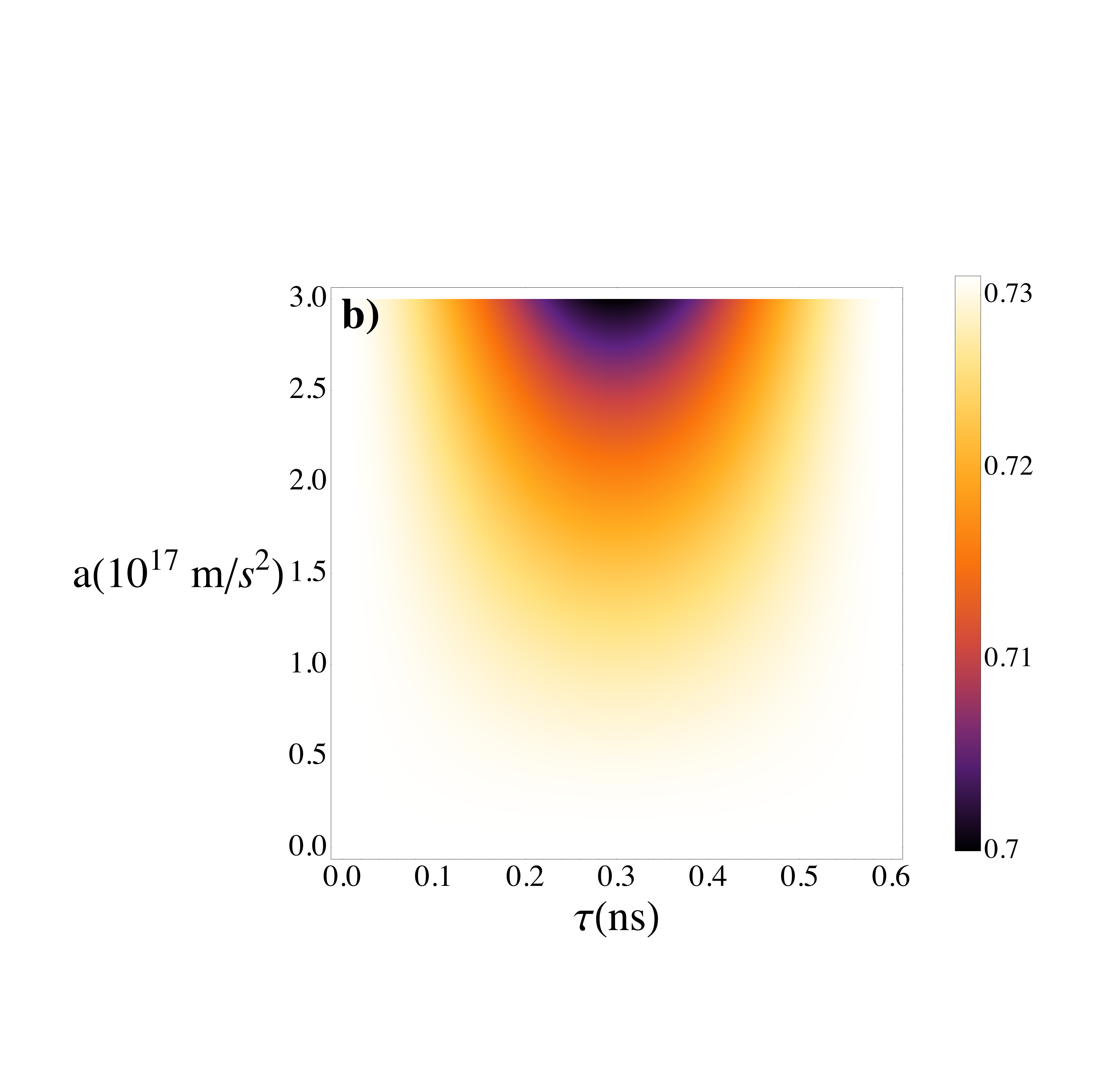}
\caption[Teleportation Fidelity]{The fidelities $\tilde{\mathcal{F}}$ and $\tilde{\mathcal{F}}_{\mathrm{opt}}$ are plotted in Fig.~\ref{fig:results}~a) and~\ref{fig:results}~b) respectively as functions of Rob's proper time $\tau$ and the acceleration $a$. The plots are shown for modes $k=1$ (Alice) and $k\pr=3$ (Rob). For the cavity length we use a typical value of $L_{\mathrm{eff}}=1.2\,\operatorname{cm}$. Together with the speed of light $c=1.2\cdot10^8\,\mathrm{m/s}$ we obtain a fundamental frequency $\omega_{k}=2\pi\times c/(2L)=2\pi\times 5\mathrm{GHz}$ and $\omega_{k\pr}=3\omega_{k}$. The squeezing parameter is $r=\tfrac{1}{2}$ and the maximum value of the perturbative parameter is $h^{2}=0.06$. Once the effect of the free time evolution in Fig.~\ref{fig:results}~a) is removed, the correction due to acceleration can be isolated in Fig.~\ref{fig:results}~b).}
\label{fig:results}
\end{figure}
Consider the initial state between Alice's mode $k$ and Rob's mode $k'$
\begin{eqnarray}
\boldsymbol{\Gamma}_{kk'}=\left(\begin{array}{cc}
\boldsymbol{\sigma}_{kk} & \boldsymbol{\sigma}_{kk'} \\
\boldsymbol{\sigma}_{kk'}^{\mathrm{tp}} & \boldsymbol{\sigma}_{k'k'}
\end{array} \right).
\end{eqnarray}
The reduced state between Alice and Rob after Rob's motion is given by
\begin{eqnarray}
\label{eq:alice bob entangled resource}
  \tilde{\boldsymbol{\Gamma}}_{kk'}&=&\left(\begin{tabular}{ c c }
    $\boldsymbol{O}_{kk}\boldsymbol{\sigma}_{kk}\boldsymbol{O}_{kk}^{tp}$ & $\boldsymbol{O}_{kk}\boldsymbol{\sigma}_{kk'}\boldsymbol{s}_{k'k'}^{tp}$ \\
    $\boldsymbol{s}_{k'k'}\boldsymbol{\sigma}_{kk'}^{tp}\boldsymbol{O}_{kk}^{tp}$ & $\boldsymbol{\sigma}_{B}$ \\
  \end{tabular}\right),
\end{eqnarray}
where
\begin{eqnarray}
\boldsymbol{\sigma}_{B}=\sum_{j\ne k'}\boldsymbol{s}_{jk'}\boldsymbol{s}_{jk'}^{\mathrm{tp}}+\boldsymbol{s}_{k'k'}\boldsymbol{\sigma}_{k'k'}\boldsymbol{s}_{k'k'}^{\mathrm{tp}}.
\end{eqnarray}
Here, $\boldsymbol{O}_{kk}$ are \emph{orthogonal} symplectic transformations which represent the phase rotations induced in Alice's mode. See appendix~(\ref{app:teleportation-state}) for details on the derivation of the reduced state~(\ref{eq:alice bob entangled resource}). The symplectic transformation matrices $\boldsymbol{s}_{mn}$ are defined via~\cite{friis2012-3}
\begin{eqnarray}
\boldsymbol{s}_{mn}=\left(\begin{array}{cc}
\mathrm{Re}[A_{mn}-B_{mn}] & \mathrm{Im}[A_{mn}+B_{mn}] \\
-\mathrm{Im}[A_{mn}-B_{mn}] & \mathrm{Re}[A_{mn}+B_{mn}]
\end{array} \right),
\end{eqnarray}
and the coefficients $A_{mn}, B_{mn}$ are the Bogoliubov coefficients for the travel scenario. As we are considering an initial two-mode squeezed state between Alice's mode $k$ and Rob's mode $k'$, we have
\begin{subequations}
\begin{align}
\boldsymbol{\sigma}_{kk}=\boldsymbol{\sigma}_{k'k'}&=\cosh(2r)\boldsymbol{I},\\
\boldsymbol{\sigma}_{kk'}&=\sinh(2r)\boldsymbol{\sigma}_{3}.
\end{align}
\end{subequations}
We wish to analyse the fidelity of teleportation when the entangled state between Alice and Rob is given by~(\ref{eq:alice bob entangled resource}). As in the previous section, we can compute the Bogoliubov transformation of our travel scenario as a power series in the small parameter $h$, see Eq.~(\ref{eqn:fundamental-bogos-matrix})~and~Eq.(\ref{eqn:fundamental-bogos}). As it is only Rob who as undergone non-trivial motion, only his transformation matrix $\boldsymbol{s}_{B}$ needs to be expanded in powers of $h$. We find that the first non-trivial order is $O(h^{2})$ and we can write the covariance matrix as, see appendix~(\ref{app:teleportation-state}) for details, 
\begin{eqnarray}
\tilde{\boldsymbol{\Gamma}}_{kk'}=\tilde{\boldsymbol{\Gamma}}_{kk'}^{(0)}+\tilde{\boldsymbol{\Gamma}}_{kk'}^{(2)}h^{2}.
\end{eqnarray}
First, inserting the motion-transformed covariance matrix $\tilde{\boldsymbol{\Gamma}}_{kk'}$ into Eq.~(\ref{eq:fidelity}) we find the perturbative expansion of the teleportation fidelity, i.e.,
\begin{align}
\tilde{\mathcal{F}}  &=  \tilde{\mathcal{F}}^{\raisebox{-1.7pt}{\scriptsize{$(0)$}}}- \tilde{\mathcal{F}}^{\raisebox{-1.7pt}{\scriptsize{$(2)$}}}\,h^{2}+O(h^{4})\,,
\label{eq:fidelity series expansion}
\end{align}
where the expansion coefficients are given by
\begin{subequations}
\begin{align}
\tilde{\mathcal{F}}^{\raisebox{-1.7pt}{\scriptsize{$(0)$}}}    &=
\bigl(1+\operatorname{Cosh}(2 r)- \operatorname{Cos}(\phi)\operatorname{Sinh}(2r)\bigr)^{-1}, \\
\tilde{\mathcal{F}}^{\raisebox{-1.7pt}{\scriptsize{$(2)$}}}    &=
\bigl(\tilde{\mathcal{F}}^{\raisebox{-1.7pt}{\scriptsize{$(0)$}}}\bigr)^{2} \bigl(1+e^{-2r}\bigr)\bigl(f^{\beta}_{k\pr}+f^{\alpha}_{k\pr}\,\operatorname{Tanh}(2r)\bigr)\,,
\end{align}
\label{eq:fidelity series expansion terms h2}
\end{subequations}
and $\phi=\omega_{k}t+\omega_{k\pr}\tau$. The additional expressions $f^{\alpha}_{k\pr}=\tfrac{1}{2}\sum_{n}|A_{nk\pr}^{\raisebox{0.7pt}{\tiny{$\,(1)$}}}|^{2}$ and $f^{\beta}_{k\pr}=\tfrac{1}{2}\sum_{n}|B_{nk\pr}^{\raisebox{0.7pt}{\tiny{$\,(1)$}}}|^{2}$ in Eq.~(\ref{eq:fidelity series expansion terms h2}) also depend on $\tau$. Note that $A_{kk\pr}$ and $B_{kk\pr}$ are the \emph{general} Bogoliubov transformations of the previous section. Due to the dependence of $\phi$ on proper time, there is a degradation effect on the fidelity as shown in Fig.~(\ref{fig:results}~a). 
In particular, this also occurs in the inertial case since the free evolution continuously rotates Alice's and Rob's modes affecting the optimal performance of the protocol. Inertial motion has no effects on entanglement involving observable degrees of freedom \cite{SaldanhaVedral2012a,SaldanhaVedral2012b,PeresScudoTerno2002,GingrichAdami2002,PachosSolano2003, LamataMartinDelgadoSolano2006,FriisBertlmannHuberHiesmayr2010}. 

However, the fidelity of teleportation does not only depend on entanglement. To correct the effects due to the time dependence of the phase, Alice and Rob must apply local rotations which depend only on their local proper times or choose times such that $\phi=2\pi n$. The periodicity of the modes is due to the massless nature of the field. If we introduce a bare mass for the field or extra spatial dimensions of the cavity then the perfect periodicity of the result would be lost. 

We wish to compute the optimum teleportation fidelity for our transformed state. This can be done via the smallest symplectic eigenvalue of the partial transposition of our entangled state. We can compute the symplectic eigenvalues of Eq.~(\ref{eq:alice bob entangled resource}) in the usual manner. Working to order $O(h^{2})$ we find the perturbed smallest symplectic eigenvalue of~(\ref{eq:alice bob entangled resource}) is
\begin{eqnarray}
\tilde{\nu}_{-}=e^{-2r}+(1+e^{-2r})\left[f^{\beta}_{k'}+f^{\alpha}_{k'}\mathrm{Tanh}(2r)\right]h^{2}.
\end{eqnarray}
This expression is valid only for $r>0$. In the case $r=0$, the original state shared between Alice and Rob does not contain any entanglement. The symplectic structure in this case is degenerate and the smallest symplectic eigenvalue is simply $\tilde{\nu}_{-}=+1$.

In the case $h=0$ the maximal, optimal fidelity $\mathcal{F}_{\mathrm{corr}}=\mathcal{F}_{\mathrm{opt}}=1/\bigl(1+\exp(-2r)\bigr)$ can be recovered. Remarkably, if the acceleration is nonzero the dependence of $\tilde{\mathcal{F}}^{\raisebox{-1.7pt}{\scriptsize{$(2)$}}}$ on the local phase $\phi$ can be removed by exactly the same local rotations as in the inertial case.  The degradation of the fidelity $\tilde{\mathcal{F}}_{corr}$ can then be attributed solely to the acceleration. Moreover, the protocol including the local phase rotations turns out to be optimal. In other words, the motion transformed upper bound~\cite{mari2008} on the fidelity of Eq.~(\ref{eq:optimalfidelity}) is achieved by $\tilde{\mathcal{F}}_{corr}$. Put simply, we can either time the trajectory correctly or adjust the phases locally.
In that case, we have 
\begin{eqnarray}
\tilde{\mathcal{F}}_{\mathrm{corr}}=\tilde{\mathcal{F}}_{\mathrm{opt}}=
\tilde{\mathcal{F}}^{\raisebox{-1.9pt}{\tiny{\,$(0)$}}}_{\mathrm{opt}}-
\tilde{\mathcal{F}}^{\raisebox{-1.9pt}{\tiny{\,$(2)$}}}_{\mathrm{opt}}\,h^2 +O(h^4),
\end{eqnarray}
where $\tilde{\mathcal{F}}^{\raisebox{-1.9pt}{\tiny{\,$(0)$}}}_{\mathrm{opt}}=\mathcal{F}_{\mathrm{opt}}=1/\bigl(1+\exp(-2r)\bigr)$ is the optimal expression for $h=0$ above and
\begin{align}
\tilde{\mathcal{F}}^{\raisebox{-1.9pt}{\tiny{\,$(2)$}}}_{\mathrm{opt}}  &=
\tilde{\mathcal{F}}^{\raisebox{-1.9pt}{\tiny{\,$(0)$}}}_{\mathrm{opt}}
(f^{\beta}_{k\pr}+f^{\alpha}_{k\pr}\,\operatorname{Tanh}(2r))\,.
\label{eq:corroptfid}
\end{align}
In Fig.~(\ref{fig:results}~b) we plot $\tilde{\mathcal{F}}_{\mathrm{opt}}$, allowing us to identify a regime of strength and duration of acceleration at which the corrections due to motion amount to $4 \%$ of the total fidelity. As we explain below, this regime is well within experimental reach with current technology. Furthermore, the effect can be amplified by selecting more complicated trajectories \cite{bruschi2012-2}.

\subsection{Physical Implementation: SQUIDS}

Let us now discuss the details of the experimental setup that we propose to test our predictions. It should be pointed that the first verification of Gaussian state teleportation (in particular, a coherent state) was Furusawa \emph{et. al.}~\cite{furusawa1998}. Our hope is to combine existing experimental techniques with results in relativistic quantum information to produce realisations of theoretical work. We propose to use state of the art technology in circuit quantum electrodynamics. Two-mode squeezed states in the microwave regime have been produced in the laboratory with squeezing parameter $r=\operatorname{Log}2$, see Ref.~\cite{EichlerEtal2011,FlurinRochMalletDevoretHuard2012,MenzelmEtal2012}. Beam splitters for propagating photons with frequencies around $5~\operatorname{GHz}$ based on superconducting circuit architectures are also available~\cite{HoffmannEtal2010}. Therefore, we believe that the standard continuous variables teleportation protocol may be realised experimentally. Obviously the most demanding aspect of our proposal is the implementation of highly accelerated motion. To this end we will take advantage of the technology developed for the experiment verifying the dynamical Casimir effect~\cite{WilsonDynCasNature2012}. The cavities of our setting can be engineered as a coplanar microwave waveguide terminated by two dc-SQUIDs placed at a distance $L_{0}$ from each other, see Fig.~(\ref{fig:microwave cavity setup}).

The SQUIDs generate boundary conditions for the 1-dimensional quantum field along the waveguide~\cite{JohanssonJohanssonWilsonNori2010}, producing a rigid cavity of effective length $L\neq L_0$. The boundary condition depends on the external magnetic flux threading the SQUID. The time variation of this flux amounts to a time variation of $d_{\pm}$ to produce the different  effective accelerations of the boundaries, which will keep the cavity length fixed in its rest frame. Therefore, by applying external drive fields on both SQUIDS with appropriate time profiles, the system becomes equivalent to a rigid cavity in motion. This setup has already been implemented in the laboratory~\cite{SvenssonMScThesis2012} and the cavity accommodates a few modes below the natural cutoff provided by the plasma frequency of the SQUID. The profile of the driving fields can be adjusted to mimic constant accelerated motion during a finite interval of time. Taking as a reference the oscillating motion of a single mirror in~\cite{WilsonDynCasNature2012} with a driving frequency $\omega_{D}= 2\pi\times 10\, \mathrm{GHz}$ and an amplitude of $0.1\,\mathrm{mm}$ for the effective motion, the maximum acceleration achieved was $4\cdot10^{17}\,m/s^{2}$. These realistic values are enough to observe our predictions, since they give rise to values of $h$ larger than the ones considered in Fig. (\ref{fig:results}b). We mention again that this correction can be accounted for by appropriately planning the cavity's trip. 

\subsection{Discussion}

To summarise our results, we have analysed the effect of relativistic motion on the fidelity of the standard continuous variable protocol for quantum teleportation. The effects of non-uniform acceleration on the  fidelity can be isolated by applying proper-time dependent local operations which remove the effects of time evolution. We have shown that the degradation of the fidelity due to acceleration is sizeable for realistic experimental parameters. We have further suggested a particular experimental set-up with superconducting cavities that is well within reach of state-of-the-art technology.  The origin of the fidelity loss is the same physical mechanism\textemdash particle generation due to motion\textemdash underlying the dynamical Casimir effect and the Unruh-Hawking radiation. Therefore, its observation would also shed light on these phenomena. Moreover, via the equivalence principle, our results suggest the existence of observable effects of gravity on quantum information set-ups, which may be relevant for space-based experiments~\cite{rideout2012}. Finally, it is possible that theoretical predictions derived with a similar formalism, e.g., the implementation of quantum gates by cavity motion, may be realisable in similar experiments. We believe that the effects studied in relativistic quantum information scenarios will finally leave the realm of theoretical gedanken experiments. The analysis of relativistic effects on quantum information can now be extended by empirical tests. Furthermore, low-cost experimental set-ups to test relativistic aspects of quantum communication, such as the one proposed here, will inform future space-based, high-risk experiments.

\section{Conclusions}

In this chapter we investigated the effects of motion and, via the equivalence principle, gravity on entanglement. Using a perturbative analysis, we saw how entanglement generating gates can be implemented via suitable trajectories of a cavity. These gates proved to be equivalent to two-mode squeezing operations on the modes contained within the cavities. This has interesting consequences as two-mode squeezing operations coupled with other continuous variable gates, such as those of a beam splitter found in~\cite{bruschi2013}, and the generalisation to multipartite gates~\cite{friis2012-4} can be used to implement continuous variable quantum communication.

We also analysed the effects of motion on quantum teleportation. Starting with an initial entanglement resource, we found a degradation in the teleportation fidelity due to the motion of one of the observers. The degradation can be allocated to two independent sources. One was individual mode phase rotations induced due to the total time of the motion and the other was due to the magnitude of the acceleration of the moving observer. Interestingly, the effect of the induced phases could be accounted for by suitable local operations on the teleported state. This allowed us to isolate purely acceleration effects. Isolating these purely acceleration effects allowed us to propose a realistic experimental set-up which could verify our predictions. Such a verification would have interesting consequences for any quantum system which is under the influence of motion or gravitation.

As a final comment, cavities in relativistic quantum information have proven to be very useful tools for investigating entanglement in non-inertial reference frames. It is, therefore, hoped a cavity's motion through spacetime can be viewed as a resource for quantum entanglement, communications and computing.

\chapter{Moving Cavities for Relativistic Quantum Information: Fermions}\label{chapter:moving-cavities}


\section{Introduction}

One of the fundamental problems in the emerging field of relativistic
quantum information is the degradation of correlations caused
by accelerated motion.  Studies of uniform acceleration in Minkowski
spacetime (see
\cite{alsing2003,funentes-schuller2005,alsing2006,bruschi2010,martinmartinez2011,friis2011}
for a small selection and \cite{martinmartinezthesis,alsing2012} for recent
reviews) have revealed significant differences in the degradation that
occurs for Bosonic and Fermionic fields. Our main motivation for investigating Dirac fields is the
clear qualitative differences in the Bosonic versus Fermionic entanglement \cite{martinmartinez2011,friis2011}.

In this chapter, we shall undertake the first steps of investigating
Fermionic entanglement in accelerated cavities by adapting the scalar
field analysis of the previous sections to a Dirac
Fermion. We are interested in the differences that arise in the cavity mode entanglement due to the Fermionic, rather than Bosonic, nature of the field. Conceptually, one new issue with Fermions is that the
presence of positive and negative charges allows a broader range of
initial Bell-type states to be considered. New technical issues arise from the
boundary conditions that are required to keep the Fermionic field
confined in the cavities.

We focus this Chapter on a massless Fermion in $(1+1)$-dimensions. In
this setting another new technical issue arises from a zero mode that
is present in the cavity under boundary conditions that may be
considered physically preferred. This zero mode needs to be
regularised in order to apply usual Fock space techniques.

We shall find that the entanglement behaviour of the massless Dirac
Fermion is broadly similar to that found for the massless scalar
in~\cite{alphacentauri2012}, in particular in the periodic
dependence of the entanglement on the durations of the individual
accelerated and inertial segments, and in the property that
entanglement degradation caused by accelerated segments can be
cancelled in the leading order in the small acceleration expansion by
fine-tuning the durations of the inertial segments. We shall however
find that the charge of the Fermionic excitations has a quantitative
effect on the entanglement, and there is in particular interference
between excitations of opposite charge.

We begin in Section~(\ref{sec:staticcavity}) by quantising a massless
Dirac field in an inertial cavity and in a uniformly-accelerated cavity
in $(1+1)$ dimensional Minkowski spacetime. We pay special attention to
the boundary conditions that are required for maintaining unitarity
and to the regularisation of a zero mode that arises under an arguably
natural choice of the boundary conditions.
Section~(\ref{sec:bogo transformation}) develops the Bogoliubov
transformation technique for grafting inertial and uniformly
accelerated trajectory segments, presenting the general building block
formalism and giving detailed results for a trajectory where initial
and final inertial segments are joined by one uniformly accelerated
segment. The evolution of initially maximally entangled states is
analysed in Section~(\ref{sec:fermionic state trafo}), and the results
for entanglement are presented in
Section~(\ref{sec:degradation of entanglement}).
A one-way-trip travel scenario, in which the accelerated cavity undergoes both
acceleration and deceleration, is analysed in Section~(\ref{sec:onewaytrip}). Section~(\ref{sec:conclusion}) presents a brief discussion and concluding
remarks.

\section{Fermionic Cavities}\label{sec:fermions}

\subsection{\label{sec:staticcavity}Cavity Construction}

In this section we quantise the massless Dirac field in an inertial cavity
and in a uniformly accelerated cavity, establishing the notation
and conventions for use in the later sections.

Let $(t,z)$ be standard Minkowski coordinates in
\mbox{$(1+1)$} dimensional Minkowski space,
and let $\eta_{\mu\nu}$ denote the components of the Minkowski metric,
$ds^2 = \eta_{\mu\nu} \, dx^\mu \, dx^\nu = -dt^2 + dz^2$.
From Eq.~(\ref{eqn:diracequation}), the massless Dirac equation reads
\begin{align}
i\left(\boldsymbol{\gamma}^{0}\partial_{t}+\boldsymbol{\gamma}^{3}\partial_{z}\right)\psi=\,0\ ,
\label{eq:massless Dirac eq}
\end{align}
where the $4\times4$ matrices $\boldsymbol{\gamma}^{\mu}$ form the algebra 
$\left\{\boldsymbol{\gamma}^{\mu},\boldsymbol{\gamma}^{\nu}\right\}\,=\,2\,\eta^{\mu\nu}\boldsymbol{I}$. In the massless case, all solutions to the Dirac equation decouple. This means that left/ right movers and left/ right handed particles can be treated independently. Using the representation defined in Section~(\ref{sec:dirac-field}), we define a plane wave basis of solutions as
\begin{align}
\psi_{\omega,\epsilon,\sigma}(t,z)
\,=\,
A_{\omega,\epsilon,\sigma}
\,e^{-i\omega(t-\epsilon z)}\,U_{\epsilon,\sigma}
\, ,
\end{align}
where $\omega\in\BbbR$,
$\epsilon\in\{1,-1\}$,
$\sigma\in\{1,-1\}$,
the constant spinors
$U_{\epsilon,\sigma}$
satisfy
\begin{subequations}
\begin{align}
\boldsymbol{\alpha}_{3} U_{\epsilon,\sigma}
&= \epsilon U_{\epsilon,\sigma} \ ,
\\
\boldsymbol{\gamma}^{5} U_{\epsilon,\sigma}
&= \sigma U_{\epsilon,\sigma} \ ,
\\
U_{\epsilon,\sigma}^{\dagger}
U_{\epsilon^{\prime},\sigma^{\prime}}
& =
\delta_{\epsilon\epsilon^{\prime}}\delta_{\sigma\sigma^{\prime}}\, ,
\label{eq:spinor normalization condition}
\end{align}
\end{subequations}
and $A_{\omega,\epsilon,\sigma}$ is a normalisation constant. In our particular representation the spinors $U_{\epsilon,\sigma}$ are
\begin{eqnarray}
U_{+,+}=\begin{pmatrix} 1 \\ 0 \\ 1 \\ 0 \end{pmatrix},U_{+,-}=\begin{pmatrix} 0 \\ 1 \\ 0 \\ 1 \end{pmatrix},U_{-,+}=\begin{pmatrix} 1 \\ 0 \\ -1 \\ 0 \end{pmatrix},U_{-,-}=\begin{pmatrix} 0 \\ 1 \\ 0 \\ -1 \end{pmatrix}.
\end{eqnarray}
Physically, $\omega$ is the
frequency with respect to the timelike Killing vector field $\partial_{t}$, the eigenvalue
$\epsilon$ of the operator $\boldsymbol{\alpha}_{3}=\boldsymbol{\gamma}^{0}\boldsymbol{\gamma}^{3}$
indicates whether the solution is
a right-mover ($\epsilon=1$) or a left-mover
\mbox{($\epsilon=-1$)}, and $\sigma$ is the eigenvalue of the
helicity/chirality operator
$\boldsymbol{\gamma}^{5}=i\boldsymbol{\gamma}^{0}\boldsymbol{\gamma}^{1}\boldsymbol{\gamma}^{2}\boldsymbol{\gamma}^{3}$~\cite{srednicki2007}.
The right-handed $(\sigma=+1)$ and left-handed $(\sigma=-1)$
solutions are decoupled because
(\ref{eq:massless Dirac eq}) does not contain a mass term.

We encase the field in the inertial cavity
$a \le z \le b$, where $a$ and $b$ are positive
parameters satisfying $a<b$.
The Minkowski coordinate inner product reads
\begin{align}
\left(\,\psi_{(1)},\psi_{(2)}\right)
\,=\,
\int\limits_{a}^{b}\!dz\,\psi_{(1)}^{\dagger}\,\psi_{(2)}
\ ,
\label{eq:mink-ip}
\end{align}
where the integral is
evaluated on a surface of constant~$t$.
To ensure unitarity of the time evolution, so that
the inner product (\ref{eq:mink-ip}) is conserved in time, the Hamiltonian
must be defined as a self-adjoint operator by introducing suitable boundary conditions at
$z=a$ and $z=b$~\cite{simon1975,bonneau2001}.
We specialise to boundary conditions that do not
couple right-handed and left-handed spinors.
For concreteness, we consider from now on
only left-handed spinors,
and we drop the index~$\sigma$.
The analysis for right-handed spinors is similar.

We seek solutions to the Dirac equation of the form
\begin{align}
\psi_{\omega}(t,z)\,=\,A_\omega\,e^{-i\omega(t-z)}\,U_{+}
\,+\,B_\omega
\,e^{-i\omega(t+z)}\,U_{-}\, ,
\label{eq:in-psiraw}
\end{align}
where $A_\omega$ and $B_\omega$ are complex-valued constants and we have used the notation $U_{\pm}\equiv U_{\pm,+}$.
We wish to regard the cavity as an interval
with two spatially separated endpoints hence we specialise
to boundary conditions that ensure
vanishing of the probability current independently at each wall.

The standard conserved current associated with the Dirac field is defined as~\cite{birrell1982,crispino2008}
\begin{eqnarray}
j^{\mu}=\psi^{\dag}\boldsymbol{\gamma}^{0}\boldsymbol{\gamma}^{\mu}\psi.
\end{eqnarray}
The boundary condition on the eigenfunctions thus reads
\begin{align}
\left.\,\psi_{\omega}^{\dag}\boldsymbol{\gamma}^{0}\boldsymbol{\gamma}^{3}\psi_{\omega'}\,\right|_{z=a}
\,=\,
0
\,=\,
\left.\,\psi_{\omega}^{\dag}\boldsymbol{\gamma}^{0}\boldsymbol{\gamma}^{3}\psi_{\omega'}\,\right|_{z=b}.
\label{eq:in-bc}
\end{align}
Following the procedure of~\cite{simon1975,bonneau2001}, we find from
Eq.~(\ref{eq:in-psiraw}) and Eq.~(\ref{eq:in-bc}) that the field's Hamiltonian is specified by two independent phases,
characterising the phase shifts of reflection from the two walls.  We
encode these phases in the parameters $\theta \in [0,2\pi)$ and
$s\in[0,1)$, which can be understood respectively as the normalised
sum and difference of the two phases. The quantum theories then fall
into two qualitatively different cases, the generic case
$0<s<1$ and the special case $s=0$.

In the generic case $0<s<1$, the orthonormal eigenfunctions are
\begin{subequations}
\label{eq:in-psis}
\begin{align}
\psi_{n}(t,z)
& =
\frac{e^{-i\omega_{\!n}t}\left[e^{+i\omega_{\!n}(z-a)}\,U_{+}\,+\,e^{i\theta}
e^{-i\omega_{\!n}(z-a)}\,U_{-}\right]}{\sqrt{2\cavlength}}
\, ,
\label{eq:in-psis-psi}
\\
\omega_{n}
&= \frac{(n+s)\pi}{\cavlength}
\ ,
\end{align}
\end{subequations}
where $n\in\BbbZ$ and $\cavlength := b-a$.
Note that $\omega_{n} \ne0$ for all~$n$,
and positive (or negative)
frequencies are obtained for $n\ge0$ ($n<0$). 
A~Fock space quantisation can be performed in
a standard manner~\cite{srednicki2007}. Following the free field case~(\ref{eqn:minkowski-car}), the canonical procedure for quantising the Dirac field is by imposing the equal time anti-commutation relations
\begin{eqnarray}
\label{eq:dirac car}
\left\lbrace\hat{\psi}_{\alpha}(t,z),\hat{\psi}^{\dag}_{\beta}(t,z')\right\rbrace=\delta_{\alpha\beta}\delta(z-z'),
\end{eqnarray}
where $\hat{\psi}_{\alpha}$ denotes the $\alpha$ component of the field $\hat{\psi}$ (not to be confused with the mode solutions $\psi_{n}$) and all other anti-commutators vanishing.

The special case $s=0$ corresponds to assuming that the two walls are
of identical physical build. In this case $\omega_{n} \ne0$ for
$n\ne0$ but $\omega_{0}= 0$. In what follows we
consider the $s=0$ quantum theory to be defined by first quantising
with $s>0$ and at the end taking the limit $s\to0_+$.  All our
entanglement measures will be seen to remain well defined in this
limit.

Coordinates convenient for the accelerated cavity are the
Rindler coordinates $(\eta,\chi)$,
defined in the right Rindler wedge $z > |t|$ by
\begin{align}
t=\chi\sinh\eta
\ , \
z=\chi\cosh\eta\ ,
\label{eq:mink-in-rindler}
\end{align}
where $0<\chi<\infty$ and $-\infty<\eta<\infty$.
The metric reads $ds^2 = -\chi^2 d\eta^2 + d\chi^2$.
A uniformly accelerated cavity, as described by a comoving observer, sits in the interval $a\le\chi\le b$,
and the boost Killing vector field along which
the cavity is dragged takes the form $\partial_\eta$. For more details see Section~(\ref{app:rindler-coords}).

In the inertial frame, the cavity walls are
on the worldlines
$z = \sqrt{a^2 + t^2}$
and
$z = \sqrt{b^2 + t^2}$, where the notation is as above.
The proper accelerations of the ends are $1/a$ and~$1/b$ respectively,
and the cavity as a whole is static in the sense that
it is dragged along the boost Killing vector field
$\partial_{\eta}=z\partial_t + t\partial_z$.
At $t=0$ the accelerated cavity overlaps precisely
with the inertial cavity.

In Rindler coordinates the massless Dirac equation (\ref{eqn:rindler-dirac-equation})
takes the form
\cite{birrell1982,mcmahon2006}
\begin{eqnarray}
i\partial_{\eta}\widehat{\psi}_{\Omega,\epsilon}=-i\left(\boldsymbol{\alpha}_{3}/2+\boldsymbol{\alpha}_{3}\chi\partial_{\chi}\right)\widehat{\psi}_{\Omega,\epsilon},
\end{eqnarray}
and the inner product for a field encased in
the accelerated cavity reads
\begin{align}
\bigl(\,{\widehat{\psi}}_{(1)},{\widehat{\psi}}_{(2)}\bigr)
\,=\,
\int\limits_{a}^{b} d\chi\,{\widehat{\psi}}_{(1)}^{\dagger}\,{\widehat{\psi}}_{(2)}
\ ,
\label{eq:rindler-ip}
\end{align}
where the integral is evaluated on a surface of constant~$\eta$.
We find that the orthonormal energy eigenfunctions are
\begin{subequations}
\label{eq:rind-psis}
\begin{align}
{\widehat{\psi}}_{n}(\eta ,\chi)
&=
\frac{\displaystyle e^{-i\Omega_{n}\eta}\left[\left(\frac{\chi}{a}\right)^{i\Omega_{n}}U_{+}+
e^{i\theta}\left(\frac{\chi}{a}\right)^{-i\Omega_{n}}U_{-}\right]}{\sqrt{2\chi\ln(b/a)}} ,
\\
\Omega_{n}
&=
\frac{(n\,+\,s)\pi}{\ln(b/a)}\ ,
\label{eq:rindler solutions frequency}
\end{align}
\end{subequations}
where $n\in\BbbZ$.
The parameters $\theta$ and $s$ have the same meaning and values as above:
we consider the microphysical build of the cavity
walls not to be affected by their acceleration.
In analogy with the inertial case, we impose the same anti-commutation relations~(\ref{eqn:rindler-car}) for the Rindler coordinate Dirac field.
For $s=0$ the mode $n=0$ is again a zero mode, and we consider the
$s=0$ quantum theory to be defined as the limit $s\to0_+$.

For simplicity, in the following analysis of the Dirac field, we drop our previous notation of denoting operators with hats.

\subsection{\label{sec:bogo transformation}Non-uniform Motion}

We now turn to a cavity whose
trajectory consists of inertial and uniformly accelerated segments.

\begin{figure}[t]
\begin{center}
\includegraphics[width=0.7\textwidth]{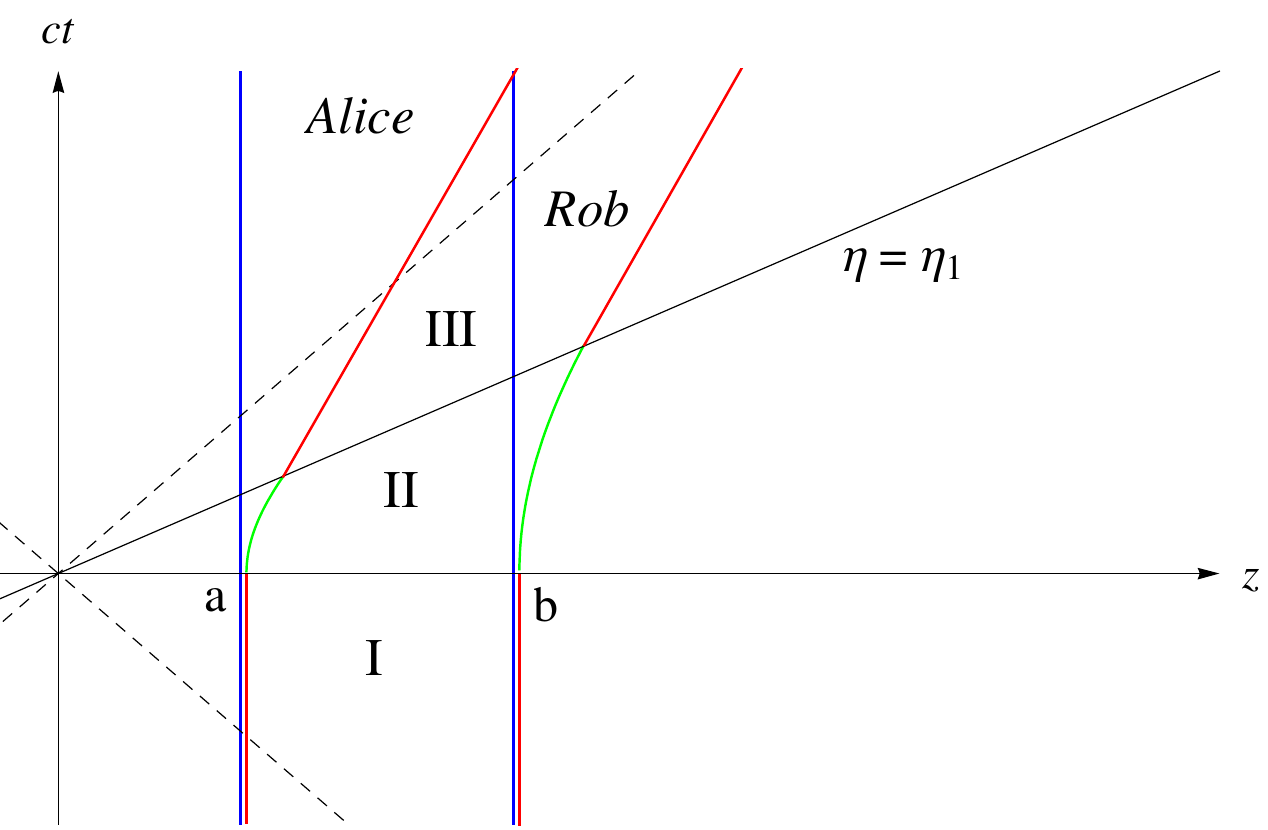}
\caption[Fermion Cavity Set-up]{Space-time diagram of cavity motion is shown.
Rob's cavity is at rest initially (Region~$\mathrm{I}$),
then undergoes a period of uniform acceleration from $t=0$ to $\eta=\eta_{1}$
(Region ($\mathrm{I\!I}$)) and is thereafter
again inertial (Region ($\mathrm{I\!I\!I}$)).
Alice's cavity overlaps with
Rob's cavity in Region $\mathrm{I}$
and remains inertial throughout.}
\label{fig:bogobogoboxes}
\end{center}	
\end{figure}

The prototype cavity configuration is shown in Fig.~(\ref{fig:bogobogoboxes}).
Two cavities, referred to as Alice and Rob, are initially
inertial. At $t=0$, Rob's cavity begins to accelerate to the right,
following the Killing vector field $K = \partial_\eta$.
We found that it was more convenient to compute the entanglement in terms of the proper time and proper acceleration at the centre of our cavities. Thus from now on, any proper time and proper acceleration will be that at the centre of a cavity unless stated otherwise. The acceleration ends at Rindler time $\eta = \eta_1$,
and the duration of the acceleration in proper time measured at the
centre of the cavity is
$\tau_1 := \tfrac12 (a+b)\eta_1$.
We refer to the three segments of Rob's trajectory as
Regions ($\mathrm{I}$), ($\mathrm{I\!I}$) and~($\mathrm{I\!I\!I}$).
Alice remains inertial throughout.
We shall discuss the evolution in Rob's cavity in two steps:
first from Region $\mathrm{I}$ to Region ($\mathrm{I\!I}$)
and then from Region ($\mathrm{I\!I}$) to Region ($\mathrm{I\!I\!I}$).
We then use the evolution to relate the operators and the vacuum of Region
($\mathrm{I}$) to those in Region~($\mathrm{I\!I\!I}$).

Consider the Dirac field in Rob's cavity. In Regions
($\mathrm{I}$) and ($\mathrm{I\!I}$) we may expand the field using
the solutions (\ref{eq:in-psis}) and (\ref{eq:rind-psis}) respectively as
\begin{subequations}
\label{eq:regionsI-IIquantization}
\begin{align}
\mathrm{I}:\ \ \
\psi	&=	\sum\limits_{n\geq0}a_{n}\,\psi_{n}
\,+\,
\sum\limits_{n<0}b_{n}^{\dagger}\,\psi_{n}
\ ,
\label{eq:region I quantization}\\[2mm]
\mathrm{I\!I}:\ \ \
\psi	&=	\sum\limits_{m\geq0}\widehat{a}_{m}\,\widehat{\psi}_{m}
\,+\,
\sum\limits_{m<0}\widehat{b}_{m}^{\dagger}\,\widehat{\psi}_{m}
\ ,
\label{eq:region II quantization}
\end{align}
\end{subequations}
and due to the quantisation rule~(\ref{eq:dirac car}) the nonvanishing anticommutators are
\begin{subequations}
\label{eq:nonvanishing anticommutators}
\begin{align}
\mathrm{I}:\ \ \
&\left\{a_{m},a_{n}^{\dagger}\right\}=
\left\{b_{m},b_{n}^{\dagger}\right\}=\delta_{mn} \, ,
\\
\mathrm{I\!I}:\ \ \
&\left\{\widehat{a}_{m},\widehat{a}_{n}^{\dagger}\right\}=
\left\{\widehat{b}_{m},\widehat{b}_{n}^{\dagger}\right\}=\delta_{mn} \, .
\end{align}
\end{subequations}
Note in the above we have chosen the convention to denote \emph{anti-particles} mode functions with $\psi_{n}$ and $\hat{\psi}_{n}$ with the understanding $n<0$.

The critical step in our procedure is relating the ``in" inertial field operator and the final ``out" accelerated field operator. It is this transformation which affects entanglement and allows us to model the motion of a cavity through spacetime. The standard method of calculating these transformations is through \emph{Bogoliubov transformations}.

Following Chapter~(\ref{chapter:bosons}), we match the two Dirac field expansions~(\ref{eq:regionsI-IIquantization})
at $t=0$, and define the Bogoliubov transformation
\begin{align}
\label{eq:change of basis}
\widehat{\psi}_{m}
= \sum\limits_{n=-\infty}^{\infty}A_{mn}\,\psi_{n}
\ , \ \ \
\psi_{n}
= \sum\limits_{m=-\infty}^{\infty}\bar{A}_{mn}\,\widehat{\psi}_{m}\, ,
\end{align}
where the elements of the
Bogoliubov coefficient matrix $A_{mn}$
are given by
\begin{align}
A_{mn}\,:=\,\bigl(\,\psi_{n},\widehat{\psi}_{m}\,\bigr)|_{t=0}
\label{eq:bogo components from I to II}
\end{align}
and the inner product in (\ref{eq:bogo components from I to II})
is evaluated on the hypersurface $t=0$. Using the orthonormality and completeness relation of the field modes, it is easy to show
\begin{eqnarray}
\left(\boldsymbol{A}\boldsymbol{A}^{\dag}\right)_{nm}&\equiv&\delta_{nm}
\end{eqnarray}
thus $\boldsymbol{A}^{\dag}=\boldsymbol{A}^{-1}$. This relation will be useful for calculating inverse transformations later, however it is \emph{not} the symplectic transformation we found in Chapter~(\ref{chapter:bosons}).

To compute the Bogoliubov coefficients, we shall work perturbatively in the limit
where the acceleration of Rob's cavity
is small. To implement this, we follow
the Bosonic case of Chapter~(\ref{chapter:bosons}) and
introduce the dimensionless parameter
$h:=2\cavlength/(a+b)$, satisfying $0<h<2$.
Physically, $h$ is the product of the cavity's
length $\cavlength$ and the acceleration at
the centre of the cavity.
Expanding
Eq.~(\ref{eq:bogo components from I to II})
in a Maclaurin series in~$h$,
we find
\begin{align}
\boldsymbol{A}=\boldsymbol{A}^{(0)}\,+\,\boldsymbol{A}^{(1)}h\,+\,\boldsymbol{A}^{(2)}h^{2}\,+\,{O}(h^{3})\ ,
\label{eq:bogo coefficient expansion}
\end{align}
where the superscript indicates
the power of $h$ and
the explicit expressions for
$\boldsymbol{A}^{(0)}$, $\boldsymbol{A}^{(1)}$ and $\boldsymbol{A}^{(2)}$ read
\begin{subequations}
\label{eq:bogo coeff pert}
\begin{align}
A_{mn}^{(0)}
&=
\delta_{mn}\,,\label{eq:region I to II bogo order zero}
\\[1ex]
A_{nn}^{(1)}
&= 0 \,,
\\[1ex]
A_{mn}^{(1)}
&=
\frac{\bigl[(-1)^{m+n}-1\bigr](m+n+2s)}{2\pi^{2}(m-n)^{3}}\, \,,
\ \ \ (m\ne n)
\\
A_{nn}^{(2)}
&= -
\left(\frac{1}{96}+\frac{\pi^{2}(n+s)^2}{240}\right)\,,
\\[1ex]
A_{mn}^{(2)}
&=
\frac{\bigl[(-1)^{m+n}+1\bigr]}{8\pi^{2}(m-n)^{4}}
\bigl[(m+s)^{2}+3(n+s)^{2}
\notag
\\[.5ex]
&
\hspace{9ex}
+8(m+s)(n+s) \bigr]\,.
\ \ \ (m\ne n)
\end{align}
\end{subequations}
The procedure for the derivation of the above Bogoliubov coefficients can be found in appendix~(\ref{appendix:dirac-bogo}).
The perturbative unitarity of $\boldsymbol{A}$ persists in the
limit $s\to0_+$.

In Region ($\mathrm{I\!I\!I}$), we expand the Dirac field in Rob's cavity as
\begin{align}
\mathrm{I\!I\!I}:\ \ \
\psi	&=	\sum\limits_{n\geq0}\tilde{a}_{n}\,\tilde{\psi}_{n}
\,+\,
\sum\limits_{n<0}\tilde{b}_{n}^{\dagger}\,\tilde{\psi}_{n}
\ ,
\label{eq:region III quantization}
\end{align}
where the mode functions $\tilde{\psi}_{n}$ are as in
Eq.~(\ref{eq:in-psis}) but $(t,z)$ are replaced by the
Minkowski coordinates $(\tilde{t},\tilde{z})$
adapted to the cavity's new rest frame,
with the surface $\tilde{t}=0$ coinciding with $\eta=\eta_1$.
The nonvanishing anticommutators are
\begin{align}
\mathrm{I\!I\!I}:\ \ \
&\left\{\tilde{a}_{m},\tilde{a}_{n}^{\dagger}\right\}=
\left\{\tilde{b}_{m},\tilde{b}_{n}^{\dagger}\right\}=\delta_{mn} \, .
\label{eq:III-anticommutators}
\end{align}
The Bogoliubov transformation between the
Region
($\mathrm{I}$)
and Region
($\mathrm{I\!I\!I}$)
modes can then be written as
\begin{align}
\label{eq:IIIchange of basis}
\tilde{\psi}_{m}
= \sum\limits_{n}\mathcal{A}_{mn}\,\psi_{n}
\ , \ \ \
\psi_{n}
= \sum\limits_{m}\bar{\mathcal{A}}_{mn}\,\tilde{\psi}_{m}\, ,
\end{align}
where the coefficient matrix $\mathcal{A}_{mn}$
has the decomposition
\begin{align}
\mathbfcal{A}
=
\boldsymbol{A}^\dagger\,\boldsymbol{G}(\eta_{1})\,\boldsymbol{A},
\label{eq:bogo from I to III}
\end{align}
and $\boldsymbol{G}(\eta_1)$
is the diagonal matrix whose elements are
\begin{align}
G_{nn}(\eta_1) = G_n :=
\exp(i\,\Omega_{n}\,\eta_{1})
\,.
\label{eq:free Rindler evolution diagonal matrix}
\end{align}
During coasting periods of either constant velocity or constant acceleration, the states undergo free evolution i.e. there are no interactions between the modes. Therefore, the role of $\boldsymbol{G}(\eta_{1})$ in Eq.~(\ref{eq:bogo from I to III})
is to compensate for the phases that the modes $\widehat{\psi}_{m}$
develop between $\eta=0$ and $\eta=\eta_1$. 

As parametrised via the proper time and acceleration at the centre of the cavity, the Rindler phases are given by
\begin{eqnarray}
\Omega_{n}(h)\eta_{1}(\tau)=\omega_{n}\tau+O(h^{2}),
\end{eqnarray}
thus the power expansion of $\boldsymbol{G}(\eta_{1})$ is given by
\begin{eqnarray}
\boldsymbol{G}(\eta_{1})=\boldsymbol{G}^{(0)}(\tau_{1})+\boldsymbol{G}^{(2)}h^{2}+O(h^{3}),
\end{eqnarray}
where $G^{(0)}_{nm}=e^{i\omega_{n}\tau_{1}}\delta_{nm}$ and $\omega_{n}$ is the Minkowski frequency of mode $n$.

The matrix $\boldsymbol{A}^\dagger = \boldsymbol{A}^{-1}$
in Eq.~(\ref{eq:bogo from I to III})
arises from matching
Region
($\mathrm{I\!I}$)
to Region
($\mathrm{I\!I\!I}$) at $\eta=\eta_1$. Expanding $\mathbfcal{A}$ in a Maclaurin series in $h$ as
\begin{align}
\mathbfcal{A}=\mathbfcal{A}^{(0)}\,+\,\mathbfcal{A}^{(1)}h
\,+\,
\mathbfcal{A}^{(2)}h^{2}\,+\,{O}(h^{3})\ ,
\label{eq:IIIbogo coefficient expansion}
\end{align}
where the superscript again indicates
the power of $h$, we obtain from
Eq.~(\ref{eq:bogo coefficient expansion}) and Eq.~(\ref{eq:bogo from I to III})
\begin{subequations}
\label{eq:region I to III bogos three orders}
\begin{align}
\mathbfcal{A}^{(0)}
&=
\boldsymbol{G}^{(0)}
\,,
\label{eq:region I to III bogo order zero}
\\[1.5mm]
\mathbfcal{A}^{(1)}
&=
\boldsymbol{G}^{(0)}\,\boldsymbol{A}^{(1)} + \bigl(\boldsymbol{A}^{(1)}\bigr)^\dagger \boldsymbol{G}^{(0)}
\,,
\label{eq:region I to III bogo order one}
\\[1.5mm]
\mathbfcal{A}^{(2)}
&=
\boldsymbol{G}^{(0)}\,\boldsymbol{A}^{(2)}
+ \bigl(\boldsymbol{A}^{(2)}\bigr)^\dagger \boldsymbol{G}^{(0)}
+ \bigl(\boldsymbol{A}^{(1)}\bigr)^\dagger \boldsymbol{G}^{(0)} \,\boldsymbol{A}^{(1)}
+\bigl(\boldsymbol{A}^{(0)}\bigr)^\dagger \boldsymbol{G}^{(2)} \,\boldsymbol{A}^{(0)}
\,.
\label{eq:region I to III bogo order two}
\end{align}
\end{subequations}
Note that as the diagonal elements of $\boldsymbol{A}^{(1)}$ are vanishing and $\boldsymbol{G}$ is diagonal,
the diagonal elements of $\mathbfcal{A}^{(1)}$ also vanish.
Unitarity of $\mathbfcal{A}$ implies the perturbative relations
\begin{subequations}
\begin{align}
\boldsymbol{0} &= \bar{\boldsymbol{G}}^{(0)}\mathbfcal{A}^{(1)}
+
\bigl(\mathbfcal{A}^{(1)}\bigr)^\dagger \boldsymbol{G}^{(0)}
\,,
\label{eq:first order unitarity condition}
\\[1.5mm]
\boldsymbol{0} &= \bar{\boldsymbol{G}}^{(0)} \mathbfcal{A}^{(2)}
+
\bigl(\mathbfcal{A}^{(2)}\bigr)^\dagger \boldsymbol{G}^{(0)}
+ \bigl(\mathbfcal{A}^{(1)}\bigr)^\dagger \mathbfcal{A}^{(1)}
\,,
\label{eq:second order unitarity condition}
\end{align}
\end{subequations}
which will be useful below.

We denote the Fock vacua in Regions
($\mathrm{I}$) and ($\mathrm{I\!I\!I}$)
by $\left|\,0\,\right\rangle$ and $\left|\,\tilde{0}\,\right\rangle$ respectively.
To relate the two,
we mimic the Bosonic case and make the ansatz~\cite{fabbri2005}
\begin{eqnarray}
\left|\,0\,\right\rangle\,=\,N e^W \left|\,\tilde{0}\,\right\rangle \, ,
\label{eq:vacuum trafo}
\end{eqnarray}
where
\begin{eqnarray}
W\,=\,\sum\limits_{p\ge 0}\sum\limits_{q<0}\,V_{pq}\,\tilde{a}_{p}^{\dagger}\,\tilde{b}_{q}^{\dagger},
\label{eq:W vacuum trafo}
\end{eqnarray}
and the coefficient matrix $V_{pq}$
and the normalisation constant $N$ are to be determined.
Note that the two indices of $V$ take values in disjoint sets.

Using the same concepts that were introduced in Section~(\ref{sec:bogo-qft}), it follows from
Eq.~(\ref{eq:region III quantization})
and
Eq.~(\ref{eq:IIIchange of basis})
that the creation and annihilation operators
in Regions
($\mathrm{I}$) and ($\mathrm{I\!I\!I}$) are related by
\begin{subequations}
\begin{align}
n\geq0:\
a_{n} &=\left(\,\psi_{n},\psi\,\right)
= \sum\limits_{m\geq0}\tilde{a}_{m}\,\mathcal{A}_{mn}\,+\,\sum\limits_{m<0}\tilde{b}_{m}^{\dagger}\,\mathcal{A}_{mn}\ ,
\label{eq:bogo decomposition of a sub n full}
\\
n<0:\ b_{n}^{\dagger} &= \left(\,\psi_{n},\psi\,\right)
=
\sum\limits_{m\geq0}\tilde{a}_{m}\,\mathcal{A}_{mn}\,+\,\sum\limits_{m<0}\tilde{b}_{m}^{\dagger}\,\mathcal{A}_{mn}\ .
\label{eq:bogo decomposition of bdag sub n full}
\end{align}
\end{subequations}
Using Eq.~(\ref{eq:vacuum trafo}) and Eq.~(\ref{eq:bogo decomposition of a sub n full}),
the vacuum annihilation condition $a_{n}\left|\,0\,\right\rangle=0$ reads
\begin{align}
\Biggl(\,
\sum\limits_{m\geq0}\tilde{a}_{m}\,\mathcal{A}_{mn}
+\sum\limits_{m<0}\tilde{b}_{m}^{\dagger}\,\mathcal{A}_{mn}
\Biggr)
e^{W} \left|\,\tilde{0}\,\right\rangle=0\,.
\label{eq:using vacuum definition}
\end{align}
From the anticommutators
(\ref{eq:III-anticommutators})
it follows that
\begin{subequations}
\label{eq:aW-commutators}
\begin{align}
& \left[\,W\,,\,\tilde{a}_{m}\,\right]\,=\,-\sum\limits_{q<0}V_{mq}\,\tilde{b}_{q}^{\dagger}\,,
\hspace{5mm}(m\ge 0)\\
& \left[\,W\,,\,\left[\,W\,,\,\tilde{a}_{m}\,\right]\,\right]\,=\,0\,\hspace{5mm}(m\ge 0).
\end{align}
\end{subequations}
Upon multiplying on the left by $e^{-W}$, using Eq.~(\ref{eq:aW-commutators}) and the
Hadamard lemma,
\begin{align}
e^{A}\,Be^{-A}\,=\,B\,+\,\left[\,A\,,\,B\,\right]\,+\,\tfrac{1}{2}\left[\,A\,,\,\left[\,A\,,\,B\,\right]\,\right]\,+\,\ldots
,
\label{eq:hadamard lemma}
\end{align}
Eq.~(\ref{eq:using vacuum definition}) reduces to
\begin{align}
\sum\limits_{m\geq0}\mathcal{A}_{mn}\,V_{mq}\,=\,-\,\mathcal{A}_{qn}
\ \ \ (n\,\geq\,0\,,\ q\,<\,0)\ .
\label{eq:first relation between V and A}
\end{align}
A similar computation shows that
the condition $b_{n}\,\left|\,0\,\right\rangle=0$
reduces to
\begin{align}
\sum\limits_{m<0}\bar{\mathcal{A}}_{mn}\,V_{pm}\,=\,\,\bar{\mathcal{A}}_{pn}
\ \ \ (n\,<\,0\,,\ p\,\geq\,0)\ .
\label{eq:second relation between V and A}
\end{align}
Working perturbatively in~$h$, the invertibility
assumptions hold, and using Eq.~(\ref{eq:IIIbogo coefficient expansion})
and Eq.~(\ref{eq:region I to III bogos three orders}) we find, with $\mathbfcal{A}_{nm}^{(0)}$ being diagonal, $\boldsymbol{V}^{(0)}$ identically vanishes and
\begin{align}
\boldsymbol{V}\,=\,\boldsymbol{V}^{(1)}h\,+\,{O}(h^{2})
\label{eq:expansion of V}
\end{align}
where
\begin{align}
V_{pq}^{(1)}\,=\,-\mathcal{A}_{qp}^{(1)}\,\bar{G}_{p}
\,=\,
\bar{\mathcal{A}}_{pq}^{(1)}\,G_{q}\,\ \ (p\ge 0, \ q<0).
\label{eq:leading order of V}
\end{align}
We shall show in
Section~(\ref{sec:fermionic state trafo}) that the normalisation
constant $N$ has the small $h$ expansion
\begin{align}
N\,=\,1\,-\,\tfrac{1}{2}\sum\limits_{p,q}|V_{pq}|^{2}
\ +{O}(\boldsymbol{V}^{3})\,.
\label{eq:normalization constant to order h squared}
\end{align}

\subsection{\label{sec:fermionic state trafo}Evolution of entangled states}

In this Section we apply the results of Section~(\ref{sec:bogo transformation})
to the evolution of Bell-type quantum states between the two cavities which are initially maximally
entangled. There are two questions we wish to address. The first is do Bosons and Fermions behave in a similar manner? In other words, do we see degradation affects on entangled states due to non-uniform motion. The second question is how does the charge of a spin-$1/2$ particle affect and change the degradation? Identifying any differences (or similarities) between Bosons and Fermions could allow us to propose new quantum information protocols which exploit the nature of the fields we are considering.

We shall work perturbatively to quadratic order in~$h$. Focusing first on Rob's cavity only,
we write out the Region ($\mathrm{I}$) vacuum and the Region ($\mathrm{I}$) states with a
single (anti-) particle in terms of Region ($\mathrm{I\!I\!I}$)
excitations on the Region ($\mathrm{I\!I\!I}$) vacuum.
Next, we address
an entangled state where one field mode is controlled by Alice and one by Rob.
Finally we investigate
a state of the type analysed in~\cite{martinmartinez2011}
where the entanglement between Alice and Rob
is in the charge of the field modes.

\subsection*{\label{subsec:vacuum and single particle states}Rob's cavity: vacuum and single-particle states}

Consider the Region ($\mathrm{I}$) vacuum $\left|\,0\,\right\rangle$ in
Rob's cavity.
We shall use Eq.~(\ref{eq:vacuum trafo}) to express this state in terms of
Region ($\mathrm{I\!I\!I}$) excitations over the
Region ($\mathrm{I\!I\!I}$) vacuum~$\left|\,\tilde{0}\,\right\rangle$. From now on, we assume $p,i\ge 0$ and $q,j<0$ unless stated otherwise.

We expand the exponential in Eq.~(\ref{eq:vacuum trafo}) as
\begin{equation}
\begin{aligned}	 		
e^W
&=
\mathds{1}\,+\,\sum\limits_{p,q}V_{pq}\,\tilde{a}_{p}^{\dagger}\,\tilde{b}_{q}^{\dagger}
\\
&\hspace{5ex}
+ \tfrac{1}{2}\,\sum\limits_{p,q,i,j}V_{pq}\,V_{ij}\,
\tilde{a}_{p}^{\dagger}\,\tilde{b}_{q}^{\dagger}\,\tilde{a}_{i}^{\dagger}\,\tilde{b}_{j}^{\dagger}
\,+\,{O}(\boldsymbol{V}^{3})\,.
\label{eq:expansion of expW}
\end{aligned}
\end{equation}
In Region ($\mathrm{I\!I\!I}$), we adopt the notation
\begin{align}
\left|\tilde{1}_{p}\right\rangle^{\!+}\ldots\left|\tilde{1}_{i}\right\rangle^{\!+} \left|\tilde{1}_{q}\right\rangle^{\!-}\ldots\left|\tilde{1}_{j}\right\rangle^{\!-}&:= \tilde{a}_{p}^{\dagger}\ldots\,\tilde{a}_{i}^{\dagger}\,\tilde{b}_{q}^{\dagger}\ldots\,\tilde{b}_{j}^{\dagger}\left|\,\tilde{0}\,\right\rangle
\end{align}
and the superscript $\pm$ indicates the sign of the charge. From Eq.~(\ref{eq:expansion of expW}) we obtain
\begin{equation}
\begin{aligned}
e^{W}\,\left|\,\tilde{0}\,\right\rangle
&=
\left|\,\tilde{0}\,\right\rangle
\,+\,\sum\limits_{p,q}V_{pq}\,
\left|\,\tilde{1}_{p}\right\rangle^{+}\,\left| \tilde{1}_{q}\,\right\rangle^{-}			
\\[.5ex]
&\hspace{5mm}
-\tfrac{1}{2}\sum\limits_{p,q,i,j}V_{pq}V_{ij}\varphi_{pi}\varphi_{qj}
\left|\tilde{1}_{p}\right\rangle^{\!+}
\left|\tilde{1}_{i}\right\rangle^{\!+}
\left|\tilde{1}_{q}\right\rangle^{\!-}
\left|\tilde{1}_{j}\right\rangle^{\!-}
+{O}(\boldsymbol{V}^{3})\,,
\label{eq:expansion of expW acting on tilde0}
\end{aligned}
\end{equation}
where we have employed the notation
\begin{eqnarray}
\varphi_{mn}:=(1-\delta_{mn}),
\end{eqnarray}
which encodes the Pauli exclusion principle of our Fermions into the summations. The overall negative sign in the third term of the r.h.s of Eq.~(\ref{eq:expansion of expW acting on tilde0}) is due to our choice of basis. Imposing vacuum normalisation~$\left\langle 0|0\right\rangle=+1$ and using
\begin{eqnarray}
\left\langle\tilde{1}_{q}\right|^{\!-}\left\langle\tilde{1}_{p}\right|^{\!+}\cdot\left|\tilde{1}_{i}\right\rangle^{\!+}\left|\tilde{1}_{j}\right\rangle^{\!-}=\delta_{pi}\delta_{qj}
\end{eqnarray}
it follows that the normalisation constant $N$ is given
by Eq.~(\ref{eq:normalization constant to order h squared}),
and Eq.~(\ref{eq:vacuum trafo}) gives
\begin{equation}
\begin{aligned}
\left|\,0\,\right\rangle
&=
\Bigl(1\,-\,\tfrac{1}{2}\sum\limits_{p,q}|V_{pq}|^{2}\Bigr)
\left|\,\tilde{0}\,\right\rangle
\,+\,\sum\limits_{p,q}V_{pq}\,
\left|\,\tilde{1}_{p}\right\rangle^{+}\,\left| \tilde{1}_{q}\,\right\rangle^{-}			
\\[.5ex]
&\hspace{5mm}
-\tfrac{1}{2}\sum\limits_{p,q,i,j}V_{pq}V_{ij}\varphi_{pi}\varphi_{qj}
\left|\tilde{1}_{p}\right\rangle^{\!+}
\left|\tilde{1}_{i}\right\rangle^{\!+}
\left|\tilde{1}_{q}\right\rangle^{\!-}
\left|\tilde{1}_{j}\right\rangle^{\!-}
+{O}(\boldsymbol{V}^{3})\,.
\label{eq:region I vacuum state to order h squared}
\end{aligned}
\end{equation}
Consider in Rob's cavity the state with exactly one
Region ($\mathrm{I}$) particle,
$\left|1_{k}\right\rangle^{\!-} := {b}_{k}^{\dagger} \left|\,0\,\right\rangle$ for $k<0$
or
\mbox{$\left|1_{k}\right\rangle^{\!+} := {a}_{k}^{\dagger} \left|\,0\,\right\rangle$} for $k\ge0$.
Acting on the Region ($\mathrm{I}$) vacuum (\ref{eq:region I vacuum state to order h squared})
by Eq.~(\ref{eq:bogo decomposition of bdag sub n full})
and the Hermitian conjugate of Eq.~(\ref{eq:bogo decomposition of a sub n full}) respectively,
we find
\begin{equation}
\begin{aligned}
k<0:
\ \ \
\left|1_{k}\right\rangle^{\!-}
&=
\sum\limits_{p,q}V_{pq}\mathcal{A}_{pk}\left|\tilde{1}_{q}\right\rangle^{\!-}+
\sum\limits_{m<0}\mathcal{A}_{mk}
\Biggl[
\Bigl(1-\tfrac{1}{2}\sum\limits_{p,q}|V_{pq}|^{2}\Bigr)
\left|\tilde{1}_{m}\right\rangle^{\!-}\\
&\hspace{3mm}+\sum\limits_{p,q}V_{pq}\varphi_{mq}
\left|\tilde{1}_{p}\right\rangle^{\!+}\!
\left|\tilde{1}_{q}\right\rangle^{\!-}\!
\left|\tilde{1}_{m}\right\rangle^{\!-} \\
&\hspace{3mm} - \tfrac{1}{2}\sum\limits_{p,q,i,j}\hspace*{-1mm}V_{pq}V_{ij}
\varphi_{pi}\varphi_{qj}\varphi_{mq}\varphi_{mj}
\left|\tilde{1}_{p}\right\rangle^{\!+}\!
\left|\tilde{1}_{i}\right\rangle^{\!+}\!
\left|\tilde{1}_{q}\right\rangle^{\!-}\!
\left|\tilde{1}_{j}\right\rangle^{\!-}\!
\left|\tilde{1}_{m}\right\rangle^{\!-}
\Biggr]\\
& \hspace{20ex}+{O}(\boldsymbol{V}^{3})\,,
\label{eq:region I single antiparticle state to order h squared}
\end{aligned}
\end{equation}
\begin{equation}
\begin{aligned}
\label{eq:region I single particle state to order h squared}
k>0: \ \ \
\left|1_{k}\right\rangle^{\!+}
&=
-\sum\limits_{p,q}V_{pq}\bar{\mathcal{A}}_{qk}
\left|\tilde{1}_{p}\right\rangle^{\!+}+
\sum\limits_{m\geq0}\bar{\mathcal{A}}_{mk}
\Biggl[
\Bigl(1-\tfrac{1}{2}\sum\limits_{p,q}|V_{pq}|^{2}\Bigr)
\left|\tilde{1}_{m}\right\rangle^{\!+}\\
& \hspace{3mm}+\sum\limits_{p,q}V_{pq}\,\varphi_{mp}
\left|\tilde{1}_{m}\right\rangle^{\!+}\!
\left|\tilde{1}_{p}\right\rangle^{\!+}\!
\left|\tilde{1}_{q}\right\rangle^{\!-}\\
& \hspace{3mm} - \tfrac{1}{2}\sum\limits_{p,q,i,j}V_{pq}V_{ij}
\varphi_{pi}\varphi_{qj}\varphi_{mp}\varphi_{mi}
\left|\tilde{1}_{m}\right\rangle^{\!+}\!
\left|\tilde{1}_{p}\right\rangle^{\!+}\!
\left|\tilde{1}_{i}\right\rangle^{\!+}\!
\left|\tilde{1}_{q}\right\rangle^{\!-}\!
\left|\tilde{1}_{j}\right\rangle^{\!-}
\Biggr]\\
& \hspace{20ex}+{O}(\boldsymbol{V}^{3})\,.
\end{aligned}
\end{equation}

\subsection*{\label{subsec:two mode entangled states}Entangled two-mode states}

We wish to consider a Region ($\mathrm{I}$) state where
one field mode is controlled by Alice and one by Rob.
Concretely, we take
\begin{equation}
\begin{aligned}
\label{eq:initial pure state rob both}
\left|\,\phi^{\pm}_{\text{init}}\,\right\rangle_{\zeta}
&=
\tfrac{1}{\sqrt{2}} \left(\,
\left|\,0\,\right\rangle_{A}\left|\,0\,\right\rangle_{R}\,\pm\,
\left|\,1_{\hat{k}}\,\right\rangle_{A}^{\kappa}\left|\,1_{k}\,\right\rangle_{R}^{\zeta}\,\right),
\end{aligned}
\end{equation}
where the subscripts $A$ and $R$ refer to the cavity and the
superscripts $\kappa$ and $\zeta$ indicate whether the mode has positive or negative
frequency, so that
$\kappa\hspace{1mm}(\zeta) =+$ for $\hat{k}\hspace{1mm}(k)\ge0$ and 
$\kappa\hspace{1mm}(\zeta) =-$ for $\hat{k}\hspace{1mm}(k)<0$.
Furthermore, we consider the two particle basis state of the two mode Hilbert space,
corresponding to one excitation each in the modes $\hat{k}$ in Alice's cavity and~$k$ in
Rob's cavity. As pointed out in
Ref.~\cite{montero2011}, making such a choice can lead to ambiguities in the
entanglement. It should be pointed out that the ambiguity only arises due to the inability to map Fermionic states to qubit states in a consistent way. However, if one is to define states in terms of creation operators and work with those then no problems arise. In our case, the ambiguity amounts to a relative phase shift of $\pi$, i.e., a sign change, in Eq.~(\ref{eq:initial pure state rob both}), which does not affect the amount
of entanglement. In other words, the states (\ref{eq:initial pure state rob both})
are pure, bipartite, maximally entangled states of mode
$\hat{k}$ in Alice's cavity and mode $k$ in Rob's cavity.

We form the density matrix for each of the states~(\ref{eq:initial pure state rob both}),
express the density matrix in terms of Rob's Region ($\mathrm{I\!I\!I}$) basis to order $h^2$ using Eq.~~(\ref{eq:region I vacuum state to order h squared}), Eq.~(\ref{eq:region I single antiparticle state to order h squared}) and Eq.~\ref{eq:region I single particle state to order h squared}), the perturbative Bogoliubov expansions and take the partial trace over all of Rob's modes except the reference mode~$k$.
All of Rob's modes except $k$ are thus regarded as an environment,
to which information is lost due to the acceleration.

The partial tracing of Fermionic modes has been at the heart of much discussion and controversy in the literature. Different choices of basis elements corresponding to different mode operator orderings and can lead to differences in entanglement~\cite{montero2011}. Here we adopt the so-called ``tracing inside-out" method which corresponds to fixing any possible sign switches by comparing the full trace of a full density matrix with the full trace of a reduced density matrix. For details on the mapping of Fermionic states to qubit states in a consistent way and avoiding possible ambiguities see~\cite{friis2013}. For explicit details of the rules for Fermionic partial tracing see appendix~(\ref{app:fermion-tracing}).

We will from now on drop the subscript $R$ from Rob's states. The relevant partial traces of Rob's matrix elements depend on the sign of the mode label~$k$ which is indicated by $\zeta$. We find
\begin{subequations}
\label{eq:par-tr k pos all}
\begin{align}
\tr_{\lnot k}\left|\,0\right\rangle\left\langle\,0\right|
&=
(1-f_{k}^{-\zeta}h^{2})\left|\,\tilde{0}\right\rangle\left\langle\tilde{0}\right|
+f_{k}^{-\zeta}h^{2}\left|\tilde{1}_{k}\right\rangle^{\!\zeta\zeta\!}\!\left\langle\tilde{1}_{k}\right|,
\label{eq:par-tr k pos 0 0}
\\
\tr_{\lnot k}\left|\,0\right\rangle^{\zeta\,\,}\!\!\left\langle1_{k}\right|
&=
\Bigl(G_{k}+\mathcal{A}_{kk}^{(2)}h^{2}\Bigr)^{(\zeta *)}
\left|\,\tilde{0}\right\rangle^{\zeta\,\,}\!\!\left\langle\tilde{1}_{k}\right|,
\label{eq:par-tr k pos 0 1+}
\\
\tr_{\lnot k}\left|1_{k}\right\rangle^{\!\zeta\zeta\!}\!\left\langle1_{k}\right|
&=
(1-f_{k}^{\zeta}h^{2})
\left|\tilde{1}_{k}\right\rangle^{\!\!\zeta\zeta\!\!}\!\left\langle\tilde{1}_{k}\right|+
f_{k}^{\zeta}h^{2}\left|\,\tilde{0}\right\rangle\left\langle\tilde{0}\right|,
\label{eq:par-tr k pos 1+ 1+}
\end{align}
\end{subequations}
where we have used Eq.~(\ref{eq:first order unitarity condition})
and Eq.~(\ref{eq:leading order of V}). Note that in the above we have defined the symbol $(\zeta *)$ to represent complex conjugation if $\zeta=-$ (anti-particles) and no conjugation otherwise (particles) and have introduced the abbreviations
\begin{equation}
f_{k}^{+}
:=
\sum\limits_{p\geq0}\bigl|\mathcal{A}_{pk}^{(1)}\bigr|^{2}
\,, \ \ \ \
f_{k}^{-}
:=
\sum\limits_{q<0}\bigl|\mathcal{A}_{qk}^{(1)}\bigr|^{2}
\,.
\end{equation}

\subsection*{\label{subsec:particle antiparticle entangled states}States with entanglement between opposite charges}

We finally consider the Region ($\mathrm{I}$) state
\begin{align}
\left|\,\chi^{\pm}_{\text{init}}\,\right\rangle\,=\,
\tfrac{1}{\sqrt{2}}
\left(\,
\left|\,1_{k}\,\right\rangle_{A}^{\!+}\left|\,1_{k^{\prime}}\,\right\rangle_{R}^{\!-}\,\pm\,
\left|\,1_{k^{\prime}}\,\right\rangle_{A}^{\!-}\left|\,1_{k}\,\right\rangle_{R}^{\!+}\,\right),
\label{eq:initial particle antiparticle entangled state}
\end{align}
where the meaning of the subscripts and superscripts is as described for Eq.~(\ref{eq:initial pure state rob both}), indicating that
$k\geq0$ and $k^{\prime}<0$. In this state Alice and Rob each have access to both of the modes
$k$ and $k^{\prime}$, and
the entanglement is in the \emph{charge\/} of the field modes,
similarly to the states considered in~\cite{martinmartinez2011}.

We form the reduced density matrix to order $h^2$
as in Section~(\ref{subsec:two mode entangled states}),
but now the partial tracing over Rob's modes excludes both mode $k$ and mode~$k^{\prime}$.
The relevant matrix elements take the form
\begin{equation}
\begin{aligned}
\tr_{\lnot k,k^{\prime}}\left|1_{k^{\prime}}\right\rangle^{\!\!--\!\!}\!\left\langle1_{k^{\prime}}\right|\ &=\
f_{k^{\prime}}^{-}h^{2}
\left|\,\tilde{0}_{k}\right\rangle^{\!\!+}\!\left|\,\tilde{0}_{k^{\prime}}\right\rangle^{\!\!--\!\!}\!
\left\langle\tilde{0}_{k^{\prime}}\right|^{+\!\!}\!\left\langle\tilde{0}_{k}\right| \\
&\hspace{5mm}+\,\bigl(1\!-\!f_{k^{\prime}}^{-}h^{2}\!-\!f_{k}^{-}h^{2}\!+\!\bigl|\mathcal{A}_{kk^{\prime}}^{(1)}\bigr|^{2}h^{2}\bigr)
\left|\,\tilde{0}_{k}\right\rangle^{\!\!+}\!\left|\,\tilde{1}_{k^{\prime}}\right\rangle^{\!\!--\!\!}\!
\left\langle\tilde{1}_{k^{\prime}}\right|^{+\!\!}\!\left\langle\tilde{0}_{k}\right| \\
&\hspace{5mm}+\,\bigl(f_{k}^{-}h^{2}\!-\!\bigl|\mathcal{A}_{kk^{\prime}}^{(1)}\bigr|^{2}h^{2}\bigr)
\left|\,\tilde{1}_{k}\right\rangle^{\!\!+}\!\left|\,\tilde{1}_{k^{\prime}}\right\rangle^{\!\!--\!\!}\!
\left\langle\tilde{1}_{k^{\prime}}\right|^{+\!\!}\!\left\langle\tilde{1}_{k}\right| \\
&\hspace{5mm}+\,\Bigl(\sum\limits_{q<0}\bar{G}_{k}G_{k^{\prime}}\bar{\mathcal{A}}_{qk}^{(1)}\mathcal{A}_{qk^{\prime}}^{(1)}h^{2}
\left|\,\tilde{0}_{k}\right\rangle^{\!\!+}\!\left|\,\tilde{0}_{k^{\prime}}\right\rangle^{\!\!--\!\!}\!
\left\langle\tilde{1}_{k^{\prime}}\right|^{+\!\!}\!\left\langle\tilde{1}_{k}\right| +\,\mathrm{h.c.}\Bigr)\,,
\label{eq:par-tr k,kprime 1- 1-}
\end{aligned}
\end{equation}
\begin{equation}
\begin{aligned}
\tr_{\lnot k,k^{\prime}}\left|1_{k}\,\right\rangle^{\!\!++\!\!}\!\left\langle1_{k}\,\right|\ &=\
f_{k}^{+}h^{2}
\left|\,\tilde{0}_{k}\right\rangle^{\!\!+}\!\left|\,\tilde{0}_{k^{\prime}}\right\rangle^{\!\!--\!\!}\!
\left\langle\tilde{0}_{k^{\prime}}\right|^{+\!\!}\!\left\langle\tilde{0}_{k}\right| \\
&\hspace{5mm}+\,\bigl(1\!-\!f_{k^{\prime}}^{+}h^{2}\!-\!f_{k}^{+}h^{2}\!+\!\bigl|\mathcal{A}_{kk^{\prime}}^{(1)}\bigr|^{2}h^{2}\bigr)
\left|\,\tilde{1}_{k}\right\rangle^{\!\!+}\!\left|\,\tilde{0}_{k^{\prime}}\right\rangle^{\!\!--\!\!}\!
\left\langle\tilde{0}_{k^{\prime}}\right|^{+\!\!}\!\left\langle\tilde{1}_{k}\right| \\
&\hspace{5mm}+\,\bigl(f_{k^{\prime}}^{+}h^{2}\!-\!\bigl|\mathcal{A}_{kk^{\prime}}^{(1)}\bigr|^{2}h^{2}\bigr)
\left|\,\tilde{1}_{k}\right\rangle^{\!\!+}\!\left|\,\tilde{1}_{k^{\prime}}\right\rangle^{\!\!--\!\!}\!
\left\langle\tilde{1}_{k^{\prime}}\right|^{+\!\!}\!\left\langle\tilde{1}_{k}\right| \\
&\hspace{5mm}-\,\Bigl(\sum\limits_{p\geq0}\bar{G_{k}}G_{k^{\prime}}\bar{\mathcal{A}}_{pk}^{(1)}\mathcal{A}_{pk^{\prime}}^{(1)}h^{2}
\left|\,\tilde{0}_{k}\right\rangle^{\!\!+}\!\left|\,\tilde{0}_{k^{\prime}}\right\rangle^{\!\!--\!\!}\!
\left\langle\tilde{1}_{k^{\prime}}\right|^{+\!\!}\!\left\langle\tilde{1}_{k}\right|+\,\mathrm{h.c.}\Bigr)\,,
\label{eq:par-tr k,kprime  1+ 1+}
\end{aligned}
\end{equation}
\begin{equation}
\begin{aligned}
\tr_{\lnot k,k^{\prime}}\left|1_{k}\,\right\rangle^{\!\!+-\!\!}\!\left\langle1_{k^{\prime}}\,\right|
&=
\Bigl(
\bar{G}_{k}\bar{G}_{k^{\prime}}
\bigl|\mathcal{A}_{kk^{\prime}}^{(1)}\bigr|^{2}h^{2}
&+
\bar{\mathcal{A}}_{kk}\bar{\mathcal{A}}_{k^{\prime}k^{\prime}}
\Bigr)
\left|\,\tilde{1}_{k}\right\rangle^{\!\!+}\!\left|\,\tilde{0}_{k^{\prime}}\right\rangle^{\!\!--\!\!}\!
\left\langle\tilde{1}_{k^{\prime}}\right|^{+\!\!}\!\left\langle\tilde{0}_{k}\right|\,,
\label{eq:par-tr k,kprime  1+ 1-}
\end{aligned}
\end{equation}

where in Eq.~(\ref{eq:par-tr k,kprime  1+ 1-}),
$\bar{\mathcal{A}}_{kk}\bar{\mathcal{A}}_{k^{\prime}k^{\prime}}$
is kept only to order $h^2$ in the small $h$ expansion
\begin{align}
\bar{\mathcal{A}}_{kk}\bar{\mathcal{A}}_{k^{\prime}k^{\prime}}\,=\,
\bar{G}_{k}\bar{G}_{k^{\prime}}+
\bar{G}_{k^{\prime}}\bar{\mathcal{A}}_{kk}^{(2)}h^{2}+
\bar{G}_{k}\bar{\mathcal{A}}_{k^{\prime}k^{\prime}}^{(2)}h^{2}+{O}(h^{3})\,.
\label{eq:Akk times Akprimekrprime expansion}
\end{align}
Note that due to the diagonal elements of $\mathcal{A}^{(1)}_{mn}$ identically vanishing, there are no terms proportional to $h$ in the above expansions.

\subsection{\label{sec:degradation of entanglement}Entanglement degradation}

Our aim is to investigate how non-uniform motion affects an initially maximally entangled Fermionic state. This is in direct analogy of the Bosonic case of~\cite{alphacentauri2012} and the previous chapter~(\ref{chapter:bosons}). In particular, we wish to analyse particles which are entangled between their charge degrees of freedom are effected by non-uniform motion.

Consider the states
$\left|\,\phi^{\pm}_{\text{init}}\,\right\rangle_{+}$ and
$\left|\,\phi^{\pm}_{\text{init}}\,\right\rangle_{-}$, Eq.~(\ref{eq:initial pure state rob both}), in which
Alice and Rob can access one mode each. Even though the reduced states are bipartite in two Fermionic modes $k,k'$, the state between Alice and Rob is mixed. Thus, the entropy of entanglement~(\ref{eqn:entropy-of-entanglement}) is not guaranteed to quantify all quantum correlations. We therefore turn to a more easily computable measure. We shall quantify the entanglement by the negativity
\cite{vidal2002,audenaert2003,plenio2007}.

The negativity $\mathcal{N}(\rho)$ quantifies how strongly the partial transpose of a density operator
$\rho$ fails to be positive. From Section~(\ref{sec:negs}) we define the negativity as
\begin{align}
\mathcal{N}(\rho)\,=\,\sum\limits_{j}\,\left|\lambda_{j}^{-}\right|\ ,
\label{eq:negativity}
\end{align}
where partial transposition as been performed on Rob's state (without loss of generality) and $\lambda_{j}^{-}$ are the negative eigenvalues of $\rho^{\mathrm{tp}_{R}}$. 

We work perturbatively in~$h$. We first write the full state of the system as
\begin{equation}
\begin{aligned}
\rho_{\zeta}^{\pm}&=\frac{1}{2}\left| 0\right\rangle_{\! AA\!}\!
\left\langle 0\right|\otimes\left| 0\right\rangle\left\langle 0\right|
\pm\frac{1}{2} 
\left| 0\right\rangle_{\! AA\!}^{\phantom{\kappa}\kappa\!\!}\!
\left\langle 1_{\hat{k}}\right|\otimes\left| 0\right\rangle^{\phantom{\zeta}\zeta}\!
\left\langle 1_{k}\right|\\
& \hspace{4mm}\pm\frac{1}{2}\left| 1_{\hat{k}}\right\rangle_{\! AA\!}^{\kappa\phantom{\kappa}\!\!}\!
\left\langle 0\right|\otimes\left| 1_{k}\right\rangle^{\zeta\phantom{\zeta}}\!
\left\langle 0\right|
+\frac{1}{2}
\left| 1_{\hat{k}}\right\rangle_{\! AA\!}^{\kappa\kappa\!\!}\!
\left\langle 1_{\hat{k}}\right|\otimes\left| 1_{k}\right\rangle^{\zeta\zeta}\!
\left\langle 1_{k}\right|
\end{aligned}.
\end{equation}
Using the partial tracing rules found in appendix~(\ref{app:symplectic-matrix-derivation}), we can write the reduced state between Alice and Rob as
\begin{equation}
\begin{aligned}
\tr_{\lnot k}\rho_{\zeta}^{\pm}
&=\frac{1}{2}\left| 0\right\rangle_{\!\! AA\!}\!
\left\langle 0\right|\otimes\left[\left(1-f^{-\zeta}_{k}h^{2}\right)\left|\tilde{0}\right\rangle\!
\left\langle\tilde{0}\right| +f^{-\zeta}_{k}h^{2}\left| \tilde{1}_{k}\right\rangle^{\zeta\zeta}\!
\left\langle \tilde{1}_{k}\right|\right]\\
& \hspace{4mm}\pm\frac{1}{2}\left(G_{k}+\mathcal{A}^{(2)}_{kk}h^{2}\right)^{(\zeta *)}\left| 0\right\rangle_{\! AA\!}^{\phantom{\kappa}\kappa\!\!}\!
\left\langle 1_{\hat{k}}\right|\otimes\left| \tilde{0}\right\rangle^{\phantom{\zeta}\zeta}\!
\left\langle\tilde{1}_{k}\right| +\text{h.c.}\\
& \hspace{4mm}+\frac{1}{2}\left| 1_{\hat{k}}\right\rangle_{\!\! AA\!}^{\kappa\kappa\!\!}\!
\left\langle 1_{\hat{k}}\right|\otimes\left[\left(1-f^{\zeta}_{k}h^{2}\right)\left|\tilde{1}_{k}\right\rangle^{\zeta\zeta}\!
\left\langle\tilde{1}_{k}\right| +f^{\epsilon}_{k}h^{2}\left| \tilde{0}\right\rangle\!
\left\langle \tilde{0}\right|\right]
\end{aligned}.
\end{equation}
In matrix form, with respect to the basis 
\begin{eqnarray}
\lbrace\left| 0\right\rangle_{A}\left| 0\right\rangle_{R},\left| 0\right\rangle_{A}\left| \tilde{1}_{k}\right\rangle^{\zeta}_{R},\left| \tilde{1}\right\rangle^{\kappa}_{A}\left| 0\right\rangle_{R},\left| 1_{\hat{k}}\right\rangle^{\kappa}_{A}\left| \tilde{1}_{k}\right\rangle^{\zeta}_{R}\rbrace ,
\end{eqnarray}
and denoting the matrix form of $\tr_{\lnot k}\rho_{\zeta}^{\pm}$ as $\boldsymbol{\rho}_{\zeta,k}^{\pm}$ we have
\begin{equation}
\begin{aligned}
\boldsymbol{\rho}_{\zeta,k}^{\pm}&=
\frac{1}{2}\left(
\begin{array}{cccc}
1 & 0 & 0 & \pm G_{k}^{(\zeta *)}\\
0 & 0 & 0 & 0\\
0 & 0 & 0 & 0\\
\pm \bar{G}_{k}^{(\zeta *)} & 0 & 0 & 1
\end{array}
\right)
\\
& \hspace{5mm}
+\frac{1}{2}
\left(
\begin{array}{cccc}
-f_{k}^{-\zeta} & 0 & 0 & \pm\mathcal{A}_{kk}^{(2)(\zeta *)}\\
0 & f_{k}^{-\zeta} & 0 & 0\\
0 & 0 & f_{k}^{\zeta} & 0\\
\pm\bar{\mathcal{A}}_{kk}^{(2)(\zeta *)} & 0 & 0 & -f_{k}^{\zeta}
\end{array}
\right)h^{2}.
\end{aligned}
\end{equation}
In order to check the properties of this state are correct, such as positive definiteness, we need to compute its eigenvalues. This can be done via perturbation theory. As the Ferimon case is more involved than the Boson case of Chapter~(\ref{chapter:bosons}), a quick review of perturbation theory will also be useful when computing the eigenvalues for the negativity. The eigenvalues and normalised eigenvectors of the unperturbed matrix $\boldsymbol{\rho}_{\zeta,k}^{\pm(0)}$ are
\begin{eqnarray}
\text{Eig}(\boldsymbol{\rho}_{\zeta,k}^{\pm(0)})=\lbrace 1,0,0,0\rbrace,
\end{eqnarray}
\begin{eqnarray}
\text{Vec}(\boldsymbol{\rho}_{\zeta,k}^{\pm(0)})=\lbrace 
\frac{1}{\sqrt{2}}\left(\begin{array}{c} G_{k}^{(\zeta *)} \\ 0 \\ 0 \\ 1 \end{array}\right),
\frac{1}{\sqrt{2}}\left(\begin{array}{c} -G_{k}^{(\zeta *)} \\ 0 \\ 0 \\ 1 \end{array}\right),
\left(\begin{array}{c} 0 \\ 1 \\ 0 \\ 0 \end{array}\right),
\left(\begin{array}{c} 0 \\ 0 \\ 1 \\ 0 \end{array}\right)
\rbrace
\end{eqnarray}
For convenience we will denote the unperturbed eigenvalues above as $\lambda_{j}^{(0)}$ and the unperturbed eigenvectors as $v_{j}^{(0)}$. As we are perturbing the matrix $\boldsymbol{\rho}_{\zeta,k}^{\pm(0)}$ by a small parameter, played here by $h^{2}$, we can compute the corrections to the unperturbed eigenvalues via the relation
\begin{eqnarray}
\lambda_{j}=\lambda_{j}^{(0)}+v_{j}^{(0) \dag}\boldsymbol{\rho}_{\zeta,k}^{\pm(2)}v_{j}^{(0)}h^{2}+O(h^{3}).
\end{eqnarray}
Note the above formula is only valid for unperturbed eigenvalues $\lambda_{j}^{(0)}$ which are \emph{non-degenerate}, with corresponding eigenvectors $v_{j}^{(0)}$. Thus, for the case at hand we find the correction to the first eigenvalue $\lambda_{1}^{(0)}=+1$ is
\begin{eqnarray}
\lambda_{1}^{(2)}=-\frac{1}{2}f_{k}h^{2}
\end{eqnarray}
where we have denoted $f_{k}=f^{+}_{k}+f^{-}_{k}$ and used the identity $2\text{Re}(\bar{G}_{k}\mathcal{A}_{kk}^{(2)})=-f_{k}$. However, we have the triply degenerate eigenvalue $\lambda_{j}^{(0)}=0$. Thus, the standard perturbation method needs to be replaced by the degenerate case. In our case, we find the perturbed eigenvalues of $\lambda_{j}^{(0)}=0$ by finding the eigenvalues of its degenerate subspace matrix $\boldsymbol{W}$. $\boldsymbol{W}$ is defined as~\cite{bransden2000}
\begin{eqnarray}
W_{mn}=v_{m}^{\dag (0)}\boldsymbol{\rho}_{\zeta,k}^{\pm(2)}v_{n}^{(0)},
\end{eqnarray}
where $v_{n}^{(0)}$ are the eigenvectors of the degenerate eigenvalues. Computing $\boldsymbol{W}$ and finding its eigenvalues we have
\begin{eqnarray}
\text{Eig}(\boldsymbol{W})=\lbrace \frac{1}{2}f^{+}_{k},\frac{1}{2}f^{-}_{k},0\rbrace ,
\end{eqnarray}
where again the use of $2\text{Re}(\bar{G}_{k}\mathcal{A}_{kk}^{(2)})=-f_{k}$ is needed. Hence collecting the full set of perturbed eigenvalues to $O(h^{2})$ we have
\begin{eqnarray}
\text{Eig}(\boldsymbol{\rho}_{\zeta,k}^{\pm})=\lbrace 1-(1/2)f_{k}h^{2},(1/2)f_{k}^{+}h^{2},(1/2)f_{k}^{-}h^{2},0\rbrace .
\end{eqnarray}
Note that due to the symmetry of the perturbations, the dependence on $\zeta$ has vanished. We can also remark that all eigenvalues are positive and sum to zero which is needed for the positive definiteness of the state $\rho_{\zeta,k}^{\pm}$ and its normalisation $\text{tr}\rho_{\zeta,k}^{\pm}=+1$. Although this is a useful check that our perturbation expansion holds, we want to quantify the entanglement in the system using the negativity. We therefore need to compute the perturbed eigenvalues of the partially transposed state of Alice and Rob. Denoting partial transposition with respect to Rob's subsystem as $(\mathds{1}\otimes \mathrm{tp}_{R})$, the partially transposed state reads
\begin{equation}
\begin{aligned}
(\mathds{1}\otimes \mathrm{tp}_{R})\boldsymbol{\rho}_{\zeta,k}^{\pm}&=\frac{1}{2}\left(
\begin{array}{cccc}
1 & 0 & 0 & 0\\
0 & 0 & \pm G_{k}^{(\zeta *)} & 0\\
0 & \pm \bar{G}_{k}^{(\zeta *)} & 0 & 0\\
0 & 0 & 0 & 1
\end{array}
\right)\\
& \hspace{5mm}+\frac{1}{2}
\left(
\begin{array}{cccc}
-f_{k}^{-\zeta} & 0 & 0 & 0\\
0 & f_{k}^{-\zeta} & \pm\mathcal{A}_{kk}^{(2)(\zeta *)} & 0\\
0 & \pm\bar{\mathcal{A}}_{kk}^{(2)(\zeta *)} & f_{k}^{\zeta} & 0\\
0 & 0 & 0 & -f_{k}^{\zeta}
\end{array}
\right)h^{2}.
\end{aligned}
\end{equation}
The eigenvalues and eigenvectors of the unperturbed state $(\mathds{1}\otimes \mathrm{tp}_{R})\rho_{\zeta}^{0}$ are
\begin{eqnarray}
\text{Eig}((\mathds{1}\otimes \mathrm{tp}_{R})\boldsymbol{\rho}_{\zeta,k}^{\pm (0)})=\lbrace -1/2,1/2,1/2,1/2\rbrace ,
\end{eqnarray}
\begin{eqnarray}
\text{Vec}((\mathds{1}\otimes \mathrm{tp}_{R})\boldsymbol{\rho}_{\zeta,k}^{\pm (0)})=\lbrace 
\frac{1}{\sqrt{2}}\left(\! \begin{array}{c} 0 \\ -G_{k}^{(\zeta *)} \\ 1 \\ 0 \! \end{array}\right),
\frac{1}{\sqrt{2}}\left(\! \begin{array}{c} 0\\ G_{k}^{(\zeta *)} \\ 1 \\ 0 \! \end{array}\right),
\left(\begin{array}{c} 1 \\ 0 \\ 0 \\ 0 \end{array}\right),
\left(\begin{array}{c} 0 \\ 0 \\ 0 \\ 1 \end{array}\right)
\rbrace .
\end{eqnarray}
We notice again we have a triply degenerate eigenvalue $1/2$. However, we are interested in the negative eigenvalues of this state as this is what quantifies our entanglement. Following the same procedure as before, we arrive at the perturbed eigenvalues
\begin{eqnarray}
\text{Eig}((\mathds{1}\otimes T_{R})\boldsymbol{\rho}_{\zeta,k}^{\pm})=\lbrace -1/2+1/2f_{k}h^{2},1/2-1/2f_{k}^{+}h^{2},1/2-1/2f_{k}^{-}h^{2},1/2\rbrace .
\end{eqnarray}
Thus our positive eigenvalues remain positive and the negative eigenvalue remains negative. Thus the leading order correction to the
negativity comes in~$O(h^2)$, and to this order the
negativity formula reads
\begin{eqnarray}
\label{eq:negativity of rhoARplusminus}
\mathcal{N}(\rho_{\zeta,k}^{\pm})\,=\,\tfrac{1}{2} \left( 1 - f_{k}h^{2}\right).
\end{eqnarray}
Using the relation $\mathcal{A}_{nk}^{(0)}=(G^{(0)}_{n}-G^{(0)}_{k})A_{nk}^{(1)}$, which results from our Bogoliubov identities and the particular form of $A_{nm}$, we can rewrite $f_{k}$ as
\begin{eqnarray}
\begin{aligned}
f_{k}
&
=
\sum\limits_{p=-\infty}^{\infty}
\bigl|E_1^{k-p} - 1\bigr|^2
\bigl| A_{kp}^{(1)}\bigr|^{2}
\label{eq:fk-def}
\end{aligned},
\end{eqnarray}
and
\begin{eqnarray}
E_1 := \exp \!
\left(\frac{i\pi\tau_1}{\cavlength}\right).
\end{eqnarray}
\begin{figure}[b!]
\centering
\includegraphics[width=0.7\textwidth]{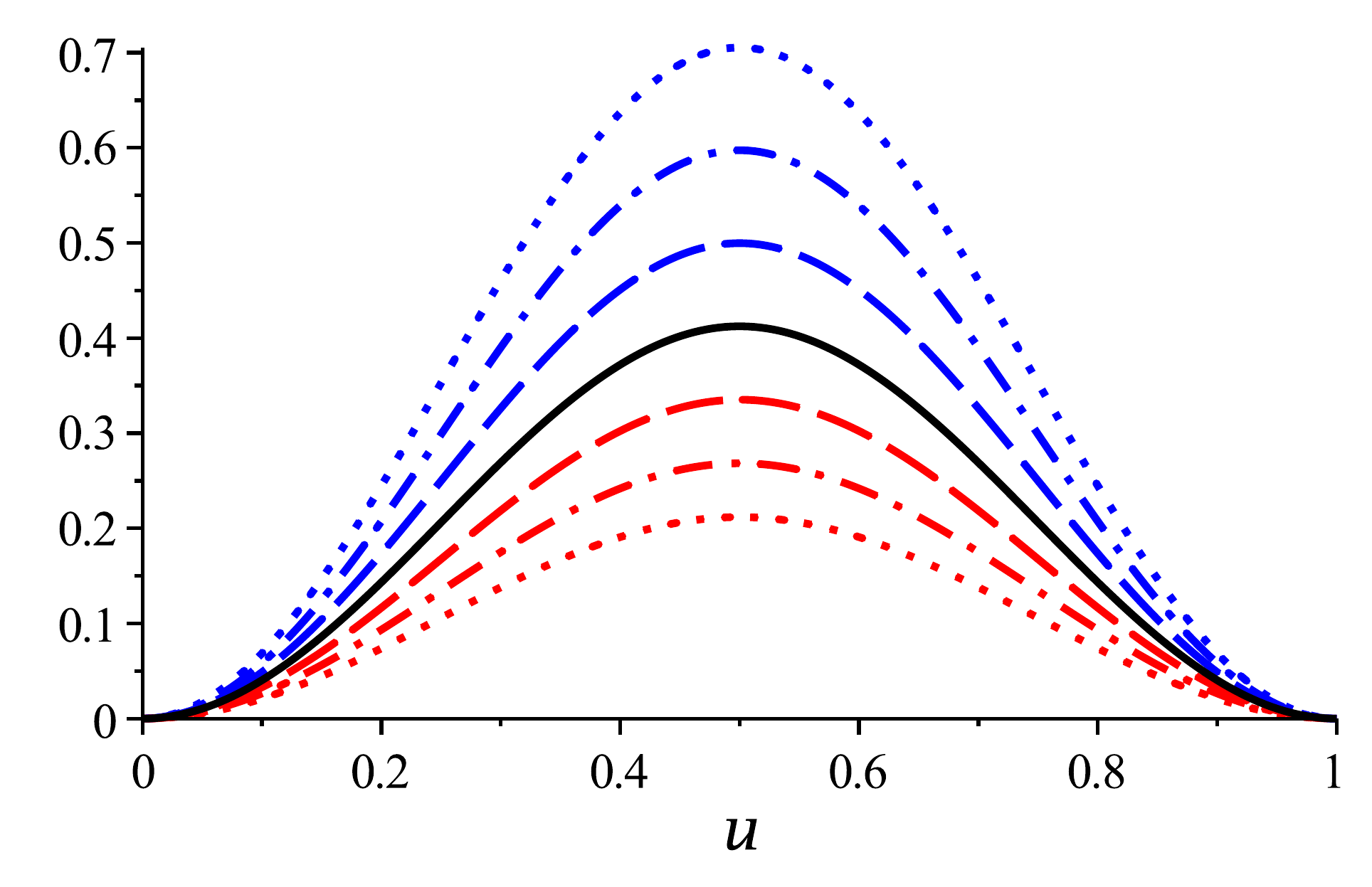}
\caption[Fermion Negativity Correction]{The plot shows $f_{k}$
as a function of
$u:=\tfrac12\eta_{1}/\ln(b/a) =
\tau_1 /2\cavlength$,
over the full period $0\le u \le 1$.
The solid curve (black) is for $s=0$ with $k=\pm1$.
The dashed, dash-dotted and dotted
curves are respectively for
$s=\frac14$, $s=\frac12$ and $s=\frac34$,
for $k=1$ (blue) above the solid curve and
for $k=-1$ (red) below the solid curve.
\label{fig:correction term I to III}}
\end{figure}
We see from Eq.~(\ref{eq:negativity of rhoARplusminus}) that acceleration
does degrade the initial maximal entanglement,
and the degradation is determined
by the function $f_{k}$~(\ref{eq:fk-def}).
$f_{k}$~is periodic in $\tau_1$,
which is the proper time measured at the centre of Rob's
cavity between sending and recapturing a light ray that
is allowed to bounce off each wall once.
$f_{k}$~is non-negative, and it
vanishes only when it is a integer multiple of $\delta$.
$f_{k}$~is not even in $k$ for generic values of~$s$, but
it is even in $k$ in the limiting case
$s=0$ in which the spectrum is symmetric
between positive and negative charges.
$f_{k}$ diverges at large $|k|$ proportionally to~$k^2$,
and the domain of validity of our
perturbative analysis is $|k| h \ll 1$.
Plots for $k=\pm1$ are shown in
Fig.~\ref{fig:correction term I to III}.

We finally turn to the entanglement between opposite charges in the state
$\left|\,\chi^{\pm}_{\text{init}}\,\right\rangle$~(\ref{eq:initial particle antiparticle entangled state}). Expressing the density matrix in the Region $\mathrm{I\!I\!I}$ basis,
tracing over Rob's unobserved modes and working perturbatively to
order~$h^2$, we find that the only nonvanishing elements of the
reduced density matrix are within an $8\times8$ block, which decomposes further into two
$3\times3$ blocks that correspond respectively to
Eq.~(\ref{eq:par-tr k,kprime 1- 1-}) and Eq.~(\ref{eq:par-tr k,kprime 1+ 1+}) and a
$2\times2$ block. Explicitly we write
\begin{eqnarray}
\boldsymbol{\rho}=\frac{1}{2}\boldsymbol{\rho}_{1}\oplus\boldsymbol{\rho}_{2}\oplus\boldsymbol{\rho}_{3},
\end{eqnarray}
where
\begin{subequations}
\begin{align}
\boldsymbol{\rho}_{1}&=
\begin{pmatrix}
f_{k}^{-}h^{2} & 0 & \sum\limits_{q< 0}G_{k}\bar{G}_{k'}\bar{\mathcal{A}}_{qk'}^{(1)}\mathcal{A}_{qk}^{(1)}h^{2} \\
0 & 1-f^{-}_{k'}h^{2}-f^{-}_{k}h^{2}+\left|\mathcal{A}_{k'k}^{(1)}\right|^{2}h^{2} & 0 \\
\sum\limits_{q< 0}\bar{G}_{k}G_{k'}\mathcal{A}^{(1)}_{qk'}\bar{\mathcal{A}}_{qk}^{(1)}h^{2} & 0 & f^{-}_{k'}-|\mathcal{A}^{(1)}_{k'k}|^{2}h^{2}
\end{pmatrix}\\
\boldsymbol{\rho}_{2}&=
\begin{pmatrix}
f_{k}^{+}h^{2} & 0 & -\sum\limits_{p\ge 0}G_{k}\bar{G}_{k'}\bar{\mathcal{A}}_{pk'}^{(1)}\mathcal{A}_{pk}^{(1)}h^{2} \\
0 & 1-f^{+}_{k'}-f^{+}_{k}+\left|\mathcal{A}_{k'k}^{(1)}\right|^{2}h^{2} & 0 \\
-\sum\limits_{p\ge 0}\bar{G}_{k}G_{k'}\mathcal{A}^{(1)}_{pk'}\bar{\mathcal{A}}_{pk}^{(1)}h^{2} & 0 & f^{+}_{k'}-|\mathcal{A}^{(1)}_{k'k}|^{2}h^{2}
\end{pmatrix}\\
\boldsymbol{\rho}_{3}&=
\pm \begin{pmatrix}
0
&
\hspace*{-6ex}
G_{k}G_{k^{\prime}}\bigl|\mathcal{A}_{kk^{\prime}}^{(1)}\bigr|^{2}h^{2} +
\mathcal{A}_{kk}\mathcal{A}_{k^{\prime}k^{\prime}}
\\[1.5mm]
\bar{G}_{k}\bar{G}_{k^{\prime}}\bigl|\mathcal{A}_{kk^{\prime}}^{(1)}\bigr|^{2}h^{2} +
\bar{\mathcal{A}}_{kk}\bar{\mathcal{A}}_{k^{\prime}k^{\prime}}
&
\hspace*{-6ex}0\\
\end{pmatrix} ,
\label{eq:nonzero two by two block}
\end{align}
\end{subequations}
and all components are kept only to
order $h^2$ in their small $h$ expansion, see Eq.~(\ref{eq:Akk times Akprimekrprime expansion}).

The only negative eigenvalue comes from the
$2\times2$ block~(\ref{eq:nonzero two by two block}).
We find that the negativity is given by
\begin{equation}
\begin{aligned}
\mathcal{N}(\rho_{\chi}^{\pm})
&=
\tfrac{1}{2} \,-\,
\tfrac{1}{4}\,\sum\limits_{p\neq k^{\prime}}\bigl|\mathcal{A}_{kp}^{(1)}\bigr|^{2}h^{2}\,-\,
\tfrac{1}{4}\,\sum\limits_{p\neq k}\bigl|\mathcal{A}_{k^{\prime}p}^{(1)}\bigr|^{2}h^{2}
\\
&=
\tfrac12
\,-\,
\tfrac14
\left(
f_k + f_{k'}
\right)
+ \tfrac12
\bigl|E_1^{k-k'} - 1\bigr|^2
\bigl| A_{k k'}^{(1)}\bigr|^{2} h^{2} .
\label{eq:negativity of rhochiplusminus}
\end{aligned}
\end{equation}
The entanglement is hence again degraded by the acceleration, and the
degradation has the same periodicity in $\tau_1$ as in the cases
considered above. The degradation now depends however on $k$ and $k'$ not just
through the individual functions $f_k$ and $f_{k'}$ but also through
the term proportional to $\bigl| A_{k k'}^{(1)}\bigr|^{2}$
in Eq.~(\ref{eq:negativity of rhochiplusminus}): 
this interference term is
nonvanishing if, and only if, $k$ and $k'$ have different parity, and when it is
nonvanishing, it diminishes the degradation effect. In the charge-symmetric
special case of $s=0$ and $k=-k'$, the degradation coincides with that
found in Eq.~(\ref{eq:negativity of rhoARplusminus}) for the two-mode
states~(\ref{eq:initial pure state rob both}).

\subsection{\label{sec:onewaytrip}One-way journey}

Our analysis for the Rob trajectory that comprises Regions
($\mathrm{I}$), ($\mathrm{I\!I}$) and~($\mathrm{I\!I\!I}$) can be
generalised in a straightforward way to any trajectory obtained by
grafting inertial and uniformly-accelerated segments, with arbitrary
durations and proper accelerations. The only delicate point is that the
phase conventions of our mode functions distinguish the left boundary
of the cavity from the right boundary, and in the previous sections
we set up the Bogoliubov transformation from Minkowski to Rindler
assuming that the acceleration is to the right. It follows
that the Bogoliubov transformation from Minkowski to
leftward-accelerating Rindler is obtained by performing the replacement $h\rightarrow -h$ in the expansions
(\ref{eq:bogo coeff pert}).

\begin{figure}[t]
\centering
\vspace*{-4ex}%
\includegraphics[width=0.7\textwidth]{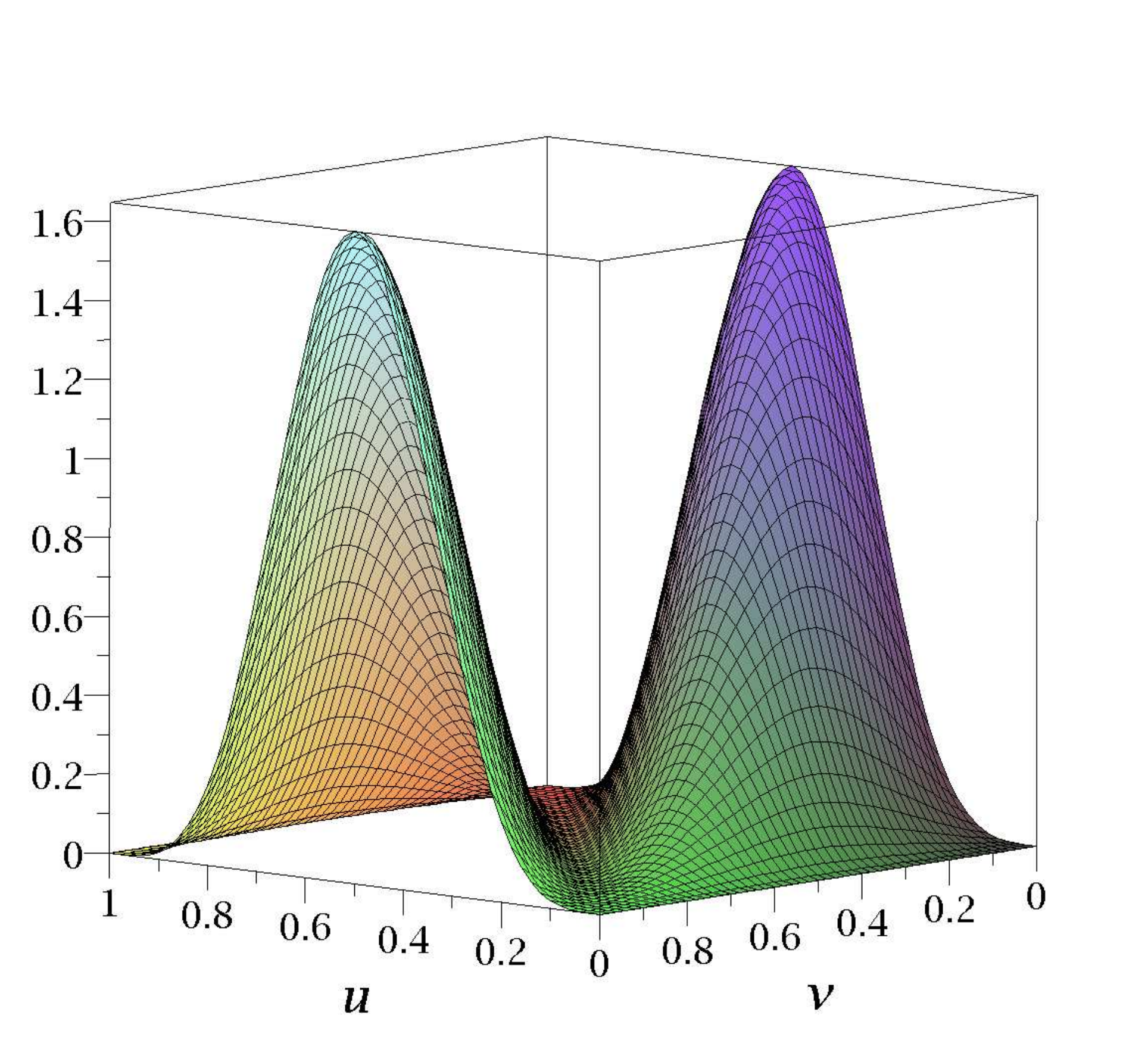}
\caption[Fermion One-way Trip]{The plot shows ${\widetilde{\!\widetilde f}}_k$
as a function of
$u:= \tau_{1}/2\delta$
and
$v:= \tau_2 /(2\cavlength)$
over the full period $0\le u \le 1$ and  $0\le v \le 1$,
for $s=0$ and $k=1$.  Note the zeroes at $u\equiv0 \mod 1$ and at
$u+v \equiv 0 \mod 1$.
\label{fig:onewaytripfigure}}
\end{figure}
As an example, consider the Rob cavity trajectory that starts
inertial, accelerates to the right for proper time $\tau_1$ as above,
coasts inertially for proper time $\tau_2$ and finally performs a
braking manoeuvre that is the reverse of the initial acceleration,
ending in an inertial state that has vanishing velocity with respect
to the initial state. Denoting the mode functions in the final
inertial state by ${\widetilde{\!\widetilde\psi}}_n$, and writing
\begin{align}
\label{eq:onewaychange-of-basis}
{\widetilde{\!\widetilde\psi}}_m
= \sum\limits_{n}\mathcal{B}_{mn}\,\psi_{n}
\ ,
\end{align}
we can find the entanglement of a general travel scenario by simply replacing the Bogoliubov coefficients  $\mathcal{A}_{mn}$ in our entanglement measures with new coefficients $\mathcal{B}_{mn}$. For the one-way journey we find
\begin{align}
\bigl| \mathcal{B}_{mn}^{(1)}\bigr|^{2}
=
\bigl|E_1^{m-n} - 1\bigr|^2
\bigl|{(E_1 E_2)}^{m-n} - 1\bigr|^2
\bigl| A_{mn}^{(1)}\bigr|^{2},
\end{align}
where
$E_2 := \exp(i\pi \tau_2/\cavlength)$.
For the two-mode initial
states
$\left|\,\phi^{\pm}_{\text{init}}\,\right\rangle_{+}$ and
$\left|\,\phi^{\pm}_{\text{init}}\,\right\rangle_{-}
$~(\ref{eq:initial pure state rob both}),
the negativity reads
\begin{subequations}
\begin{align}
\mathcal{N}(\rho^{\pm}_{AR\pm})
& \,=\,
\tfrac{1}{2} \Bigl( 1 - {\widetilde{\!\widetilde f}}_k h^{2}\Bigr) ,
\label{eq:onewaynegativity of rhoARplusminus}
\end{align}
\end{subequations}
where
\begin{align}
{\widetilde{\!\widetilde f}}_k
&
=
\sum\limits_{p=-\infty}^{\infty}
\bigl| \mathcal{B}_{kp}^{(1)}\bigr|^{2}.
\label{eq:onewayfk-def}
\end{align}
The negativity in the state
$\left|\,\chi^{\pm}_{\text{init}}\,\right\rangle$
(\ref{eq:initial particle antiparticle entangled state}) reads
\begin{align}
\mathcal{N}(\rho_{\chi}^{\pm})
&=
\tfrac12
\,-\,
\tfrac14
\Bigl(
{\widetilde{\!\widetilde f}}_k h^{2}
+ {\widetilde{\!\widetilde f}}_{k'}h^{2}
\Bigr)
\notag
\\
&
\hspace{2ex}
+ \tfrac12
\bigl|E_1^{k-k'} - 1\bigr|^2
\bigl|{(E_1E_2)}^{k-k'} - 1\bigr|^2
\bigl| A_{k k'}^{(1)}\bigr|^{2}  .
\label{eq:onewaynegativity of rhochiplusminus}
\end{align}
The degradation caused by acceleration is thus periodic in
$\tau_1$ and $\tau_2$. The degradation
vanishes if, and only if, $E_1=1$ or $E_1 E_2=1$, so that any degradation caused by
the accelerated segments can be cancelled by fine-tuning the duration
of the inertial segment, to the order $h^2$ in which we are working.
A~plot of ${\widetilde{\!\widetilde f}}_k$
is shown in Fig.~(\ref{fig:onewaytripfigure}).

\section{Conclusions}\label{sec:conclusion}

We have analysed the entanglement degradation for a massless Dirac
field between two cavities in $(1+1)$-dimensional Minkowski spacetime, one
cavity inertial and the other moving along a trajectory that
consists of inertial and uniformly accelerated segments. Working in
the approximation of small accelerations but arbitrarily long travel
times, we found that
the degradation is qualitatively similar to that found in
\cite{alphacentauri2012}, and Chapter~(\ref{chapter:bosons}), for a scalar field with
Dirichlet boundary conditions. The degradation is
periodic in the durations of the individual inertial and accelerated
segments, and we identified a travel scenario where the degradation
caused by accelerated segments can be undone by fine-tuning the
duration of an inertial segment. The presence of charge allows however
a wider range of initial states of interest to be analysed. As an
example, we identified a state where the entanglement degradation
contains a contribution due to interference between excitations of
opposite charge.

Our analysis contained two limitations. First, while our
Bogoliubov transformation technique can be applied to
arbitrarily complicated graftings of inertial and
uniformly accelerated cavity trajectory segments,
the treatment is perturbative in the small dimensionless parameter $h$.
We were thus not able to address the large $h$ limit,
in which striking qualitative differences between Bosonic and
Fermionic entanglement have been found for field modes that are not
confined in
cavities~\cite{funentes-schuller2005,alsing2006,bruschi2010,martinmartinez2011,FriisBertlmannHuberHiesmayr2010}. However, it should be pointed out that for a given fixed cavity length of the order $1$m, the magnitudes of acceleration accommodated by our approach can be as large as $10^{14}\mathrm{ms}^{-2}$.

Second, a massless Fermion in a $(1+1)$-dimensional cavity is unlikely
to be a good model for systems realisable in a laboratory. A~Fermion
in a linearly-accelerated rectangular cavity in $(3+1)$-dimensions can
be reduced to the $(1+1)$-dimensional case by separation of variables,
but for generic field modes the transverse quantum numbers then
contribute to the effective $(1+1)$-dimensional mass; further, any
foreseeable experiment would presumably need to use Fermions that have
a positive mass already in $(3+1)$-dimensions before the reduction. However, progress in experiments using Graphene have opened up the possibility of implementing quantum information protocols with particles which are effectively massless Dirac spinors~\cite{lukyanchuck2004}.


However, there is an encouraging physical result we have learned from studying Fermionic entanglement is the role of a particles charge. Although we have seen that Bell state entanglement is qualitative similar to the Boson case, we found that considering states which are symmetric in charge are more robust against the effects of non-uniform motion. To be more precise, for a given trajectory, even if we are unable to time the accelerations perfectly such that entanglement is not ideally stored, we can choose the initial entangled state such that the effect is mitigated. Physically, this is a consequence of imposing boundary conditions on the flow of probability in and out of the cavity. This gives us the freedom of splitting the energy spectrum into ``unequal" parts. Breaking the symmetry of the energy spectrum between particles and anti-particles allow us to construct states that can be more robust against acceleration effects.

The particular choice of boundary conditions imposed on the field also opens up further questions. As an example, is there a natural way of picking a fixed value for $s$? The natural choice would seem to be $s=0$, which results in a symmetric energy spectrum, but we have seen this leads to a zero mode which, \emph{a priori}, could pose problems. It would therefore be interesting to investigate other types of boundary conditions which have a different physical interpretation. One such model could be the so-called MIT bag model~\cite{alberto1996}. The MIT bag model effectively traps a quantum mechanical particle in a finite well potential. In the limit the depth of the well goes to infinity, we recover a scenario which resembles perfectly reflective cavity walls.

We started this chapter by emphasising that a cavity localises the
quantum degrees of freedom in the worldtube of the cavity, and our
assumption of inertial initial and final trajectory segments localises
the acceleration effects in a finite interval of the cavity's proper
time. We should perhaps end by emphasising that we are not attempting
to localise measurements of the field at more precise spacetime
locations within the worldtube of the cavity, and we are hence not
proposing cavities as a fundamental solution to the open conceptual
issues of a quantum measurement theory in relativistic
spacetime~\cite{sorkin1993}. A~cavity can however reduce the
measurement ambiguities from, say, megaparsecs to centimetres, which
may well suffice to resolve the conceptual issues in specific
experimental settings of interest, gedanken or otherwise.

\chapter{Spatially Extended Particle Detectors}\label{chapter:fat-detectors}

\section{Introduction}
The field of quantum information aims at understanding how to store, process, transmit, and read information efficiently exploiting quantum resources \cite{nielsen2010}. In the standard quantum information scenarios, observers may share entangled states, employ quantum channels, quantum operations, classical resources and perhaps more advanced devices such as quantum memories and quantum computers to achieve their goals.
In order to implement any quantum information protocol, all parties must be able to \textit{locally} manipulate the resources and systems which are being employed. Although quantum information has been enormously successful at introducing novel and efficient ways of processing information, it still remains an open question to what extent relativistic effects can be used to enhance current quantum technologies and give rise to new relativistic quantum protocols.

The novel and exciting field of relativistic quantum information has recently gained increasing attention within the scientific community. An important aim of this field is to understand how the state of motion of an observer and gravity affects quantum information tasks. For a review on developments in this direction see~\cite{alsing2012}.
Recent work has focussed on developing mathematical techniques to describe localised quantum fields to be used in future relativistic quantum technologies. The systems under investigation include fields confined in moving cavities~\cite{alphacentauri2012} and wave-packets \cite{dragan2012,downes2013}. Moving cavities in spacetime can be used to generate observable amounts of bipartite and multipartite entanglement \cite{friis2012-2,friis2012-4}. In Chapter~(\ref{chapter:bosons}), it was shown that the relativistic motion of these systems can be used to implement quantum gates, thus bridging the gap between relativistic-induced effects and quantum information processing. In addition, references~\cite{bruschi2012-2,friis2012-4} employed the covariance matrix formalism within the framework of continuous variables and showed that most of the gates necessary for universal quantum computation could be obtained by simply moving the cavity through especially tailored trajectories~\cite{bruschi2013}. This result pioneers on the implementation of quantum gates in relativistic quantum information.

A third local system that has been considered for relativistic quantum information processing is the well known Unruh-DeWitt detector~\cite{unruh1976}, a point-like  quantum system which follows a classical trajectory in spacetime and interacts locally with a global free quantum field. Such a system has been employed with different degrees of success in a variety of scenarios, such as  in the work unveiling the celebrated Unruh effect~\cite{unruh1976} or to extract entanglement from the vacuum state of a bosonic field~\cite{reznik2003}.  Unruh-DeWitt detectors seem convenient for relativistic quantum information processing. However, the mathematical techniques involved, namely perturbation theory, become extremely difficult to handle even for simple quantum information tasks such as teleportation~\cite{lin2012}. 

The main aim of our research program is to develop detector models which are mathematically simpler to treat so they can be used in relativistic quantum information tasks. A first step in this direction was taken in~\cite{dragan2011} where a model to treat analytically a finite number of harmonic oscillator detectors interacting with a finite number of modes was proposed exploiting techniques from the theory of continuous variables. The covariance matrix formalism was employed to study the Unruh effect and extraction of entanglement from quantum fields without perturbation theory.  The techniques introduced in~\cite{dragan2011}  are restricted to simple situations in which the time evolution is trivial.  To show in detail how the formalism introduced was applied, the authors presented simplified examples using detectors coupled to a single mode of the field which is formally only applicable when the field can be decomposed into a discrete set of modes with large frequency separation. This situation occurs, for example, when the detectors are inside a cavity. The detector model introduced in this work generalises the model presented in~\cite{dragan2011} to include situations in which the time evolution is non-trivial. 

The chapter is structured as follows: we will begin by introducing the standard, two-level system, Unruh-DeWitt detector model. In particular, we will investigate spatial profiles which are constant in time, but not point-like. In particular, we show that a Gaussian shaped profile allows a detector to couple to a peaked distribution of field modes when undergoing different forms of motion. These considerations will give us the necessary knowledge to consider more general interactions between quantum mechanical objects and a field. Next, we replace the two-level system with a harmonic oscillator and analyse the case where the coupling strength varies in space and time. It is shown that the detectors effectively couple to a time evolving frequency distribution of plane waves that can be described by a single mode. We introduce the mathematical techniques required to solve the time evolution of a detector which couples to an arbitrary \emph{time-dependent frequency distribution} of modes. The interaction of the detector with the field is purely quadratic in the operators and, therefore, we can employ the formalism of continuous variables taking advantage of the powerful mathematical techniques that have been developed in the past decade~\cite{alsing2012}. These techniques allow us to obtain the explicit time dependent expectation value of relevant observables, such as mean excitation number of particles. As a concrete example, we employ our model to analyse the response of a detector, which moves along an arbitrary trajectory and is coupled to a time-dependent frequency distribution of field modes. 

The techniques we will present simplify the Hamiltonian and an exact time dependent expression for the number operators can be obtained. We also discuss the extent of the impact of the techniques developed in this chapter: in particular, we stress that they can be successfully applied for a finite number of detectors following arbitrary trajectories. The formalism is also applicable when the detectors are confined within cavities. In this last case, the complexity of our techniques further simplify due to the discrete structure of the energy spectrum. We also note that the model can be generalised to the case where the detector is a quantum field itself. 

We close the chapter by combining the Unruh-DeWitt detector with the cavities of Chapter~(\ref{chapter:bosons}). We investigate the potential for using point-like objects to generate entanglement between two spatially distinct cavities in relative motion. Entangling the modes of two separate cavities has significant consequences as it can be used as a resource for relativistic quantum information based experiments.

\section{Constant Spatial Profile}\label{sec:free-detector}

\subsection{Introduction}

In this section we propose the use of finite-size detectors \cite{takagi1986,grove1983,schlicht2004}, i.e. detectors with a position dependent coupling strength, which are not only more realistic but also have the advantage of coupling to peaked distributions of modes. We design Gaussian-type spatial profiles such that a uniformly accelerated detector naturally couples to peaked distributions of  Rindler modes. By expanding the field in terms of what are known as Unruh modes, we show that the accelerated detector couples simultaneously to two peaked distribution of modes corresponding to left and right Unruh modes. As expected, the same detector interacts with a Gaussian distribution of Minkowski modes when it follows an inertial trajectory. In the Minkowski vacuum, the response of the detector has a thermal signature when it is uniformly accelerated and the temperature depends on the proper acceleration of the detector.

In the prototypical studies of quantum entanglement in non-inertial frames, observers are assumed to analyse states involving sharp frequency modes \cite{alsing2003,funentes-schuller2005}. In particular, recent works analysing the entanglement degradation between global modes seen by uniformly accelerated observers consider states of modes labelled by Rindler frequencies~\cite{bruschi2010,bruschi2012,martinmartinez2010}. Our analysis provides further insight into the physical interpretation of the particle states which were analysed in these works. We show that the response of the finite-size detector when the state of the field has a single particle labelled by a Unruh frequency has a thermal term plus a correction with noise that depends on acceleration. Therefore, a degradation of global mode entanglement in non-inertial frames as a function of acceleration should be detected by uniformly accelerated finite-size detectors. Although global mode entanglement cannot be detected directly it can be extracted by Unruh-DeWitt type detectors becoming useful for quantum information tasks.

\subsection{Unruh-DeWitt Particle Detector model}

From Section~(\ref{sec:qft-unruh-detector}), we start with the action which describes an Unruh-DeWitt detector interacting with the real Klein-Gordon field. Parametrised via appropriately chosen comoving coordinates, $(\tau,\boldsymbol{\zeta})$, we can write
\begin{eqnarray}
S_{I}&=&\int d\tau M(\tau)\tilde{\phi}(\tau),
\end{eqnarray}
where the field the detector couples to is given by,
\begin{eqnarray} 
\label{eq:Mfield}
\tilde{\phi}(\tau):=\int d^{3}\boldsymbol{\zeta}\sqrt{-g(\boldsymbol{\zeta)}}f(\boldsymbol{\zeta})\hat{\phi}(\tau,\boldsymbol{\zeta})
\end{eqnarray}
and $\text{det}(g_{\mu\nu})=g$ is the determinant of the metric tensor.  We assume that the center of the detector follows a classical trajectory in spacetime and the spatial profile $f(\boldsymbol{\zeta})$ determines how the detector couples to the field along the trajectory. This function, which must be real to ensure that the action is Hermitian, can be interpreted as a position dependent coupling strength. Consider $u_{\boldsymbol{k}}(\boldsymbol{\zeta}(\tau))$ to be field solutions to the Klein-Gordon equation evaluated along a point-like worldline parametrised by $\tau$ corresponding to the center point of the detector. The momentum $\boldsymbol{k}$ of the modes are determined by an observer comoving with the center of the detector. The Hamiltonian in terms of these modes takes the form,
\begin{eqnarray}
\label{hamiltonian}
\hat{H}_{I}&=&\hat{M}(\tau)\cdot\int d^{3}\boldsymbol{k} \tilde{f}(\boldsymbol{k})\left(u_{\boldsymbol{k}}(\boldsymbol{\zeta}(\tau))\hat{a}_{\boldsymbol{k}}+\mathrm{h.c.}\right),
\end{eqnarray}
where $\hat{a}^{\dagger}_{\boldsymbol{k}}$ and $\hat{a}_{\boldsymbol{k}}$ are creation and annihilation operators associated to the field modes of momentum $\boldsymbol{k}$. The frequency distribution $\tilde{f}(\boldsymbol{k})$ corresponds to a transformation of  $f(\boldsymbol{\zeta})$ into frequency space. In the ideal case, where the detector is considered to be point-like, the spatial profile is $f(\boldsymbol{\zeta})=\delta^{3}(\boldsymbol{\zeta}-\boldsymbol{\zeta}')$ and the detector couples locally to the field and the coupling strength is uniformly equal for all frequency modes. Here $ \delta^{3}(\boldsymbol{\zeta}-\boldsymbol{\zeta}'):=\delta(\zeta_{1}-\zeta_{1}')\delta(\zeta_{2}-\zeta_{2}')\delta(\zeta_{3}-\zeta_{3}')$ is the three dimensional Dirac delta distribution. When we model a finite-size detector, which corresponds to a more realistic situation, the detector couples naturally to a distribution of field modes. The frequency distribution will be determined by the spatial profile. In this sense the field $\tilde{\phi}(\tau)$ corresponds to a {\it window} of frequencies.

\subsection{Inertial Trajectory}

Now we specify a trajectory for the detector.  When the detector follows an inertial trajectory it is convenient to use Minkowski coordinates $(t,\boldsymbol{x})$ where  $\boldsymbol{x}:=(x,\boldsymbol{y}_{\perp})$. In this case, the proper time of a comoving observer is $\tau=t$ and we can also write the comoving spatial coordinates as $\boldsymbol{\zeta}=\boldsymbol{x}$. The solutions to the Klein-Gordon equation correspond to plane waves,
\begin{eqnarray}
\phi^{M}_{\boldsymbol{k}}(t,\boldsymbol{x})&:=&\frac{1}{\sqrt{2(2\pi)^{3}\omega_{\boldsymbol{k}}}}e^{-i\omega t+i\boldsymbol{k}\cdot\boldsymbol{x}},
\end{eqnarray}
where the frequency of the mode $\omega_{\boldsymbol{k}}\equiv\sqrt{\boldsymbol{k}\cdot\boldsymbol{k}+m^{2}}$ is strictly positive and $\boldsymbol{k}\in\mathbb{R}^{3}$ denotes the momentum  $\boldsymbol{k}:=(k_x,\boldsymbol{k}_{\perp})$. Canonical quantisation proceeds in the same manner as detailed in Chapter~(\ref{chapter:qft}) and the previous chapters.

The field can be expanded in Minkowski modes as
\begin{equation}
\hat{\phi}(t,\boldsymbol{x})=\int d^{3}\boldsymbol{k}\left(\hat{a}_{\boldsymbol{k}}\phi^{M}_{\boldsymbol{k}}(t,\boldsymbol{x})+\text{h.c.}\right).
\end{equation}
From this, the frequency distribution expressed in Minkowski modes is
\begin{eqnarray}
\label{eqn:freq-distro}
\tilde{f}(\boldsymbol{k})&=&\int d^{3}\boldsymbol{x} f(\boldsymbol{x})e^{+i\boldsymbol{k}\cdot\boldsymbol{x}},
\end{eqnarray}
which is the Fourier transform of the spatial profile $f(\boldsymbol{x})$ and dictates the specific form of Eq.~(\ref{hamiltonian}).

We now design a spatial profile tailored so that the corresponding frequency detection window of the detector is a Gaussian distribution of modes peaked around a Minkowski momentum vector $\boldsymbol{\lambda}$. This choice is motivated by early works on relativistic entanglement
where the states analysed involved sharp frequencies $\Omega$ and $\Omega'$. For example, the Bell state,
\begin{equation}
|\phi^{+}\rangle=\frac{1}{\sqrt{2}}(|0\rangle_{\Omega}|0\rangle_{\Omega'}+|1\rangle_{\Omega}|1\rangle_{\Omega'}),
\end{equation}
which was analysed in a flat (1+1)-dimensional space in the charged  \cite{bruschi2012} and uncharged massless bosonic case \cite{bruschi2010}. Entanglement for Bell states in non-inertial frames was also discussed for Dirac fields \cite{martinmartinez2009,martinmartinez2010,martinmartinez2012-2}. 
Sharp frequency states, $|1_{k}\rangle=\hat{a}_{k}^{\dag}|0\rangle$, are an idealisation of Gaussian wave-packets of the form
\begin{equation}
\label{Eq:particlesmearing}
|1_{\lambda}\rangle=\int dk \Phi(\lambda,k)\hat{a}_{k}^{\dag}|0\rangle,
\end{equation}
where $\Phi(\lambda,k)$ is a Gaussian distribution centred around $\lambda$ \cite{bruschi2010}. Our detector model will be useful to investigate questions of entanglement in non-inertial frames from an operational perspective and extract entanglement for relativistic quantum information processing. An interesting question is how much entanglement can be extracted by our detectors from field mode entangled states as a function of the detector's acceleration. However, we shall not pursue these questions here.
\begin{figure}[t]
\centering
\includegraphics[scale=0.7]{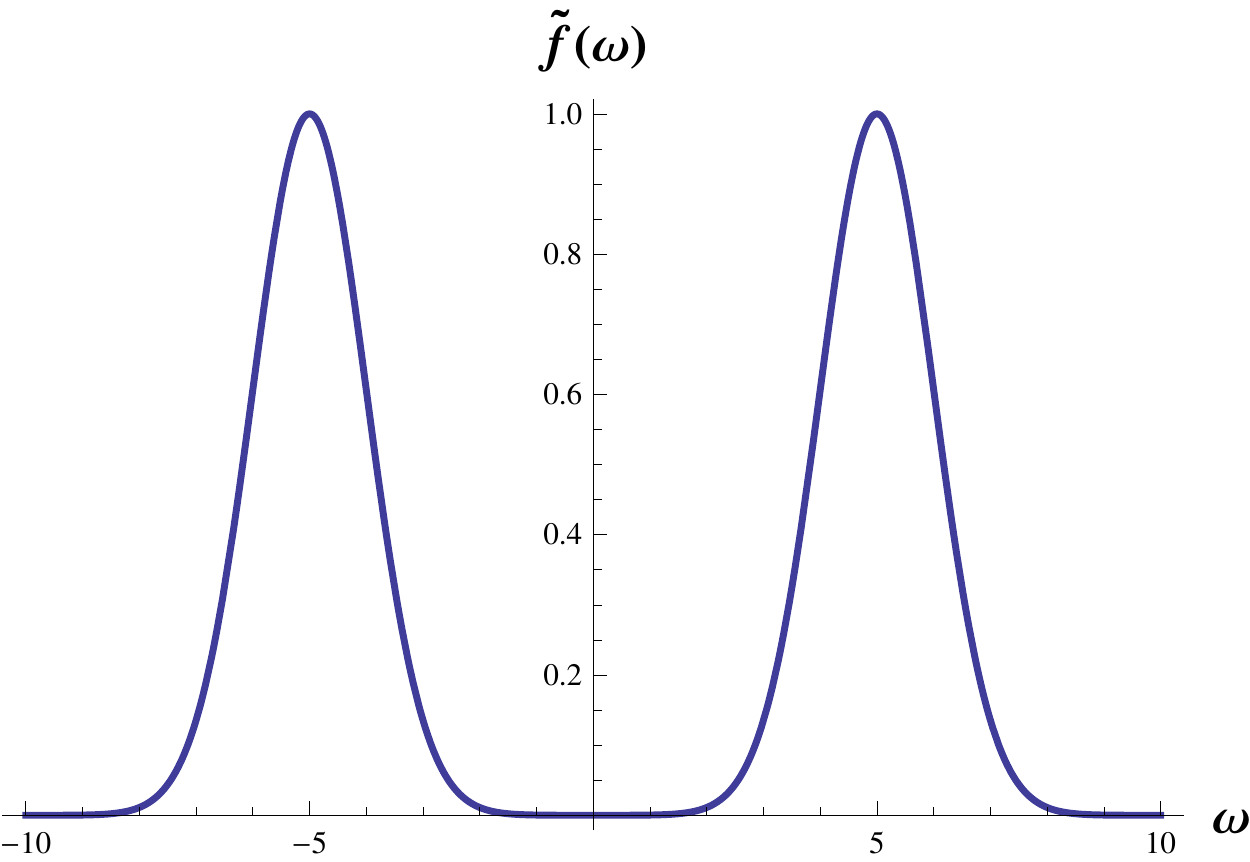}
\caption[Unruh-DeWitt Detector Gaussian Peaking]{\label{gaussianPeakingfig} $(1+1)$ dimensional example of a frequency distribution peaked around $\pm\lambda$ given by Eq.(\ref{gaussianPeaking}) for $\sigma=1$ and $\lambda=5$. This frequency distribution peaks around the desired frequency $\lambda$ but has a double peaking due to the two exponential terms. In the $(1+1)$ massless case, the field is expanded as an integral over $\omega>0$ and so the peak in the $\omega<0$ region does not contribute.} 
\end{figure}
We find that a Gaussian frequency window of width $\sigma$ centred around frequency $\boldsymbol{\lambda}$ as shown in Fig.(\ref{gaussianPeakingfig}), can be engineered by choosing the following spatial profile,
\begin{eqnarray}
\label{gaussianprof}
f(\boldsymbol{x})&=&n_{\sigma}e^{-\frac{1}{2}\sigma^{-2}\boldsymbol{x}\cdot\boldsymbol{x}}\left(e^{-i\boldsymbol{\lambda}\cdot\boldsymbol{x}}+e^{+i\boldsymbol{\lambda}\cdot\boldsymbol{x}}\right),
\end{eqnarray}
which corresponds to a Gaussian multiplied by a superposition of planes waves of opposite momentum $\boldsymbol{\lambda}$. $n_{\sigma}$ is a normalisation constant. The spatial profile is therefore transformed into, via Eq.~(\ref{eqn:freq-distro}), the momentum distribution 
\begin{eqnarray}
\label{gaussianPeaking}
\tilde{f}(\boldsymbol{k})=e^{-\frac{1}{2}\sigma^{2}(\boldsymbol{k}-\boldsymbol{\lambda})\cdot(\boldsymbol{k}-\boldsymbol{\lambda})}+e^{-\frac{1}{2}\sigma^{2}(\boldsymbol{k}+\boldsymbol{\lambda})\cdot(\boldsymbol{k}+\boldsymbol{\lambda})}.
\end{eqnarray}
This means that in order to couple our detector to a peaked Gaussian distribution of modes centred around $\boldsymbol{\lambda}$ it is necessary to engineer a field-detector coupling strength which not only is peaked around the atom's trajectory but also oscillates with position. Sharp frequency modes $\tilde{f}(\boldsymbol{k})=\delta^{3}(\boldsymbol{k}-\boldsymbol{\lambda})+\delta^{3}(\boldsymbol{k}+\boldsymbol{\lambda})$ are obtained when  $f(\boldsymbol{x})\sim\text{exp}(-i\boldsymbol{\lambda}\cdot\boldsymbol{x})+\text{exp}(+i\boldsymbol{\lambda}\cdot\boldsymbol{x})$. In the massless $1+1$-dimensional case the frequency distribution obtained from a given spatial profile is defined as a function of $\omega\ge 0$ only. Therefore, given the window profile peaks are sufficiently narrow and separated, the second peaking corresponding to the $\omega<0$ region, does not contribute to the frequency window in this case. The field to which the detector couples given by Eq.(\ref{eq:Mfield}) is therefore,
\begin{eqnarray}
\tilde{\phi}(t)=\int\limits_{0}^{\infty}d\omega N_{\omega}e^{-\frac{1}{2} \sigma^{2}({\lambda} -\omega )^2}\left[e^{-i\omega t}\hat{a}_{\omega}+e^{+i\omega t}\hat{a}^{\dagger}_{\omega}\right].
\end{eqnarray}
In the general case, the frequency window is peaked around two modes corresponding to negative and positive momentum.  

Now that we have defined our spatial profile and found the smeared field operator, we would like to know how the detector behaves when travelling on an inertial trajectory. We do this, as mentioned in Section~(\ref{sec:qft-unruh-detector}), by analysing the instantaneous transition rate of the detector. To first order in perturbation theory, the transition rate is a function of the detector's energy gap $\Delta$ given by~\cite{schlicht2004}
\begin{eqnarray}
\label{transitionrate}
\dot{F}_{\tau}(\Delta)&:=&2\int_{0}^{\infty}ds Re\left[e^{-i\Delta s}W(t,t-s)\right],
\end{eqnarray}
where $W(t,t'):=\bra{\psi}\phi(t)\phi(t')\ket{\psi}$ is the so-called Wightman-function and $\ket{\psi}$ denotes the state of the field.  Note that we have assumed that the detector is turned on in the distant past.

The transition rate is the time derivative of the probability amplitude of the detector undergoing a change in its internal state. In other words, it is the average time the detector ``clicks" over some time interval. A straight forward derivation can be found in~\cite{schlicht2004}.
\begin{figure}[t]
\centering
\includegraphics[scale=0.7]{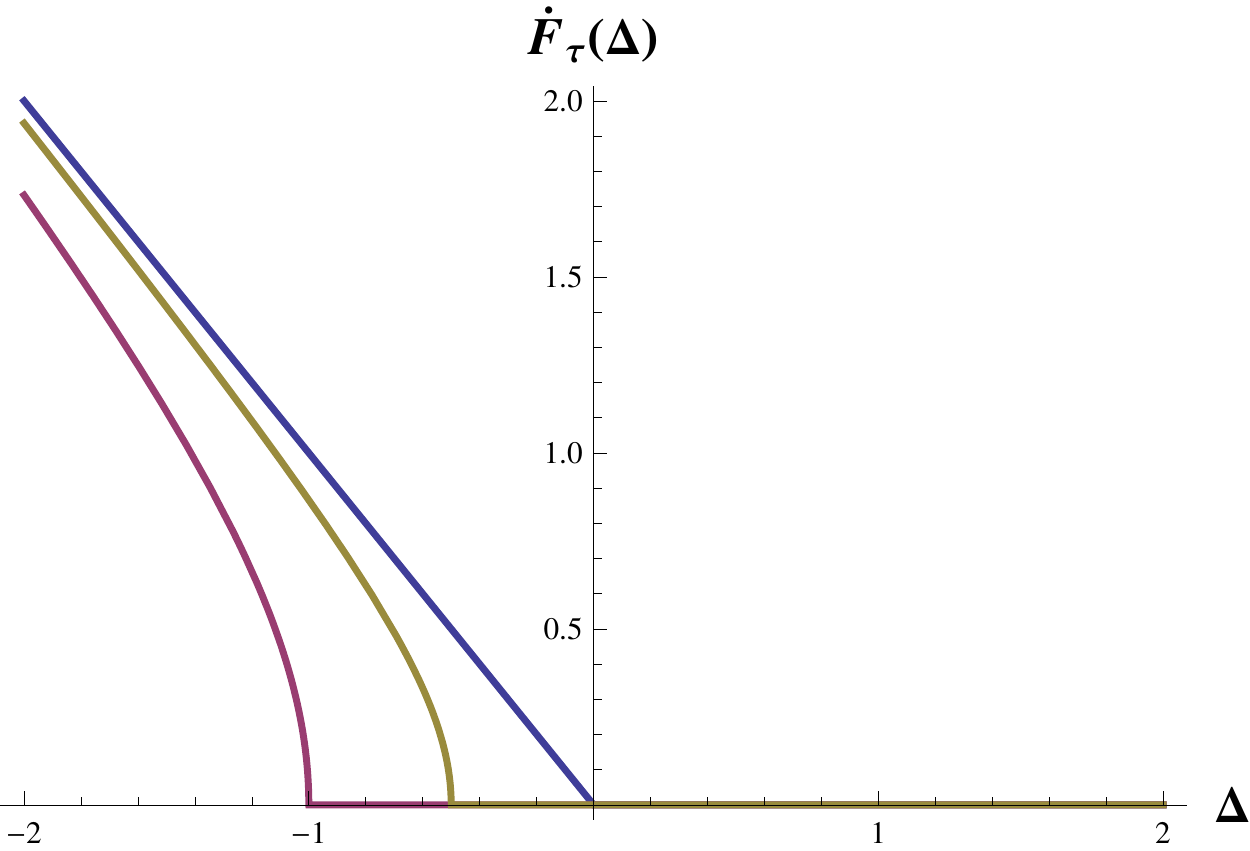}
\caption[Unruh DeWitt Detector Inertial Transition Rate]{\label{inertialresponse} The transition rate of a point-like inertial detector probing the massive Minkowski vacuum~(\ref{eqn:inertial-response}). Here we see that for massive fields, the detector only undergoes emission when it has an energy gap greater than the bare mass of the field. We plotted for mass values $m=0,0.5,1$ which are represented by the purple, mustard and blue lines respectively.} 
\label{fig:inertial-response}
\end{figure}
Expanding the field in terms of Minkowski modes we find the Minkowski vacuum transition rate for a stationary detector is
\begin{eqnarray}
\label{eqn:inertial-response}
\dot{F}_{\tau}(\Delta)=-\Theta(-\Delta-m)\Xi(\Delta),
\end{eqnarray}
where 
\begin{eqnarray}
\Xi(\Delta)=\sqrt{(-\Delta)^{2}-m^{2}}|\tilde{f}(-\Delta)|^{2}
\end{eqnarray}
and $\Theta(x)$ is the Heavisde theta function defined as 
\begin{eqnarray}
\Theta(x)&=&\left\{\begin{array}{ccl} 0 & : & x<0\\ 1 & : & x\ge 0\end{array}\right. .
\end{eqnarray}
Note in the above result we have explicitly assumed $\tilde{f}(\boldsymbol{k})=\tilde{f}(|\boldsymbol{k}|)$ i.e. the Fourier transform of the spatial profile $f(\boldsymbol{x})$ depends on the magnitude of $\boldsymbol{k}$ only. The result is explicitly independent of the time parameter $t$. Further, it is trivial to see as $f(\boldsymbol{x})\rightarrow\delta(\boldsymbol{x})$ then $|\tilde{f}|\rightarrow +1$ and we recover the standard literature result of a stationary detector probing the Minkowski vacuum~\cite{schlicht2004}. It is interesting to note that the transition rate of the detector is tempered only by the square of the frequency distribution $\tilde{f}$. See Fig.~(\ref{fig:inertial-response}) for plots of Eq.~(\ref{eqn:inertial-response}) for mass values of $m=0,0.5,1$.

We are also interested in analysing the response of the detector when the field has a single Minkowski excitation, \begin{eqnarray}
\ket{1_{\boldsymbol{\lambda}}}&:=&\int d^{3}\boldsymbol{k} \Phi(\boldsymbol{\lambda},\boldsymbol{k})\hat{a}_{\boldsymbol{k}}^{\dag}\ket{0},
\end{eqnarray}
where we define a delta normalised state to have the property $\int d^{3}\boldsymbol{k} |\Phi(\boldsymbol{\lambda},\boldsymbol{k})|^{2}=\delta^{3}(\boldsymbol{0})$ and a properly normalised state to have the property $\int d^{3}\boldsymbol{k} |\Phi(\boldsymbol{\lambda},\boldsymbol{k})|^{2}=1$. This state is the generalisation of Eq.(\ref{Eq:particlesmearing}) to three spatial dimensions. The Wightman-function in this case is 
\begin{eqnarray}
W(t,t')&:=& \bra{1_{\Phi}}\tilde{\phi}(t)\tilde{\phi}(t')\ket{1_{\Phi}},
\end{eqnarray}
Writing the states and the field in terms of the Minkowski modes and normal ordering the associated operators we find that
\begin{equation}
\begin{aligned}
\label{wightmann1p}
W(t,t')&=\bra{0}\tilde{\phi}(t)\tilde{\phi}(t')\ket{0}\cdot\int d^{3}\boldsymbol{k} |\Phi(\boldsymbol{\lambda},\boldsymbol{k})|^{2}\\
& \hspace{5mm}+2\text{Re}\left[\bar{I}_{\boldsymbol{\lambda}}(t)I_{\boldsymbol{\lambda}}(t')\right],
\end{aligned}
\end{equation}
where  
\begin{eqnarray}
I_{\boldsymbol{\lambda}}(t)=\int d^{3}\boldsymbol{k}\Phi(\boldsymbol{\lambda},\boldsymbol{k})\tilde{f}(\boldsymbol{k})\frac{1}{\sqrt{\omega_{\boldsymbol{k}}}}e^{-i\omega_{\boldsymbol{k}} t}.
\end{eqnarray}  
We notice there are two terms in the Wightman-function. The first one corresponds to the vacuum state and the second is the contribution from the particle present in the field. 

The single particle contribution factorises into two independent functions of $t$ and $t'$. This allows us to analyse the transition rate with relative ease. Substituting Eq.~(\ref{wightmann1p}) into Eq.~(\ref{transitionrate}) we find the usual vacuum transition rate is modified by an integral expression which contains a term proportional to
\begin{eqnarray}
\label{iota}
\iota_{t}(\Delta)&:=&\int_{0}^{\infty} ds e^{-is\Delta}I_{\boldsymbol{\lambda}}(t-s).
\end{eqnarray}
This integral is effectively a Fourier transform of $I_{\boldsymbol{\lambda}}(t-s)$  in the $s$ variable and can be computed, either analytically or numerically, for specifically chosen $\tilde{f}$ and $\Phi$.  Employing the Riemann-Lebesgue lemma, which can only be used for functions which are integrable on the real line, one can show that $I_{\boldsymbol{\lambda}}(t)\rightarrow 0$ as $t\rightarrow\pm\infty$ as long as $\tilde{f}$ and $\Phi$ are well-behaved. $I_{\boldsymbol{\lambda}}(t)$ vanishes in the distant past and future where the detector is responding only to vacuum fluctuations. In other words, the detector only observes a constant spectrum in these asymptotic regions. In the intermediate regions the oscillatory response is due to the presence of the particle.

\subsection{Accelerated Trajectory}

We now consider a detector following a uniformly accelerated trajectory.  Conformally flat Rinder coordinates $(\rho,\boldsymbol{\xi})=(\rho,\xi,\boldsymbol{y}_{\perp})$ are a convenient choice in this case.  The transformation between Rindler and Minkowski coordinates in this case is given by
\begin{equation}
\label{eq:rindler}
\begin{aligned}
t & =  a^{-1}e^{a\xi}\sinh(a\rho)\\ 
x & =  a^{-1}e^{a\xi}\cosh(a\rho)
\end{aligned}
\end{equation}
where $a$ is a positive parameter and the $\boldsymbol{y}_{\perp}$ are unchanged.  This transformation holds for the spacetime region $|t|>x$, i.e. the right Rindler wedge. The coordinate transformation for $|t|>-x$ (left region) differs from Eq.~\eqref{eq:rindler} by an overall sign in the $x$ coordinate. The coordinates are tailored specifically to the trajectory $\boldsymbol{\xi}=\boldsymbol{0}$ so that an observer travelling along this worldline will measure a proper acceleration $\sqrt{-A^{\mu}A_{\mu}}=a$ and the proper time is parametrised by the coordinate time $\rho$.
The Klein-Gordon equation for a massive bosonic field in a (3+1)-dimensional flat spacetime in this case takes the form
\begin{eqnarray}
\partial_{\rho\rho}\phi-\left[\partial_{\xi\xi}+e^{2a\xi}(\partial_{yy}+\partial_{zz})-m^{2}e^{2a\xi}\right]\phi=0 ,
\end{eqnarray}
and the solutions are the Rindler modes \cite{crispino2008}
\begin{equation}
\begin{aligned}
u_{\Omega,\boldsymbol{k}_{\perp},\alpha}(\rho,\boldsymbol{\xi})&:= N_{\Omega/a}K_{i\Omega/a}\left(\kappa a^{-1}e^{a\xi}\right)e^{-i\Lambda_{\alpha}(\rho,\boldsymbol{\xi})},\\
\Lambda_{\alpha}(\rho,\boldsymbol{\xi})&:=\alpha \rho \Omega-\boldsymbol{k}_{\perp}\cdot\boldsymbol{x}_{\perp},
\end{aligned}
\end{equation}
with $\kappa=\sqrt{\boldsymbol{k}_{\perp}\cdot\boldsymbol{k}_{\perp}+m^{2}}$ and  $N_{\Omega/a}$ is the mode normalisation constant. The functions $K_{i\Omega/a}(R)$ are modified Bessel functions of the second kind. Here $\boldsymbol{x}_{\perp}:=(y,z)$ and $\boldsymbol{k}_{\perp}:=(k_{y},k_{z})$ are position and momentum vectors perpendicular to the direction of acceleration. $\Omega$ is strictly positive and denotes the \emph{dimension-full} Rindler frequency and  $\alpha=+1(-1)$ corresponds to right (left) Rindler wedges,  respectively. The canonical orthonormality relations for the $3+1$ massive field are,
\begin{equation}(u_{\Omega,\boldsymbol{k}_{\perp},\alpha},u_{\Omega',\boldsymbol{k}^{\prime}_{\perp},\alpha'})=\delta(\Omega-\Omega')\delta^{2}(\boldsymbol{k}_{\perp}-\boldsymbol{k}_{\perp}')\delta_{\alpha\alpha'},
\end{equation}and commutation relations satisfy
\begin{equation}[\hat{a}_{\Omega,\boldsymbol{k}_{\perp},\alpha},\hat{a}^{\dag}_{\Omega'\boldsymbol{k}^{\prime}_{\perp},\alpha'}]=\delta(\Omega-\Omega')\delta^{2}(\boldsymbol{k}_{\perp}-\boldsymbol{k}_{\perp}')\delta_{\alpha\alpha'}.
\end{equation} 
From our coordinate definitions Eq.~\eqref{eq:rindler}, we choose the detector to be travelling in the right Rindler wedge. Thus, our comoving coordinates can be parametrised as $\tau=\rho$ and $\boldsymbol{\zeta}=\boldsymbol{\xi}$. The field expansion in terms of the parametrised Rindler modes is
\begin{equation}
\label{eq:rinderexpansion}
\hat{\phi}(\rho,\boldsymbol{\xi})=\int d\Omega d^{2}\boldsymbol{k}_{\perp}\left[u_{\Omega,\boldsymbol{k}_{\perp},+}(\rho,\boldsymbol{\xi})\hat{a}_{\Omega,\boldsymbol{k}_{\perp},+}+\mathrm{h.c.}\right].
\end{equation}
Note the left Rindler modes do not appear in Eq.~(\ref{eq:rinderexpansion}) as the detector is assumed to be moving in the right Rindler wedge. The explicit form of the accelerated detectors frequency distribution is
\begin{eqnarray}
\label{Eq:acceleratedsmearingexpression}
\tilde{f}(\Omega,\boldsymbol{k}_{\perp})=\int d^{3}\boldsymbol{\xi} e^{2a\xi}f(\boldsymbol{\xi})K_{i\Omega/a}\left(\kappa a^{-1}e^{a\xi}\right)e^{+i\boldsymbol{k}_{\perp}\cdot\boldsymbol{x}_{\perp}}.
\end{eqnarray}
The most significant difference between the inertial and the accelerated frequency distributions is the appearance of a non-trivial metric factor and the Bessel function. Note also for both massless and massive fields, the Rindler modes are defined as an integral over $\Omega\in\mathbb{R}^{+}$, unlike the Minkwoski mode case. Eq.(\ref{Eq:acceleratedsmearingexpression}) is a Fourier transform in the $y$ and $z$ coordinates, however, it is a non-standard integral transformation in the $\xi$ coordinate. Reminiscent of a Hankel transformation, we expect our desired properties of arbitrary mode peaking to still hold. Using the integral representation of the Modified Bessel function of the second kind~\cite{NIST:DLMF}
\begin{eqnarray}
K_{i\Omega/a}(R)=\frac{\sqrt{\pi}\left(\frac{1}{2} R\right)^{i\Omega/a}}{\Gamma(i\Omega/a +1/2)}\int_{0}^{\infty}dt\frac{\left(\sinh(t)\right)^{2i\Omega/a +1}}{e^{R\cosh(t)}},
\end{eqnarray}
valid for $\Omega/a>0$ and $R>0$, we can write the frequency distribution as a Fourier type integral that takes the form
\begin{eqnarray}
\tilde{f}(\boldsymbol{k})=\int d^{3}\boldsymbol{\xi} \beta(\boldsymbol{\xi})e^{+i\boldsymbol{\xi}\cdot\boldsymbol{k}},
\end{eqnarray}
where now $\boldsymbol{k}=(\Omega+\delta,\boldsymbol{k}_{\perp})$ and
\begin{eqnarray}
\beta(\boldsymbol{\xi})=\sqrt{\pi}\frac{\left(\frac{1}{2}\frac{M}{a}\right)^{i\Omega/a}}{\Gamma(i\Omega/a +1/2)}f(\boldsymbol{\xi})\int_{0}^{\infty}dr \frac{(\sinh(r))^{2i\Omega/a}}{e^{\frac{M}{a}e^{a\xi}\cosh(r)}}.
\end{eqnarray}
$\delta$ is a phase that is acquired from the integral representation of the modified Bessel function. This shows that, in principle, the standard properties of the Fourier transformation can be used to design a detector profile such that we obtain a peaked distribution in momentum space. For a concrete example, we shall consider the massless $(1+1)$ field case. The appropriate transformation, in terms of Rindler modes, is given by
\begin{eqnarray}
\label{1plus1fourier}
\tilde{f}(\Omega)=\int d\xi e^{2a\xi} f(\xi)e^{i\Omega\xi}.
\end{eqnarray}
and the spatial profile we propose in this case is 
\begin{eqnarray}
\label{Eq:modifedspatialprofile}
f(\xi)&=&N_{\sigma}e^{-2a\xi}e^{-\frac{1}{2}\sigma^{-2}{\xi^2}}\left(e^{-i{\lambda}{\xi}}+e^{+i{\lambda}{\xi}}\right)
\end{eqnarray}
which includes the conformal metric factor that arises from the Rindler coordinate transformation. Here $N_{\sigma}$ is a normalization constant.
This profile reduces the integral transformation in Eq.~(\ref{1plus1fourier}) to a standard Fourier transformation and the resulting frequency distribution is
\begin{eqnarray}
\tilde{f}(\Omega)=e^{-\frac{1}{2} \sigma^{2}(\lambda -\Omega )^2}+e^{-\frac{1}{2} \sigma ^2(\lambda +\Omega )^2}.
\end{eqnarray}
Substituting this frequency distribution into Eq.(\ref{eq:Mfield}), we obtain
\begin{eqnarray}
\tilde{\phi}(\rho)=\int d\Omega N_{\Omega}e^{-\frac{1}{2} \sigma^{2}(\lambda -\Omega )^2}\left[e^{-i\Omega\rho}\hat{a}_{\Omega,I}+\mathrm{h.c.}\right],
\end{eqnarray}
noting again that the left Rindler modes are absent. In the limiting case where the acceleration is zero $\Omega\rightarrow\omega$ and the spatial profile reduces to the $(1+1)$ version of the Minkowski profile given by Eq.({\ref{gaussianprof}}).  

We would now like to expand the field in terms of Unruh modes which play an important role in the literature. These modes are given by \cite{bruschi2010,crispino2008},
\begin{eqnarray}
\begin{array}{ccl} 
\label{eqn:unruh-modes}
u_{\Omega,\boldsymbol{k}_{\perp},R}&:=&\cosh(r_{\Omega/a})u_{\Omega,\boldsymbol{k}_{\perp},+}+\sinh(r_{\Omega/a})\bar{u}_{\Omega,\boldsymbol{k}_{\perp},-}\\
u_{\Omega,\boldsymbol{k}_{\perp},L}&:=&\cosh(r_{\Omega/a})u_{\Omega,\boldsymbol{k}_{\perp},-}+\sinh(r_{\Omega/a})\bar{u}_{\Omega,\boldsymbol{k}_{\perp},+}
\end{array},
\end{eqnarray}
where $\tanh(r_{\Omega/a})=e^{-\pi\Omega/a}$. The subscript $R/L$ in Eqns.~(\ref{eqn:unruh-modes}) denote what are known as either ``right (R)" or ``left (L)" Unruh modes. The Unruh modes are simply a different basis of Klein-Gordon field solutions we can use. They are convenient as their quantised field operator admits the same vacuum as the Minkowski quantisation. This is useful as the Unruh annihilation operators therefore annihilate the Minkowski vacuum. Further, as can be seen from Eq.~(\ref{eqn:unruh-modes}), the Bogoliubov transformations between Rindler modes and Unruh modes is simple. This transformation will allows us to derive simple expressions for the case of a detector in uniform acceleration.

Upon parametrising the Unruh modes with our accelerated comoving coordinates, i.e.~$(\rho,\boldsymbol{\xi})=(\tau,\boldsymbol{\xi})$, and noting that the left Rindler modes have no support in the right Rindler wedge, we find that the Unruh modes reduce to
\begin{eqnarray}
\begin{array}{ccl} 
u_{\Omega,\boldsymbol{k}_{\perp},R}(\rho,\boldsymbol{\xi})&=&\cosh(r_{\Omega/a})u_{\Omega,\boldsymbol{k}_{\perp},+}(\rho,\boldsymbol{\xi}),\\
u_{\Omega,\boldsymbol{k}_{\perp},L}(\rho,\boldsymbol{\xi})&=&\sinh(r_{\Omega/a})\bar{u}_{\Omega,\boldsymbol{k}_{\perp},+}(\rho,\boldsymbol{\xi}).
\end{array}
\end{eqnarray}
Expanding the field in terms of Unruh modes in the right wedge results, we find for the field and its smeared counterpart
\begin{subequations}
\begin{align}
\label{Eq:unruhfieldop}\hat{\phi}(\rho,\boldsymbol{\xi})&=\int d\Omega d^{2}\boldsymbol{k}_{\perp}N_{\Omega/a}K_{i\Omega/a}\left(\frac{\kappa}{a}e^{a\xi}\right)\left[ \mathrm{ch}_{\Omega}e^{-i\Omega\rho +i\boldsymbol{k}_{\perp}\cdot\boldsymbol{x}_{\perp}}\hat{a}_{\Omega,\boldsymbol{k}_{\perp},R}\right.\notag\\
& \hspace{3mm}\left. +\mathrm{sh}_{\Omega}e^{+i\Omega\rho -i\boldsymbol{k}_{\perp}\cdot\boldsymbol{x}_{\perp}}\hat{a}_{\Omega,\boldsymbol{k}_{\perp}L} +\mathrm{h.c.}\right],\\
\tilde{\phi}(\rho)&=\int d\Omega d^{2}\boldsymbol{k}_{\perp}N_{\Omega/a}\tilde{f}(\Omega,\boldsymbol{k}_{\perp})\left[\mathrm{ch}_{\Omega}e^{-i\Omega\rho}\hat{a}_{\Omega,\boldsymbol{k}_{\perp},R}\right.\notag\\
& \left.\hspace{5mm}+\mathrm{sh}_{\Omega}e^{+i\Omega\rho}\hat{a}_{\Omega,\boldsymbol{k}_{\perp},L}+\mathrm{h.c.}\right],
\end{align}
\end{subequations}
where we have used
\begin{subequations}
\begin{align}
\mathrm{ch}_{\Omega}&\equiv\cosh(r_{\Omega/a}),\\
\mathrm{sh}_{\Omega}&\equiv\sinh(r_{\Omega/a}).
\end{align}
\end{subequations}
As uniform acceleration is also a stationary orbit of flat spacetime, we expect a time independent vacuum transition rate.
Using the parametrised Unruh modes, we can calculate the transition rate of the accelerated detector. We find that for the field in its vacuum state, the detectors transition rate is
\begin{figure}[t]
\centering
\includegraphics[scale=0.7]{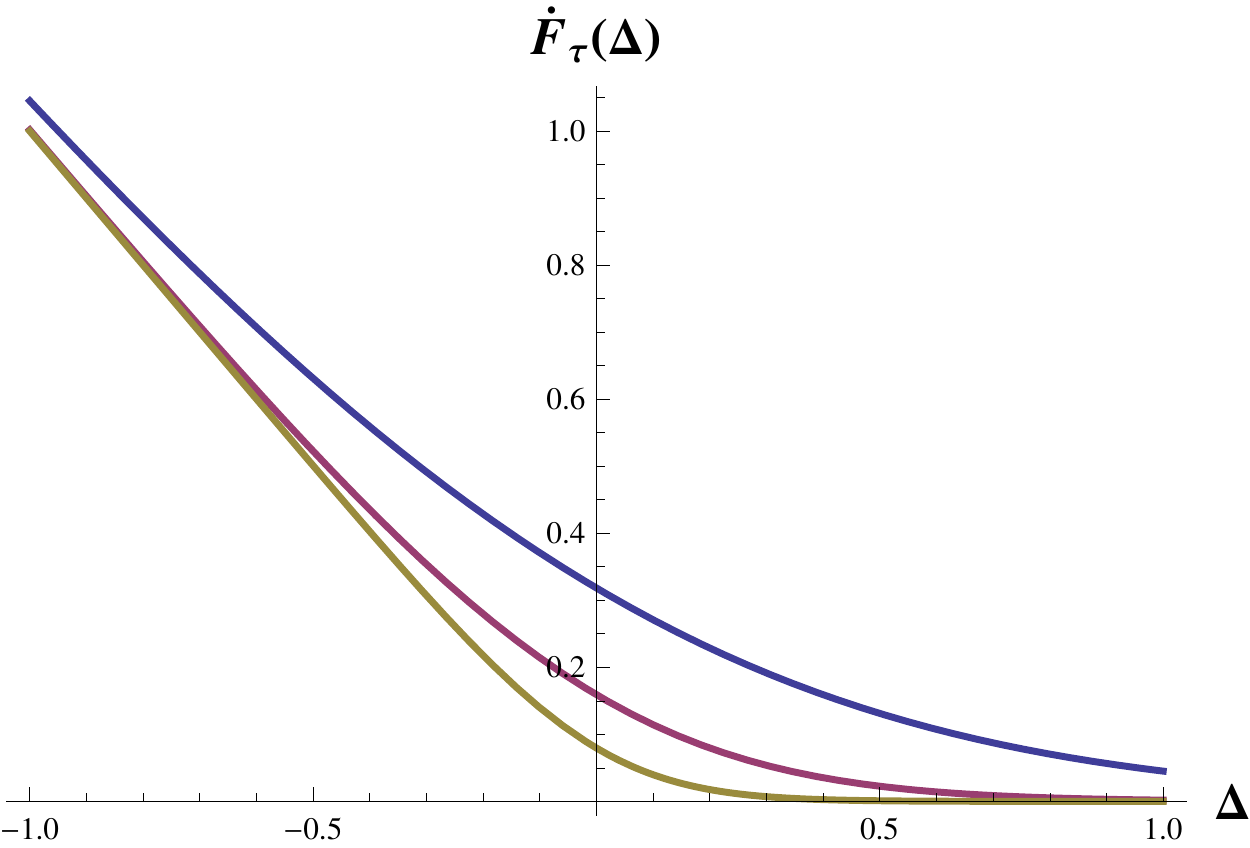}
\caption[Unruh-DeWitt Detector Accelerated Transition Rate]{The transition rate of a point-like uniformly accelerated detector probing the massless Minkowski vacuum in $3+1$-dimensions~(\ref{acceleratedresponse}). We plotted for acceleration values $a=.5,1,2$ which are represented by the mustard, purple and blue lines respectively.}
\label{fig:acceleratedresponse}
\end{figure}
\begin{eqnarray}
\label{acceleratedresponse}
\dot{F}_{\tau}(\Delta)&=&\frac{1}{e^{2\pi\Delta/a}-1}\Xi\left(\Delta\right),
\end{eqnarray}
where
\begin{equation}
\Xi\left(\Delta\right):=\int d^{2}\boldsymbol{k}_{\perp}\left[N^{2}_{\boldsymbol{\Delta}}|\tilde{f}(\boldsymbol{\Delta})|^{2}\Theta(\Delta) - N^{2}_{-\boldsymbol{\Delta}}|\tilde{f}(-\boldsymbol{\Delta})|^{2}\Theta(-\Delta)\right],
\end{equation}
with $\pm\boldsymbol{\Delta}:=(\pm\Delta,\boldsymbol{k}_{\perp})$ and $N_{\boldsymbol{\Delta}}$ denotes the appropriate normalisation for the Rindler modes. We can see immediately that the transition rate of the detector is the expected thermal distribution, where the temperature is inversely proportional to the acceleration parameter $a$, but again modified by the smeared field operator. We also note that Eq.~\eqref{acceleratedresponse} satisfies the Kubo-Martin-Schwinger condition \cite{hodgkinson2012}
\begin{eqnarray}
\dot{F}_{\tau}(\Delta)=e^{-\frac{2\pi}{a}\Delta}\dot{F}_{\tau}(-\Delta).
\end{eqnarray}
which is used as a criterion for the state of the detector being a thermal state. See Fig.~(\ref{fig:acceleratedresponse}) for a plot of the point-like limit, i.e. $\Xi\left(\Delta\right)\rightarrow 1$, of Eq.~(\ref{acceleratedresponse}) for accelerations $a=0.5,1,2$.

The transition rate is, as expected, independent of time due to the stationarity of the trajectory and the invariance of the vacuum state. Now we shall analyse the response of our accelerated detector model when the field contains a single Unruh particle. In the literature, well analysed states of the field correspond to maximally entangled Bell states, see for example, \cite{alsing2003,friis2011,datta2009}. These states contain both the vacuum and a single Unruh particle. Our starting point will again be the Wightman-function which, for the one particle state, takes the form
\begin{eqnarray}
W(\rho,\rho')&:=&\bra{1_{p}}\tilde{\phi}(\rho)\tilde{\phi}(\rho')\ket{1_{p}},
\end{eqnarray}
where $\ket{1_{p}}$ is a one Unruh particle state defined as 
\begin{eqnarray}
\ket{1_{p}}:=\int d\Omega  d^{2}\boldsymbol{k}_{\perp} \Phi(\Omega,\boldsymbol{k}_{\perp})\hat{a}^{\dag}_{\Omega,\boldsymbol{k}_{\perp},p}\ket{0},
\end{eqnarray}
for $p=R,L$. Continuing in the exact same fashion as the Minkowski one particle state we find 
\begin{equation}
\label{Eq:wightmann1paccelerated}
W(\rho,\rho')=\bra{0}\phi(\rho)\phi(\rho')\ket{0}\cdot\int d\Omega d^{2}\boldsymbol{k}_{\perp} |\Phi(\boldsymbol{k})|^{2}+2\text{Re}\left[\bar{I}_{p}(\rho)I_{p}(\rho')\right],
\end{equation}
where $I_{p}(\rho):=\int d\Omega d^{2}\boldsymbol{k}_{\perp}\tilde{f}(\Omega,\boldsymbol{k}_{\perp})\Phi(\Omega,\boldsymbol{k}_{\perp})U_{\Omega,p}(\rho)$ with
\begin{eqnarray}
   U_{\Omega,p}(\rho) = N_{\Omega/a}\left\{
     \begin{array}{rcl}
        \cosh(r_{\Omega/a})e^{-i\Omega\rho} & : & p=R\\
        \sinh(r_{\Omega/a})e^{+i\Omega\rho} & : & p=L
     \end{array}
   \right. .
\end{eqnarray}
We see that the Wightman function in Eq.~(\ref{Eq:wightmann1paccelerated}) oscillates as a function of time. It is these oscillations that contribute to the undulatory behaviour of the detectors transition rate. For the Unruh state, the corresponding accelerated expression for Eq.~(\ref{iota}) again has a time dependent oscillatory integral. It is clear, for appropriately behaved functions $\tilde{f}$ and $\Phi$, the same analysis can be applied here as for the Minkwoski particle. The Riemann-Lebesgue lemma can be used to show that in the asymptotic past and future, the response of the detector is the same as the vacuum and hence has a thermal signature. As with the Minkowski particle, the intermediate regions between past and future asymptotic times give rise to an oscillatory response function. In the limit of high acceleration, the second terms in Eq.~\eqref{Eq:wightmann1paccelerated} become negligible and the state tends to the maximally mixed state. Thus, the single particle state of the field is dominated by the vacuum fluctuations.

\subsection{Discussion}

We have introduced a detector model which naturally couples to peaked frequency distributions of Minkowski, Unruh and Rindler modes. This detector model is suitable for studies of entanglement extraction in non-inertial frames. In the $(3+1)$-dimensional case, the frequency window of the detector peaks around positive and negative momentum inducing a double peaking. In the massless $(1+1)$-dimensional case, frequency distributions naturally peak around a single frequency. We obtain analytical results for the instantaneous transition rates of detectors undergoing inertial and uniformly accelerated motion. In particular, the transition rate of the accelerated detector is the expected thermal distribution modified by a smearing function that arises from the detectors spatial profile. We also showed that the well studied single Unruh particle states produce an oscillatory response that is only thermal in the asymptotic past and future. As the accelerated detector observes a thermal-type noise for both the Minkowski vacuum and the single Unruh particle state, it would be reasonable to assume that any entangled states between global Unruh modes will degraded due to the Unruh effect.
\section{Time Varying Spatial Profile}\label{sec:time-ordering}

The techniques employed to solve the interaction of a detector and a quantum field commonly require perturbative methods. We introduce mathematical techniques to solve the time evolution of an arbitrary number of detectors interacting with a quantum field moving in space-time while using non-perturbative methods. Our techniques apply to harmonic oscillator detectors and can be generalised to treat detectors modelled by quantum fields. Since the interaction Hamiltonian we introduce is quadratic in creation and annihilation operators, we are able to draw from continuous variable techniques commonly employed in quantum optics.

\subsection{Interacting systems}

From the precious section, the interaction Hamiltonian $\hat{H}_I(t)$ between a quantum mechanical system (detector) interacting with a Bosonic quantum field $\hat{\Phi}(t,\boldsymbol{x})$ in $4$-dimensional~spacetime~is given by
\begin{eqnarray}
\hat{H}_{I}(t)=\, \hat{M}(t)\int d^3\boldsymbol{x} \sqrt{-g}f(t,\mathbf{x})\hat{\Phi}(t,\boldsymbol{x}),\label{general:interaction:hamiltonian}
\end{eqnarray}
where $(t,\boldsymbol{x})$ are a suitable choice of coordinates for the spacetime (not necessarily Minkowski), $\hat{M}(t)$ is the monopole moment of the detector and $g$ denotes the determinant of the metric tensor \cite{fabbri2005}. The function $f(t,\boldsymbol{x})$ is the spatial profile of the detector and describes the effective interaction strength between the detector and the field. Notice that the spatial profile is now dependent on time. As we will see, this has useful properties for reducing the complexity of the detector-field interaction. When written in momentum space, it describes how the internal degrees of freedom of the detector couple to a time dependent distribution of the field modes.

The field $\hat{\Phi}$ can be expanded in terms of a particular set of solutions to the field equation $\phi_{\boldsymbol{k}}(t,\boldsymbol{x})$ as
\begin{eqnarray}
\label{discretedecomp}
\hat{\Phi}=\sum_{\boldsymbol{k}}\left[\hat{D}_{\boldsymbol{k}}\phi_{\boldsymbol{k}}+\mbox{h.c.}\right],
\end{eqnarray}
where the variable $\boldsymbol{k}$ is a set of discrete parameters and $\hat{D}_{\boldsymbol{k}}$ are bosonic operators that satisfy the time independent canonical commutation relations $[\hat{D}_{\boldsymbol{k}},\hat{D}_{\boldsymbol{k'}}^{\dag}]=\delta_{\boldsymbol{k}\boldsymbol{k'}}$. We refer to the solutions $\phi_{\boldsymbol{k}}$ as field modes. We emphasise that the modes $\phi_{\boldsymbol{k}}$ need \textit{not} be standard solutions to the field equations (i.e. plane waves in the case of a scalar field in Minkowski spacetime) but can also be wave-packets formed by linear superpositions of plane waves.

We can engineer the function $f(t,\boldsymbol{x})$ such that 
\begin{eqnarray}
\int d^{3}\boldsymbol{x} \sqrt{-g}f(t,\boldsymbol{x})\hat{\Phi}(t,\boldsymbol{x})=h(t)\hat{D}_{\boldsymbol{k}_{*}}+\mbox{h.c.},
\end{eqnarray}
where one mode, labelled via $\boldsymbol{k}_{*}$, has been selected out of the set $\lbrace\phi_{\boldsymbol{k}}\rbrace$, which in turn implies
\begin{eqnarray}
\hat{H}_{I}(t)= \hat{M}(t)\left[h(t)\hat{D}_{\boldsymbol{k}_{*}}+\mbox{h.c.}\right].\label{single:mode:interaction:hamiltonian}
\end{eqnarray}
Therefore, the coupling strength has been specially designed to make the detector couple to a single mode,  in this case labelled by $\boldsymbol{k}_{*}$. In the case of a free $(1+1)$-dimensional relativistic scalar field, the mode the detector couples to corresponds to a time dependent frequency distribution of plane waves. In the following we clarify, using a specific example, what we mean by a time-dependent frequency distribution. The $(1+1)$ massless scalar field $\hat{\Phi}(t,x)$ obeys the standard Klein-Gordon equation $(-\partial_{tt}+\partial_{xx})\hat{\Phi}(t,x)=0$. It can be expanded in terms of standard Minkowski modes (plane waves) as \cite{takagi1986,crispino2008}
\begin{eqnarray}
\hat{\Phi}(t,x)=\int_{-\infty}^{+\infty} \frac{dk}{\sqrt{2\pi |k|}}\left[\hat{a}_k e^{-i(|k| t-kx)}+\hat{a}^{\dagger}_k e^{i(|k| t-kx)}\right],\label{Minkowski:field}
\end{eqnarray}
where the momentum $k\in\mathbb{R}$ and $k>0$ labels right moving modes while $k<0$ labels left moving modes and each particle has energy $\omega_{k}=|k|$. The creation and annihilation operators satisfy the canonical commutation relations $[\hat{a}_{k},\hat{a}_{k^{'}}^{\dag}]=\delta(k-k^{'})$. We substitute Eq.~(\ref{Minkowski:field}) into~(\ref{general:interaction:hamiltonian}), assuming for simplicity a flat spacetime, i.e. $\sqrt{-g}=1$, and by inverting the order of integration we obtain
\begin{eqnarray}
\hat{H}_{I}(t)=\, \hat{M}(t)\int_{-\infty}^{+\infty} \frac{dk}{\sqrt{2\pi |k|}}\left[\hat{a}_k e^{-i|k| t}\tilde{\bar{f}}(t,k)+\hat{a}^{\dagger}_k e^{i|k|t}\tilde{f}(t,k)\right],\label{equation:fourier}
\end{eqnarray}
where we have defined the spatial Fourier transform $\tilde{f}(t,k)$ of the function $f(t,x)$ as
\begin{eqnarray}
\label{eqn:tdfd}
\tilde{f}(t,k):=\int_{-\infty}^{+\infty} dx f(t,x)e^{-ikx}.
\end{eqnarray}
Note that Eq.~(\ref{eqn:tdfd}) is the time dependent generalisation of~(\ref{eqn:freq-distro}). We call the function~(\ref{eqn:tdfd}) a \emph{time dependent frequency distribution}. Thus given a general interaction strength, the momenta contained within the field that interacts with the detector will be modified in a time dependent way.

We should add that our detector model given by Eq.~(\ref{general:interaction:hamiltonian}) extends the well-known pointlike Unruh-DeWitt detector which has been extensively studied in the literature \cite{DeWitt1979,takagi1986,hu2012}. When the spatial profile approximates a delta function $f(t,x)=\delta(p(t)-x)$, the detector approximates a point-like system following a classical trajectory $p(t)$ \cite{takagi1986,schlicht2004}. 

In our analysis we have considered the detector to be a harmonic oscillator. By doing this we will be able to draw from continuous variables techniques in quantum optics that will simplify our computations. However, the original Unruh-DeWitt detector consists of a two-level system. The excitation rate of a harmonic oscillator has been shown to approximate well that of a two-level system at short times \cite{lin2010,hu2012}. However, for long interaction times the difference becomes significant and the models cannot be compared directly.

In the following, we explain how to solve the time evolution of an arbitrary number of detectors interacting with an arbitrary number of fields when the interaction Hamiltonian is of a purely quadratic form given by Eq.~(\ref{single:mode:interaction:hamiltonian}).

\subsection{Time evolution of N interacting bosonic systems}

We start this section by reviewing Lie algebra theory and techniques from symplectic geometry. By combining these techniques we will derive equations that govern the evolution of a quantum system. The generalisation of the quadratic Hamiltonian given by Eq.~(\ref{single:mode:interaction:hamiltonian}) to $N$ interacting bosons is 
\begin{eqnarray} \label{eq:hamil}
\hat{H}(t)=\sum_{j=1}^{N(2N+1)}\lambda_j(t)\hat{G}_j,\label{interaction:hamiltonian:expansion}
\end{eqnarray}
where the functions $\lambda_{j}$ are real and the operators $\hat{G}_{j}$ are Hermitian and quadratic combinations of the harmonic creation and annihilation operators $\lbrace(\hat{D}_{j},D^{\dagger}_{j})\rbrace$. For example, $\hat{G}_{1}=\hat{D}_{1}^{\dag}\hat{D}_{2}^{\dag}+\hat{D}_{1}\hat{D}_{2}$. The summation is over the total number of independent, purely quadratic, operators which for $N$ modes is $N(2N+1)$. The operators $\hat{G}_i$ form a closed Lie algebra with Lie bracket
\begin{equation}
\label{eqn:structureconstant}
[\hat{G}_i,\hat{G}_j\bigr]=c_{ijk}\hat{G}_k.
\end{equation}
As we saw in Section~(\ref{sec:symplectic-representations}), the algebra generated by the $N(2N+1)$ operators $\hat{G}_{j}$ is the algebra generated by \textit{all} possible linearly independent quadratic combinations of creation and annihilation bosonic operators. The set of operators $\left\{\hat{G}_{j}\right\}$ can be divided into four subsets, where $N$ operators generate phase rotations, $2N$ single mode squeezing operations, $N^{2}-N$ independent beam splitting operations and $N^{2}-N$ two mode squeezing operations. Phase rotations and beam splitting together form the well known set of passive transformations \cite{wolf2003}. There are  $(N^{2}-N)+N=N^2$ generators of passive transformations which, excluding the total number operator $\sum \hat{D}^{\dag}_{i}\hat{D}_{i}$ that commutes with all passive generators, form the well known sub algebra $SU(N)$ of the total algebra of our model, where $\mathrm{dim}(SU(N))=N^2-1$ \cite{puri1994}.

The complex numbers $c_{ijk}$ are the structure constants of the algebra generated by the operators $\hat{G}_{j}$. In general they form a tensor that is antisymmetric in its first two indices only. Moreover, the values taken by the $c_{ijk}$ explicitly depend on the choice of representation for the $\hat{G}_{j}$”.

We wish to find the time evolution of our interacting system. In the general case, the Hamiltonian $\hat{H}(t)$ does not commute with itself at different times $[\hat{H}(t),\hat{H}(t')\bigr]\neq0$. Therefore, the time evolution is induced by the unitary operator
\begin{eqnarray}
\hat{U}(t)=\overleftarrow{T}e^{-i\int_{0}^{t} dt' \hat{H}(t')}\label{evolution:operator}
\end{eqnarray}
where $\overleftarrow{T}$ stands for the time ordering operator \cite{greiner2007}. We can employ techniques from Lie algebra and symplectic geometry \cite{wilcox1967,berndt2001,hall2004} to explicitly find a solution to Eq.~(\ref{evolution:operator}). The unitary evolution of the Hamiltonian can be written as \cite{puri2001},
\begin{eqnarray}
\label{eeo}
\hat{U}(t)=\prod_{j}\hat{U}_j(t)=\prod_{j}e^{-iF_j(t)\hat{G}_j},
\end{eqnarray}
where the functions $F_j(t)$ associated with generators $\hat{G}_{j}$ are real and depend on time.  By equating Eq.~(\ref{evolution:operator}) with Eq.~(\ref{eeo}), differentiating with respect to time and multiplying on the right by $\hat{U}^{-1}(t)$ we find a sum of similarity transformations
\begin{eqnarray}
\hat{H}(t)&=&\dot{F}_{1}(t)\hat{G}_{1}+\dot{F}_{2}(t)\hat{U}_{1}\hat{G}_{2}\hat{U}^{-1}_{1}+\dot{F}_{3}(t)\hat{U}_{1}\hat{U}_{2}\hat{G}_{3}\hat{U}^{-1}_{2}\hat{U}^{-1}_{1}+\ldots \label{hamiltonianequation}
\end{eqnarray}
In this way, we obtain a set of $N(2N+1)$ coupled, non-linear, first order ordinary differential equations of the form
\begin{eqnarray}
\label{odes}
\sum_{j}\alpha_{ij}(t)\dot{F}_{j}(t)+\sum_{j}\beta_{ij}F_{j}(t)+\gamma_{i}(t)=0\label{general:ode},
\end{eqnarray}
where the coefficients $\alpha_{ik}(t)$ and $\beta_{ik}(t)$, which are \emph{not} Bogoliubov coefficients, will in general be functions of the $F_{j}(t)$ and $\lambda_j(t)$. The form of the Hamiltonian and the initial conditions $F_{j}(0)=0$ completely determine the unitary time evolution operator~(\ref{evolution:operator}).

The equations can be re-written using the continuous variable ideas of Chapter~(\ref{chapter:cv}). To this end, we define the vector of operators
\begin{eqnarray}
\hat{\boldsymbol{\xi}}:=\left(\hat{D}_{1},\ldots ,\hat{D}_{N},\hat{D}_{1}^{\dagger}, \ldots ,\hat{D}_{N}^{\dagger}\right),
\end{eqnarray}
which is of course the complex form basis of Chapter~(\ref{chapter:cv}), Section~(\ref{sec:symplectic-representations}). In this formalism, successive applications of the Baker-Campbell-Hausdorff formula which are required in the similarity transformations of Eq.~(\ref{hamiltonianequation}) will be replaced by simple matrix multiplications reducing the problem from a tedious Hilbert space computation to simple linear algebra. We write
\begin{eqnarray}
\hat{U}_{j}(t)\,\hat{G}_{k}\,\hat{U}^{-1}_{j}(t)&=&\hat{\boldsymbol{\xi}}^{\dag}\cdot \boldsymbol{S}_{j}(t)^{\dag}\cdot \boldsymbol{G}_{k} \cdot \boldsymbol{S}_{j}(t) \cdot \hat{\boldsymbol{\xi}}\label{useful:equation}, 
\end{eqnarray}
where we have used the identity $\hat{U}_{j}(t)\,\hat{\boldsymbol{\xi}}\,\hat{U}^{-1}_{j}(t)\equiv\boldsymbol{S}_{j}(t)\cdot\hat{\boldsymbol{\xi}}$ and $\boldsymbol{G}_{j}$ is the matrix representation of $\hat{G}_{j}$, defined via $\hat{G}_{j}:=\hat{\boldsymbol{\xi}}^{\dag}\cdot\mathbf{G}_{j}\cdot\hat{\boldsymbol{\xi}}$. The dynamical transformation of the vector of operators $\hat{\boldsymbol{\xi}}$ generated by the interaction Hamiltonian $\hat{G}_{j}$ is given by the symplectic matrix~\cite{luis1995}
\begin{eqnarray}
\boldsymbol{S}_{j}:=e^{-iF_{j}(t)\boldsymbol{K}\boldsymbol{G}_{j}}
\end{eqnarray}
where $F_{j}(t)$ are real functions associated with the generator $\hat{G}_{j}$ and $K_{mn}:=[\hat{\xi}_m,\hat{\xi}_{n}^{\dag}]$ is the symplectic form. A symplectic matrix $\boldsymbol{S}$ satisfies $\boldsymbol{S}\,\boldsymbol{K}\,\boldsymbol{S}^{\dag}=\boldsymbol{K}$. In this formalism, we can use Eq.~(\ref{useful:equation}) and the identity $\hat{H}=\hat{\boldsymbol{\xi}}^{\dag}\cdot\boldsymbol{H}\cdot \hat{\boldsymbol{\xi}}$ to obtain the matrix representation of the Hamiltonian $\hat{H}$,
\begin{eqnarray}
\label{matrixequations}
\boldsymbol{H}(t)&=&\dot{F}_{1}(t)\,\boldsymbol{G}_{1}+\dot{F}_{2}(t)\,\boldsymbol{S}_{1}(t)^{\dag}\cdot\boldsymbol{G}_{2}\cdot\boldsymbol{S}_{1}(t)\nonumber\\
& &\hspace{4mm}+\dot{F}_{3}(t)\,\boldsymbol{S}_{1}(t)^{\dag}\cdot\boldsymbol{S}_{2}(t)^{\dag}\cdot\boldsymbol{G}_{3}\cdot\boldsymbol{S}_{2}(t)\cdot\boldsymbol{S}_{1}(t)+\ldots\label{final:equation}
\end{eqnarray}
It is necessary to explicitly compute the matrix products of the form $\boldsymbol{S}_{k}^{\dag}(t)\cdot\boldsymbol{G}_{j}\cdot\boldsymbol{S}_{k}(t)$ in order to re-write Eq.~(\ref{matrixequations}) in terms of the generators ${\bf G}_i$.
By equating the coefficients of Eq.~\ref{matrixequations}) to the coefficients $\lambda_j(\tau)$ in the matrix representation of Eq.~\ref{eq:hamil}) we obtain a set of coupled $N(2N+1)$ ordinary differential equations. Solving for the functions $F_j(t)$, we obtain the explicit expression for the time evolution of the system as described by Eq.~(\ref{eeo}). The final expression is
\begin{eqnarray}
\label{cm:evolution:operator}
\boldsymbol{S}(t)=\prod_{j}\boldsymbol{S}_j(t)=\prod_{j}e^{-iF_{j}(t)\boldsymbol{K}\boldsymbol{G}_{j}},
\end{eqnarray}
which corresponds to the time evolution of the whole system. Systems of great interest are those of Gaussian states which are common in quantum optics and relativistic quantum theory~\cite{alsing2012}. In this case the state of the system is encoded by the first moments $\langle \hat{\xi}_{j}\rangle$ and a covariance matrix $\boldsymbol{\Gamma}(t)$ defined by 
\begin{eqnarray}
\Gamma_{ij}=\langle \hat{\xi}_{i}\hat{\xi}_{j}^{\dag}+\hat{\xi}_{j}^{\dag}\hat{\xi}_{i}\rangle -2\langle \hat{\xi}_{i}\rangle\langle \hat{\xi}_{j}^{\dag}\rangle.
\end{eqnarray} 
From Section~(\ref{chapter:cv}), the time evolution of the initial state $\boldsymbol{\Gamma}(0)$ is given by the symplectic transformation
\begin{eqnarray}
\label{eqn:symplectictimeevolution}
\boldsymbol{\Gamma}(t)=\boldsymbol{S}(t)\boldsymbol{\Gamma}(0)\boldsymbol{S}^{\dag}(t).
\end{eqnarray}

\subsection{Application: Time evolution of a detector coupled to a field}

We now apply our formalism to describe a situation of great interest to the field of relativistic quantum information: a single detector following a general trajectory and interacting with a quantum field via a general time and space dependent coupling strength.
We therefore return to the $(1+1)$ massless scalar field. The standard plane-wave solutions  to the field equation in Minkowski coordinates  are
\begin{equation}
\phi_{k}^{M}(t,x)=\frac{1}{2\pi\sqrt{|k|}}e^{-i|k| t+kx}\label{minkowski:modes}
\end{equation} 
which are (Dirac delta) normalized as $ \bigl(\phi_{k}^{M},\phi_{k'}^{M}\bigr)=\delta(k-k')$ through the standard conserved inner product $\bigl(\cdot,\cdot\bigr)$ \cite{crispino2008}. The mode operators associated with these modes, $\hat{a}_{k}$, define the Minkowski vacuum via $\hat{a}_{k}\left| 0\right>_{M}=0$ for all $k$.

The field expansion in Eq.~(\ref{Minkowski:field}) contains both right and left moving Minkowski plane waves. In general, given an arbitrary trajectory of the detector and an arbitrary interaction strength, the detector couples to both right and left moving waves. However, for the sake of simplicity, in Section~\ref{concrete:example:section} we will consider an example where the detector follows an inertial trajectory.  In the 1+1 dimensional case right and left moving waves decouple, therefore it is reasonable to assume that the detector couples only to right moving waves. This situation could correspond to a photodetector which points in one particular direction \cite{downes2013}.

The degrees of freedom of the detector which we assume to be a harmonic oscillator are described by the bosonic operators $\hat{d},\hat{d}^{\dagger}$ that satisfy the usual time independent commutation relations $\bigl[\hat{d},\hat{d}^{\dagger}\bigr]=1$. The ground state $\bigl|0\bigr>_d$ of the detector is defined by $\hat{d}\bigl|0\bigr>_d=0$. Therefore, the vacuum $\bigl|0\bigr>$ of the \textit{non-interacting} theory takes the form  $\bigl|0\bigr>:=\bigl|0\bigr>_d\otimes\bigl|0\bigr>_M$. 

In the interaction picture, we use Eq.~(\ref{general:interaction:hamiltonian}) and assume the detector couples to the field via the interaction Hamiltonian 
\begin{eqnarray}
\hat{H}_{I}(t)= \hat{M}(t)\int dx\, \sqrt{-g}\, f(t,x)\int_{-\infty}^{+\infty} dk\left[\hat{a}_{k}\phi_{k}^{M}(t,x)+\hat{a}^{\dagger}_{k}\bar{\phi}_{k}^{M}(t,x)\right],\label{example:interaction:hamiltonian}
\end{eqnarray}
Using the Hamiltonian~(\ref{example:interaction:hamiltonian}), we parametrise the interaction via a suitable set of coordinates, $(\tau,\xi)$, that describe a frame comoving with the detector. A standard choice, as in the previous section, is to use the so-called Fermi-Walker coordinates~\cite{takagi1986,schlicht2004}. This amounts to expressing $(t,x)$ within the integrals of~(\ref{example:interaction:hamiltonian}) as the functions $(t(\tau,\xi),x(\tau,\xi))$. In the comoving frame, the monopole moment of the detector takes the form
\begin{eqnarray}
\hat{M}(\tau)=e^{-i\tau\Delta}\hat{d}+e^{i\tau\Delta}\hat{d}^{\dagger}.
\end{eqnarray}
In momentum space the detector couples to a time-dependent frequency distribution of Minkowski plane-wave field modes. Here we consider a coupling strength that can be designed to couple the detector to a time-varying wave-packet $\psi(\tau,\xi)$. It is therefore more convenient to decompose the field not in the plane-wave basis but in a special decomposition 
\begin{equation}
\hat{\Phi}(\tau,\xi)=\hat{D}_{k_{*}}\psi(\tau,\xi)+\hat{D}_{k_{*}}^{\dagger}\bar{\psi}(\tau,\xi)+\hat{\Phi}'(\tau,\xi),\label{discrete:field:decomposition}
\end{equation}
where $\psi$ is the mode the detector couples to, which corresponds to a time dependent frequency distribution of plane waves, and $k_{*}$ represents a particular mode we wish to distinguish in the field expansion. Note that for the case of a field contained within a cavity, where the set of modes is discrete, our methods apply without an explicit need to form discrete wave packets. The operators $\hat{D}_{k_{*}},\hat{D}_{k_{*}}^{\dagger}$ are time independent and satisfy the canonical time independent commutation relations $\bigl[\hat{D}_{k_{*}},\hat{D}_{k_{*}}^{\dagger}\bigr]=1$. The field $\hat{\Phi}'$ includes all the modes orthogonal to $\psi$ and we will assume them to be countable. Once expressed in the comoving coordinates, the Hamiltonian~(\ref{example:interaction:hamiltonian}) takes on the single mode form~(\ref{single:mode:interaction:hamiltonian}) when the following conditions are satisfied
\begin{eqnarray}
\label{eqn:singlemodeconditions}
h(\tau) = \int d\xi\,f(\tau,\xi)\psi(\tau,\xi),\quad
\int d\xi\,f(\tau,\xi)\hat{\Phi}'(\tau,\xi) =0\,\,\forall \tau\label{mode:choice:condition}.
\end{eqnarray}
The decomposition~(\ref{discrete:field:decomposition}) can always be formed from a complete orthonormal basis (an example of which can be found in \cite{takagi1986}). In general, the operator $\hat{D}_{k_{*}}$ does not annihilate the Minkowski vacuum $\bigl|0\bigr>_M$. This observation is a consequence of fundamental ideas that lie at the foundation of quantum field theory, where different and inequivalent definitions of particles can coexist. Such concepts are, for example, at the very core of the Unruh effect~\cite{unruh1976} and the Hawking effect~\cite{hawking1975}. 

The operator $\hat{D}_{k_{*}}$ will annihilate the vacuum $\bigl|0\bigr>_D$. Note that, in general, the vacuum state $\bigl|0\bigr>_I$ of this interacting system is different from the vacuum state $\bigl|0\bigr>$ of the noninteracting theory, i.e. $\bigl|0\bigr>\neq\bigl|0\bigr>_I$. 

Under the conditions~(\ref{eqn:singlemodeconditions}), the interaction Hamiltonian takes a very simple form
\begin{eqnarray}
\hat{H}_{I}(\tau)=\hat{M}(\tau)\cdot \left[h(\tau)\hat{D}_{k_{*}}+\bar{h}(\tau)\hat{D}_{k_{*}}^{\dagger}\right],\label{interaction:hamiltonian:example}
\end{eqnarray}
which describes the effective interaction between the internal degrees of freedom of a detector following a general trajectory and coupling to a \textit{single mode} of the field described by $\hat{D}_{k_{*}}$. The time evolution of the system can be solved in this case by employing the techniques we introduced in the previous section.  However, this formalism is directly applicable to describe the interaction of N detectors with the field. In that case, our techniques yield differential equations which can be solved numerically. We choose here to demonstrate our techniques with the single detector case since it is possible to compute a simple expression for the expectation value of the number of particles in the detector. 

Let the detector-field system be in the ground state $\bigl|0\bigr>_D$ at $\tau=0$. We design a coupling such that we obtain an interaction of the form (\ref{interaction:hamiltonian:example}). In this case, the covariance matrix only changes for the detector and our preferred mode. This subsystem, described by $\hat{d},\hat{D}_{k_{*}}$, is always separable from the rest of the non-interacting modes. The covariance matrix of the vacuum state $\bigl|0\bigr>_D$ is represented by the $4\times 4$ identity matrix, i.e. $\boldsymbol{\Gamma}(0)=\boldsymbol{I}$. From Eq.~(\ref{eqn:symplectictimeevolution}), the final state $\boldsymbol{\Gamma}(\tau)$ therefore takes the simple form of $\boldsymbol{\Gamma}(\tau)=\boldsymbol{S}\boldsymbol{S}^{\dag}$. The final state provides the information we need to compute the time dependent expectation value of the detector $N_d(\tau) := \bigl<\hat{d}^{\dagger}\hat{d}\bigr>(\tau)$.

From the definition of the covariance matrix $\boldsymbol{\Gamma}(\tau)$, one finds that $N_d(\tau)$ is related to $\boldsymbol{\Gamma}(\tau)$ by
\begin{eqnarray}
\label{elements:of:the:state:heisenberg:picture}
\Gamma_{11}(\tau) &=& 2\left<\hat{d}^{\dagger}\hat{d}\right>(\tau)-2\left<\hat{d}^{\dagger}\right>(\tau)\left<\hat{d}\right>(\tau)+1.
\end{eqnarray}
We also choose to work with states that have first moments zero, i.e $\langle\xi_{j}\rangle=0$. In this case, since our interaction is quadratic, the first moments will always remain zero \cite{weedbrook2012}. Therefore we are left with equation
\begin{eqnarray}\label{elements:of:the:state:heisenberg:picture:final}
\Gamma_{11}(\tau) &=& 2\left<\hat{d}^{\dagger}\hat{d}\right>(\tau)+1.
\end{eqnarray}
Our expressions hold for detectors moving along an arbitrary trajectory and coupled to an arbitrary wave-packet. Given a scenario of interest, one can solve the differential equations, obtain the functions $F_{j}(\tau)$ and, by using the decomposition in Eq.~(\ref{cm:evolution:operator}), one can obtain the time evolution of the system. We can find the expression for the average number of excitations in the detector at time $\tau$, which reads
\begin{eqnarray}
\label{eqn:numberexpectation}
N_d(\tau)&=&\frac{1}{2}\left[\mathrm{ch}_{1}\mathrm{ch}_{2}\mathrm{ch}_{3}\mathrm{ch}_{4}-1\right].
\end{eqnarray}
where we have adopted the notation $\mathrm{ch}_{j}\equiv\cosh(2F_{j}(\tau))$. For our choice of initial state we find that the functions $F_{j}(\tau)$ are associated with the generators $\hat{G}_{1}=\hat{d}^{\dag}\hat{D}_{k_{*}}^{\dag}+\hat{d}\hat{D}_{k_{*}},\hat{G}_{2}=-i(\hat{d}\hat{D}_{k_{*}}-\hat{d}^{\dag}\hat{D}_{k_{*}}^{\dag}),\hat{G}_{3}=\hat{d}^{\dag 2}+\hat{d}^{2}$ and $\hat{G}_{4}=-i(\hat{d}^{\dag 2}-\hat{d}^{2})$,
respectively. The appearance of these functions can be simply related to the physical interpretation of the operators $\hat{G}_{j}$. In fact, the generators $\hat{G}_{1}$ and $\hat{G}_{2}$ are nothing more than the two-mode squeezing operators. Such operations generate entanglement and are known to break particle number conservation. The  two generators $\hat{G}_{3}$ and $\hat{G}_{4}$ are related to the single-mode squeezing operators for the mode $d$. The generators $\hat{G}_{1}\ldots \hat{G}_{4}$, together with the generators $\hat{G}_{5}=\hat{D}_{k_{*}}^{\dag 2}+\hat{D}_{k_{*}}^{2}$ and $\hat{G}_{6}=-i(\hat{D}_{k_{*}}^{\dag 2}-\hat{D}_{k_{*}}^{2})$ which represent single mode squeezing for the mode $\hat{D}_{k_{*}}$,  form the set of active transformations of a two mode Gaussian state and do not conserve total particle number \cite{arvind1995}. The remaining operators, whose corresponding functions are absent in Eq.~(\ref{eqn:numberexpectation}), form the passive transformations for Gaussian states. These transformations are also known as the generalised beam splitter transformation \cite{luis1995}; they conserve the total particle number of a state and hence do not contribute to equation~(\ref{eqn:numberexpectation}).

\subsection{Example: Inertial Detector\label{concrete:example:section}}

\begin{figure}[b!]
\includegraphics[scale=0.5]{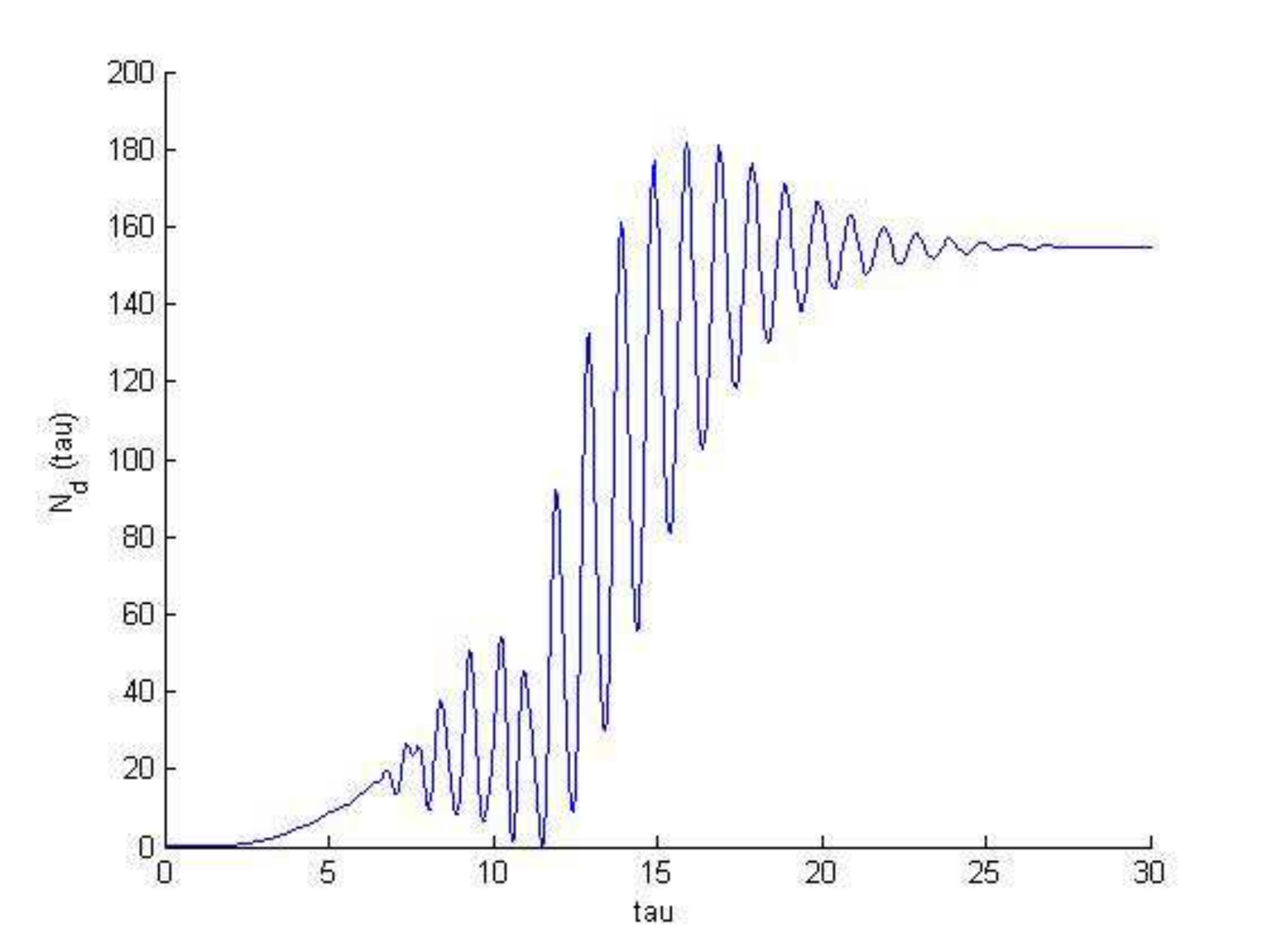}
\centering
\caption[Unruh-DeWitt Detector Number Expectation Value]{\label{figure} Mean number of particles, $N_{d}(\tau)$, as a function of time $\tau$. Here we used (without loss of generality) $\lambda=1$, $T^2=80$ and $\Delta=2\pi$.}
\end{figure}

To further specify our example we consider the detector stationary and interacting with a localised time dependent frequency distribution of plane waves. The free scalar field is decomposed into wave packets of the form \cite{takagi1986}
\begin{eqnarray}
\tilde{\phi}_{ml}:=\int dk f_{ml}(k)\phi_{k},
\end{eqnarray}
where the distributions $f_{ml}(k)$ are defined as
\begin{equation}
f_{ml}(k):=\left\lbrace\begin{array}{cl} \epsilon^{-1/2}e^{-2i\pi lk/\epsilon} & \epsilon m<k<\epsilon(m+1)\\ 0 & \mathrm{otherwise}\end{array}\right.,
\end{equation}
with $\epsilon>0$ and $\lbrace m,l\rbrace$ running over all integers. Note that if $m\ge 0$ the frequency distribution is composed exclusively of right moving modes defined by $k>0$.  The mode operators associated with these modes are defined as
\begin{eqnarray}
\hat{D}_{ml}:=\int dk\, \bar{f}_{ml}(k)\,\hat{a}_{k}.
\end{eqnarray}
Notice that for our particular choice of wave-packets, the general operators $\hat{D}_{j}$ obtain two indices. The distributions $f_{ml}(k)$ satisfy the completeness and orthogonality relations
\begin{equation}
\label{packetrelations}
\sum_{m,l}f_{ml}(k)\bar{f}_{ml}(k')=\delta(k-k'),\,\,\,\,\,
\int dk f_{ml}(k)\bar{f}_{m'l'}(k)=\delta_{mm'}\delta_{ll'}.
\end{equation}
The wave packets are normalised as $(\tilde{\phi}_{ml},\tilde{\phi}_{m'l'})=\delta_{mm'}\delta_{ll'}$ and the operators satisfy the commutation relations $[\hat{D}_{ml},\hat{D}_{m'l'}^{\dag}]=\delta_{mm'}\delta_{ll'}$.
The scalar field can then be expanded in terms of these wave packets as
\begin{equation}
\hat{\Phi}=\sum_{m,l}\left[\tilde{\phi}_{ml}\hat{D}_{ml} +h.c.\right].
\end{equation}
Following \cite{schlicht2004}, we consider an inertial detector and we can parametrise our interaction via $t=\tau$ and $x=\xi$. We now construct the spatial profile of the detector to be
\begin{equation}
\label{stationaryprofile}
f(\tau,\xi):=h(\tau)\int dk \bar{f}_{ML}(k)\phi_{k}(\tau,\xi)
\end{equation}
where $h(\tau)$ is now an arbitrary time dependent function which dictates when to switch on and off the detector. Physically, this corresponds to a detector interaction strength that is changing in time to match our preferred mode, labelled by $M,L$ i.e. $\hat{D}_{k_{*}}=\hat{D}_{ML}$. We point out that any other wave packet decomposition could be chosen as long as it satisfies completeness and orthogonality relations of the form (\ref{packetrelations}). The form of the spatial profile to pick out these modes is therefore general. Inserting the profile (\ref{stationaryprofile}) into our interaction Hamiltonian (\ref{general:interaction:hamiltonian}), we obtain
\begin{eqnarray}
\hat{H}_{I}(\tau)=\left(\hat{d} e^{-i\tau\Delta}+\hat{d}^{\dag}e^{i\tau\Delta}\right)\left(h(\tau)\hat{D}_{ML}+\bar{h}(\tau)\hat{D}_{ML}^{\dag}\right)
\end{eqnarray}
We choose the switching on function to be $h(\tau)=\lambda\tau^{2}e^{-\tau^2/T^2}$, where $T$ modulates the interaction time and $\lambda$ quantifies the interaction strength. The interaction Hamiltonian is then
\begin{equation}
\hat{H}_{I}(\tau)=\lambda\tau^{2}e^{-\frac{\tau^2}{T^2}}\left(\hat{d} e^{-i\tau\Delta}+\hat{d}^{\dag}e^{i\tau\Delta}\right)\left(\hat{D}_{LM}+\hat{D}_{LM}^{\dag}\right)
\end{equation}
which written in generator form (see Eq.(\ref{eq:hamil})) is
\begin{eqnarray}
\label{eqn:hamops}
\hat{H}_{I}(\tau)&=\lambda\tau^{2}e^{-\frac{\tau^2}{T^2}}\left[\cos\left(\tau\Delta\right)\hat{G}_{1}+\sin\left(\tau\Delta\right)\hat{G}_{2}\right.\nonumber\\
&\hspace{7mm}+\left.\cos\left(\tau\Delta\right)\hat{G}_{7}+\sin\left(\tau\Delta\right)\hat{G}_{8}\right]
\end{eqnarray}
where $\hat{G}_{7}=\hat{d}^{\dag}\hat{D}_{ML}+\hat{d}\hat{D}_{ML}^{\dag}$, $\hat{G}_{8}=-i(\hat{d}\hat{D}_{ML}^{\dag}-\hat{d}^{\dag}\hat{D}_{ML})$ and the other operators defined similarly. The matrix representation of $\hat{H}_{I}$ is
\begin{eqnarray}
\label{eqn:hammatrix}
\boldsymbol{H}_{I}(\tau)=\lambda\frac{\tau^{2}e^{-\frac{\tau^2}{T^2}}}{2}\left[\begin{array}{cccc} 
0 & e^{i\tau\Delta} & 0 & e^{i\tau\Delta} \\
e^{-i\tau\Delta} & 0 & e^{i\tau\Delta} & 0 \\
0 & e^{-i\tau\Delta} & 0 & e^{-i\tau\Delta} \\
e^{-i\tau\Delta} & 0 & e^{i\tau\Delta} & 0
\end{array}\right]
\end{eqnarray}
Equating (\ref{eqn:hammatrix}) and (\ref{matrixequations}), or equivalently (\ref{eqn:hamops}) and (\ref{interaction:hamiltonian:expansion}), gives us the ordinary differential equations we need to find the functions $F_{j}$ for this specific example. Here we solve the equations for $F_{j} (\tau)$ numerically and we plot the average number of detector excitations $N_{d}(\tau)$ as a function of time in Fig.~(\ref{figure}).

We find that the number expectation value of the detector grows and oscillates as a function of time while the detector and field are coupled. This can be expected since the time dependence of the Hamiltonian comes in through complex exponentials that will induce phase rotations in the state and hence oscillations in the number operator. Finally, the number expectation value reaches a constant value after the interaction is turned off. Once the interaction is switched off, the free Hamiltonian does not account for emissions of particles from the detector.

\subsection{Discussion}

It is of great interest to solve the time evolution of interacting Bosonic quantum systems since they are relevant to quantum optics, quantum field theory and relativistic quantum information, among many other research fields. In most cases, it is necessary to employ perturbative techniques which assume a weak coupling between the Bosonic systems. In relativistic quantum information, perturbative calculations used to study tasks such as teleportation and extraction of vacuum entanglement \cite{lin2012} become very complicated already for two or three detectors interacting with a quantum field. In cases where the computations become involved, physically motivated or \emph{ad hoc} approximations can aid, however, in most cases, powerful numerical methods must be invoked and employed to study the time evolution of quantities of interest.

We have provided mathematical methods to derive the differential equations that govern the time evolution of N interacting Bosonic modes coupled by a purely quadratic interaction. The techniques we introduce allow for the study of such systems beyond perturbative regimes. The number of coupled differential equations to solve is $N(2N+1)$,  therefore making the problem only polynomially hard.  Symmetries, separable subsets of interacting systems, among other situations can further reduce the number of differential equations. 

The Hamiltonians in our method are applicable to a large class of interactions. In this section, as a simple example, we have applied our mathematical tools to analyse the time evolution of a single harmonic oscillator detector interacting with a quantum field.  However, our techniques are readily applied to $N$ detectors following any trajectory while interacting with a finite number of wave-packets through an arbitrary interaction strength $f(t,x)$. When confined to a cavity, the detector can couple to a single mode of the field in a time independent way as, in principle, no discrete mode decomposition needs to be enforced. Therefore, the single mode interaction Hamiltonian~(\ref{single:mode:interaction:hamiltonian}) can arise in a straightforward fashion.

We have further specified our example to analyse the case of an inertial detector interacting with a time-dependent wave-packet. We showed how to engineer a coupling strength such that the interaction Hamiltonian can be descried by an effective single field mode. However, the field mode is not a plane wave but a time dependent frequency distribution of plane waves. In this case we have solved the differential equations numerically and showed the number of detector excitations oscillates in time while the detector is on. 

Finally, we comment on the physical realisation of the detectors presented in this work. A space and time dependent coupling strength can be engineered by placing the quantum system in an external potential which is also time and space dependent. These tuneable interactions have been produced in ion traps~\cite{thompson1990,miller2005}, cavity QED~\cite{walther1006} and superconducting circuits~\cite{peropadre2010,gambetta2011,srinivasan2011,sabin2012}. In an ion trap, the interaction of the ion with its vibrational modes can be modulated by a time and spatial dependent classical driving field, such as a laser~\cite{haffner2008}. Moreover, in cavity QED, time and space dependent coupling strengths are used to engineer an effective coupling between two cavity modes~\cite{imamoglu1997,guzman2006}.

Work in progress includes using these detectors to extract field entanglement and perform quantum information tasks. 
\section{Massive Cavity Entanglement Generation}\label{sec:mceg}

\subsection{Introduction}

Early results in the field of relativistic quantum information showed that entanglement of global modes is degraded from the perspective of observers moving in uniform acceleration~\cite{funentes-schuller2005,martinmartinez2009}. However, such states are not useful to perform quantum information tasks because Alice and Rob must be able to store information in local systems which they can manipulate. In this final section, we show how to use the Unruh-DeWitt detector to entangle moving cavities in the case where the cavities have two spatial dimensions. Finding ways to generate entanglement in relativistic settings is necessary for relativistic quantum processing. 

As we mentioned before, moving cavities promise to be good candidates for storing quantum information in relativistic quantum information. It has been shown that entanglement can be generated between two $(1+1)$-dimensional cavities, one inertial and the second in uniform acceleration, by letting an Unruh-DeWitt detector interact with the modes of the cavities~\cite{downes2011}. In this section, we consider the entanglement generated between the modes of moving $(2+1)$-dimensional Bosonic cavities. The transverse dimensions of the box play the role of an effective mass in the field equation. We are interested in this scenario for two reasons. One is that it is a more physically realistic set-up. The second is that the presence of mass (or effective mass) has important effects on the entanglement between cavity modes. For example, the degradation of entanglement between inertial and accelerated field modes is increased by several orders of magnitude when the fields are massive~\cite{alphacentauri2012}. It was also shown that the probability of the excitation of an atom moving through a cavity is lower for massive fields. This effect can be used to distinguish between inertial and non-inertial frames~\cite{dragan2011}. We find that the entanglement generated by an atom interacting with the field of an inertial and an accelerated cavity is robust against acceleration once it has been created but our ability to entangle the cavities is lower when when the fields are massive, given some fixed cavity size.

\subsection{Physical Set-up}

In this section, we consider a set-up to entangle two cavities introduced by Browne~\cite{browne2003}. This consists of an atom used to entangle the Bosonic modes of two different cavities. We modify the setting by considering that the pair of cavities are in relative motion and that they have two spatial dimensions. We assume one of the cavities undergoes uniform acceleration in one direction only. This means its other spatial dimension will remain inertial and hence unaffected by its motion. An atom is then passed through the two cavities and entangle them by emitting an excitation into one of the cavities.

The stationary cavity is at rest and is described by the standard Minkowski spacetime coordinates $(t,x,y)$. We shall assume that this cavity, as described by an observer at at the origin of the coordinates, has boundaries at $x_{\pm}$ and $y_{\pm}$ in the $x$- and $y$-directions respectively. We denote the length of a cavity wall by $x^{i}_{+}-x^{i}_{-}:=L_{i}$ where $i=1,2$ and denotes the spatial components of the coordinate $3$-vector. 

Since consider that the second cavity undergoes uniform acceleration in the $x$-direction, the most convenient choice of coordinates are the Rindler coordinates $(\eta,\chi,y)$ defined via
\begin{subequations}
\begin{align}
\label{MinkowskiTrans}
t &= \chi\sinh\left(\eta\right), \\
x &= \chi\cosh\left(\eta\right),
\end{align}
\end{subequations}
and the $y$ coordinate is the same as the standard Minkowski $y$ coordinate. See Chapter~(\ref{chapter:qft}), Section~(\ref{app:rindler-coords}) for details on Rindler coordinates. Analogously to the inertial cavity, we define the accelerating cavity walls by $\chi_{\pm}$ and $\tilde{y}_{\pm}$. Again, we denote the length of a cavity wall $\chi^{i}_{+}-\chi^{i}_{-}:=\tilde{L}_{i}$ in each spatial direction. The two cavity mirrors $\chi_{\pm}$ follow uniformly accelerated trajectories. They therefore accelerate with a proper acceleration of $1/\chi_{\pm}$. In our scenario, we set all cavity lengths to be equal, i.e. $L_{i}=\tilde{L}_{i}=L$. Further, we choose $x_{\pm}=\chi_{\pm}$, $y_{-}=-3L/2$, $y_{+}=\tilde{y}_{-}=-L/2$ and $\tilde{y}_{+}=+L/2$. These coordinates mean that at the instant $t=\eta=0$ the two cavities are aligned. Their $x$-coordinates overlap while their $y-$coordinates are positioned such that the cavities are side-by-side.

Finally, we consider the trajectory of the atom. We choose the atom to be always located at the centre of the cavities $x-$coordinates while passing through $y=0$ at $t=0$ with constant speed $v$ in the $y$-direction. This can be written in $3$-vector notation as $x^{\mu}_{a}(t)=(t,X_{a},vt)$ where $X_{a}=(x_{+}-x_{-})/2$. For the dynamics of the fields, it will be more useful to parametrise the cavities and trajectories in terms of the atoms proper time. The relation between the atoms proper time, which we denote as $\tau$, and its coordinate time $t$ is given by $t=\gamma\tau$ where $\gamma=1/\sqrt{1-v^2}$. The parametrisation of the atoms trajectory is then $x^{\mu}_{a}(\tau)=(\gamma\tau,X_{a},v\gamma\tau)$. 

\begin{figure}[t]
        \centering
        \begin{subfigure}[b]{0.45\textwidth}
                \centering
                \includegraphics[width=\textwidth]{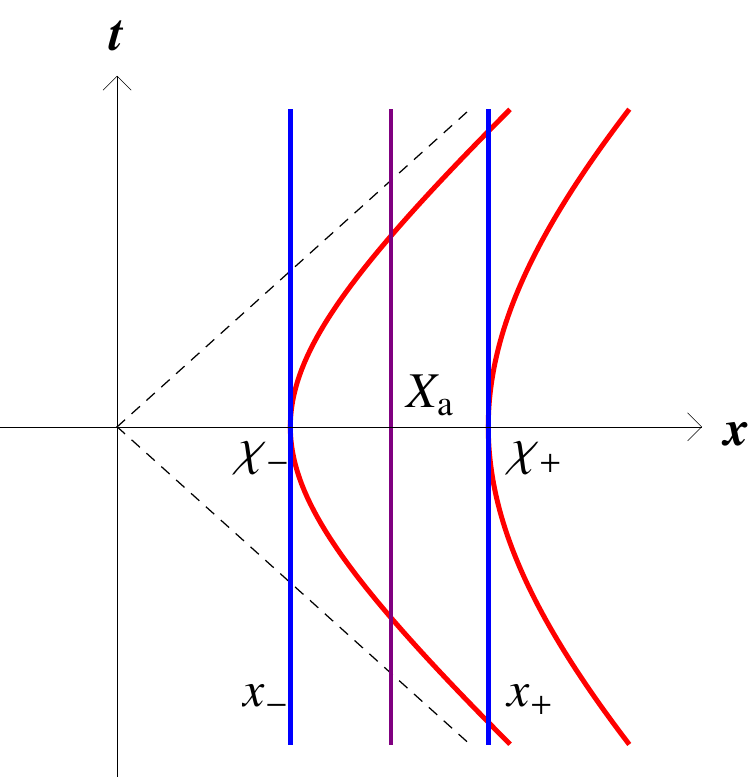}
                \caption{$t-x$ schematic}
                \label{fig:gull}
        \end{subfigure}%
\hspace{5mm}
        \begin{subfigure}[b]{0.45\textwidth}
                \centering
                \includegraphics[width=\textwidth]{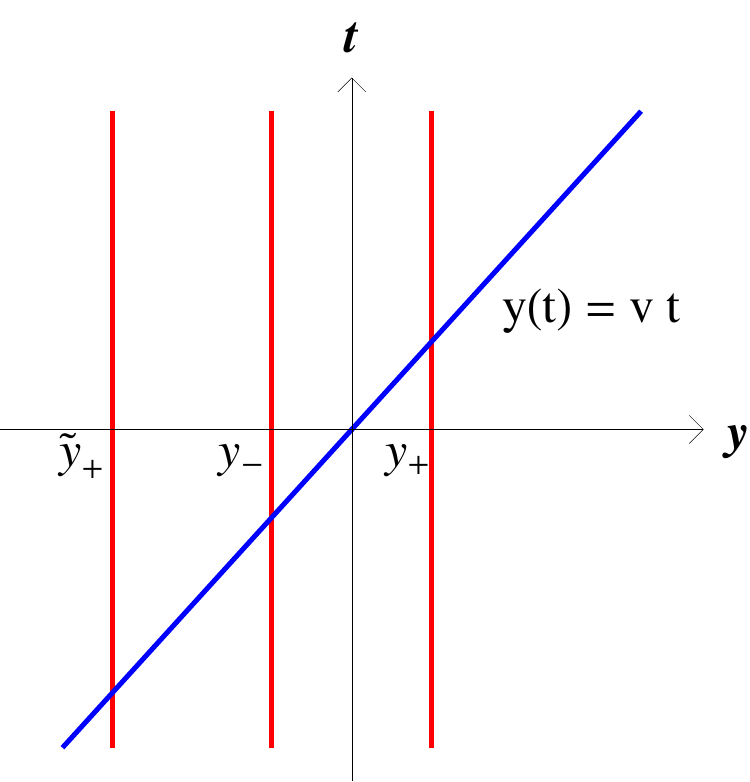}
                \caption{$t-y$ schematic}
                \label{fig:tiger}
        \end{subfigure}
        \caption[Entanglement Generation Cavity Set-up]{(\ref{fig:rindler-diagram}~a) is a schematic diagram showing the spacetime motion of Alice's and Rob's cavities with the detector. We see that Alice's cavity and the $x-$coordinated of the atom are at fixed spatial positions. Rob's cavity walls however follow hyperbolic trajectories. At the instant $t=0$ both Alice's and Rob's cavity width is equal. (\ref{fig:rindler-diagram}~b) Shows the $y$-dimension, both Alice's and Rob's cavity remain fixed while the atom follows an inertial trajectory with velocity $v$.}\label{fig:rindler-diagram}
\end{figure}

Fig.~(\ref{fig:rindler-diagram}~a) is the spacetime diagram for the $x-$coordinates which shows the atoms constant position at the centre of the cavities. Fig.~(\ref{fig:rindler-diagram}~b) shows the cavities $y-$dimension side-by-side and the atom passing through the inertial (left most) cavity first and then the accelerating (right most) cavity.
Having examined the kinematics of the cavities, let us now describe the quantum fields that are carried by them.

\subsection{Field dynamics}

We consider the field to vanish at the edges of the cavities. The fields are initially in the vacuum state according to the observers co-moving with each cavity (Alice for the inertial cavity and Rob for the accelerated one). The Bogoliubov transformation of the field between Minkowski and Rindler reference frames is highly non-trivial \cite{bruschi2010}. It is well known that Alice's cavity, according to Rob, has some excitations and vice-versa. Nonetheless the initial state of the two cavities is separable regardless of the coordinates used to describe it.

We describe the atom as a two-level system with ground state $\ket{g}$ and excited state $\ket{e}$. The cavity modes can become entangled by their interaction with the atom that passes through the cavities.  To achieve this the atom is initially prepared in its exited state. The atom passes through the cavities having a non-zero probability of emitting an excitation in one of them. The state of the atom is subsequently measured. If the atom is found to be in the ground state the cavity modes will be entangled. It is impossible to discriminate which cavity field has been excited by the atom without further measurements; therefore, the final state of the two cavities must be a superposition of both possibilities.

We use the standard massive Klein-Gordon field in two-dimensions for the dynamics of the fields contained within the cavities. We can expand the field operator in Alice's cavity as
\begin{equation}
\label{MinkowskiSolutions2}
\hat{\phi}_{A}(t,x,y)=\sum_{n,m}N_{nm}u_{n}(x)u_{m}(y)e^{-i\omega_{nm}t}\hat{a}_{nm}+\mathrm{h.c.},
\end{equation}
where the relevant Minkowski mode solutions are
\begin{subequations}
\begin{align}
u_{k}(x^{i})&=\sin\left[\frac{k\pi}{L_{i}}\left(x^{i}-x^{i}_{-}\right)\right],\\
N_{nm}&=\frac{\sqrt{2}}{\sqrt{\omega_{nm}L_{x}L_{y}}},\\
\omega_{nm}^{2}&=\left(\frac{n\pi}{L_{x}}\right)^{2}+\left(\frac{m\pi}{L_{y}}\right)^{2}+\kappa^{2}.
\end{align}
\end{subequations}
Similarly, we can expand the field contained within Rob's cavity as
\begin{equation}
\label{RindlerSolutions}
\hat{\phi}_{R}(\eta,\chi,y)=\sum_{n,m}\tilde{N}_{nm}\tilde{u}_{nm}(\chi)\tilde{u}_{m}(y)e^{-i\tilde{\Omega}_{nm}\eta}\hat{b}_{nm}+\mathrm{h.c.}
\end{equation}
Note that due to the trivial coordinate transformation in the $y$-direction then $\tilde{u}_{m}(y)=u_{m}(y)$. The $\chi$ spatial function on the other hand is highly non-trivial and obeys the modified Bessel equation. 

The boundary conditions for the accelerated cavity walls are defined by $\chi_{\pm}$. This gives for the mode solutions in the accelerated direction 
\begin{subequations}
\begin{align}
\tilde{u}_{nm}(\chi)&=I_{-i\tilde{\Omega}_{nm}}(\kappa_{m}\chi_{-})I_{i\tilde{\Omega}_{nm}}(\kappa_{m}\chi)-I_{i\tilde{\Omega}_{nm}}(\kappa_{m}\chi_{-})I_{-i\tilde{\Omega}_{nm}}(\kappa_{m}\chi)\\
\kappa_{m}^{2}&=\left(\frac{m\pi}{\tilde{L}_{y}}\right)^{2}+\kappa^{2}.
\end{align}
\end{subequations}
The $I_{\pm\Omega}(z)$ are the modified Bessel functions of the first kind~\cite{NIST:DLMF,Olver:2010:NHMF}. The quantities $\tilde{N}_{nm}$ and $\tilde{\Omega}_{nm}$ are functions of the quantum numbers $(n,m)$, the acceleration of Rob's box and the bare mass of the field. They are only analytically closed functions for the massless $(1+1)$-dimension case of \cite{downes2011}. We shall evaluate them numerically for a specified acceleration and bare mass interval.

\subsection{Atom-field interaction}

The atom is modelled by a two-level system with a characteristic frequency of $\Delta$ with raising and lowering operators $\sigma_{+}$ and $\sigma_{-}$ respectively, i.e. the Unruh-DeWitt detector. The monopole moment of the atom is given by~(\ref{eqn:internal-degrees-of-freedom})
\begin{eqnarray}
\hat{M}(\tau)=\sigma_{-}e^{-i\tau\Delta}+\sigma_{+}e^{i\tau\Delta}.
\end{eqnarray}
The interaction Hamiltonians, written in the interaction picture \cite{greiner2007}, is given by  
\begin{eqnarray}
\label{UnruhDeWittHamiltonian}
\hat{H}_{A/R}(\tau)&=&\epsilon_{A/R}(\tau)\hat{M}(\tau)\cdot\hat{\phi}_{A/R}(\tau),
\end{eqnarray}
where $\hat{\phi}_{A}$ and $\hat{\phi}_{R}$ are field operators given by Eq.(\ref{MinkowskiSolutions2}) and Eq.~(\ref{RindlerSolutions}) which are evaluated along the world line of the atom and $\tau$ is the proper time along the atoms world line. The coupling functions $\epsilon_A$ and $\epsilon_R$ represent the strength of interaction and in general can be time-dependent. We choose them to be sine functions of the detectors proper time $\tau$
\begin{eqnarray}
\epsilon(\tau)&=&\epsilon\sin^{2}\left(2\pi v\gamma\tau\right).
\end{eqnarray}
For simplicity we have considered the two coupling functions to be equal and the interaction between the atom and the cavities is smooth. In particular, when the atom is at the edge of a cavity the interaction strength is zero.

The evolution for the entire system can be written as $\ket{\psi}=\hat{U}_{R}\hat{U}_{A}\ket{\psi(0)}$ where $\ket{\psi(0)}$ is the initial state of the system. $\hat{U}_{A}$ and $\hat{U}_{R}$ are the unitary operators that evolve Alice's and Rob's subsystems respectively. This corresponds to the state evolving under the interaction of Alice's Hamiltonian and then Rob's. Assuming the cavities are initially in the vacuum state and the atom is excited, i.e. $\ket{\psi(0)}=\ket{0}_{A}\ket{0}_{R}\ket{e}$, the evolution of the state to first order in perturbation theory becomes
\begin{equation}
\ket{\psi}=\left(\hat{I}-i\int d\tau \left(\hat{H}_{A}^{I}(\tau)+\hat{H}_{R}^{I}(\tau)\right)\right)\ket{0}_{A}\ket{0}_{R}\ket{e},
\end{equation}
where the integration is over the interaction time of the atom and the cavities. To specify these interaction times we must concretely define the positions of the cavities. We consider all cavity wall lengths to be equal to unity in their rest frames i.e. $L_{i}=\tilde{L}_{i}=1$. We also define the proper acceleration at the centre of Rob's cavity to be $h$ such that $\chi_{\pm}=1/h\pm 1/2$. The position of the cavities gives the interaction time interval for Alice and Rob's cavities as $\tau\in\left[-3/2v\gamma,-1/2v\gamma\right]$ and $\tau\in\left[-1/2v\gamma,+1/2v\gamma\right]$ respectively, see Fig.~(\ref{fig:rindler-diagram}~b).

In our scheme we will be interested in post-selected events where the atom has been detected in the ground state after passing through the cavity. The remaining state of the field $|\Phi\rangle =\langle g|\psi\rangle $ takes the form
\begin{eqnarray}
\ket{\Phi} = -i\int d\tau\bra{g}\sigma_{-}e^{-i\tau\Delta}\ket{e}\epsilon(\tau)\cdot\lbrace\hat{\phi}_{A}(\tau)+\hat{\phi}_{R}(\tau)\rbrace\ket{0}_{A}\ket{0}_{R}
\end{eqnarray}
We observe the only non-zero contributions from this inner product come from the $\sigma_{-}\hat{a}^{\dag}$ and $\sigma_{-}\hat{b}^{\dag}$ terms of the Hamiltonians \eqref{UnruhDeWittHamiltonian}.

\subsection{Cavity Entanglement}

Let us determine the degree of entanglement shared by the cavities after the interaction has taken place. The state of the fields of the cavities is pure, because we only take into account the post-selected events, when the atom is found in the ground state after the interaction. Therefore we can use the entropy of entanglement~(\ref{eqn:entropy-of-entanglement}) as a valid measure of quantum correlations. 

We can write the state of the system after the atom has passed through the cavities as
\begin{eqnarray}
\label{eqn:alive-rob-cavity-state}
\ket{\Phi}&=&\displaystyle\sum_{n,m}\left[F_{nm}^{A}\hat{a}^{\dag}_{nm}+F_{nm}^{R}\hat{b}^{\dag}_{nm}\right]\ket{0}_{A}\ket{0}_{R},
\end{eqnarray}
where we have defined
\begin{subequations}
\begin{align}
F_{nm}^{A}&=N_{nm}\sin(n\pi/2)\!\!\int\limits_{-3T}^{-T}\! d\tau\Lambda(\tau) e^{+i\omega_{nm}\gamma\tau},\\
F_{nm}^{R}&=\tilde{N}_{nm}\int\limits_{-T}^{T}\! d\tau\Lambda(\tau)\tilde{u}_{nm}(\chi(\tau))e^{+i\Omega_{nm} \text{atanh}(h\gamma\tau)}.
\end{align}
\end{subequations}
Here, $\chi(\tau)=\sqrt{1/h^{2}-\gamma^{2}\tau^{2}}$, $T:=1/2v\gamma$ and we have denoted
\begin{eqnarray*}
\Lambda(\tau)&=&-i\epsilon(\tau)\sin\left[m\pi\left(v\gamma\tau-1/2\right)\right]e^{-i\tau\Delta}.
\end{eqnarray*}
It should be noted for a given atom velocity $v$, the proper acceleration at the centre of Rob's cavity has a maximum value. From $\chi(\tau)$, we can write $h\le 1/\gamma\tau$ which is minimised when $\tau=1/2v\gamma$. This gives as an upper limit of acceleration $h\le 2v$. This implies, remembering we have set all lengths to unity, the maximum acceleration at the centre of Rob's cavity is $h=2$ when $v=1$.

We first find the reduced density matrix $\hat{\rho}_{R}=\text{tr}_{A}(\ket{\Phi}\bra{\Phi})$ of Robs's cavity. Since $_A\!\bra{0}\hat{a}^{\dag}_{nm}\ket{0}_{A}=\,_A\!\bra{0}\hat{a}_{nm}\ket{0}_{A}=0$ and $_A\!\bra{0}\hat{a}_{nm}\hat{a}^{\dag}_{ij}\ket{0}_{A}=\delta_{ni}\delta_{mj}$, and denoting $\hat{b}^{\dag}_{nm}\ket{0}_{R}=\ket{1_{nm}}_{R}$, we have
\begin{figure}[t]
        \centering
        \begin{subfigure}[b]{0.45\textwidth}
                \centering
                \includegraphics[width=\textwidth]{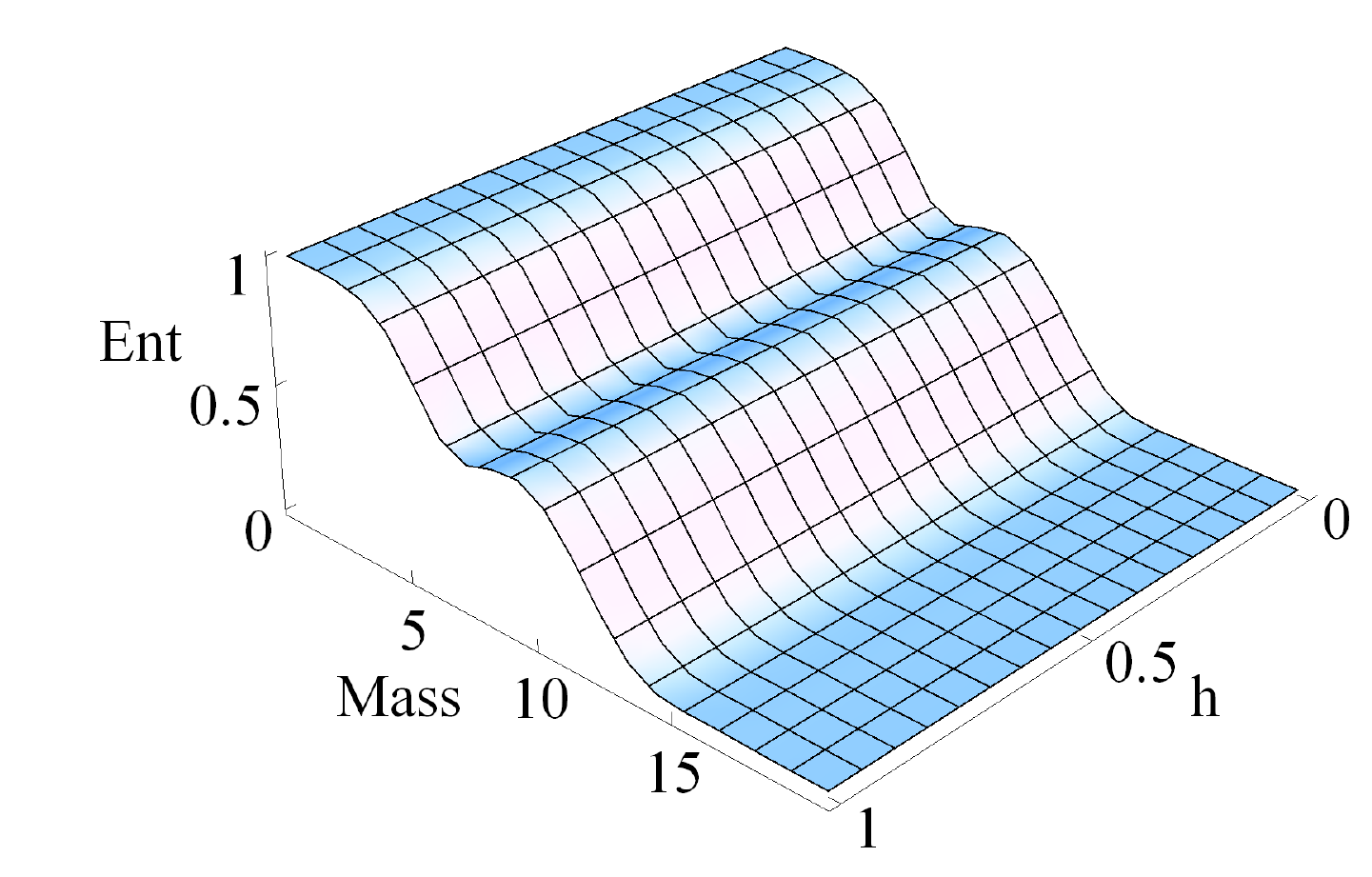}
                \caption{Entanglement plot}
                \label{fig:gull2}
        \end{subfigure}%
\hspace{5mm}
        \begin{subfigure}[b]{0.45\textwidth}
                \centering
                \includegraphics[width=\textwidth]{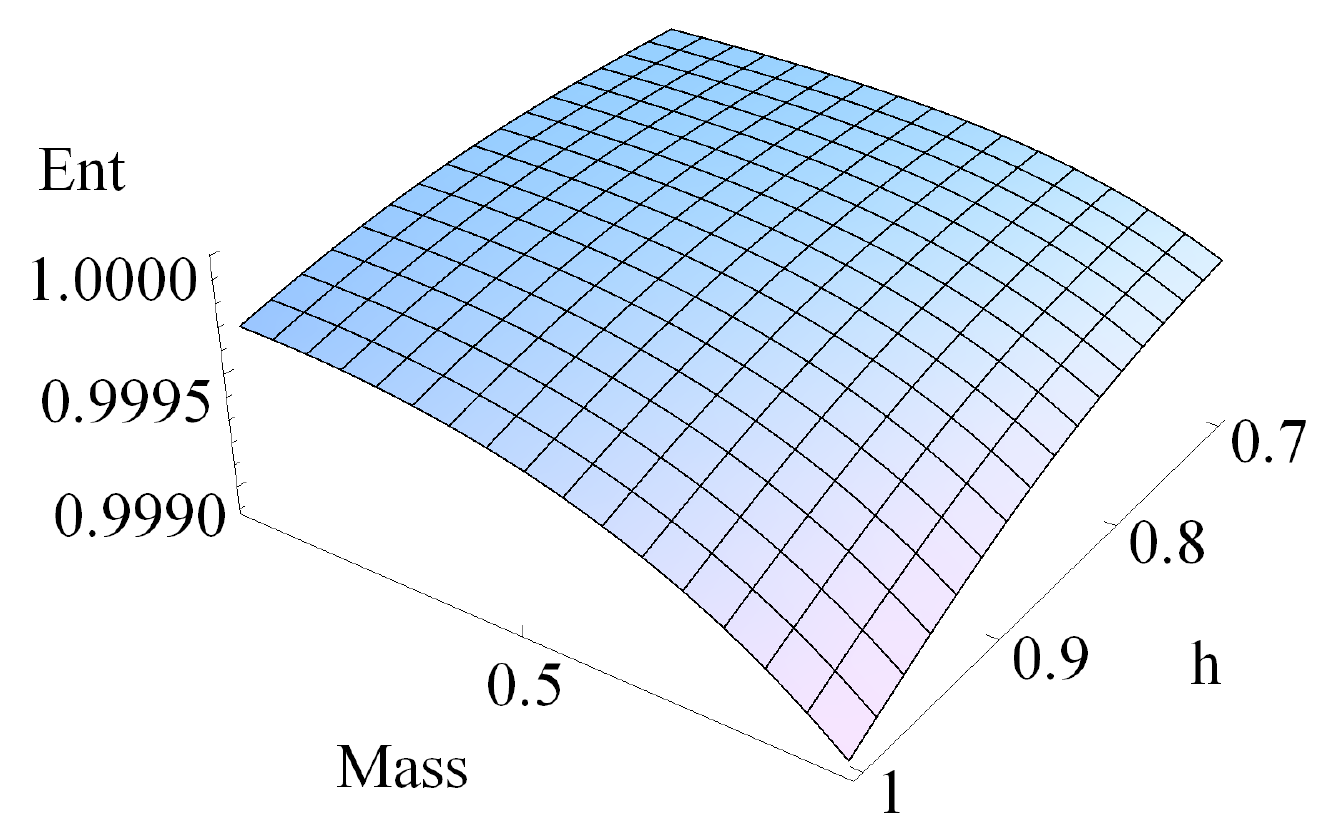}
                \caption{Enlargement}
                \label{fig:tiger2}
        \end{subfigure}
        \caption[Entanglement Generation Entropy of Entanglement]{(\ref{big2dplot}~a): Entanglement generated between the two cavitites as a function of acceleration at the centre of Rob's cavity, $h$ ,and the bare mass of his field $\kappa$. Here we set all cavity lengths to unity, $\Delta=\sqrt{2}\pi$ and $v=1/2$. (\ref{big2dplot}~b): An enlargement for $h$ from 0.7 to 1 and $\kappa$ from 0 to 1.}\label{big2dplot}
\end{figure}
\begin{eqnarray*}
\hat{\rho}_{R}&=&\sum_{n,m}|F_{nm}^{A}|^{2} \ket{0}\bra{0}+\sum_{nmij}F_{nm}^{R}\bar{F}_{ij}^{R}\ket{1_{nm}}\bra{1_{ij}}
\end{eqnarray*}
We numerically analyse the entanglement shared between Alice and Rob. To evaluate the entanglement, a truncation in the infinite summations contained within Eq.~(\ref{eqn:alive-rob-cavity-state}) is needed. Thus, we cut-off the summation after an appropriate number of terms $N$, which is found numerically also. Arranging the mode integrals into a vector as 
\begin{eqnarray}
\boldsymbol{F}=\left(F_{11}^{R},\ldots, F_{1N}^{R},\ldots, F_{N1}^{R},\ldots, F_{NN}^{R}\right),
\end{eqnarray}
will help us write the final state in a more compact way. From this vector we can construct the matrix representation of Rob's state $\boldsymbol{\rho}_{R}:=\left(\sum_{nm}|F_{nm}^{A}|^{2}\right)\oplus\boldsymbol{F}\cdot\boldsymbol{F}^{\dagger}$ where the dot denotes the outer product of the two vectors. Our newly constructed matrix then takes the form
\begin{equation}
\label{densitymatrix}
 \boldsymbol{\rho}_{R} =
 \begin{pmatrix}
  \displaystyle\sum_{n,m}|F_{nm}^{A}|^{2} & 0 & \cdots & 0 \\
  0 & |F_{11}^{R}|^{2} & \cdots & F_{11}^{R}\bar{F}_{NN}^{R} \\
  \vdots  & \vdots  & \ddots & \vdots  \\
  0 & F_{NN}^{R}\bar{F}_{11}^{R} & \cdots & |F_{NN}^{R}|^{2}
 \end{pmatrix}.
\end{equation}
Here we have defined the first element of the matrix to be the coefficient of $\ket{0}_{RR\!}\bra{0}$. After the renormalisation $\hat{\rho}_{R}\rightarrow\hat{\rho}_{R}/\text{tr}(\hat{\rho}_{R})$ where $\text{tr}(\hat{\rho}_{R})=\sum_{n,m}\left[|F_{nm}^{A}|^{2}+|F_{nm}^{R}|^{2}\right]$ the state can be used to evaluate its Von Neumann entropy $\mathrm{S}(\hat{\rho})=-\text{tr}\left(\hat{\rho}\log\hat{\rho}\right)$ and hence find the entropy of entanglement for the system. We find the eigenvalues of the truncated matrix~(\ref{densitymatrix}) and compute the Von Neumann entropy via $\mathrm{S}(\rho_{R})=-\sum_{k}\lambda_{k}\log(\lambda_{k})$ where $\lambda_{k}$ are the eigenvalues of $\boldsymbol{\rho}_{R}$.

The entanglement generated between the cavities decreases monotonically as a function of the proper acceleration at the centre of Rob's cavity and oscillates as a function of the bare mass of Rob's field, see Fig.~(\ref{big2dplot}~a). From Fig.~(\ref{big2dplot}~b), one can see that the generated entanglement is robust against the acceleration of the cavity, i.e. acceleration does not greatly effect the generated entanglement. To understand the oscillations in the plot we can consider the case of zero acceleration i.e. $h=0$. The integrals $F_{nm}^{R}$ can be explicitly calculated in this case. The resulting expression takes the form
\begin{eqnarray}
F_{nm}^{R}=f_{nm}(\kappa)\left(1-(-1)^{n}e^{i g_{nm}(\kappa)}\right),
\end{eqnarray}
where
\begin{eqnarray}
g_{nm}(\kappa)=\frac{1}{v}\left(\Delta\sqrt{1-v^{2}}-\sqrt{\pi^{2}n^{2}+\pi^{2}m^{2}+\kappa^{2}} \right),
\end{eqnarray}
and $f_{nm}(\kappa)$ is a polynomial in $\kappa$ of order $\mathcal{O}(\kappa^{-6})$. Thus the overall behaviour of the mode integrals is to decrease as bare mass increases. There are, however, points of constructive resonances where $1-(-1)^{n}e^{i g_{nm}(\kappa)}=2$. It is these resonances that contribute to the local maxima observed in the entanglement between the cavities. Physically, this corresponds to a resonance between the internal energy gap of the detector and the energy of the quanta contained within Rob's cavity.

\subsection{Discussion}

We have examined the entanglement generated between the modes of two cavities in relative motion when interacting with a two-level system. It was found that the acceleration of one of the cavities degrades the ability to prepare entangled states between the two cavities. Moreover, we found that the bare mass of the field contained within the accelerated cavity also degrades the ability to generate entangled states. We showed that the kinematical set-up of the cavities dictates the maximum acceleration of Rob's cavity i.e. infinite accelerations cannot be achieved arbitrarily. To access higher accelerations, and hence generating smaller amounts of entanglement, the velocity of the two-level system must be increased. In the limit the two-level system's velocity approaches the speed of light, the maximum acceleration at the centre of the cavity is $h=2$. This implies that one side of Rob's cavity approaches the Rindler horizon and therefore its proper acceleration diverges. It should also be pointed out that by varying the lengths of the inertial cavity we can compensate for the acceleration and field mass of Rob's cavity. This compensation allows us to maximise the entanglement generated between the cavities for a given acceleration and bare field mass.

We also note the behaviour of entanglement as a function of the accelerated cavity's bare mass. A damped undulatory behaviour is observed. This can be traced back to a $\kappa$ dependent phase that can constructively interfere with the state of the field. It was also shown that once the generation of entanglement is robust against the effects of acceleration. It is this feature that could be exploited to generate entangled states between an inertial and an accelerated for use in quantum information protocols.

Unruh-DeWitt detectors have been shown to be useful in generating entanglement between two parties. Future directions of work could be to extend the work of scalar fields to Dirac fields and to use the detectors themselves for entanglement extraction of a relativistic quantum field.

\section{Conclusions}

In this chapter, we investigated the use of Unruh-DeWitt detectors in relativistic quantum information. Given their localised nature, they are ideally suited to model systems that Alice and Rob can access. We first analysed how considering detectors with a spatially dependent profile influences their interaction with a quantum field. It was shown that the response of detector deviates from the purely thermal response of the literature and therefore changes, in principle, the experimental observation of the Unruh effect. In particular, a Gaussian shape profile naturally allowed both inertial and accelerated detectors to couple to peaked distributions of field modes. The deviation from the standard Unruh effect is due to the finite size of the detector inducing an effective, non-uniform, coupling to the field. We then went on to introduce spatial profiles that depend on both position and time. Physically, the detector's profile was changing in time to ``match" the mode it was probing. Using these profiles, we engineered simplified detector-field interactions. Using ideas from symplectic geometry, we derived equations of motion that govern the evolution of a state that allowed for non-perturbative calculations. These non-perturbative calculations allow for exact calculations of quantities such as the average number of particles in the field and even the entanglement between the detector and the field. Going beyond perturbation calculations is important as it allows one to analyse contributions from higher order effects.

Finally, we used the Unruh-DeWitt detector to generate entanglement between moving cavities. It was show that the total amount of entanglement generated between an inertial and an accelerated cavity is degraded as the proper acceleration at the centre of the moving cavity increases. Physically, this is due to the particle creation in the accelerated cavity. As the detector will observe a thermal-type bath of particles in the moving cavity, the probability of it undergoing spontaneous emission or excitation will be increased. Therefore, the probability of it emitting a particle in the stationary cavity will be reduced and hence the interaction would generate a weakly entangled state. Further, we showed how extra spatial dimensions, which act as an effective mass of the field, degrade the entanglement generated between the two cavities. This can be traced back to the fact that a massive mode needs more energy to become excited. Thus, for a low velocity Unruh-DeWitt detector, the interaction is not energetic enough to excite the massive modes. This has important consequences as realistic experimental proposals, which would in based in $(3+1)$-dimensions, would need to take into both effects from acceleration and extra dimensions.

\chapter*{Part III}
\addcontentsline{toc}{chapter}{Part III}

\chapter{Conclusion and Discussion}

In this final chapter we will summarise the results presented in this thesis. Current work in the field of relativistic quantum information will be commented on and finally a discussion of possible new directions of research will be mentioned.

\section{Results}

\subsection{Moving Cavities for Relativistic Quantum Information}

Our main objective was to describe a localised quantum field suitable for quantum information tasks (issue (1) in the thesis introduction). We implemented this localisation by imposing boundary conditions on the quantum field being considered. Modelling the motion of a cavity via Bogoliubov transformations, we were able to describe how entanglement was generated and affected by acceleration and gravitational fields. We developed a versatile framework in which to pose and answer questions in relativistic quantum information. 

\subsubsection{Bosons}

In Chapter~(\ref{chapter:bosons}), we confined the quantum Klein-Gordon field to a cavity described mathematically as ideal boundary conditions. We modelled non-uniform trajectories of the cavity as different periods of inertial and uniformly accelerated motion. By using Bogoliubov transformations, we were able to describe very general cavity trajectories. Further, we were able to draw from the field of Gaussian state quantum information to investigate how entanglement can be generated within a moving cavity.

We showed that for a single cavity whose initial state is separable, the general motion of a cavity through spacetime will generate entanglement between modes of the field contained within the cavity. Moreover, we found that by repeating a particular travel scenario, a remarkable linear growth of entanglement was observed. Physically, this was due to a resonance between the cavity modes and the total time of the travel scenario. It was further discussed how these entangling trajectories can be viewed as quantum gates. This tantalising realisation opens up the possibility of implementing quantum computing via cavity motion.

We also investigated the protocol of Gaussian state quantum teleportation. Starting with two entangled cavities, we showed that, the when one of the cavities undergoes non-uniform motion, acceleration reduces the ability to perform the teleportation of a continuous variable coherent state efficiently. Physically, this is due to particle creation by the dynamics of spacetime. Particle creation underpins the Unruh effect and introduces noise to a system, therefore, reducing the entanglement between the cavities modes. However, we were able to show how to compensate for this negative effect by carefully choosing the trajectory of the cavity and performing local operations on the teleported state. Remarkably, this allowed the fidelity of the teleportation to obtain its optimal value. Finally, we described a possible experimental implementation of our results using superconducting circuits and SQUIDS. Using state of the art technology, we proposed a realistic set-up which could be one of the first verifications of the theoretical work in relativistic quantum information. 

\subsubsection{Fermions}

In Chapter~(\ref{chapter:moving-cavities}), we confined the quantum Dirac field to a cavity described mathematically as ideal boundary conditions. However, in the case of massless Fermions, the boundary conditions we imposed resulted in a so-called zero mode. This, essentially, is a mode of the field which has a non-zero wavefunction but physically has no energy. We therefore had to introduce a method of regularising this zero mode to obtain physical results. We went on to analyse different types of state that the Dirac field admits such as standard Bell-type states between different modes and different particle charges. We found that an entanglement degradation effect occurs for modes which undergo non-uniform motion. It was shown that the qualitative behaviour of Fermions is similar to previous results of Bosons. However, they differed in two significant ways. One was that the spin-statistics of the Fermions prevented highly populated modes. This presented a technical simplification of our analysis as compared to Bosons. This is because the partial tracing of Fermions only involves zero and single particle states. When dealing with Bosons, once must trace over states which, in principle, contain an infinite amount of particles. On the other hand, a technical complication arose in possible ambiguities associated with a choice of basis. This required us to adopt a particular procedure for consistently tracing over the Fermionic modes as to avoid any such problems. Finally, we analysed the degradation of entanglement when Rob went on a way-one trip to another location. It was shown that the degradation of entanglement could be compensated for by appropriately constructing the trajectory of Rob's cavity. Also, we found that for appropriate boundary conditions, some Fermionic states were actually more robust against the motion than others. Finally, we pointed out a possible realisation of an accessible massless Fermion particle using recent experimental progress using Graphene. 

\subsubsection{Final Comments}

Even though perturbative, the framework we have developed provides new tools to investigate entanglement in quantum field theory. We were able to derive a very general formula for the entanglement generated by non-uniform motion. The generality of the result is clear, one need only compute the Bogoliubov coefficients for a particular scenario, which could involve arbitrary cavity trajectories or non-trivial spacetimes, to compute how entanglement changes. We hope these tools will be used for future investigations for motion through spacetime. 

\subsection{Unruh-DeWitt Detectors}

Our main objective was to investigate how Unruh-DeWitt detectors could be used for quantum information tasks (issue (2) in the thesis introduction). By exploiting their properties, we introduced new techniques, based on symplectic geometry, which can be used to evolve quantum systems in a non-perturbative way. We also used Unruh-DeWitt detectors to investigate a possible experimental scenario in relativistic quantum information. 

\subsubsection{Modified Spatial Profiles}

In Chaper~(\ref{chapter:fat-detectors}), Sections~(\ref{sec:free-detector},\,\,\ref{sec:time-ordering}), we discussed how to model a finite size Unruh-DeWitt detector via a spatial profile. By considering arbitrarily shaped spatial profiles, we showed that the standard Unruh-DeWitt detector transition rate is modified in a potentially significant way. Moreover, the particular realisation of an Unruh-DeWitt detector, which can introduce different spatial profiles, will influence any experimental observations. Hence, the physical implementation of an Unruh-DeWitt detector will have strong ramifications in an experiment. Further, we also showed how particles present in the field also affect the response of an Unruh-DeWitt detector in a non-trivial way. Physically, this indicates that the noise due the presence of a particle could play an important role in the verification of the Unruh effect. 

We went on to allow the size, or spatial profile, to change in a time dependent way. Using these time dependent spatial profiles, we constructed simplified, single mode, detector-field interaction Hamiltonians. Physically, these time and space dependent profiles can be understood as the detector ``changing" in time to match the mode it is trying to probe. By reducing the complexity of the interaction, we were able to go beyond the usual methods of perturbation theory in quantum field theory. This is useful as investigating quantities in quantum information, such as the entropy of entanglement which depends on the Von Neumann entropy of a state, do not always admit a perturbative expansion (see Chapter~(\ref{chapter:bosons})). By utilising concepts from symplectic geometry, we derived the equations of motion which governs the evolution of a quantum state. In a non-perturbative way, we concretely analysed how the state of an Unruh-DeWitt detector (modelled by a harmonic oscillator) changed when coupled to a single mode of a quantum field. We showed that the average number of excitations in the internal degrees of freedom of the detector became populated in an oscillatory way. Therefore, the non-perturbative tools we have presented offer a useful way to explore quantum information concepts in quantum field theory. 

\subsubsection{Entanglement Generation}

In our final chapter, Section~(\ref{sec:mceg}), we looked at how the Unruh-DeWitt detector and cavities in quantum field theory can be combined. We showed that robust entanglement can be generated between two spatially distinct cavities. It was also shown that extra spatial cavity dimensions, which act as an effective mass of a field, degrade the ability to entangle the two cavities. While the robustness against acceleration is an encouraging prospect for experiments in relativistic quantum information, any realistic experiment will likely involve fields with more than one spatial dimension and, therefore, possible effects which arise due to an effective mass could pose problems in verifying theoretical predictions. Although, including mass terms in our theory could provide useful ways to investigate gravitational effects on entanglement. As a final comment, understanding interference effects will allow us to develop more accurate experimental proposals. 

\subsubsection{Final Comments}

Modified Unruh-DeWitt detectors offer us an encouraging tool to investigate entanglement between moving objects. In particular, the reduction of common quantum field theory type interactions to those found in quantum optics allows one to use a vast array of tools developed in the past decade, such as Gaussian state information theory. By bringing together tools and techniques from these two different fields, Unruh-DeWitt detectors could provide an ideal model to produce new ideas and directions of research in relativistic quantum information. 

\section{Current Work}

We now discuss ideas that are currently being investigated. It is hoped that these ideas will provide further insight into relativistic quantum information and help build a solid foundation on which to continue investigating new concepts and scenarios.

\begin{enumerate}

  \item The work presented in Chapters~(\ref{chapter:bosons},\,\,\ref{chapter:moving-cavities}) relied on perfect boundaries for the cavities. This means that there is no possibility of particles leaking in or out of the system. In a realistic scenario, dissipative effects play an important role in quantum systems~\cite{weiss1999}. Therefore taking into account these effects is an important extension to our work. The motivation is two-fold. One is that experiments are never ideal and so a more accurate description how they behave is crucial. The second is that if any possible entanglement that is produced within a cavity needs to be accessed by an external agent. This is impossible if the cavity walls allow no interaction with the environment outside. We would therefore like a framework in which to describe the interaction of entangled cavity modes and an external source.
    
  \item Another limitation of the work presented in Chapters~(\ref{chapter:bosons},\,\,\ref{chapter:moving-cavities}) was the perturbative treatment of the model. In particular, early theoretical work which identified significant differences between Bosons and Fermions relied on a non-perturbative, high acceleration analysis. Extending our work to accommodate for any possible value of cavity length and acceleration will allow us to verify the differences between Bosons and Fermions found in early literature. A possible route to achieving this goal would be a numerical analysis of the needed Bogoliubov transformations.
  
  \item We also mentioned, in Chapter~(\ref{chapter:fat-detectors}), using Unruh-DeWitt detectors to extract entanglement from a quantum field. One can consider two Unruh-DeWitt detectors which are separated by some space-like distance. If the detectors interact with a quantum field locally, by the properties of quantum mechanics, they will become entangled after some finite amount of time even though they have never interacted directly. While such configurations have been considered~\cite{reznik2005} our equations for the dynamics of a quantum system could provide new insight into the exact dynamics quantum entanglement in relativistic settings.

\end{enumerate}

\section{Possible Directions}

In this final section, we propose some possible, and no doubt ambitious, directions for new research. 

\begin{enumerate}

  \item In our analyses of entanglement in quantum field theory we have relied heavily on the ability to decompose a quantum field into definite momentum eigenfunctions, also known as a Fourier decomposition. This was based exclusively on the concept of a timelike Killing vector. As mentioned earlier, a generic spacetime does not admit timelike Killing vectors in a global manner. Another interesting scenario where definite momentum Fourier decompositions are difficult to find is interacting quantum field theory~\cite{srednicki2007}. In the case of interacting quantum field theory on flat spacetime, the concept of \emph{path integrals} have proven to be a very powerful tool for computing expectation values of operators. 
    
  An initial step in combining path integrals and quantum information resulted in the so-called ``entropy-area" law~\cite{srednicki1993}. By considering a Klein-Gordon field which has internal degrees of freedom within some finite volume of spacetime, the entanglement between these degrees of freedom and those in the rest of the spacetime is proportional to the area of the finite volume. This elegant result opens up a plethora of interesting questions: \emph{does self-interaction affect the entanglement of a field?, how does entanglement change when the field interacts with another field? can path integrals extend our notions of entanglement to curved spacetimes?} A tentative start could be to reformulate our results in the language of path integrals. Even though these questions are rather speculative, partial answers to quantum information theoretic considerations in quantum field theory have already been addressed. In particular, the analogous result for the entropy area law for Dirac fields has been investigated~\cite{casini2005} and multipartite correlations have also been considered~\cite{hayden2013}.
  
    \item It would also be interesting to investigate how other important information theoretic quantities behave when quantum field theory is introduced. Such a quantity is the \emph{channel capacity} of quantum map. A quantum map is a transformation that takes one given state to another and its channel capacity quantifies the rate at which information can be transmitted. A simple example of a quantum map is the time evolution operator. There are, of course, more complicated maps and an interesting question is how much information can be sent (or even how much information is lost) when a state is sent through a relativistic quantum channel. Partial answers to such questions have already been proposed in the context of analysing Rindler observers~\cite{bradler2012} and developments in Gaussian state quantum channels~\cite{monras2010} could provide a workable framework for incorporating relativity. Quantum channels have many implications in different fields of quantum information such as metrology (accurate estimation of parameters in a quantum system) and quantum communication in general. 

  \item There has been recent interest in the experimental validation of predictions in relativistic quantum information~\cite{friis2013-2,rideout2012}. As current experiments are pushing towards space-based scenarios, the need to understand fundamental phenomena in these situations is critical. A possible way to prepare for currently unknown problems is by testing mathematical theories in experiments which are easily accessible. This is done via so-called \emph{analogue systems}. One attractive field is that of analogue gravity. Gravitational physics, such as small effects due to general relativity, is notoriously difficult to test reliably. Thus, people have turned to systems which can be easily set-up experimentally to investigate the mathematical description of gravity. One example is the use of Bose-Einstein condensates to model different spacetime metrics. By investigating weak gravitational effects in Bose-Einstein condensates, we can attempt to verify the mathematical description of true spacetime-gravitation effects.

  \item Another interesting possibility is to explore correlations in quantum gravity. New models have been put forward that could be the solution to a quantum description of gravity~\cite{engle2008,freidel2008}. In particular, models for black holes have also been investigated in these situations~\cite{ashtekar2004}. Along similar lines, a model for black hole evaporation which used purely information theoretic concepts has been put forward~\cite{braunstein2011}. It would be worth investigating how these two new models can be combined to shed possible light on the black hole information paradox~\cite{hawking2005}. 

  \item Other interesting topics that combine relativity and quantum information are: a consistent measurement theory for quantum field theory~\cite{benicasa2012,lin2012-2}, the ability to encode information into a quantum field for later use~\cite{olson2011} (so-called ``time-like" teleportation) and more general quantum correlations than entanglement (e.g. quantum discord)~\cite{adesso2012}. 
  
\end{enumerate}

\appendix

\chapter{Symplectic Matrix Derivation}\label{app:symplectic-matrix-derivation}
We begin by defining the vector of Bosonic annihilation and creation operators
\begin{eqnarray}
\hat{\boldsymbol{\xi}}=(\hat{a}_{1},\ldots,\hat{a}_{N},\hat{a}^{\dag}_{1},\ldots,\hat{a}^{\dag}_{N}).
\end{eqnarray}
The commutation relations of these operators can be compactly written as
\begin{eqnarray}
K_{mn}=[\hat{\xi}_{m},\hat{\xi}_{n}^{\dag}]\Rightarrow\boldsymbol{K}=\left(
\begin{array}{cc}
\boldsymbol{I} & \boldsymbol{0} \\
\boldsymbol{0} & -\boldsymbol{I}
\end{array}\right).
\end{eqnarray}
Consider next the action of a unitary operator whose exponent, which is anti-Hermitian, is purely quadratic in Bosonic operators. The most general Hermtian combination of Bosonic operators can be written in the form
\begin{eqnarray}
\label{eqn:quad-ham}
\hat{H}=\hat{\boldsymbol{\xi}}^{\dag}\cdot\boldsymbol{H}\cdot\hat{\boldsymbol{\xi}},
\end{eqnarray}
where the matrix representation of $\hat{H}$ takes the form
\begin{eqnarray}
\boldsymbol{H}=\left(\begin{array}{cc}
\boldsymbol{U} & \boldsymbol{V} \\
\bar{\boldsymbol{V}} & \bar{\boldsymbol{U}}
\end{array}\right),
\end{eqnarray}
with the specific conditions $\boldsymbol{U}=\boldsymbol{U}^{\dag}$ and $\boldsymbol{V}=\boldsymbol{V}^{\mathrm{tp}}$. The conditions on $\boldsymbol{U}$ and $\boldsymbol{V}$ ensure the Hermiticity of $\boldsymbol{H}$. Next consider the unitary transformation of the vector of operators $\hat{\boldsymbol{\xi}}$ such that
\begin{eqnarray}
\label{eqn:unitary-transformation}
e^{-i\hat{H}}\hat{\xi}_{m}e^{+i\hat{H}}=S_{mn}\hat{\xi}_{n},
\end{eqnarray}
where $S_{mn}$ will be identified with a symplectic matrix. One could use the Hadamard lemma
\begin{eqnarray}
e^{-\hat{X}}\hat{Y}e^{+\hat{X}}=\hat{Y}-[\hat{X},\hat{Y}]+\frac{1}{2!}[\hat{X},[\hat{X},\hat{Y}]]+\ldots
\end{eqnarray}
to find the explicit form for each transformation however using the commutation relations of our Bosonic operators we can find a link between a given quadratic unitary operator and its symplectic counter part. Fully writing out the expression for the operator $\hat{H}$ from Eqn.~(\ref{eqn:quad-ham}) in terms of Bosonic operators we have
\begin{eqnarray}
\hat{H}=U_{mn}\hat{a}^{\dag}_{m}\hat{a}_{n}+V_{mn}\hat{a}^{\dag}_{m}\hat{a}^{\dag}_{n}
+\bar{V}_{mn}\hat{a}_{m}\hat{a}_{n}+\bar{U}_{mn}\hat{a}_{m}\hat{a}_{n}^{\dag}.
\end{eqnarray}
Using the Bosonic commutation relations $[\hat{a}_{m},\hat{a}^{\dag}_{n}]=\delta_{mn}$ it is straightforward to show Eqn~(\ref{eqn:unitary-transformation}) can be written as
\begin{eqnarray}
\left[-i\hat{H},\hat{a}_{k}\right]&=&-i\left(U_{km}a_{m}+V_{km}a_{m}^{\dag}\right),\\
\left[-i\hat{H},\hat{a}_{k}^{\dag}\right]&=&+i\left(\bar{V}_{km}a_{m}^{\dag}+\bar{U}_{km}a_{m}\right),
\end{eqnarray}
which is conveniently written in matrix form as
\begin{eqnarray}
[-i\hat{H},\hat{\boldsymbol{\xi}}]=-i\boldsymbol{K}\boldsymbol{H}\hat{\boldsymbol{\xi}}.
\end{eqnarray}
From there it is trivial to show the Hadamard lemma gives
\begin{eqnarray}
e^{-i\hat{H}}\hat{\boldsymbol{\xi}}e^{+i\hat{H}}=e^{-i\boldsymbol{K}\boldsymbol{H}}\hat{\boldsymbol{\xi}},
\end{eqnarray}
and hence we can identify a quadratic unitary operator with a sympletic matrix as
\begin{eqnarray}
\hat{U}=e^{-i\hat{\boldsymbol{\xi}}^{\dag}\cdot\boldsymbol{H}\cdot\hat{\boldsymbol{\xi}}}\rightarrow\boldsymbol{S}=e^{-i\boldsymbol{K}\boldsymbol{H}}.
\end{eqnarray}

\chapter{Partial Tracing for Fermions}\label{app:fermion-tracing}
Here we state, for reference purposes, the partial tracing rules used to derive the reduced state of Fermionic modes. All notation follows that of section~(\ref{sec:staticcavity}). 

Consider a Fermionic state which contains a reference mode $k\ge 0$. The partial tracing of the state, leaving the mode $k\ge 0$ untouched, is achieved by using the following rules:
\begin{subequations}
\begin{align}
\tr_{\lnot k}\ket{0}\bra{0}&=\ket{0}\bra{0},\\
\tr_{\lnot k}\ket{1_{p}}\bra{1_{i}}&=\begin{cases}
\delta_{pi}\ket{0}\bra{0} & p\neq k,\,\,i\neq k,\\
\ket{1_{k}}\bra{1_{k}} & p=i=k,\\
0 & \mathrm{otherwise},
\end{cases}\\
\tr_{\lnot k}\ket{1_{p}}\bra{0}&=\delta_{kp}\ket{1_{k}}\bra{0},\\
\tr_{\lnot k}\ket{1_{p}}\ket{1_{q}}\bra{1_{j}}\bra{1_{i}}&=\delta_{qj}\delta_{pi}\ket{0}\bra{0}\,\,p\neq k,\,\,i\neq k,
\end{align}
\end{subequations}
with all other partial traces vanishing. The corresponding expressions for $k<0$ can be found by the replacements $(+\rightarrow -)$, $p\rightarrow q$ and $i\rightarrow j$. 

Next we define the relevant partial traces leaving the modes $k\ge 0$ and $k'<0$ untouched:
\begin{subequations}
\begin{align}
\tr_{\lnot k,k'}\ket{0}\ket{0}\bra{0}\bra{0}&=\ket{0}\ket{0}\bra{0}\bra{0},\\
\tr_{\lnot k,k'}\ket{1_{p}}\ket{1_{q}}\bra{0}\bra{1_{i}}&=\begin{cases}
\ket{1_{k}}\ket{1_{k'}}\bra{0}\bra{1_{k}} & p=i=k,\,\,q=k',\\
0 & \mathrm{otherwise},
\end{cases}\\
\tr_{\lnot k,k'}\ket{0}\ket{1_{q}}\bra{0}\bra{0}&=\begin{cases}
\ket{0}\ket{1_{k'}}\bra{0}\bra{0} & q=k', \\
0 & \mathrm{otherwise},
\end{cases}\\
\tr_{\lnot k,k'}\ket{1_{p}}\ket{0}\bra{0}\bra{0}&=\begin{cases}
\ket{1_{k}}\ket{0}\bra{0}\bra{0} & p=k, \\
0 & \mathrm{otherwise},
\end{cases}\\
\tr_{\lnot k,k'}\ket{1_{p}}\ket{0}\bra{1_{j}}\bra{0}&=\begin{cases}
\ket{1_{k}}\ket{0}\bra{1_{k'}}\bra{0} & p=k,\,\,j=k',\\
0 & \mathrm{otherwise},
\end{cases}\\
\tr_{\lnot k,k'}\ket{0}\ket{1_{q}}\bra{1_{j}}\bra{0}&=\begin{cases}
\delta_{qj}\ket{0}\ket{0}\bra{0}\bra{0} & q\neq k',j\neq k',\\
\ket{0}\ket{1_{k'}}\bra{1_{k'}}\bra{0} & q=j=k',\\
0 & \mathrm{otherwise},
\end{cases}\\
\tr_{\lnot k,k'}\ket{1_{p}}\ket{1_{q}}\bra{1_{j}}\bra{1_{i}}&=\begin{cases}
\delta_{pi}\delta_{qj}\ket{0}\ket{0}\bra{0}\bra{0} & p\neq k,i\neq k, q\neq k', j\neq k',\\
\ket{1_{k}}\ket{1_{k'}}\bra{1_{k'}}\bra{1_{k}} & p=i=k,q=j=k',\\
\delta_{qj}\ket{1_{k}}\ket{0}\bra{0}\bra{1_{k}} & q\neq k', j\neq k', p=i=k,\\
\delta_{pi}\ket{0}\ket{1_{k'}}\bra{1_{k'}}\bra{0} & p\neq k, i\neq k, q=j=k',\\
0 & \mathrm{otherwise},
\end{cases}
\end{align}
\end{subequations}
where all other traces are given by either the Hermitian conjugates of the above or vanish.

\chapter{Derivation of two-mode transformation}\label{app:Derivation of two-mode transformation}
In section~(\ref{sec:resonance}) we saw the state between two modes, $k,k'$, depends only on the Bogoliubov transformation between those two modes. In other words, we can effectively truncate all states and transformations to the two modes we are interested in. This is, however, only true when working to first order in our perturbation parameter $h$. To show this statement is true, it is more convenient to use the real representation of the symplectic group defined with respect to the basis $\hat{\boldsymbol{X}}=(\hat{x}_{1},\hat{p}_{1},\ldots)$. We first write the general Bogoliubov transformation of a single cavity as
\begin{eqnarray}
\label{eq:bogo-matrix}
  \boldsymbol{S}&=&\left(\begin{tabular}{ c c c }
    $\boldsymbol{s}_{kk}$ & $\boldsymbol{s}_{kk'}$ & $\boldsymbol{s}_{kE}$ \\
    $\boldsymbol{s}_{k'k}$ & $\boldsymbol{s}_{k'k'}$ & $\boldsymbol{s}_{k'E}$ \\
    $\boldsymbol{s}_{Ek}$ & $\boldsymbol{s}_{Ek'}$ & $\boldsymbol{s}_{EE}$ 
  \end{tabular}\right),
\end{eqnarray}
where we have arranged the matrix such that the two modes we are interested in, labelled by $k,k'$, are separated from the rest of the system, which we label with $E$ for environment. To be clear, the matrices which involve the subscript $E$ are infinite dimensional as they contain the transformations between $k,k'$ and \emph{all} other modes. However, the matrices which involve only $k$ and $k'$ are just $2\times 2$ in size. The explicit expressions for the Bogoliubov matrices are found by transforming the matrix $\boldsymbol{Q}$ in~(\ref{eqn:bogo-matrices-bosons}) to the $\hat{\boldsymbol{X}}$ basis. This results in every block matirx in Eqn.~(\ref{eq:bogo-matrix}) taking the form~\cite{friis2012-3}
\begin{eqnarray}
\label{eqn:real-form-bogo-defs}
\boldsymbol{s}_{kk'}=\left(\begin{array}{cc}
\mathrm{Re}(\alpha_{kk'}-\beta_{kk'}) & \mathrm{Im}(\alpha_{kk'}+\beta_{kk'}) \\
-\mathrm{Im}(\alpha_{kk'}-\beta_{kk'}) & \mathrm{Re}(\alpha_{kk'}+\beta_{kk'})
\end{array}\right).
\end{eqnarray}

Using the identities, where the label $j\in (k,k',E)$,
\begin{eqnarray}
\boldsymbol{s}_{jj}&=&\boldsymbol{s}^{(0)}_{jj}+O(h^{2}),\\
\boldsymbol{s}_{jj'}&=&\boldsymbol{s}^{(1)}_{jj'}h+O(h^{2})\hspace{2mm}j\ne j',
\end{eqnarray}
we can decompose the full Bogoliubov matrix~(\ref{eq:bogo-matrix}) as $\boldsymbol{S}=\boldsymbol{S}^{(0)}+\boldsymbol{S}^{(1)}+O(h^{2})$. As we are considering the initial state is the vacuum, i.e. $\boldsymbol{\Gamma}_{0}=\boldsymbol{I}$, the final state after the transformation takes the form
\begin{eqnarray*}
\boldsymbol{S}\boldsymbol{S}^{tp}=\boldsymbol{I}+
\left(\begin{tabular}{ c c c }
\label{eq:firstorderstate}
    $\boldsymbol{0}$ & $\boldsymbol{A}$ & $\boldsymbol{B}$ \\
    $\boldsymbol{A}^{\mathrm{tp}}$ & $\boldsymbol{0}$ & $\boldsymbol{C}$ \\
    $\boldsymbol{B}^{\mathrm{tp}}$ & $\boldsymbol{C}^{\mathrm{tp}}$ & $\boldsymbol{0}$ 
  \end{tabular}\right)h+O(h^{2}),
\end{eqnarray*}
where we have defined the matrices
\begin{subequations}
\begin{align}
\boldsymbol{A}&=\boldsymbol{s}_{kk'}^{(1)}\boldsymbol{s}_{k'k'}^{(0)tp}+\boldsymbol{s}_{kk}^{(0)}\boldsymbol{s}_{k'k}^{(1)tp},\\
\boldsymbol{B}&=\boldsymbol{s}_{kE}^{(1)}\boldsymbol{s}_{EE}^{(0)tp}+\boldsymbol{s}_{kk}^{(0)}\boldsymbol{s}_{Ek}^{(1)tp},\\
\boldsymbol{C}&=\boldsymbol{s}_{k'E}^{(1)}\boldsymbol{s}_{EE}^{(0)tp}+\boldsymbol{s}_{k'k'}^{(0)}\boldsymbol{s}_{Ek'}^{(1)tp}.
\end{align}
\end{subequations}
We can immediately see that to $O(h)$, all sub-blocks of~(\ref{eq:firstorderstate}) contain transformations relating only to two subsystems. The reduced state of the $k,k'$ modes is thus
\begin{eqnarray}
\label{eqn:two-mode-reduced-state}
\boldsymbol{\Gamma}_{kk'}&=&\boldsymbol{I}+
\left(\begin{tabular}{c c}
    $\boldsymbol{0}$ & $\boldsymbol{A}$\\
    $\boldsymbol{A}^{\mathrm{tp}}$ & $\boldsymbol{0}$ 
  \end{tabular}\right)h+O(h^{2}).
\end{eqnarray}
This expression can also be obtained from considering two-mode transformation
\begin{eqnarray}
\boldsymbol{\Gamma}_{kk'}=\boldsymbol{S}_{kk'}\boldsymbol{S}_{kk'}^{\mathrm{tp}},
\end{eqnarray}
where the two-mode matrix $\boldsymbol{S}_{kk'}$ is defined via
\begin{eqnarray}
\boldsymbol{S}_{kk'}=\left(\begin{tabular}{ c c}
    $\boldsymbol{s}_{kk}$ & $\boldsymbol{s}_{kk'}$\\
    $\boldsymbol{s}_{k'k}$ & $\boldsymbol{s}_{k'k'}$
  \end{tabular}\right).
\end{eqnarray}
From Eq.~(\ref{eqn:two-mode-reduced-state}), it is easy to see the state is pure as $\text{det}(\boldsymbol{\Gamma}_{kk'})=1+O(h^{2})$ to first order in $h$. Thus the claim our state is pure and depends upon the modes $k,k'$ only to first order is justified.

\chapter{Teleportation State Derivation}\label{app:teleportation-state}
In this appendix we derive the state used in the quantum teleportation scheme of section~(\ref{sec:teleportation-results}). We first consider an initial state with correlations between Alice's mode $k$ and Rob's mode $k'$ and is separable with the rest of Alice and Rob's subsystems. Using the real representation of the symplectic group with basis $\hat{\boldsymbol{X}}=(\hat{x}_{k},\hat{p}_{k},\ldots,\hat{x}_{k'},\hat{p}_{k'},\ldots)$ where we have denoted Alice's quadrature operators as $(\hat{x}_{k},\hat{p}_{k})$ and Rob's quadrature operators as $(\hat{x}_{k'},\hat{p}_{k'})$. We can write our initial state as
\begin{eqnarray}
\boldsymbol{\Gamma}_{0} = \left(\begin{tabular}{ c  c | c  c }
    $\boldsymbol{\sigma}_{kk}$ &  & $\boldsymbol{\sigma}_{kk'}$ & \\ 
     & $\boldsymbol{I}$ &  &  \\ \hline
    $\boldsymbol{\sigma}_{kk'}^{tp}$ &  & $\boldsymbol{\sigma}_{k'k'}$ &  \\
     &  &  & $\boldsymbol{I}$ 
  \end{tabular}\right),
\end{eqnarray}
which indicates the state reduce states of Alice and Rob's system are, respectively, $\boldsymbol{\sigma}_{kk}$ and $\boldsymbol{\sigma}_{k'k'}$ and they are correlated trough $\boldsymbol{\sigma}_{kk'}$. All other modes are separable and in the vacuum state, represented by the identity matrix. Next we write the transformation on Alice and Rob's subsystems. Alice and Rob's cavity undergo separate evolution and so we can write the total symplectic matrix which governs their dynamics as
\begin{eqnarray}
  \boldsymbol{S}&=&\boldsymbol{O}_{A}\oplus\boldsymbol{S}_{B},
\end{eqnarray}
where $\boldsymbol{O}_{A}$ is an orthogonal matrix which represents inertial motion in Alice's cavity and $\boldsymbol{S}_{B}$ represents the non-uniform motion undergone by Rob's cavity. We can write $\boldsymbol{S}$ in the block form
\begin{eqnarray}
\boldsymbol{S}&=&\left(\begin{tabular}{ c  c |  c  c }
    $\boldsymbol{O}_{kk}$ &  & & \\ 
     & $\boldsymbol{O}_{EE}$ & & \\ \hline
 &  & $\boldsymbol{s}_{k'k'}$ & $\boldsymbol{s}_{k'E}$ \\
 &  & $\boldsymbol{s}_{Ek'}$ & $\boldsymbol{s}_{EE}$ 
  \end{tabular}\right),
\end{eqnarray}
where again the subscript $E$ denotes the environment modes of Alice and Rob's cavities and the matrices $\boldsymbol{s}_{jj'}$ are defined via~(\ref{eqn:real-form-bogo-defs}). The transformation of the whole state then goes as
\begin{eqnarray}
\boldsymbol{\Gamma}=\boldsymbol{S}\boldsymbol{\Gamma}_{0}\boldsymbol{S}^{tp}.
\end{eqnarray}
After a short computation we arrive at the final state
\begin{eqnarray}
\label{eqn:teleportation-transformed-state}
\boldsymbol{\Gamma}=\left(\begin{tabular}{ c  c |  c  c }
    $\boldsymbol{O}_{kk}\boldsymbol{\sigma}_{kk}\boldsymbol{O}_{kk}^{tp}$ &  & $\boldsymbol{O}_{kk}\boldsymbol{\sigma}_{kk'}\boldsymbol{s}_{k'k'}^{tp}$ & $\cdot$ \\ 
     & $\boldsymbol{I}_{EE}$ &  & \\ \hline
    $\boldsymbol{s}_{k'k'}\boldsymbol{\sigma}_{kk'}^{tp}\boldsymbol{O}_{kk}^{tp}$ &  & $\boldsymbol{\Gamma}_{B}$ & $\cdot$ \\
    $\cdot$ &  & $\cdot$ & $\cdot$
  \end{tabular}\right),
\end{eqnarray}
where $\cdot$ denotes a non-zero entry and
\begin{eqnarray}
\boldsymbol{\Gamma}_{B}=\boldsymbol{s}_{Ek'}\boldsymbol{s}_{Ek'}^{tp}+\boldsymbol{s}_{k'k'}\boldsymbol{\sigma}_{k'k'}\boldsymbol{s}_{k'k'}^{tp}.
\end{eqnarray}
The expression $\boldsymbol{s}_{Ek'}\boldsymbol{s}_{Ek'}^{tp}$ represents an infinite sum over all other modes contained within Rob's cavity. Written out fully we have
\begin{eqnarray}
\boldsymbol{s}_{Ek'}\boldsymbol{s}_{Ek'}^{tp}=\sum_{j\ne k'}\boldsymbol{s}_{jk'}\boldsymbol{s}_{jk'}^{tp},
\end{eqnarray}
where $\boldsymbol{s}_{jk'}$ represents the Bogoliubov transformation between the modes $j$ and $k'$. Therefore the reduced state between Alice's mode $k$ and Rob's mode $k'$ is
\begin{eqnarray}
\tilde{\boldsymbol{\Gamma}}=\left(\begin{array}{cc}
\boldsymbol{O}_{kk}\boldsymbol{\sigma}_{kk}\boldsymbol{O}_{kk}^{tp} & \boldsymbol{O}_{kk}\boldsymbol{\sigma}_{kk'}\boldsymbol{s}_{k'k'}^{tp}\\
\boldsymbol{s}_{k'k'}\boldsymbol{\sigma}_{kk'}^{tp}\boldsymbol{O}_{kk}^{tp} & \boldsymbol{\Gamma}_{B}
\end{array}\right).
\end{eqnarray}
In our case, the initial state is a two-mode squeezed state and so we have
\begin{subequations}
\begin{align}
\boldsymbol{\sigma}_{kk}&=\boldsymbol{\sigma}_{k'k'}=\cosh(2r)\boldsymbol{I},\\
\boldsymbol{\sigma}_{kk'}&=\sinh(2r)\boldsymbol{\sigma}_{1}.
\end{align}
\end{subequations}
Further, we can use the expressions
\begin{subequations}
\begin{align}
\boldsymbol{s}_{jj}&=\boldsymbol{s}^{(0)}_{jj}+\boldsymbol{s}_{jj}^{(2)}h^{2}+O(h^{3}),\\
\boldsymbol{s}_{jj'}&=\boldsymbol{s}^{(1)}_{jj'}h+O(h^{2})\hspace{3mm}j\ne j',
\end{align}
\end{subequations}
to write the transformed state between Alice and Rob as
\begin{eqnarray}
\label{eqn:power-expansion-teleportation-state}
\tilde{\boldsymbol{\Gamma}}=\tilde{\boldsymbol{\Gamma}}^{(0)}+\tilde{\boldsymbol{\Gamma}}^{(2)}h^{2}+O(h^{3}).
\end{eqnarray}
Explicitly we have for the order zero term
\begin{eqnarray}
\tilde{\boldsymbol{\Gamma}}^{(0)}=\left(
\begin{array}{cc}
\cosh(2r)\boldsymbol{I} & \sinh(2r)\boldsymbol{R}\\
\sinh(2r)\boldsymbol{R} & \cosh(2r)\boldsymbol{I}
\end{array}\right),
\end{eqnarray}
with the rotation matrix
\begin{eqnarray}
\boldsymbol{R}=\left(
\begin{array}{cc}
\cos\left(\phi_{k}+\phi_{k'}\right) & -\sin\left(\phi_{k}+\phi_{k'}\right)\\
-\sin\left(\phi_{k}+\phi_{k'}\right) & -\cos\left(\phi_{k}+\phi_{k'}\right)
\end{array}\right).
\end{eqnarray}
Here, $\boldsymbol{R}$ accounts for the free evolution of both Alice and Rob's cavities in the presence of no acceleration and $\phi_{k}:=\omega_{k}\tau$ is the phase of a mode. For the second order term in~(\ref{eqn:power-expansion-teleportation-state}) we find
\begin{eqnarray}
\tilde{\boldsymbol{\Gamma}}^{(2)}=\left(
\begin{array}{cc}
\boldsymbol{0} & \sinh(2r)\tilde{\boldsymbol{\Gamma}}_{AB}^{(2)}\\
\sinh(2r)\tilde{\boldsymbol{\Gamma}}_{AB}^{(2)\mathrm{tp}} & \tilde{\boldsymbol{\Gamma}}_{B}^{(2)}
\end{array}\right),
\end{eqnarray}
such that
\begin{subequations}
\begin{align}
\tilde{\boldsymbol{\Gamma}}_{AB}^{(2)}&=
\left(\begin{tabular}{cc}
	$\cos(\phi_{k})$ & $-\sin(\phi_{k})$ \\
    $-\sin(\phi_{k})$ & $-\cos(\phi_{k})$
  \end{tabular}\right)\left(\begin{tabular}{cc}
	$\mathrm{Re}(D_{-})$ & $-\mathrm{Im}(D_{-})$ \\
    $\mathrm{Im}(D_{+})$ & $\mathrm{Re}(D_{+})$
  \end{tabular}\right),\\
\tilde{\boldsymbol{\Gamma}}^{(2)}_{B}&=2\left(\begin{tabular}{cc}
	$\Gamma_{11}$ & $\Gamma_{12}$ \\
    $\Gamma_{12}$ & $\Gamma_{22}$
  \end{tabular}\right),
\end{align}
\end{subequations}
with $D_{\pm}\equiv\alpha^{(2)}_{k'k'}\pm\beta^{(2)}_{k'k'}$ and the elements of $\boldsymbol{\Gamma}_{B}^{(2)}$ are 
\begin{subequations}
\begin{align}
\Gamma_{11}&=f^{\alpha}_{k'}+f^{\beta}_{k'}-\text{Re}[f^{\alpha\beta}_{k'}]+\cosh(2r)(-f^{\alpha}_{k'}+f^{\beta}_{k'}-\text{Re}[f^{\alpha\beta}_{k'}]),\\
\Gamma_{12}&=\cosh(2r)(-f^{\alpha}_{k'}+f^{\beta}_{k'}-\text{Re}[e^{i\phi_{k'}}\beta^{2}_{k'k'}]),\\
\Gamma_{22}&=f^{\alpha}_{k'}+f^{\beta}_{k'}+\text{Re}[f^{\alpha\beta}_{k'}]+\cosh(2r)(-f^{\alpha}_{k'}+f^{\beta}_{k'}+\text{Re}[f^{\alpha\beta}_{k'}]).
\end{align}
\end{subequations}
These lengthy expressions are finished with the definitions
\begin{subequations}
\begin{align}
f^{\alpha}_{k'}&=\sum_{n\neq k'}|\alpha_{k'n}^{(1)}|^{2},\\
f^{\alpha}_{k'}&=\sum_{n\neq k'}|\beta_{k'n}^{(1)}|^{2},\\
f^{\alpha\beta}_{k'}&=\sum_{n\neq k'}\alpha_{k'n}^{(1)}\beta_{k'n}^{(1)}.
\end{align}
\end{subequations}

\chapter{Derivation of Dirac Bogoliubov Transformations}\label{appendix:dirac-bogo}

For the massless Dirac field in $1+1$-dimensions, we can write the Minkowksi and Rindler mode solutions, respectively, as 
\begin{subequations}
\begin{align}
\psi_{n}(t,z)&=\frac{e^{-i\omega_{\!n}t}\left[e^{+i\omega_{\!n}(z-a)}\,U_{+}\,+\,e^{i\theta}
e^{-i\omega_{\!n}(z-a)}\,U_{-}\right]}{\sqrt{2\cavlength}}\!\!,\\
\widehat{\psi}_{n}(\eta,\chi)&=\frac{\displaystyle e^{-i\Omega_{n}\eta}\left[\left(\frac{\chi}{a}\right)^{i\Omega_{n}}U_{+}+
e^{i\theta}\left(\frac{\chi}{a}\right)^{-i\Omega_{n}}U_{-}\right]}{\sqrt{2\chi\ln(b/a)}}.
\end{align}
\end{subequations}
We wish to compute the Bogolibov transformation from the Minkwoski mode solutions to the Rindler mode solutions. This involves computing the inner product
\begin{eqnarray}
\label{eqn:bogo-inner-product}
A_{mn}=\left(\psi_{m},\widehat{\psi}_{n}\right)|_{t=0},
\end{eqnarray}
where the Rindler modes $\widehat{\psi}_{n}$ have been expressed in terms of Minkwoski coordinates $(t,z)$ and the integral is taken over the constant hypersurface $t=0$. The Rindler coordinates are defined as
\begin{subequations}
\begin{align}
\chi&=\sqrt{z^{2}-t^{2}},\\
\eta&=\mathrm{arcTanh}\left[t/z\right].
\end{align}
\end{subequations}
For the computation of the Bogoliubov coefficients, it is convenient to work with the dimensionless coordinates
\begin{subequations}
\begin{align}
Z&=\frac{z-a}{\delta}-\frac{1}{2}\in\left[-1/2,+1/2\right],\\
T&=\frac{t}{\delta}\in\mathbb{R}.
\end{align}
\end{subequations}
We can identify a third dimensionless parameter to work with as
\begin{eqnarray}
h=2\delta/(b+a).
\end{eqnarray}
This will be our small expansion parameter. Using the identities
\begin{subequations}
\begin{align}
\omega_{n}t&=(n+s)\pi T,\\
\Omega_{n}\eta &=\frac{(n+s)\pi\mathrm{arcTanh}\left[\frac{T}{Z+1/h}\right]}{\ln\left[\frac{2+h}{2-h}\right]},\\
\chi &=\delta\sqrt{\left(Z+1/h\right)^{2}-T^{2}},\\
a &=\delta\left(1/h-1/2 \right),
\end{align}
\end{subequations}
we can write the Minkowski and Rindler modes as
\begin{subequations}
\begin{align}
\psi_{n}(T,Z)&=\frac{1}{\sqrt{2\delta}}e^{-i(n+s)\pi T}\left[e^{i(n+s)\pi(Z+1/2)}U_{-}+e^{i\theta}e^{-i(n+s)\pi(Z+1/2)}U_{+}\right],\\
\widehat{\psi}_{n}(T,Z)&=\frac{1}{\sqrt{2\delta}}\frac{1}{\sqrt{\ln\left[\frac{2+h}{2-h}\right]\sqrt{\left(Z+1/h\right)^{2}-T^{2})}}}e^{-i(n+s)\pi\mathrm{\frac{T}{Z+1/h}}}\notag\\
& \hspace{3mm}\times\left[e^{i(n+s)\pi\Lambda}U_{-}+e^{i\theta}e^{-i(n+s)\pi\Lambda}U_{+}\right],
\end{align}
\end{subequations}
where we have defined the function
\begin{eqnarray}
\Lambda=\frac{\ln\left[\frac{\sqrt{(Z+1/h)^{2}-T^{2}}}{1/h-1/2}\right]}{\ln\left[\frac{2+h}{2-h}\right]}.
\end{eqnarray}
It is then a matter of performing the appropriate power expansion in powers of $h$ around $h=0$ to compute the Bogoliubov coefficients. The zero order expansion (i.e. when there is no acceleration) gives trivially,
\begin{eqnarray}
\widehat{\psi}_{n}^{(0)}(T,Z)=\psi_{n}(T,Z),
\end{eqnarray}
and so the Bogoliubov coefficient $A_{mn}$ to order zero is
\begin{eqnarray}
A_{mn}=\delta_{mn},
\end{eqnarray}
where the integration is performed over the dimensionless coordinate $Z\in [-1/2,1/2]$ and we have taken the constant time hypersurface as $T=0$. A similar analysis can be done for each order of $h$ which results in the Bogoliubov expansions of Chapter~(\ref{chapter:moving-cavities}). The same procedure can of course be applied to the Klein-Gordon field to obtain the expansion~(\ref{eqn:fundamental-bogos}). However in the case of a massive field, the needed expansions are more involved and involve uniform expansions of modified Bessel functions. As the computations are lengthy not illuminating, I have omitted their derivation and just quote the results~(\ref{eqn:fundamental-bogos}).

\chapter{Notation\label{chapter:notation}}

A list of notational conventions and the definition of the Pauli matrices used throughout the thesis.

    \begin{tabular}{l  p{9cm}}
    Notation & Meaning \\ \\
    \hline \\
	 $\hat{O}$ & Hilbert space operator \\
	 $\ket{\psi}$, $\hat{\rho}$ &  pure and mixed quantum states respectively \\
	 $\langle\cdot | \cdot\rangle\in\mathbb{C}$ &  Hilbert space inner product \\
	 $\mathrm{tp}$ &  matrix transposition  \\
	 $\mathrm{tp}_{j}$ &  partial matrix transposition \\
	 $\mathrm{tr}$  & operator trace \\
	 $\mathrm{tr}_{j}$ &  partial trace over $j$-th subsystem \\
	 $\mathrm{det}$ &  matrix determinant \\
	 $\dag$ &  Hermitian adjoint or conjugate transposition \\
	 $\mathrm{Eig}(X)$ &  eigenvalues of $X$  \\
	 $\mathrm{Eig}_{+}(X)$ &  strictly positive eigenvalues of $X$ \\
	 $\mathrm{Vec}(X)$  & eigenvectors of $X$ \\
	 $\mathrm{max}(\cdot)$ &  maximum value of an expression \\
	 $\langle\hat{O}\rangle$, $\mathrm{Var}(\hat{O})$ &  expectations value and variance of an operator \\
	 $\delta_{nn}$, $\delta(x-x')$  & Kronecker delta and Dirac delta distribution \\
	  $\Theta(x)$  & Heaviside theta function \\
	 $(t,\boldsymbol{x})$ &  $4$-vector representing spacetime coordinates \\
	 $x^{\mu}$ & denotes the components of the $4$-vector \\
	 $\partial_{j}\equiv\frac{\partial}{\partial x^{j}}$  & coordinate differentiation \\
	 $g^{\mu\nu}$, $\eta^{\mu\nu}$ &  metric tensor components \\
	 $\nabla_{\mu}$ &  covariant derivative \\
	 $\sum_{j}x^{j}x_{j}\equiv x^{j}x_{j}$ & Einstein summation convention \\
	 $(\cdot,\cdot)\in\mathbb{C}$ &  field equation solution inner product \\
	 \end{tabular}

	 \begin{tabular}{l  p{9cm}}
    \hline \\
	 $\bigotimes, \otimes$  & tensor product \\
	 $\bigoplus,\oplus$ &  direct product \\
	 $\Sigma_{j}$ &  summation of $j$ \\
	 $\Pi_{j}$ &  product over $j$ \\
	 $[\hat{A},\hat{B}]\equiv\hat{A}\hat{B}-\hat{B}\hat{A}$ &  commutator \\
	 $\lbrace\hat{A},\hat{B}\rbrace\equiv\hat{A}\hat{B}+\hat{B}\hat{A}$ & anti-commutator \\
	 $\boldsymbol{M}$ &  matrix \\
	 $M_{mn}$ & components of matrix \\
	 $\boldsymbol{v}$ &  vector \\
	 $v_{m}$ & components of a vector \\
	 $\boldsymbol{v}\cdot\boldsymbol{w}\in\mathbb{C}$ &  vector inner product  \\
	 $|\cdot|$ &  complex number modulus \\
	 $\|\cdot\|_{1}$ & operator trace norm  \\
    \end{tabular}
\\ \\ \\
Pauli matrices 
\begin{eqnarray*}
\boldsymbol{\sigma}_{0}=\begin{pmatrix} 1 & 0 \\ 0 & 1 \end{pmatrix},\,\boldsymbol{\sigma}_{1}=\begin{pmatrix} 0 & 1 \\ 1 & 0 \end{pmatrix},\,\boldsymbol{\sigma}_{2}=\begin{pmatrix} 0 & -i \\ i & 0 \end{pmatrix},\,\boldsymbol{\sigma}_{3}=\begin{pmatrix} 1 & 0 \\ 0 & -1 \end{pmatrix}
\end{eqnarray*}

\renewcommand{\bibname}{References} 
\bibliographystyle{unsrt}
\bibliography{6_backmatter/references} 

\end{document}